%% file: thesis.tex
\documentstyle[twoside]{mythesis}
\begin{document}

% The title page and copyright page are unnumbered.
%\pagestyle{empty}

% The title page generated by LaTeX is a place-holder; the real title
% page must be obtained from the Graduate Degrees Section, Room 1,
% California Hall.

\include{macro}

\input{titlepage}
%\cleardoublepage

\input{abstractpage}
%\cleardoublepage
\pagestyle{plain}
\pagenumbering{roman}

\input{signaturepage}
%\cleardoublepage

\input{copyrightpage}

\input{dedication}

%\cleardoublepage

% The table of contents, list of figures, list of tables uses
% lower-case Roman numerals.

\tableofcontents
%\cleardoublepage

\listoffigures
%\cleardoublepage

\listoftables
\clearpage

\input{preface}

\clearpage

\input{acknow}
\clearpage

\pagenumbering{arabic}
\pagestyle{headings}
\include{chap1}

\include{chap2}

\include{chap3}

\include{chap4}

\include{chap4b}

\include{chap5}

\include{chap6}

\include{appendixa}
\include{appendixb}

\bibliography{references}

\bibliographystyle{plain}

\end{document}

%% file: macro.tex
\hyphenation{an-iso-tropy}
\hyphenation{brems-strah-lung}
\hyphenation{Robert-son}
\hyphenation{Fried-man}
\hyphenation{an-iso-tropies}
\hyphenation{quad-rupole}
\input epsf
\def\mycaption#1{\vskip 0.1truecm 
\ssp\rightskip=3truepc\leftskip=3truepc\baselineskip=11pt
	\noindent{\footnotesize#1}}

\def\COBE{{\it COBE}}
\def\be{\begin{equation}}
\def\bel#1{\begin{equation}\label{#1}}
\def\ee{\end{equation}}
\def\bea{\begin{eqnarray}}
\def\beal#1{\begin{eqnarray}\label{#1}}
\def\eea{\end{eqnarray}}
\def\eal{\!\!\!&=&\!\!\!}
\def\eadef{\!\!\!&\equiv&\!\!\!}
\def\eapp{\!\!\!&\approx&\!\!\!}
\def\eaprop{\!\!\!&\propto&\!\!\!}
\def\elt{\!\!\!&<&\!\!\!}
\def\eqn#1{(\ref{eq:#1})}
\def\frac#1/#2{\leavevmode\kern.1em
 .5ex\hbox{\the\scriptfont0 #1}\kern-.1em
 /\kern-.15em\lower.25ex\hbox{\the\scriptfont0 #2}}
\def\approx{\simeq}
%%%%%%%%%%%%%%%%%%%%%%%%%%%%%%%%%%%%%%%%%%%%%%%%%%%%%%%%%%%%%%%%%%%
%
%\simlt and \simgt produce > and < signs with twiddle underneath
\def\spose#1{\hbox to 0pt{#1\hss}}
\def\simlt{\mathrel{\spose{\lower 3pt\hbox{$\mathchar"218$}}
     \raise 2.0pt\hbox{$\mathchar"13C$}}}
\def\simgt{\mathrel{\spose{\lower 3pt\hbox{$\mathchar"218$}}
     \raise 2.0pt\hbox{$\mathchar"13E$}}}
%\simpropto produces \propto with twiddle underneath
\def\simpropto{\mathrel{\spose{\lower 3pt\hbox{$\mathchar"218$}}
     \raise 2.0pt\hbox{$\propto$}}}
%
%\newif\ifmathmode
%\def\mathflag#1${\mathmodetrue#1\mathmodefalse$}
%\everymath{\mathflag}
%\def\displayflag#1$${\mathmodetrue#1\mathmodefalse$$}
%\everydisplay{\displayflag}
%\mathmodefalse
\font\tenmmib=cmmib10 scaled\magstep1
%\font\tenmmib=cmmib10 
\newfam\bmitfam\textfont\bmitfam=\tenmmib
\def\bg{{\fam\bmitfam\mathchar"710D}}
\def\bx {{\bf x}}
\def\bk {{\bf k}}
\def\bv {\bf v}
\def\ie {{\it i.e.}}
\def\eg {{\it e.g.}}
\def\etal {{\it et al.}}

\def\sp {\kern 0.5pt}
\def\va {v^{(1)}}
\def\vb {v^{(2)}}
\def\vi {v^{i(1)}}
\def\vd {v \delta}
\def\P{{\bf p}}
\def\Q{{\bf q}}
\def\V{{\bf v}}
\def\K{{\bf k}}
\def\X{{\bf x}}
\def\cos{{\rm cos}}
\def\sin{{\rm sin}}
\def\exp{{\rm exp}}
\def\ct {\cos \theta}
\def\cst {\cos^2 \theta}
\def\e {{\rm e}}
\def\Cy {Compton-$y$}
\def\opc{\left( 1 + \cos^2 \beta \right)}
\def\cosb{\cos \beta}
\def\del{\delta(p-p')}
\def\ddel{\left[ {\partial \over {\partial p'}} \del \right]}
\def\dddel{\left[ {\partial^2 \over {\partial p'^2}} \del \right]}
\def\fmf{F_1(t,\X,\P,\P')}
\def\ffff{F_2(t,\X,\P,\P')}
\def\fof{F_3(\X,\P,\P')}
\def\pmp{(\P-\P')}
\def\bomb{\cosb (1-\cosb)}
\def\pv{\V_b \cdot \P}
\def\ppv{\V_b \cdot \P'}
\def\pq{\Q \cdot \P}
\def\ppq{\Q \cdot \P'}
\def\df{{\partial f \over \partial p}}
\def\dfo{{\partial f_0 \over \partial p}}
\def\ddf{{\partial^2 f \over \partial p^2}}

%spectral distortion
\def\part#1;#2 {\partial#1 \over \partial#2}
\def\deriv#1;#2 {d#1 \over d#2}
\def\obh {\Omega_b h^2}
\def\oh {\Omega_0 h^2}
\def\sc {{T_e \over T}}
\def\dndt {\part f;t }
\def\xe{x_\nu}
\def\nsc{{d\tau \over dt}}
\def\To {\Theta_{2.7}}
\def\ktmc {{T_e \over m_e}}
\def\lp {\left(}
\def\rp {\right)}
\def\e {{e}}
\def\exe {\e^{\xe}-1}
\def\de {{\delta \rho_\gamma \over \rho_\gamma}}
\def\dn {{\delta n_\gamma \over n_\gamma}}
\def\mup {\mu'(\xe)}
\def\exp {{\rm exp}}
\def\ng {n_{\gamma BE}}
\def\dde {{\delta \rho_\gamma / \rho_\gamma}}
\def\ddn {{\delta n_{\gamma} / n_{\gamma}}}
\def\ngbe {n_{\gamma BE}}
\def\sci#1 {\ \times\ 10^{#1}}
\def\standO {$\oh=0.25,$\ $\obh=0.025$}
\def\SC {Compton scattering}
\def\USC {Compton scattering}
\def\DC {double Compton scattering}
\def\UDC {Double Compton scattering}
\def\yf {(1-Y_p/2)}
\def\sec {{\rm s}}
\def\xe{x_p}

%chap4
\def\vsn{(v_x^N - v_x^S)}
\def\dotaa{{\dot a \over a}}

%% file: titlepage.tex
%
% title page is done in {titlepage} environment
%
\begin{titlepage}
%
% no page number
%
\begin{center}
\begin{large}			% title is set in \large size
{Wandering in the Background: \\
A Cosmic Microwave Background Explorer}
\end{large}
\end{center}

\begin{center}
\vspace{\stretch{0.5}}		% (rubber) vertical space after title
by\\[\stretch{0.5}]		% (rubber) vertical space
Wayne Hu \\
\vspace{\stretch{1.0}}		% (rubber) vertical space after univ. name
A.B. (Princeton University) 1990 \\
M.A. (University of California at Berkeley) 1992 \\
\vspace{\stretch{1.25}}		% (rubber) vertical space after univ. name
%
% double space ``requirements'' part
%
\renewcommand{\baselinestretch}{2.0}\large\normalsize
A thesis submitted in partial satisfaction of the\\
requirements for the degree of\\
Doctor of Philosophy\\
in\\
Physics\\%[\stretch{0.75}]	% (rubber) vertical space
in the \\%[\stretch{0.75}]
{\large GRADUATE DIVISION}\\
of the \\
{\large UNIVERSITY of CALIFORNIA at BERKELEY}
\end{center}
\vspace{\stretch{1.0}}		% (rubber) vertical space after univ. name
\vspace{\stretch{1.0}}		% (rubber) vertical space before committee
\begin{committee}
\item Professor Joseph Silk, Chair
\item Professor Marc Davis
\item Professor Hyron Spinrad
\end{committee}
\vspace{\stretch{0.25}}		% (rubber) vertical space before year
\begin{center}
{\large 1995 \par}		% set year in \large size
\end{center}
\end{titlepage}

%% file: abstractpage.tex
\pagestyle{empty}
\begin{abstitle}

\begin{center}
Wandering in the Background:\\
A Cosmic Microwave Background Explorer \\%[\stretch{0.5}]
%\vskip 0.20 in
by \\
%\vskip 0.20 in
Wayne Hu\\
\vskip 0.10 in
Doctor of Philosophy in Physics \\
University of California at Berkeley\\
Professor Joseph Silk, Chair \\
\vskip 0.20 in
\end{center}
We develop and examine the principles governing the formation of 
distortions in the cosmic microwave background.
Perturbations in the frequency or spectral distribution of the background
probe the thermal history of the universe, whereas those in the angular
temperature distribution probe its dynamics and geometry.  
Stressing model independent results, we show how the microwave background
can be used to extract information on the mass density, vacuum density, 
baryon content, radiation content, expansion rate and some aspects of 
structure formation in the universe.   To address these issues, we 
develop elements of relativistic kinetic and perturbation theory as
they become necessary for the description of the particle and gravitational
interactions of the photons.  Subtle issues such as fluctuation representation, 
or gauge, normal mode analysis in an open geometry, and second order 
effects are considered in detail. Employing analytic and numerical results, 
we construct anisotropies in a critical, open, and cosmological constant 
universe with adiabatic and/or isocurvature initial conditions allowing for
possible early reionization.  We find that
anisotropy formation is a simple process governed by the Compton scattering
of photons off electrons and their gravitational coupling to the other
particle species in the universe.

%\vskip 0.75in
%\begin{center}
%\def\xlen{4in}
%\def\xspace{0.25in}
%\def\xspacenew{0.2in}
%\begin{minipage}{\xlen}
%\begin{flushleft}
%\rule{\xlen}{1pt} \\
%{Chair \hfill Date} \\[\xspace]
%\end{flushleft}
%\end{minipage}
%\end{center}
\end{abstitle}

%% file: signaturepage.tex
\begin{signaturepage}
\thispagestyle{empty}
\vspace*{\stretch{1.0}}	% (rubber) space before sig. introduction
%
% sig. introduction is centered
%
\begin{center}
%
% double-space signature introduction
%
\renewcommand{\baselinestretch}{2.0}\large\normalsize
The thesis of Wayne Hu is approved: \\
\end{center}
\vspace{\stretch{1.0}}	% (rubber) space before sig. lines
\begin{center}
\epsfxsize=4in \epsfbox{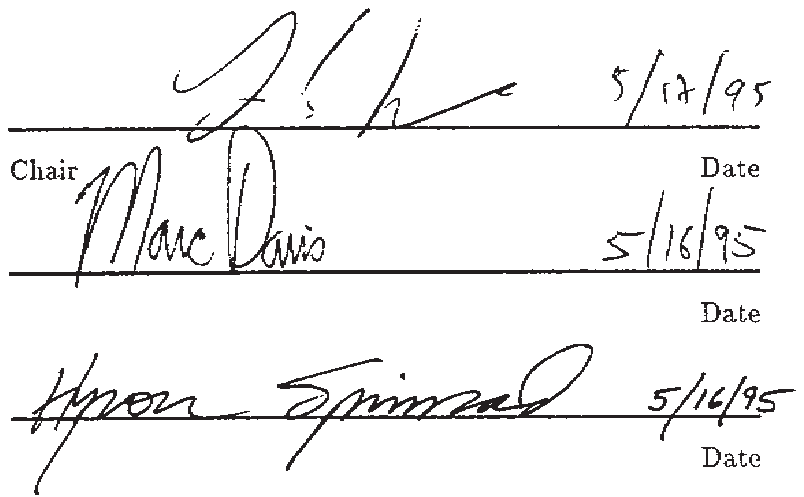}
%\def\xlen{4in}
%\def\xspace{0.25in}
%\def\xspacenew{0.2in}
%\begin{minipage}{\xlen}
%\begin{flushleft}
%\rule{\xlen}{1pt} \\
%{Chair \hfill Date} \\[\xspace]
%\rule{\xlen}{1pt} 
%{\vphantom{Chair} \hfill Date} \\[\xspace]
%\rule{\xlen}{1pt} \\
%{\hfill Date} 
%\end{flushleft}
%\end{minipage}
\end{center}
\vspace{\stretch{1.0}}	% (rubber) length before univ. & year
\begin{center}
University of California, Berkeley \\
1995
\vspace{\stretch{1.0}}	% (rubber) length before page number
\end{center}
\end{signaturepage}

%% file: copyrightpage.tex
\begin{copyrightpage}
\vspace*{\stretch{1.0}}
\begin{center}
Wandering in the Background: \\
A Cosmic Microwave Background Explorer \\
\vskip 0.25 in
{\copyright} copyright 1995 \\
by \\
Wayne Hu \\
\end{center}
\end{copyrightpage}

%% file: dedication.tex
%%%%%%%%%%%%%%%%%%%%%%%%%%%%%
% DEDICATION PAGE
 
%\begin{titlepage}
\vphantom{mark} 
\vskip1.8truein
\noindent{\large To Chuang-tzu,}
\vskip0.25truein

{\it\rightskip=3truepc\leftskip=3truepc\noindent
Said the disciple, ``After I heard your words,
one year and I ran wild, two years and I was
tame, three years and positions interchanged, four years and things
settled down, five years and things came to me \ldots"
\vskip 0.10 in
\centerline{--Chuang-tzu, 27}
}

\vskip0.5truein
\noindent{\large From Chuang-tzu,}
\vskip0.25truein

{\it\rightskip=3truepc\leftskip=3truepc\noindent
I hear that there is a sacred tortoise which has been dead for three
thousand years. His Majesty keeps it wrapped up in a box at the top 
of the hall in the ancestral shrine.  Would this tortoise rather be dead,
to be honored as preserved bones, or would it rather be alive and
dragging its tail in the mud...  Away with you!  I shall drag my tail
in the mud. 
\vskip 0.10 in
\centerline{--Chuang-tzu, 17}
} 

%\end{titlepage}
 
\newpage
 
%%%%%%%%%%%%%%%%%%%%%%%%%%%%%

%% file: preface.tex
\vphantom{mark}
\vskip 0.truein
\noindent{\Huge\bf Preface}
\addcontentsline{toc}{chapter}{Preface}
\vskip 0.3truein
\begin{quote}
\footnotesize\it
If you have a great tree and think it's a pity it's so useless,
Why not plant it in the middle of nowhere in the wilds which spread out,
and go rambling away aimlessly at its side, wander around and fall asleep
in its shade?
\vskip 0.25truecm
\centerline{--Chuang-tzu, 1 \footnote{Translations of the {\it Chuang-tzu}
throughout are adapted from \cite{Graham}.}}
\end{quote}
A mere three years ago when I started work on the cosmic microwave
background (CMB) with Joe Silk, anisotropies had not yet been discovered.
The theory of anisotropy formation was considered arcana and earned
barely a mention in the standard textbooks of the time.  
With the number of detections now in the double digits, CMB anisotropies have
joined spectral distortions, light element abundances, and large scale
structure measurements as some of our most powerful observational
probes of cosmology.  
The depth that even the interested non-specialist needs to understand the 
principles governing fluctuations in the CMB has consequently increased.  
This work begins the task of assembling the material necessary for
a modern understanding of the CMB.  Of course, the whole task is beyond
the scope of a 200 some page dissertation assembled in a month's time!  
I make no claims of completeness.  Rather, I develop
a handful of general principles that seem to me may have 
lasting interest.  As a consequence, I do not treat in any
detail CMB constraints on 
specific cosmological models, except where necessary to illustrate
general points.  Moreover, important issues of statistical analysis 
related to the current generation of experiments are not covered here.  
I happily refer the interested reader
to the excellent ``companion thesis'' by Emory Bunn \cite{Bunn}.  

Chapter 1 is provided as a qualitative and hopefully intuitive
introduction to the subject.  The formal development begins in 
chapter 2 with relativistic kinetic theory and continues in 
chapter 4 with relativistic perturbation theory. Readers who
are familiar with these subjects should skip to their applications:
spectral distortions in chapter 3 and density perturbation evolution in 
chapter 5.  Given its importance, anisotropy formation occupies the
rest of this work.  
Again, I stress robust features that may survive
the current generation of models.  I discuss how these features
may be used to probe general
cosmological issues such as the matter content, dynamics, and 
geometry of the universe.  
Advanced topics such as radiation 
feedback effects, polarization, and the
details of recombination are saved for Appendix A.   
Appendix B gathers together useful material
scattered throughout the text.

Some topics are covered at greater depth than others.  Some will be of
more interest to the specialist than to the general cosmologist.  This
thesis is nowhere near as homogeneous as the subject it purports to study
(though it may be as directionless)!  
Wander through its pages and perhaps you will find it of some use
-- if nothing else, for its soporific qualities.
\vskip 0.5truecm
{\hfill Wayne Hu} \\
\noindent
{\ssp\it
Berkeley, California \\
April 1995 }

%% file: acknow.tex
\vphantom{mark}
\vskip 0.truein
\noindent{\Huge\bf Acknowledgments}
\addcontentsline{toc}{chapter}{Acknowledgements}
\vskip 0.2truein
First and foremost I would like to thank my advisor Joe Silk for his
constant flow of ideas, support and encouragement. 
He introduced me to all the right people and helped me gain exposure
in the field. 
I would also like to thank the whole Berkeley CMB group.  This 
thesis has arisen in large part through discussions and collaborations
with them.  Specifically, Ted Bunn deserves special credit for putting
up with me as an officemate in general and my dumb statistical questions
in particular.  Douglas Scott started me out in the anisotropy game
and taught me a good part of my working knowledge of astronomy and
cosmology.  Naoshi Sugiyama devoted much time and effort to share
with me his expertise in the CMB.  Our many fruitful collaborations
form the basis of the latter half of this work.  
Martin White lent his critical skills
in helping me develop and refine the material here.   I would also
like to thank Naoshi and Martin for allowing me to use results from
their Boltzmann codes.  
Marc Davis and Hy Spinrad provided me with the ideal set of
comments on a draft of this thesis. 

My officemates
Lexi Moustakas and Dan Stevens provided me with a daily dose of
entertaining conversations
on a variety of off-the-wall subjects.  Dan, as the patron saint of
coffee, also supplied me with much needed caffeine during the writing
of this work. 
Matt Craig and David Schlegel endured many rehearsals
of my ``cute pictures'' talk.  David Weinberg earns my special thanks
for giving that talk more credit than it probably deserved!  
Marc Kamionkowski and David Spergel urged me to aim high. 
Max Tegmark threatened me with monetary
gain had I {\it not} finished this work on time.
Dan Plonsey saved one of my chapters from encryption hell.  
Eric Gawiser proofread several chapters.  Ann Takizawa,
who knows everything there is to know about UC Berkeley, saved me
on many occasions from missing important deadlines.

Tom Donnelly taught me that even physicists can ``get huge.'' 
The I-house bunch, especially Justin Bendich, Dan Krejsa 
and Raymond Yee, provided
memorable dinner conversations and an outlet for the
frustrations of the early years. Even more so than his physics
acumen, Dan's baking skills made him an 
ideal roommate.  Via email, Mike Aguilar and the rest of 
the college crew watched my progress from a writer of hat-obsessed bad
poetry to jargon-filled bad prose.  Finally, Meow Vatanatumrak 
selectively {\it impeded} progress on this work.  Her diabolical 
tactics helped me better enjoy my stay at
Berkeley.

%% file: chap1.tex
\chapter{Overview}
%\vphantom{MARKER}
\begin{quote}
\footnotesize\it
Is the azure of the sky its true color?  Or is it that the distance 
into which we are looking is infinite?  The P'eng never stops flying higher 
till everything below looks the same as above: heat-hazes, dust-storms, 
the breath which living things blow at each other \dots
\vskip 0.1truecm
\centerline{--Chuang-tzu, 1}
\end{quote}
\input chap1/firas.tex

%\vphantom{MARKER}
%\vskip 0.75truecm

\section{Cosmological Background}
\vskip 0.25truecm
With the discovery of the cosmic microwave background (CMB) by Penzias and 
Wilson in 1965 \cite{PW}, modern cosmology was born.  Long the realm of
armchair philosophers, the study of the origins and evolution of
the universe became a physical science with falsifiable theories. 
As light from
an earlier epoch, the CMB provides evidence that has 
proven many a cosmological theory wrong.   Still, cosmology has
remained a data-starved field until quite recently.  Unlike its brethren
disciplines, experimentation is not possible.  Given access to this one
universe alone, one must piece together the principles of its formation
out of what observations of it are possible.  The task is made even more
challenging due to the enormous range of physical and temporal 
scales involved.

We are now at the threshold of a new era in cosmology.  With telescopes
probing ever earlier epochs and larger volumes, 
we are making rapid progress in improving the quantity and quality of
data.  Cosmology is at last becoming a precision science.
Once again the CMB is taking a central place in this
transition.
Launched in late 1989, the {\it COBE} satellite ushered in the era of precision
cosmology. It has revealed in the CMB 
a perfect thermal or blackbody spectrum of
temperature $T_0 = 2.726 \pm 0.010$K (95\% CL), with deviations 
no more than several parts in $10^4$ \cite{Mather}, and
temperature anisotropies at the level of one part 
in $10^{5}$ \cite{Smoot}.

\vbox{
\subsection{Perfection and Its Implications}
\begin{quote}
\footnotesize\it
Observe the void -- its emptiness emits a pure light. 
\vskip 0.1truecm
\centerline{--Chuang-tzu, 4}
\end{quote}
}
The cosmic microwave background spectrum and anisotropy:
near perfection and slight imperfection.  
The implications of the former
run deep; the applications of the latter are broad.  
A thermal radiation background is 
a definite and almost unique prediction of the big bang cosmology. 
Why is the spectrum thermal at $2.7$K, a much lower temperature than most
astronomical matter in the universe?  Let us recall the basic
facts and premises upon which the big bang model is built.
Light from distant galaxies is redshifted in proportion to their
distance.  
In the big bang model, this is interpreted as 
a consequence of the universal expansion of the universe.  
Due to the light travel time, distant sources emitted their light long
ago when the universe was smaller.
During the expansion, the wavelengths of photons are stretched and
particle number densities drop leading to the low temperature and
photon density observed in the background today.  Conversely, extrapolating
backwards in time, we infer that the universe began in a 
hot dense state.  As we discuss in more
detail in \S \ref{ch-spectral}, at sufficiently high
temperatures interactions between particles were rapid enough to
bring the universe into a state of thermal equilibrium.  This and the
fact that adiabatic cooling from the expansion preserves the thermal spectrum
explains
the blackbody nature of the observed spectrum (see Fig.~\ref{fig:1firas}).  
No other
model for cosmology yet proposed can account for the stunningly
thermal spectrum.
Even in the big bang model, the lack of distortions to the spectrum
provides serious constraints on physical and astrophysical processes
that could have occurred between the thermalization 
redshift $z \approx 10^7$ and the present,
\ie\ very nearly the {\it whole} history of the universe.  

\input chap1/dmr.tex
The second pillar upon which the big bang model stands is the large scale
homogeneity and isotropy of the universe.  Originally only a hypothesis
based on simplicity and a Copernican desire not to occupy a preferred
position in the universe, this ``cosmological principle'' finds its validation
most dramatically
in the radio source catalogue of Gregory and Condon \cite{GC} and in the
extreme isotropy of the CMB.  Aside from a dipole anisotropy of $3.343 \pm
0.016 $mK (95\% CL) \cite{Smoot91},
almost certainly due to the Doppler effect from our
own motion, the CMB is isotropic at the level of one part in $10^{5}$.

In fact, the high degree of isotropy has long been a puzzle to cosmologists.
The CMB last interacted with the matter through Compton scattering as long
ago as redshift $z \approx 10^3$, when the photons no longer had the
energy to keep hydrogen photoionized, and no later than $z$ of a few tens
if hydrogen was ionized by some external source.   Our extrapolation 
backwards to this early time tells us that the patches of sky off which the CMB
last scattered should not have been in causal contact at that time.  
This seemingly
acausal isotropy of the CMB temperature is called the {\it horizon problem}.
The most promising solution to date, called the inflationary scenario,
postulates an early phase of rapid expansion that separates originally
causally connected regions by the vast distances necessary to account for the
large scale isotropy of the CMB.  Alternatively, it may be just a 
boundary condition of the universe imposed by unknown physics at the
Planck epoch.  

Potentially more troubling to cosmologists is the fact that the universe
at small scales is manifestly inhomogeneous
as the distribution of galaxies and indeed our
own existence implies.  In the big bang model, perturbations grow
by gravitational instability 
slowly due to the expansion, \ie\ power law rather than 
exponential growth (see \S
\ref{ch-perturbation}, \S\ref{ch-evolution}).  Even though the CMB bears
the imprint of an earlier and less evolved epoch, fluctuations must be present
at the $10^{-6}-10^{-5}$ level to be consistent with the simple
gravitational instability model. 
The announcement by the {\it COBE} DMR group of the first detection of
CMB anisotropies was thus met with expressions of relief and elation
by cosmologists.

\vbox{
\subsection{Imperfection and Its Applications}
\begin{quote}
\footnotesize\it
Said Hui-Shih to Chuang-tzu: `This talk of yours is big but useless.' 
\vskip 0.1truecm
\centerline{--Chuang-tzu, 1}
\end{quote}
}
As is often the case in physics, the deviations are of greater 
practical interest
than the mean.  While measurements of the 
thermal nature and isotropy of the CMB 
reveal
strong support for the general hot big bang scenario, they are shed
no light upon 
the details of the cosmological model.  Anisotropies on the other 
hand bear the imprint, filtered through the dynamics and geometry of
the expanding universe, of the fluctuations which eventually led to 
structure formation in the universe.  CMB anisotropies can therefore
shed light on  not only the mysteries of structure formation but also 
such fundamental quantities as the expansion rate, matter content and 
geometry of the universe. 
Let us briefly review the current status of some of these unresolved
issues.

Hubble's law states that the observed redshift 
scales with distance as $z = H_0 d$ due to the uniform 
expansion.  Measurement of the
proportionality constant, the so-called Hubble constant,
is notoriously difficult due to the need
to obtain absolute distances to galaxies.  
The uncertainty is usually parameterized
as $H_0 = 100 h$ km s$^{-1}$ Mpc$^{-1}$ where observations roughly require
 $ 0.5 \simgt h \simgt 1$.
High values of the Hubble constant $h \approx 0.8$ seem currently favored
by many distance scale calibrations 
(see \cite{Jacoby} for a review and \cite{Freedman}
for recent advances), but the issue is far from settled (see \eg\ \cite{Saha}).
Because $H_0$ sets the expansion time scale 
$H_0^{-1} \approx 10 h^{-1}$Gyr, its measurement is crucial in 
determining the age of the universe.  Through the theory of stellar
evolution, globular clusters are inferred to be as old as $14 \pm 2$ Gyr
\cite{Sandage,Shi} which may lead to an age crisis if $H_0$ turns out
to be in the upper range of modern measurements.  

How acute the age
crisis might be depends on the second major source of dispute: the
density of the universe.  Because mass tends to deccelerate the expansion,
a higher energy density implies a younger universe.  
The mass is usually parameterized by $\Omega_0$ which is the energy density
in units of the critical density $\rho_{crit} = 3H_0^2/8\pi G 
= 1.879 \times 10^{-29} h^2$ g cm$^{-3}$.  
There is also the possibility that vacuum energy and pressure, 
\ie\ the cosmological
constant $\Lambda$,
can provide an acceleration of the expansion leading to an arbitrarily
old universe.
A universe with $\Omega_0+\Omega_\Lambda = 1$ is special
in that it is the only one that is spatially flat.
Dynamical measurements of 
the mass in the halo of galaxies from their velocity dispersion implies that 
$\Omega_0 \simgt 0.1-0.3$.  The inequality
results from the fact that these measurements cannot probe the amount
of mass that is not clustered with galaxies. 
Large scale velocity fields can test larger regions
and though the situation to date is far from clear,
current measurements tend to yield slightly higher values for $\Omega_0$
(see \eg\ \cite{Strauss} for a recent review).  

Let us examine the constituents of the total density.  
Luminous matter in the form of stars in the central part of galaxies
only accounts for $\Omega_* \approx 0.004$ of the critical density. 
Compared with dynamical measurements, this indicates that most of
the matter in the universe is dark.
On the other hand, the CMB energy
density $\Omega_\gamma h^2= 2.38 \times 10^{-5} \Theta_{2.7}^4 $, where
$\Theta_{2.7} = T_0/2.7$K.  Although negligible today,
in the early universe it increases in importance relative to the matter
energy density $\rho_m$ since 
$\rho_\gamma/\rho_m \propto 1+z$ due to the redshift.  
With the photon density
thus fixed through the CMB temperature, primordial nucleosynthesis
and observations of the light element abundances imply that the baryon
fraction is low 
$ \Omega_b h^2 = 0.01-0.02$ \cite{Smith,Walker}.  A significant
amount of non-baryonic dark matter is apparently present in the 
universe.  The amount and nature of dark matter in the
universe has significant consequences for structure formation.  
The most crucial aspect of its nature for these purposes is the mass
of its constituent particles.
Collisionless dark matter, unlike baryonic matter,
does not suffer dissipative processes. 
Thus the particle mass determines whether their rms velocity
is high enough to escape gravitational collapse.

CMB anisotropies can provide information on all these fundamental issues
and more.  Since the issue of anisotropy formation is of such central
importance, its systematic development occupies the greater part 
of this work \S \ref{ch-perturbation}--\ref{ch-secondary}.
Gravitational and Compton coupling of the CMB represent intertwining
themes that recur throughout these chapters. 
It is therefore useful to give here a brief 
exposition
of these concepts, their importance for anisotropy formation, and
their implications for cosmology \cite{Color}.  
\\
\input chap1/cdm.tex
\section{Anisotropy Formation}
\label{sec-1anisotropy}
\nobreak
\begin{quote}
\footnotesize\it
Words are for catching ideas; once you've caught the idea, you can 
forget about the words.  Where can I find a man who knows how to forget
about words so that I might have a few words with him?
\vskip 0.1truecm
\centerline{--Chuang-tzu, 26}
\end{quote}
\nobreak
Fluctuations in the total matter density, which includes decoupled species such
as the neutrinos and possibly collisionless dark matter,
interact with the photons through the gravitational potentials they 
create.  These same fluctuations grow by gravitational attraction, \ie\
infall into their own potential wells, to eventually form large scale 
structure in the universe.  Their presence in the early universe
is also responsible for anisotropy formation.  

Before redshift $z_* \approx 1000$, the CMB was hot enough to 
ionize hydrogen.
Compton scattering off electrons, which are in turn linked to the protons
through Coulomb interactions, strongly couples the photons to the baryons 
and establishes a photon-baryon fluid.  Photon pressure resists
compression of the fluid by gravitational infall and sets up
acoustic oscillations. 
At $z_*$, 
recombination produces neutral hydrogen
and the photons last scatter.  Regions
of compression and rarefaction at this epoch represent hot and cold spots
respectively.  Photons also suffer gravitational 
redshifts from climbing out of the potentials on the last scattering
surface.  The resultant fluctuations 
appear to the observer today as anisotropies on the sky.   
By developing the simple picture outlined
above in greater detail, we show how realistic anisotropies
such as those depicted in Fig~\ref{fig:1cdm} are formed. 

\subsubsection{Notation}

Although sky maps such as Fig.~\ref{fig:1dmr} are visually impressive, the
anisotropy must be analyzed statistically.  For gaussian fluctuations, 
the statistical
content is encapsulated in the two point temperature correlation function,
or equivalently its angular decomposition into Legendre moments $C_\ell$. 
In Fig.~\ref{fig:1cdm}, we show a typical prediction for the anisotropy
power spectrum $C_\ell$ compared with the current state of observations.  

Predictions for $C_\ell$ are obtained by tracking the evolution of
temperature fluctuations. Their 
equations of motion
take on a simple form when decomposed into normal modes.  These are plane
waves for a flat geometry, referred to in this chapter as such even when
considering their open geometry 
generalization (see \S \ref{ss-4laplacian} and \cite{Harrison,Wilson}).
We represent temperature fluctuations in Newtonian form,
which simplifies concepts such as infall and redshift, 
by defining them on the spatial hypersurfaces of the
conformal Newtonian gauge (see \S \ref{sec-4gauge}).

Under the gravitational force $F$, a temperature perturbation 
$\Theta_0 = \Delta T/T$ of
comoving wavenumber $k$ evolves
almost as a  
simple harmonic oscillator before recombination \cite{HSa}
$(1+R) \ddot \Theta_0 + {k^2 \over 3} \Theta_0 \approx F$.  
The overdots represent 
derivatives with respect to conformal time $\eta = \int (1 + z) dt$ with
$c=1$ and $R=3\rho_b/4\rho_\gamma = 3.0 \times 10^{4} (1+z)^{-1} 
\Omega_b h^2$ accounts for the baryonic contribution to the
effective mass of the oscillator.  Notice that the restoring force
from photon pressure is independent of the baryon content.  The
frequency of the oscillator is constructed out of these quantities 
as $\omega = k c_s$ where the 
sound speed $c_s$, which measures the resistance of the fluid to compression, 
is $c_s \equiv \dot p / \dot \rho = 1/\sqrt{3(1+R)}$.  The oscillator
equation can thus be rewritten as $\ddot \Theta_0 + k^2 c_s^2 \Theta_0
\approx F/(1+R)$.  

Let us now consider the gravitational driving force
$F/(1+R) \approx - k^2 \Psi/3 -\ddot \Phi$, where $\Psi$
is the Newtonian gravitational potential, obtained from density 
fluctuations via the generalized Poisson equation, 
and $\Phi \approx -\Psi$ is the perturbation to the space curvature.
They also represent plane wave fluctuations in the time-time and space-space
metric components respectively.
%In all cases relevant to this analysis,
%$\Phi \approx -\Psi$ since anisotropic stress from the radiation
%quadrupoles is small.$^{\HSb}$  
The sign convention reflects the fact that overdensities
create positive space curvature and negative potentials, \ie\ potential
wells.
In real space though, a single plane wave represents
both overdense {\it and} underdense regions.  We use the former to
guide intuition since the distinction is only in sign.
%Furthermore, the decomposition of a general fluctuation requires
%a spectrum of these independent 
%plane waves oscillators.
% where $\Psi$ and $\Phi$ become functions of $k$. 

\input chap1/acoustic.tex
\subsection{Acoustic Oscillations}

Let 
us first consider temperature fluctuations before recombination
in the case of a {\it static} potential \cite{DZS,Bond,Jorgensen}.
Although only appropriate for a universe which has {\it always}
been matter 
dominated, it illustrates the general nature of the acoustic
oscillations.
In this case, $F= -k^2 (1+R) \Psi /3$ and represents 
the usual driving force of gravity that leads to infall into potential 
wells.
Since big bang nucleosynthesis implies that the baryon
density is low, $\Omega_b h^2 \approx 0.01-0.02$, 
as a first approximation
assume that $R \ll 1$ and the photons completely 
dominate the fluid $c_s \approx 1/\sqrt{3}$. 

Gravitational infall compresses the fluid until resistance from
photon pressure reverses the motion. 
Since the gravitational force is constant in this case, it merely
shifts
the zero point of the oscillation to $\Theta_0= -\Psi$.
To determine the amplitude of the oscillations, 
we must first fix the initial conditions.  
The relation between the matter density fluctuations and the potential
$\delta_m(0) = -2\Psi$ is fixed by demanding consistency with the Poisson and
Euler equations.  Let us assume adiabatic initial conditions for the
photons $\Theta_0(0) = {1 \over 3} \delta_m(0) =-{2 \over 3} \Psi$ 
and $\dot \Theta_0(0)=0$ (see Fig.~\ref{fig:1acoustic}a).  
In this case, the photons follow the matter, making
the temperature higher inside a potential well.
The effective initial displacement of
$\Theta_0(0) + \Psi = {1 \over 3}\Psi$
then evolves as $\Theta_0(\eta) 
= {1 \over 3} \Psi\cos(kc_s\eta) - \Psi$.  At last scattering $\eta_*$, 
the photons
decouple from the baryons and stream out of potential wells suffering
gravitational redshifts equal to $\Psi$.  
%They then suffer gravitational redshifts equal to $\Psi$ 
%as they climb out of the 
%potential well making the resulting fluctuation $\Theta_0(\eta_*)+\Psi 
%= {1 \over 3} \Psi\cos(kc_s\eta)$.
We thus call $\Theta_0+\Psi$ the {\it effective} temperature
fluctuation.
Here the 
redshift exactly cancels the zero
point displacement since gravitational
infall and redshift are one and the same for a photon-dominated system.  

The phase of the oscillation at last scattering determines the
effective fluctuation.  Since the oscillation frequency $\omega = kc_s$,
the critical wavenumber $k = \pi /c_s\eta_*$ is
essentially at the scale of the {\it sound horizon} 
$c_s\eta_*$ (see Fig~\ref{fig:1acoustic}). 
%Because $c_s \approx c/\sqrt{3}$, this is close to the  
%particle (causal) horizon at $c\eta_*$.  
Larger wavelengths
will not have evolved from the initial conditions and possess
${1 \over 3} \Psi$ fluctuations after gravitational 
redshift. This combination of the intrinsic 
temperature fluctuation and the gravitational redshift is the well
known Sachs-Wolfe effect \cite{SW}.  
Shorter wavelength fluctuations can be frozen
at different phases of the oscillation.  Since fluctuations
as a function of $k$ go as $\cos(kc_s\eta_*)$ at last scattering, 
there will be a harmonic
series
of temperature {\it fluctuation} 
peaks with $k_m = m\pi/c_s\eta_*$ for the $m$th peak.
Odd peaks thus represent the compression phase (temperature crests),
whereas even peaks represent the rarefaction phase (temperature troughs), 
inside the potential wells.

\subsection{Baryon Drag}

Though effectively pressureless, the baryons still
contribute to the inertial and gravitational
mass of the fluid $m_{\rm eff} = 1+R$. 
This changes the
balance of pressure and gravity as baryons drag the photons into
potential wells.  As the baryon content $R$ is increased, 
gravitational infall leads to greater compression of the 
fluid, \ie\ a further
displacement of the oscillation zero point (see Fig.~\ref{fig:1acoustic}b).  
Since the redshift is  not affected
by the baryon content, this relative shift remains after 
last scattering to enhance all peaks from compression over those
from rarefaction.
If the baryon photon ratio $R$ were constant, 
$\Theta(\eta)+\Psi = {1 \over 3}\Psi (1+3R)\cos(kc_s\eta) -R\Psi$,
with 
compressional peaks a factor of $(1+6R)$ over the $R=0$ case.
In reality, the effect is reduced since 
$R\rightarrow 0$ at early times.

Finally the {\it evolution} of the effective mass
has a effect of its own. 
In classical mechanics, the ratio of energy $E= {1 \over 2} m_{\rm eff}
\omega^2 A^2$ to frequency of 
an oscillator $\omega$ is an adiabatic 
invariant.  Thus for the slow changes in $\omega \propto (1+R)^{-1/2}$,
the amplitude of the oscillation varies as $A \propto (1+R)^{-1/4}$.  
Since $R(\eta_*) = 30\Omega_b h^2
\simlt 1$ at recombination, this is ordinarily not a strong effect.
 
\subsection{Doppler Effect}

Since the turning points are at
the extrema, the fluid velocity oscillates 90 degrees
out of phase with the density (see Fig.~\ref{fig:1acoustic}a).  
Its motion relative to the observer causes a 
Doppler shift.  
Whereas the observer velocity creates a pure dipole anisotropy
on the sky, the fluid velocity
causes a spatial temperature variation 
$V_\gamma/\sqrt{3}$ 
on the last scattering surface from its line of sight component. 
For a photon-dominated $c_s \approx 1/\sqrt{3}$ fluid,
the velocity contribution is equal in amplitude to the density 
effect \cite{DZS,Jorgensen}. 
This photon-intrinsic Doppler shift should be distinguished from the 
scattering-induced Doppler shift of reionized scenarios (see
\S\ref{ss-6regeneration} and \cite{SZ}).

The addition of baryons significantly changes
the relative velocity contribution.  As the effective mass
increases, conservation of energy requires that
the velocity decreases for the same initial temperature displacement. 
Thus the {\it relative} amplitude of the velocity scales as $c_s$.  
In the toy model of a constant baryon-photon density ratio $R$, the oscillation
becomes $V_\gamma /\sqrt{3} = {1 \over 3}\Psi (1+3R)(1+R)^{-1/2} 
\sin(kc_s\eta)$. 
Notice that velocity oscillations are symmetric around zero leading to
even more prominent compressional peaks (see Fig.~\ref{fig:1acoustic}b).
Even in a universe with $\Omega_b h^2$ given 
by nucleosynthesis, $R$ is sufficiently large
to make velocity contributions subdominant.

\input chap1/isw.tex
\subsection{Potential Evolution}

All realistic models involve potentials which are time-dependent, leading
to a non-trivial gravitational driving force that can 
greatly enhance the prominence of the acoustic peaks \cite{HSa,HSb}.
We have hitherto assumed that matter dominates the energy density. 
In reality,
radiation dominates 
above the redshift of equality 
$z_{eq}= 2.4 \times 10^4 \Omega_0 h^2$, assuming the
usual three flavors of massless neutrinos.  
The feedback from radiation perturbations into the gravitational 
potential makes the CMB sensitive to the matter-radiation 
ratio in the background {\it and} the fluctuations. 

Consider first adiabatic initial conditions as before.
Inside the sound horizon, radiation pressure prevents gravitational infall 
during radiation domination.  Energy 
density fluctuations consequently can no longer maintain
a constant gravitational potential.  
Counterintuitively, this decaying potential can actually enhance 
temperature fluctuations through its near resonant driving force.
Since the potential decays after sound horizon crossing, it mimics
$\cos(kc_s\eta)$ for $kc_s\eta \simlt \pi$.  Consequently, it 
drives the first compression without 
a counterbalancing effect on the subsequent rarefaction or gravitational
redshift.

Moreover, there is another effect.  
Recall that the space curvature perturbation follows the potential as
$\Phi \approx -\Psi$.
Since the 
forcing function $F/(1+R) \approx -\ddot \Phi - k^2 \Psi/3$, a
changing $\Phi$ also drives oscillations.  As $\Phi$ is a
perturbation to the spatial metric, its change induces 
a time-dilation effect which is wholly analogous 
to the cosmological redshift due to the 
expansion.  Heuristically, the overdensities
which establish the 
potential well ``stretch'' the space-time fabric (see Fig.~\ref{fig:1isw}a).
As the potential well decays, it re-contracts.
Photons which are caught in this contraction find their wavelength
similarly contracted, \ie\ blueshifted.
Thus a differential change in $\Phi$ leads to a dilation 
effect, $\dot \Theta_0 = -\dot \Phi$, and consequently 
a forcing effect on $\ddot \Theta_0$ of $-\ddot\Phi$
as required.

If $\Psi$ were exactly $\cos(kc_s\eta)$, then
$\ddot \Phi$ would double the driving force. Detailed calculation
shows that 
the oscillation amplitude is boosted to $\approx 5$ times the Sachs-Wolfe
effect of ${1 \over 3}\Psi$ 
(see \S \ref{ss-4acousticdriven}).
Only short wavelengths, which 
cross the sound horizon during the radiation-dominated epoch,
experience this enhancement.
For $\Omega_0 h^2 \approx 0.25$, 
the sound horizon at equality is several times smaller than 
that at 
last scattering.   
Hence delaying equality, by lowering $\Omega_0 h^2$ or increasing the number
of relativistic species, boosts the amplitude of oscillations for
the first few peaks.  Finally, the decay of the potential $\Psi$ also removes
the zero point shift and thus lifts the pattern of alternating heights for
the peaks. 

As a second example of forced oscillations, consider isocurvature perturbations.
In this case, the matter alone carries the initial fluctuations,
\ie\ $\Theta_0(0) = 0$ and since the radiation dominates the
energy density,  $\Phi(0)=0=\Psi(0)$ as well. 
However $\dot \Theta(0) \ne 0$ and is set to
counteract the gravitational attraction of the matter.   
Consequently, the
potential grows to be significant only near sound horizon crossing and
subsequently decreases if the universe is radiation dominated. 
The forcing function resembles $\sin(kc_s\eta)$ and thus drives
the sine harmonic of oscillations.  Furthermore, since fluctuations
are initially established to counter gravity, infall
enhances 
{\it even} rather than odd peaks. 
Outside the sound
horizon, dilation implies that $\Theta_0(\eta_*) = 
-\Phi(\eta_*)$, creating a Sachs-Wolfe effect
of $[\Theta_0+\Psi](\eta_*) \approx 2\Psi(\eta_*)$.
%unlike the familiar ${1 \over 3}\Psi(\eta_*)$ of the adiabatic effect.

\input chap1/diffusion.tex
\subsection{Photon Diffusion Damping}

In reality, the photons and baryons are not perfectly coupled 
since the photons
possess a mean free path in the baryons $\lambda_C$ due to Compton scattering.
As the photons random walk through the baryons, hot spots and cold
spots are mixed (see Fig.~\ref{fig:1diffusion}).  
Fluctuations thereafter remain
only in the unscattered fraction causing a near exponential
decrease in amplitude 
as the diffusion length
$\lambda_D \sim \sqrt{N}\lambda_C = \sqrt{\eta\lambda_C}$ overtakes
the wavelength \cite{Silk}.

At last scattering, the ionization fraction $x_e$ 
decreases due to recombination, 
thus
increasing the mean free path of the photons $\lambda_C \propto
(x_e n_b)^{-1}$. 
The effective diffusion scale
is therefore extremely sensitive to the ionization history in 
addition to the baryon number density $n_b$.  
Subtle effects during and even before last scattering
can have a measurable effect on the 
damping \cite{Kaiser83,HSSW}.
Moreover, if last scattering is delayed, \eg\ by early reionization, 
diffusion continues and can destroy all the acoustic peaks 
(see \S\ref{sec-6linear}). 
Assuming a standard recombination ionization history however, the
approximate scaling can be obtained from the Saha equation for
the ionization at fixed redshift or temperature,
$x_e \propto (\Omega_bh^2)^{-1/2}$.  The final 
damping length therefore approximately scales as $\lambda_D (\eta_*)
\propto \eta_*^{1/2} (\Omega_b h^2)^{-1/4}$.  For high $\Omega_b h^2$ 
models, this scaling must be modified due to the high Lyman-$\alpha$
opacity at recombination \cite{HSsmall}.

\subsection{Integrated Sachs-Wolfe Effect}

After last scattering, the photons free stream toward the observer.  Only
gravitational effects can further alter the temperature.   
The differential redshift from $\dot \Psi$ and dilation from $\dot \Phi$
discussed above must be integrated along the trajectory 
of the photons. We thus call the combination the {\it integrated}
Sachs-Wolfe (ISW) effect \cite{SW}.
%It imprints the 
%the epoch of the 
%transition from radiation to matter domination and matter to
%matter to curvature or cosmological constant $\Lambda$ domination on 
%the CMB.
For adiabatic models, it can contribute via the potential decay 
for modes that cross the sound horizon between last scattering
and full matter domination.  
In isocurvature models, potential {\it growth}
outside the sound horizon makes the ISW effect dominate
over the Sachs-Wolfe
effect for all wavelengths larger than the sound horizon at $\eta_*$
(see \S \ref{ss-5iso}). 
Because these effects are
sensitive to the radiation content and
occur primarily at early times, we call them {\it early} ISW effects.
In an open or $\Lambda$  model, the universe enters
a rapid expansion phase once matter no longer dominates the 
expansion.  We call the effect of the resultant potential 
decay the {\it late} ISW effect.  

One additional subtlety is introduced in ISW effects.  
If the potential decays while the photon is in an underdense region,
 it
will suffer an effective redshift rather than a blueshift.
Contributions from overdense and underdense regions will
cancel and damp the ISW effect 
if the decay time is much  greater than the light travel time across
a wavelength  
(see Fig.~\ref{fig:1isw}).  
The damping does not occur for the {\it early} ISW
effect.  Since it arises when the perturbations are outside
or just crossing the horizon, the time scale for the decay 
is always less
than, or comparable to, the light travel time across a wavelength.
For the late ISW effect, decay takes on the
order of an expansion time at curvature or $\Lambda$ domination independent
of the wavelength.   Thus, cancellation 
leads to a gradual damping in $k$ of contributions as the wavelength becomes
smaller than the horizon at the decay epoch.  For a fixed $\Omega_0$, 
the decay epoch occurs much later in 
flat $\Omega_\Lambda + \Omega_0 = 1$ models than open ones.  
Consequently, $\Lambda$ models will suffer
cancellation of late ISW contributions at a much larger scale than open 
models \cite{Kofman}.  In summary, 
the epoch that the universe exits the radiation $(\Omega_0 h^2)$ and  
matter-dominated phase $(\Omega_\Lambda, 1 -\Omega_0-\Omega_\Lambda)$ 
is imprinted on the CMB by the early and late
ISW effects respectively.
 
\input chap1/projection.tex
\subsection{Projection Effects}

We have been considering the generation 
of temperature fluctuations in space. 
However, what one actually observes are temperature anisotropies on 
the sky.  
The connection between the two is that a spatial fluctuation
on a distant surface, say at last scattering for the
acoustic effects, appears as an 
anisotropy on the sky.  
%Given an initial relative weighting or spectrum
%of the $k$ modes, the contributions to the anisotropy from each mode are then
%summed.  Below we consider these processes 
%in general and as applied to adiabatic initial fluctuations.
Three quantities go into this conversion: the spectrum of 
spatial fluctuations, the distance to the surface of their generation,
and curvature or lensing in light propagation to the observer
(see Fig.~\ref{fig:1projection}).

For the acoustic 
contributions, the 
$k$ modes that reach extrema in their oscillation 
at last scattering form a harmonic series of peaks related to the
sound horizon.  
This in turn is approximately $\eta_*/\{1+ C
[1+R(\eta_*)]^{1/2}\}$, where $R(\eta_*) = 30\Omega_b h^2$ and
$C \approx \sqrt{3}-1$.
Since $\Omega_b h^2$ must be low to satisfy 
nucleosynthesis constraints, 
the sound horizon will scale roughly as the particle horizon $\eta_*$.
The particle horizon at last scattering itself scales as
$\eta_* \propto (\Omega_0 h^2)^{-1/2} f_R$.
Here $f_R = [1+(24\Omega_0 h^2)^{-1}]^{1/2} - (24\Omega_0h^2)^{-1/2}$
and is near unity if the universe is matter dominated at $\eta_*$.
For low $\Omega_0 h^2$, radiation allows for more rapid early expansion
and consequently a smaller horizon scale.
In a flat $\Lambda$ universe, the distance to the
last scattering surface scales approximately as $\eta_0 \propto
(\Omega_0 h^2)^{-1/2}f_\Lambda$ with 
$f_\Lambda =1 + 0.085\ln\Omega_0$.
Notice that the two behave similarly at high $\Omega_0 h^2$. Since the
acoustic angle 
$\theta_A \propto \eta_*/\eta_0$, the leading term
$(\Omega_0 h^2)^{-1/2}$ has no effect.  Slowly 
varying corrections from $f_R/f_\Lambda$ decreases the angular 
scale somewhat as
$\Omega_0 h^2$ is lowered. 
On the other hand, the 
damping scale subtends an angle $\theta_D \approx \lambda_D/\eta_0
\propto (\Omega_0 h^2)^{1/4} (\Omega_b h^2)^{-1/4} f_R^{1/2}/f_\Lambda$.
Even in a low $\Omega_0 h^2$ universe $\theta_D$
is only weakly dependent on $h$ unlike $\theta_A$ the acoustic scale. 

By far the most dramatic effect is due to {\it background} curvature in the 
universe \cite{SG}.
If the universe is open, photons curve on their geodesics such that
a given scale subtends a much smaller angle in the sky than in a flat
universe.  In a $\Lambda=0$ universe, the angle-distance relation
yields $\theta_A \propto \eta_* \Omega_0 h$, \ie\ 
$\propto \Omega_0^{1/2} f_R$.
Likewise, the damping scale subtends an angle
$\theta_D \propto \lambda_D \Omega_0 h$, \ie\ $\propto \Omega_0^{3/4}
\Omega_b^{-1/4} f_R^{1/2}$.  
At {\it asymptotically} high and 
low $\Omega_0 h^2$, $f_R \approx 1$ and $f_R \propto (\Omega_0 h^2)^{1/2}$
respectively, so that there is a weak but different scaling with $h$ and
strong but similar scaling with $\Omega_0$ for the two angles.
The latter should be an easily 
measurable effect \cite{KSS}.

Contributions from after last scattering, such as the ISW effects, 
arise from a distance closer to us. A given scale thus subtends a larger
angle on the sky (see Fig.~\ref{fig:1projection}). 
Their later formation also implies that the
radiation correction factor $f_R$ will be smaller.  For example, 
the angle subtended by the adiabatic early ISW effect scales nearly as
$\Omega_0^{1/2}$ in a $\Lambda=0$ universe even at low $\Omega_0h^2$.

The above discussion implicitly assumes an one-to-one correspondence
of linear scale onto angle that is strictly only true if 
the wavevector is perpendicular
to the line of sight.
In reality, the orientation of the wavevector
leads to aliasing of different, in fact larger, angles for a given 
wavelength (see Fig.~\ref{fig:1projection}b).  
This is particularly important for Doppler
contributions which vanish for the perpendicular mode 
(see \S \ref{ss-6cancellation}). 
Moreover if there is a lack of long wavelength power, \eg\ in typical baryon 
isocurvature models, 
large angle anisotropies are dominated by aliasing of power
from short wavelengths.
Consequently, the angular power spectrum may be less blue than the 
spatial power spectrum (see \S \ref{ss-5iso}).
On the other hand, for so called ``scale invariant'' or equal weighting 
of $k$ modes, 
aliasing tends to smear out sharp features but does
not change the general structure of the real to angular space mapping.  
It is evident that gravitational lensing from the curvature {\it fluctuations}
of overdense and underdense regions has a similar but usually smaller
effect \cite{SeljakLensing}.

\input chap1/total.tex

\section{Anisotropy Spectrum}

Anisotropy formation is a simple process that is governed by
gravitational effects on the photon-baryon fluid and the photons alone
before and
after
last scattering respectively.   The component contributions contain
detailed information on classical cosmological parameters.
Let us now put them together to form the total
anisotropy spectrum.

%\subsection{An Adiabatic Example}

The popular 
scale invariant adiabatic models provide a useful example of how
cosmological information is encoded into the anisotropy spectrum.
Specifically by
scale invariant, we mean that the logarithmic contribution to the
gravitational potential is initially
constant in $k$.  For open
universes, this is only one of several reasonable
choices near the curvature 
scale \cite{KamSper,LS,RP,Bucher}. In Fig.~\ref{fig:1total}, 
we display a schematic representation of the anisotropy spectrum
which separates the various effects discussed above and identifies
their dependence
on the background cosmology.

Changing the overall dynamics from $\Omega_0=1$ through 
flat $\Lambda$ models to open models is similar to shifting the spectrum
in angular space toward smaller angles. 
Beginning at the largest angles, the ISW effect from late potential
decay dominates in 
$\Omega_0 \ll 1$ models.  Cancellation suppresses contributions for
wavelengths smaller than the particle horizon
at the exit from matter domination. This damping
extends to larger angles in $\Lambda$ than in open
models 
affecting even the quadrupole.
At scales much larger than the sound horizon at $\eta_*$ 
and particle horizon at equality,
the effective temperature, or Sachs-Wolfe effect, 
is $[\Theta+\Psi](\eta_*)
\approx
{1 \over 3} \Psi(\eta_*)$.  Shifting equality through $\Omega_0 h^2$
changes the redshift contribution $\Psi(\eta_*)$.  
For scales just above the sound horizon, the early ISW effect boosts 
fluctuations as the relative 
radiation content is increased by lowering $\Omega_0 h^2$. 
In sufficiently low $\Omega_0$ open models, the late and early ISW 
effects merge and entirely dominate over the last scattering surface
effects at large angles.

The first of a series of peaks from the acoustic oscillations appear 
on the sound horizon at $\eta_*$.  In the total spectrum, the
first acoustic peak merges with the early ISW effect. 
A lower $\Omega_0 h^2$ thus serves to broaden out and change the
angular scaling of
this combined feature.  
The acoustic peak heights also depend strongly on 
$\Omega_0 h^2$ for the first few peaks
due to the driving effects of infall and dilation.
Furthermore, greater infall due to the baryons allows 
more gravitational zero point shifting if
$\Omega_0 h^2$ is sufficiently high to maintain the potentials. 
Odd peaks
will thus be enhanced over the even, as well as velocity contributions,
with increasing $\Omega_b h^2$.  
The location of
the peaks is dependent on the sound horizon, distance to last scattering, and
the curvature.  In a low $\Omega_b h^2$, high $\Omega_0 h^2$ 
universe, it is sensitive only 
to the curvature $1-\Omega_0-\Omega_\Lambda$. 
Finally, the physics of recombination sets the
diffusion damping scale which cuts off the series of acoustic peaks.

\input chap1/cancelpict.tex
\section{Robustness to Initial Conditions}

How robust are anisotropies
to model changes?  
Obviously, changing the initial spectrum will significantly modify the 
spectrum.
For example, isocurvature conditions and tilt
can alter the  relative contributions
of the various effects.  
The lack of super-curvature modes in open inflationary
models can also suppress
the low order multipoles \cite{LW}.
On the other hand, they may 
be boosted by gravitational wave ISW 
contributions \cite{TWL,Crittenden}.

Acoustic oscillations however 
are unavoidable, if there are potential
perturbations before last scattering.
Even
exotic models such as defect-induced fluctuations should give rise to 
acoustic contributions of some form.
Since adiabatic and isocurvature conditions drive two
different harmonics, they
can be distinguished by the relation
between the peaks and the sound horizon at last scattering \cite{HSb}.  
The {\it locations} of the peaks are then dependent only on the background
cosmology, \ie\  mainly on the curvature but also 
on a combination of $\Omega_b h^2$, $\Omega_\Lambda$ and $\Omega_0 h^2$.  
On the other hand, the difference in heights between odd and even peaks is
a reasonably robust probe of
the baryon-photon ratio, 
 \ie\ $\Omega_b h^2$, relative to the matter-radiation ratio 
at last scattering,
 \ie\ $\Omega_0 h^2$ and possibly even the number of massless
neutrinos.  Finally,  the damping scale probes the baryon content and
the detailed physics of recombination.
If acoustic oscillations are detected in the 
anisotropy data, clearly 
we will be able to measure many parameters of classical cosmology. 

\section{Reionization} 

The one caveat to these considerations is that reionization can completely
erase the acoustic oscillations.  In a model with sufficiently early
reionization, \ie\ $z_i \gg 10$, the photon diffusion 
length grows to be the horizon
scale at the new last scattering surface and consequently damps 
all of the peaks.   
In models such as CDM, structure forms late and early reionization is
highly unlikely.   However, it is worthwhile to consider 
its general effects on the CMB in the event that structure formation
proceeded by a qualitatively different route.

\input chap1/vishpict.tex
CMB fluctuations can be regenerated once the baryons
are released from Compton drag to evolve independently $z_d = 160
(\Omega_0 h^2)^{1/5} x_e^{-2/5}$ (see \S \ref{ss-6regeneration}).  Baryonic
infall into potential wells leads to electron bulk velocities which
induce Doppler shifts in the scattered photons. If the universe
remains ionized, last scattering effectively occurs when the 
Compton scattering time exceeds the expansion time.  Thus the
thickness of the last scattering surface is on the order of the horizon
size at last scattering.  At small scales, this thickness spans 
many wavelengths of the perturbation.  Photons that last scatter from
the fore and rear of the perturbation encounter electrons with oppositely
directed infall velocities (see Fig.~\ref{fig:1cancelpict}).   
Just like the late
ISW effect, the net contribution will be cancelled at small
scales. 

Cancellation is particularly severe for the linear theory Doppler effect
(see \S \ref{ss-6cancellation}).
This implies that higher order terms in perturbation theory will dominate
the anisotropy at small scales.  As we show in \S \ref{sec-6second}, 
the dominant second order effect is due to a coupling of 
density and velocity perturbations called the Vishniac effect 
\cite{OV,Vishniac}.  
It arises since the probability of a photon scattering off an 
overdensity is higher due to the increased electron density.  
If the overdense regions are also caught in a larger scale bulk flow, 
this can yield an anisotropy on the scale of the overdensity
since a greater fraction of the photons suffer Doppler kicks 
along lines of sight that intersect overdensities
(see Fig.~\ref{fig:1vishpict}).  
Since the effect 
depends on a coupling of modes, it is extremely sensitive to the
shape and amplitude of the baryon power spectrum.  Furthermore, the
horizon size at last scattering is imprinted as the cancelled scale
of the first order effect. Thus in the case of 
early reionization, the CMB can be
used as a sensitive probe of the model for structure formation 
and the ionization history of the universe, but yields little 
model-independent information on the classical cosmological parameters.
These secondary anisotropies are thus complementary to the primary ones.
It is possible that the observed spectrum will contain an admixture of
the two if reionization occurs but is not sufficiently early.

%% file: chap1/firas.tex
\begin{figure}[t]
\centerline{ %\hskip-0.25truecm
\epsfxsize=4.5in \epsfbox{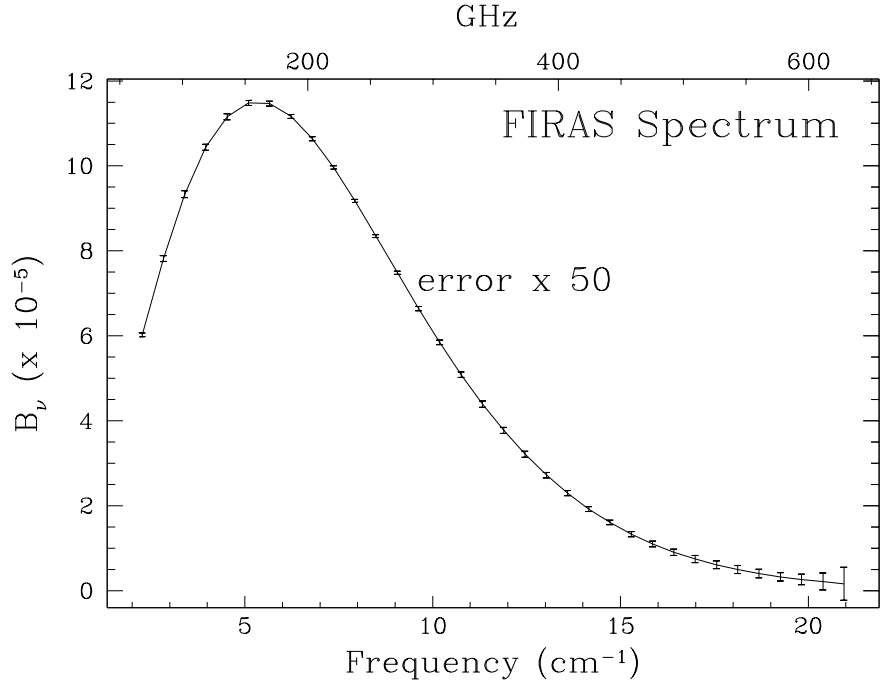}}
%\vskip -0.25truecm
\caption{FIRAS Spectral Measurement}
\mycaption{To the precision of the {\it COBE} FIRAS instrument \cite{Mather},
the CMB spectrum is a perfect blackbody with a maximum deviation of no more 
than
$3 \times 10^{-4}$ and a noise weighted rms deviation of 
under $5 \times 10^{-5}$ of
its peak intensity.  No spectral distortions have been measured to date
excluding nearly all options for its formation except in the early
stages of a hot big bang.  Plotted here is the intensity in
ergs cm$^{-2}$ s$^{-1}$ sr$^{-1}$ cm. 
}
\label{fig:1firas}
\end{figure}

%% file: chap1/dmr.tex
\begin{figure}[t]
\vphantom{marker} \vskip 0.5truecm
\centerline{ 
\epsfxsize=6.0in \epsfbox{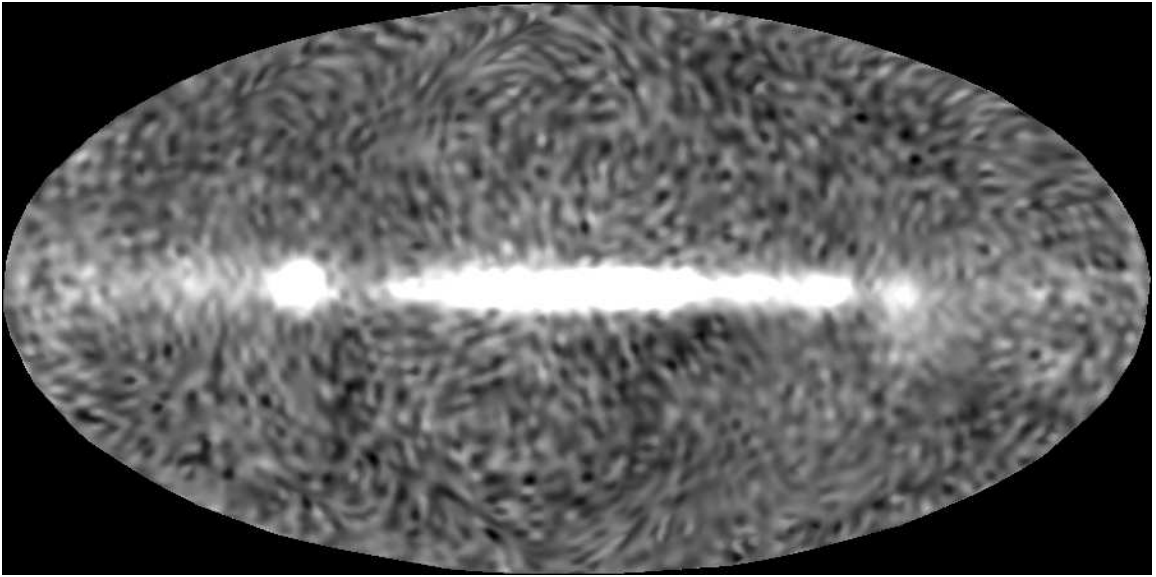}}
\vskip 0.5truecm
\caption{DMR Anisotropy Map}
\mycaption{Anisotropies in the CMB as detected by the {\it COBE} DMR 
experiment at an rms level of $\Delta T/T = {\cal O}(10^{-5})$. 
While the raw data set is noisy and suffers galactic contamination
(bright center band), filtering reveals a detection of high significance
and importance to our understanding of structure formation in the
universe.  Map courtesy of E. Bunn.
}
\label{fig:1dmr}
\end{figure}

%% file: chap1/cdm.tex
\begin{figure}[t]
\centerline{ \hskip-0.5truecm
\epsfxsize=4.5in \epsfbox{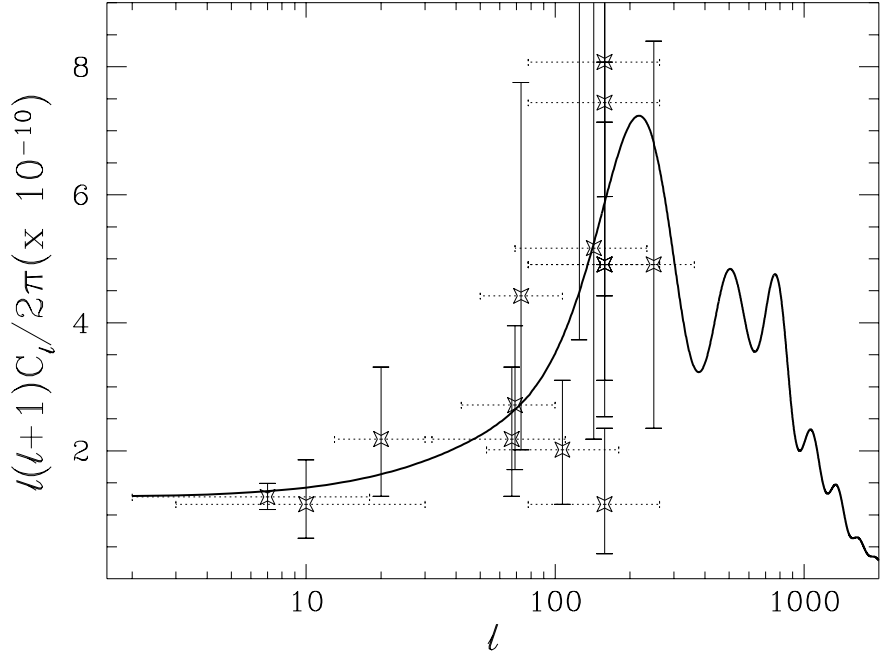}}
\vskip -0.5truecm
\caption{Anisotropies: Theory and Experiment}
\mycaption{Anisotropy data of current CMB experiments from Tab.~\ref{tab:data}
compiled by \cite{SSW}. Dotted horizontal ``error bars'' are the half
power angular range of the experiment.  
Overplotted is the predicted anisotropy power 
spectrum $C_\ell$ in a typical model: standard CDM with
$\Omega_0=1$, $h=0.5$, $\Omega_B=0.05$, scale invariant scalar initial
fluctuations, and arbitrary normalization.  
The corresponding angle on the sky is approximately
$100/\ell$ degrees.  
}
\label{fig:1cdm}
\end{figure}

%% file: chap1/acoustic.tex
\begin{figure}[p]
\centerline{ \hskip-0.5truecm
\epsfxsize=6in \epsfbox{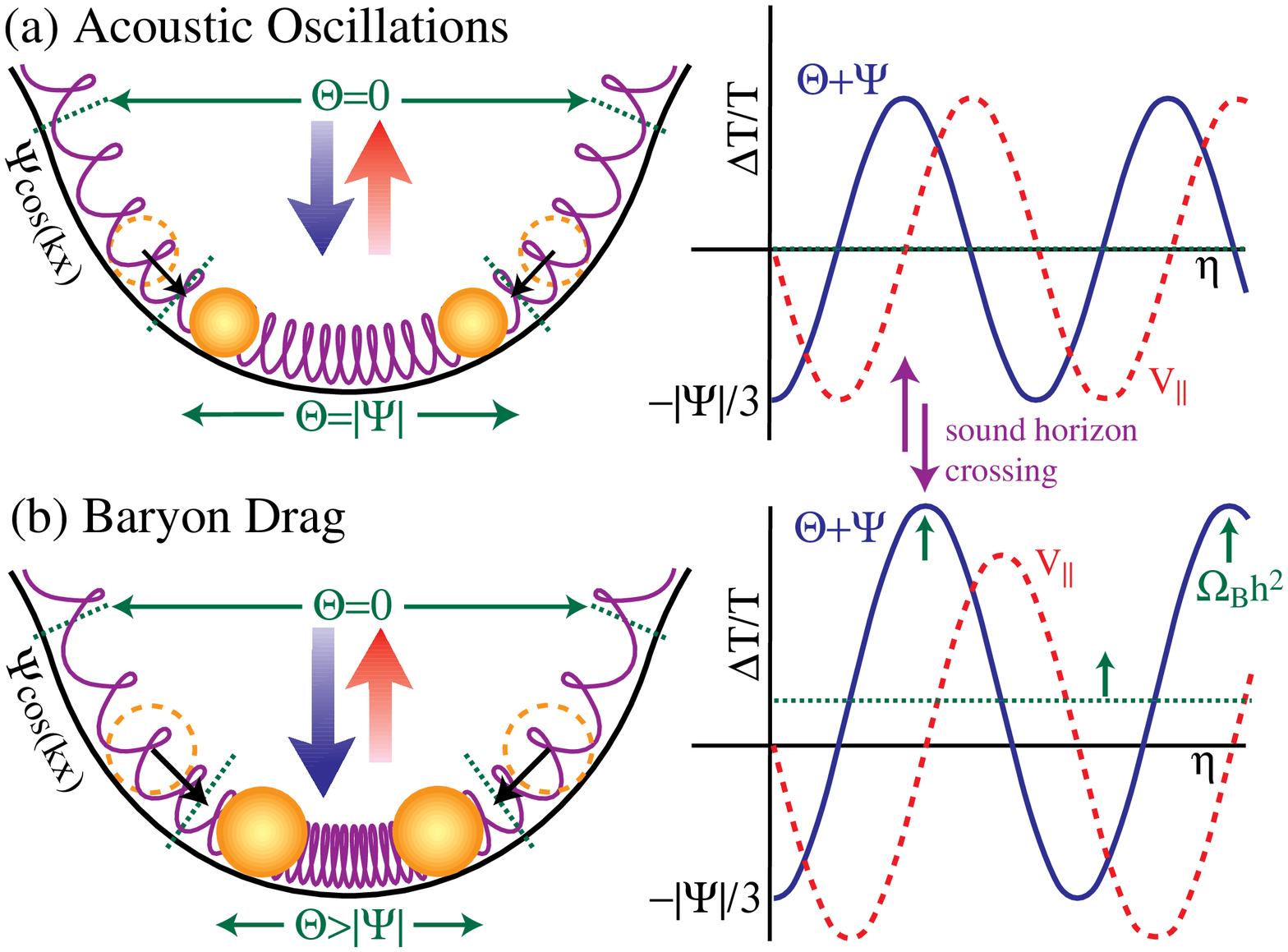}}
\vskip -0.5truecm
\caption{Acoustic Oscillations}
\mycaption{(a) Photon-dominated system.
Fluid compression through gravitational infall
is resisted by photon pressure setting up acoustic
oscillations.  Displayed here  is a potential well in real space
$-\pi/2 \simlt kx \simlt \pi/2$.
Gravity
displaces the zero point so that at the bottom of the well, the
temperature is $\Theta_0 =|\Psi| = -\Psi$ at equilibrium with $\Psi/3$
excursions.  This displacement is exactly cancelled by the redshift
$\Psi$ a photon experiences climbing out from the bottom of the potential well.
Velocity oscillations lead to a Doppler effect 90 degrees phase
shifted from the temperature perturbation.
(b) Photon-baryon system.  Baryons increase the gravitating mass, causing 
more infall and a net 
zero point displacement, even after redshift.  
Temperature crests
(compression) are enhanced 
over troughs (rarefaction) and velocity contributions.
}
\label{fig:1acoustic}
\end{figure}

%% file: chap1/isw.tex
\begin{figure}[p]
\centerline{ \hskip-0.5truecm
\epsfxsize=6in \epsfbox{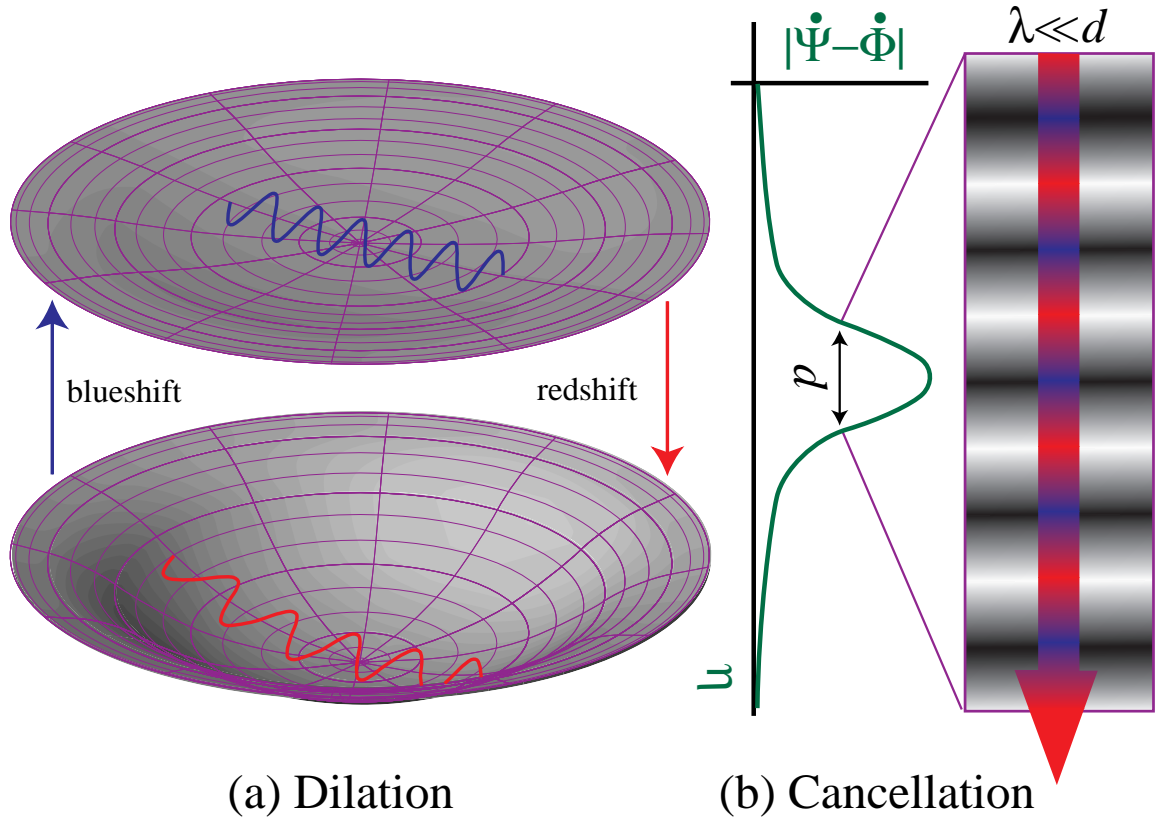}}
\vskip -0.5truecm
\caption{Differential Redshift and Dilation}
\mycaption{Gravitational redshift and dilation effects in a time dependent
potential.  Time variability occurs whenever the matter is not the sole
dynamical factor and thus probes $\Omega_0 h^2$, 
$\Omega_\Lambda$, $1-\Omega_0-\Omega_\Lambda$ and any isocurvature 
perturbations. (a) Decay of the potential $|\Psi|$ decreases the
gravitational redshift leading to an effective blueshift in
the well.
The implied 
curvature perturbation $|\Phi|$ decay represents a 
``contraction of space'' which blueshifts photons 
through time dilation, nearly doubling 
the $\Psi$ effect.  (b) In the free streaming limit after last
scattering, these two mechanisms combine to form the ISW effect.
Redshift-blueshift cancellation cuts off
contributions at small scales where the photon traverses many 
wavelengths during the decay. 
}
\label{fig:1isw}
\end{figure}

%% file: chap1/diffusion.tex
\begin{figure}[t]
%\vphantom{marker} \vskip 0.5truecm
\centerline{ \hskip -1truecm
\epsfxsize=4.0in \epsfbox{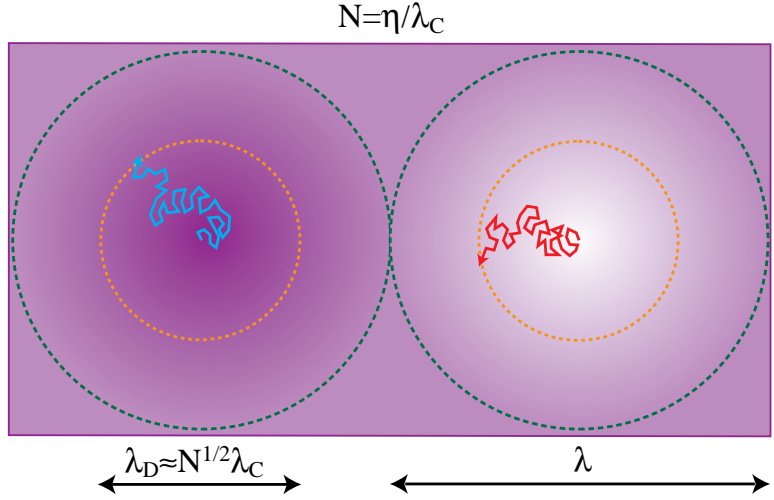}}
\vskip -0.5truecm
\caption{Photon Diffusion}
\mycaption{Photon diffusion mixes hot photons from overdense regions 
and cold photons from underdense regions as the diffusion length 
$\lambda_D$ exceeds the wavelength $\lambda$.  
Scattering averages the two
and rapidly damps anisotropies.  The diffusion length is given by
a random walk of stepsize the Compton mean free path $\lambda_C$.  
The number of steps the photon traverses in the age of the universe
$\eta$ is $\eta/\lambda_C$.  Thus the diffusion length scales as
$\lambda_D \approx N^{1/2} \lambda_C = (\eta\lambda_C )^{1/2}$.
The Compton mean free path increases near recombination causing
extensive damping at last scattering.
}
\label{fig:1diffusion}
\end{figure}

%% file: chap1/projection.tex
\begin{figure}[p]
\centerline{ \hskip-0.5truecm
\epsfxsize=6.0in \epsfbox{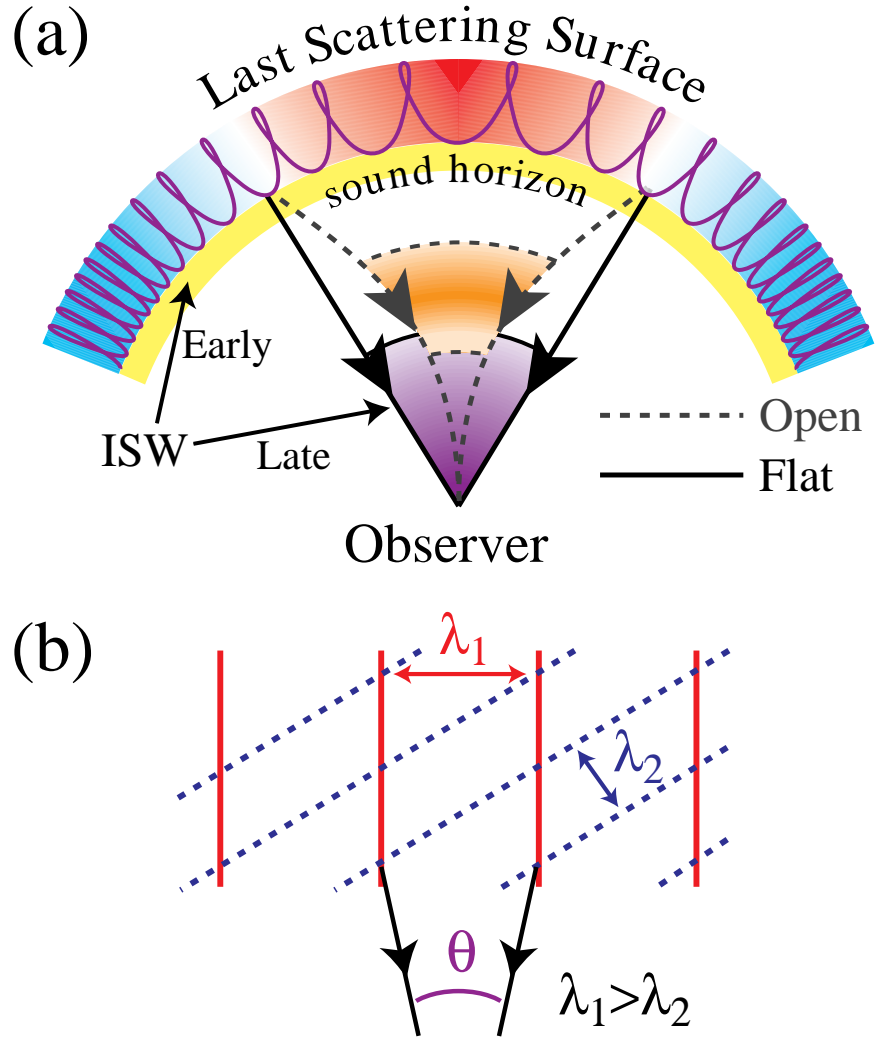}}
\vskip -0.5truecm
\caption{Projection Effect}
\mycaption{(a) Acoustic contributions 
exhibit a series of peaks with decreasing
angle beginning at the angular scale the sound horizon subtends at
last scattering.  This scale decreases significantly as the curvature
increases due to geodesic deviation.  Contributions after last scattering,
come from a smaller physical scale for the same angular scale, which
pushes the late ISW effect of flat $\Lambda$ and open models to larger angles.
(b) The orientation
of the plane wave projected on the surface of last scattering
leads to aliasing of power from shorter wavelengths onto larger angles.
This smooths out sharp features and prevents a steeply rising (blue) anisotropy
spectrum.
}
\label{fig:1projection}
\end{figure}

%% file: chap1/total.tex
\begin{figure}[p]
\centerline{ \hskip-0.5truecm
\epsfxsize=6.0in \epsfbox{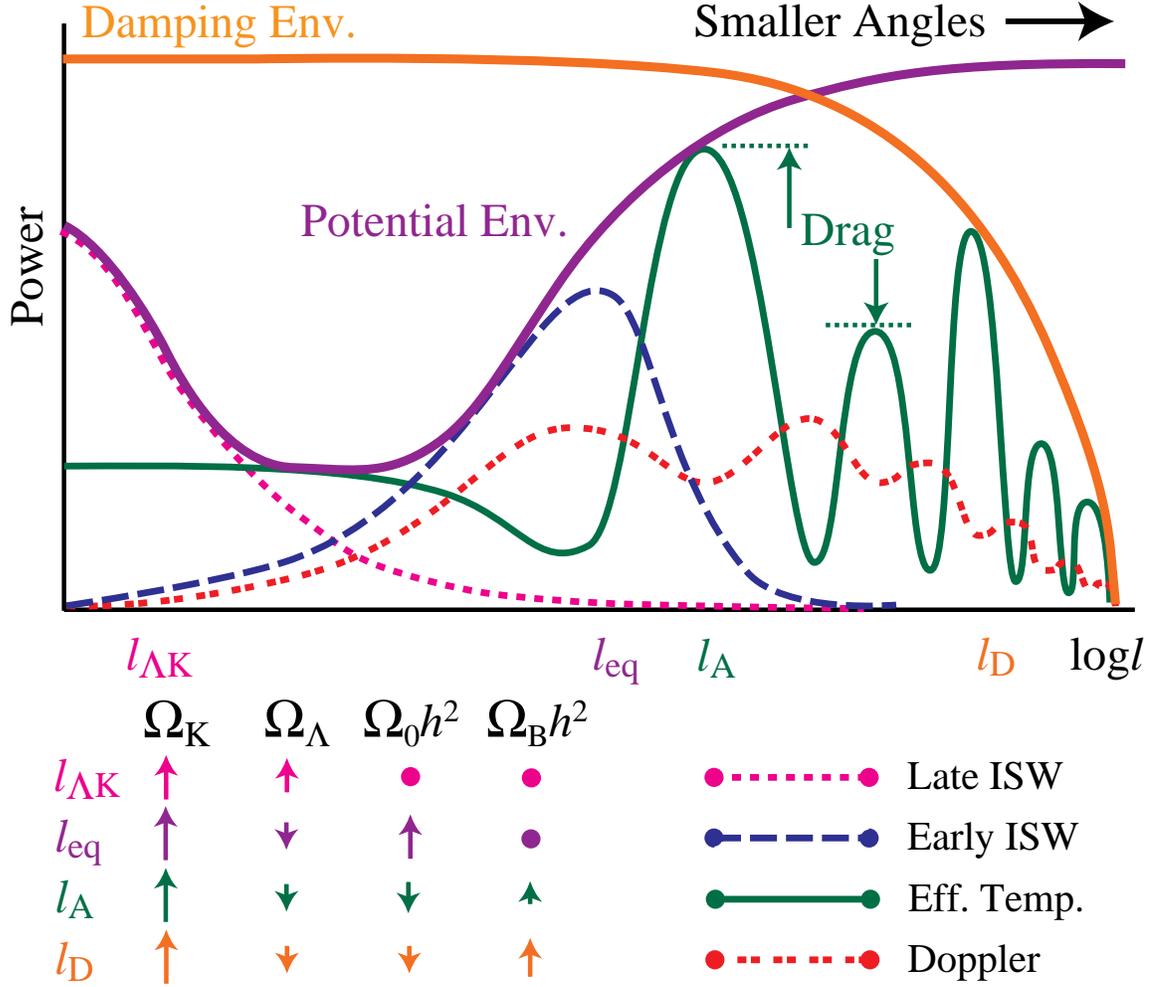}}
\vskip -0.5truecm
\caption{Total Anisotropy Spectrum}
\mycaption{A schematic representation for
scale invariant adiabatic scalar models.  
Features in open models are shifted
to significantly smaller angles compared with $\Lambda$ and $\Omega_0=1$
models, represented here as a shift in the $\ell$ axis beginning at
the quadrupole $\ell = 2$.   The  monopole and dipole fluctuations
are unobservable due to the mean temperature and peculiar velocity at the point
of observation.
The effective temperature at last scattering
$[\Theta+\Psi](\eta_*)$ includes the gravitational redshift effect
$\Psi(\eta_*)$. At large scales, the effective temperature goes
to $\Psi(\eta_*)/3$ and is called the Sachs-Wolfe (SW) contribution. 
In reality, small scale acoustic contributions
from the effective temperature and velocity
are smoothed out somewhat in $\ell$ due to projection effects 
(see Fig.~\ref{fig:1projection}).
}
\label{fig:1total}
\end{figure}

%% file: chap1/cancelpict.tex
\begin{figure}[p]
\vphantom{marker} \vskip -1.4truecm
\centerline{ \hskip -1truecm
\epsfxsize=5.5in \epsfbox{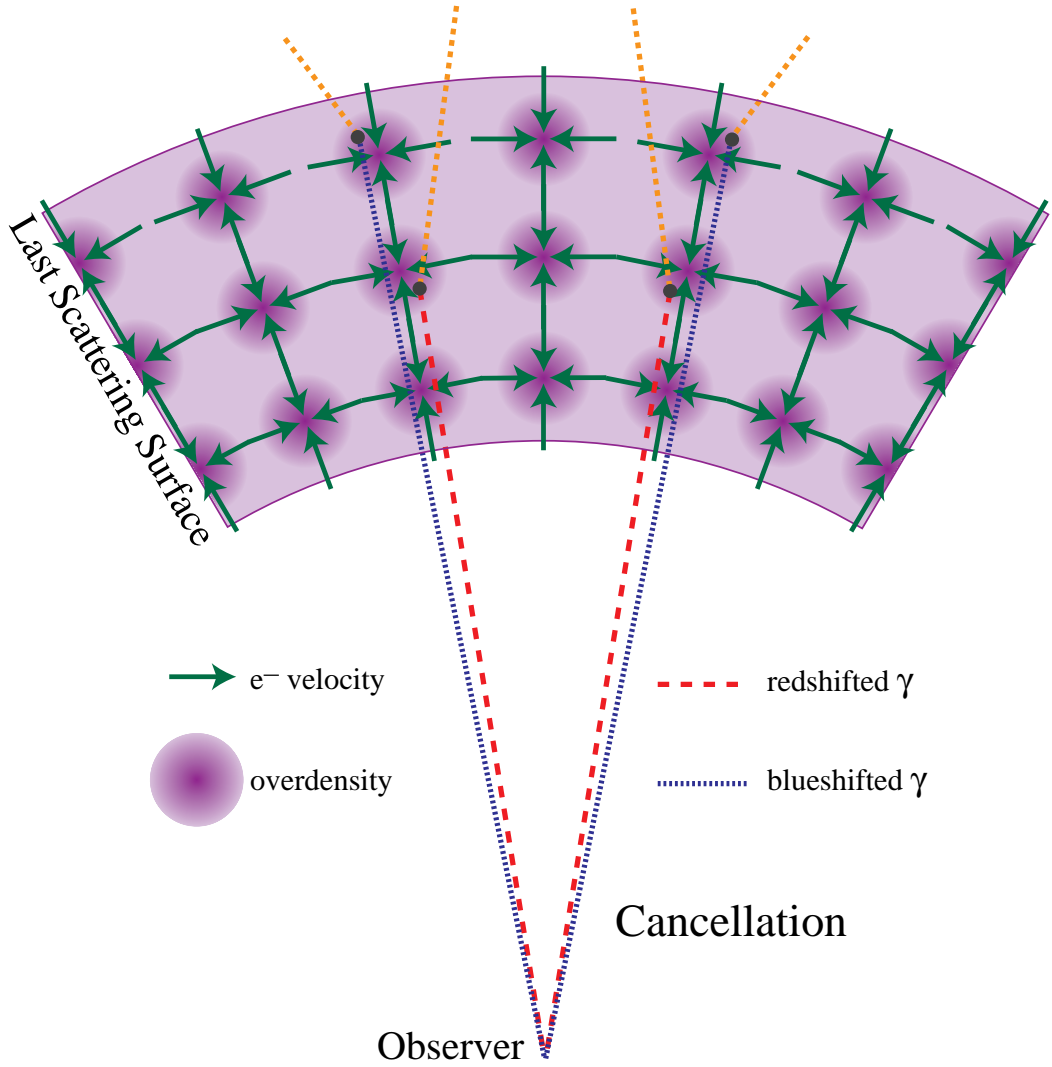}}
\vskip -0.5truecm
\caption{Cancellation Mechanism}
\mycaption{If the coherence scale, \ie\ wavelength, of the perturbation
is under the thickness of the last scattering surface, the photons suffer
alternating Doppler shifts depending on whether the photon last scattered
in the fore or rear of the perturbation.  
The small scale
Doppler effect is therefore severely cancelled.
}
\label{fig:1cancelpict}
\end{figure}

%% file: chap1/vishpict.tex
\begin{figure}[p]
%\vphantom{marker} \vskip 0.5truecm
\centerline{ \hskip -1truecm
\epsfxsize=5.5in \epsfbox{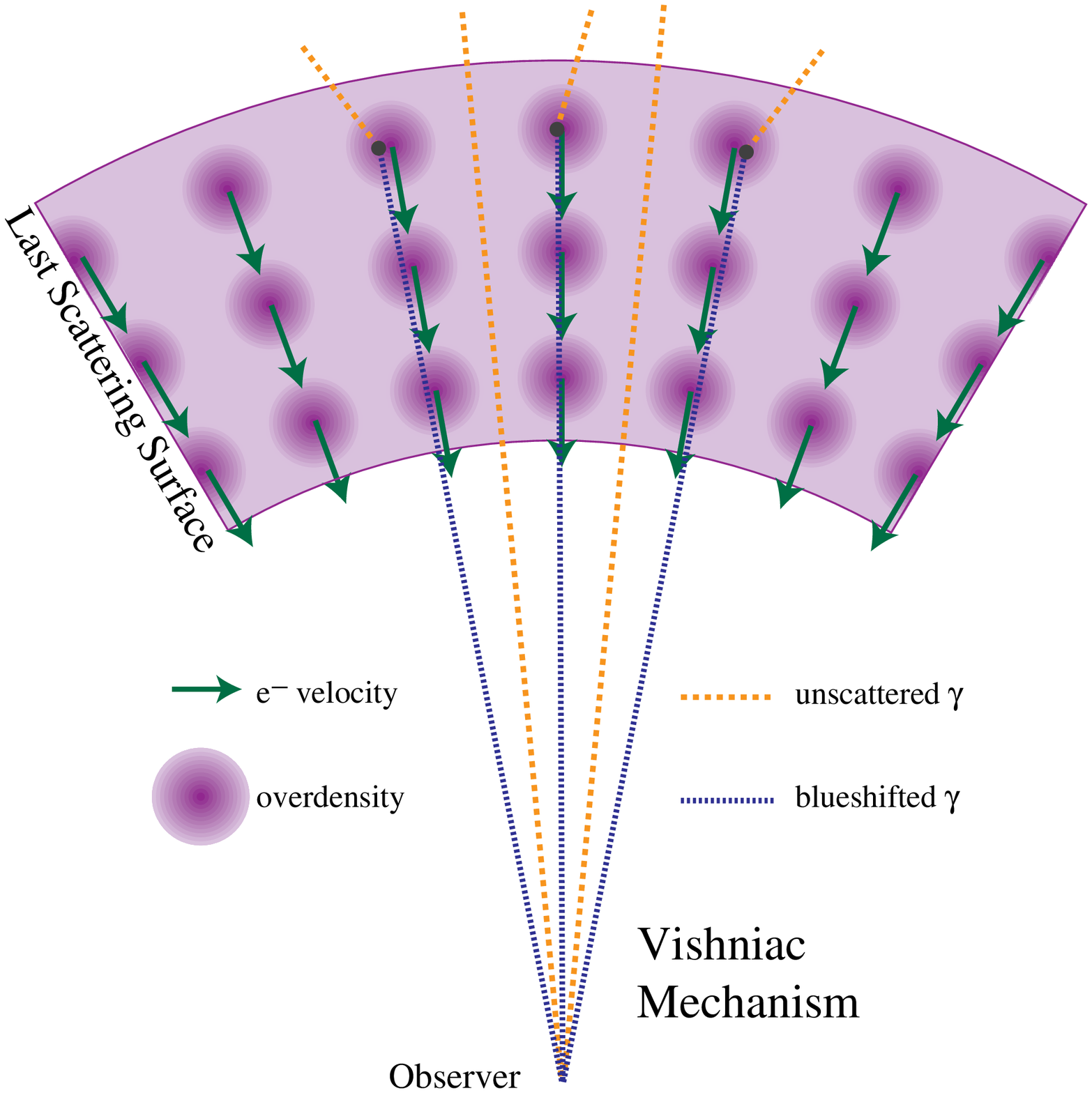}}
\vskip -0.5truecm
\caption{Vishniac Mechanism}
\mycaption{In an overdense region, the free electron density is higher.
This increases the probability of scattering.  If these overdense regions
are also caught in a large scale bulk flow, this can lead to a small 
scale variation in the temperature through preferential scattering. 
The Vishniac mechanism thus relies on a coupling of large and small 
scale perturbation modes. 
}
\label{fig:1vishpict}
\end{figure}

%% file: chap2.tex
\chapter{The Boltzmann Equation}
\label{ch-Boltzmann}
\begin{quote}
\footnotesize\it
Wonderful, the process which fashions and transforms us!  What is
it going to turn you into next, in what direction will it use
you to go?
\vskip 0.1truecm
\centerline{--Chuang-tzu, 6}

\end{quote}

The study of the formation and evolution of CMB fluctuations in both
real and frequency space begins with the
radiative transport, or  Boltzmann equation.  In this pedagogically motivated
chapter, we will examine its derivation.  
The Boltzmann equation written in abstract form as 
\bel{eq:Boltzabs}
{df \over dt} = C[f]
\ee
contains a collisionless part
$df/dt$, which deals with the effects of gravity on the
photon distribution function $f$, and collision terms $C[f]$,
which account for its interactions with other
species in the universe.
The collision terms in the Boltzmann
equation have several important effects.  Most importantly, Compton 
scattering couples the
photons and baryons, keeping the two in 
kinetic equilibrium.  This process along with interactions that create 
and destroy photons determines the extent to which the CMB can 
be thermalized.
We will examine these issues more fully in \S\ref{ch-spectral} 
where we consider spectral distortions. 
Compton scattering also governs the evolution of inhomogeneities 
in the CMB temperature which lead to anisotropies on the sky.
This will be the topic of \S\ref{ch-primary} and \S \ref{ch-secondary}.

In this chapter, we will first examine gravitational interactions and show
that the photon energy is affected by gradients in the gravitational 
potential, \ie\ the gravitational redshift, and changes in the
spatial metric, \ie\ the cosmological redshift from the scale factor
and dilation effects 
due to the space curvature perturbation.  Compton scattering
in its non-relativistic limit can be broken down in a perturbative expansion
based on the energy transfer between the photons and electrons.  
We will examine
the importance of each term in turn and derive its effects
on spectral distortions and temperature inhomogeneities in the CMB.

\section{Gravitational Interactions}
\label{sec-2collisionless}

Gravity is the ultimate source of spatial fluctuations in the photon
distribution and the cause of the adiabatic cooling of the photon
temperature from the expansion. 
Its effects are described by the collisionless Boltzmann, 
or Liouville, equation which controls the
evolution of the photon distribution $f(x,p)$ as the photons stream 
along their geodesics.  Here $x$ and $p$ are the 4-position and 
4-momentum of the photons respectively.  It is given by
\bel{eq:collisionless}
{d f \over dt} = {\partial f \over \partial x^\mu}{dx^\mu \over dt} 
+ {\partial f \over \partial p^{\mu}} {d p^\mu \over dt} = 0.
\ee
In other words, the phase space density of photons is conserved 
along its trajectory.  
The gravitational effects are hidden in the time dependence of the
photon momentum. The solution to equation \eqn{collisionless} is non-trivial
since the photons propagate in a metric distorted by the lumpy
distribution of matter.  To evaluate its effect explicitly, 
we need to examine the geodesic equation in the presence of arbitrary 
perturbations. 

\subsection{Metric Fluctuations}
\label{ss-metric}

The big bang model assumes that the universe  is homogeneous and
isotropic on the large scale.  All such cases can be described by the
Fried\-man-Robert\-son-Walker metric, where the line element takes the form
\bel{eq:FRWmetric}
ds^2 = g_{\mu\nu} dx^{\mu} dx^{\nu} = -dt^2 + (a/a_0)^2 
       \gamma_{ij} dx^{i} dx^{j},
\ee
with  $\gamma_{ij}$ as the background three-metric on a space of constant 
curvature $K = -H_0^2 (1-\Omega_0-\Omega_\Lambda)$ and the
scale factor is related to the redshift by $a/a_0 = (1+z)^{-1}$.   
We will be mainly interested in the flat $K=0$ and negatively
curved (open) $K<0$ cases.  
%The scale factor $a$ evolves according to the total density in the
%universe $\dot a /a = Ha/a_0$ where the Hubble parameter is
%\bel{eq:Hubble}
%H^2 = {\left( a_0 \over a \right)^4} {a_{eq} + a \over a_{eq} + a_0} 
%\Omega_0 H_0^2 - {\left(a_0 \over a\right)^2 K + \Omega_\Lambda H_0^2},
%\ee
%where the vacuum density $\Omega_\Lambda$ is related to the cosmological
%constant by $\Omega_\Lambda = \Lambda/3H_0^2$ and $a_{eq}$ is the epoch
%of matter-radiation equality. 
For these cases, a
convenient representation of the three-metric
which we will have occasion to use is the radial representation
\bel{eq:radial}
\gamma_{ij} dx^i dx^j = -K^{-1} [d\chi^2 + \sinh^2 \chi (d\theta^2 + 
\sin^2\theta d\phi^2)],
\ee
where the radial coordinate is scaled to the curvature length $(-K)^{-1/2}$.

Small scalar perturbations to the background
metric can in general be expressed by two spatially varying functions.
The exact form of the metric fluctuations varies with the choice of 
hypersurface on which
these perturbations are defined, \ie\ the gauge. 
We will discuss the subtleties involving the choice of gauge in
\S\ref{sec-4gauge}.
For now, let us derive the
evolution equations for the photons using the conformal Newtonian
gauge where the metric takes the form
\beal{eq:metric}
g_{00} \eal -[1+ 2\Psi(\bx,t)], \nonumber\\
g_{ij} \eal (a/a_0)^2 [1+2\Phi(\bx,t)] \gamma_{ij}.  
\eea
Note that $\Psi$ can be interpreted as a Newtonian potential. 
$\Phi$ is the fractional  perturbation to the spatial curvature as the form
of equation \eqn{radial} shows.  
As we shall see
in \S\ref{ss-4einstein}, they are related by the Einstein equations
as $\Phi = -\Psi$ when pressure
may be neglected.  We will therefore often loosely refer to both
as ``gravitational potentials.''

The geodesic equation for the photons is
\bel{eq:geodesic}
{d^2 x^\mu \over d\lambda^2} + \Gamma^{\mu}_{\ \alpha\beta} 
{d x^\alpha \over d \lambda}{d x^\beta \over d \lambda} = 0,
\ee
where $\Gamma$ is the Christoffel symbol.  The 
affine parameter $\lambda$ is chosen such that the photon
energy satisfies $p^0 = dx^0/d\lambda$.  Since the photon momentum
is given by
\bel{eq:momentum}
{p^i \over p^0} =  {dx^i \over dt},
\ee
the geodesic equation then becomes
\bel{eq:momentumev}
{d p^i \over dt} = g^{i\nu} \left( {1 \over 2} {\partial g_{\alpha\beta}
\over \partial x^\nu}- {\partial g_{\nu \alpha} \over \partial x^\beta}
	\right) {p^\alpha p^\beta \over p^0}.
\ee
This equation determines the gravitational effects on the photons
in the presence of perturbations as we shall now show.  

\subsection{Gravitational Redshift and Dilation}
\label{ss-2redshift}

Let us rewrite the Boltzmann equation in terms of the energy $p$ and 
direction of propagation of the photons $\gamma^i$ in a frame that is
locally 
orthonormal on constant time hypersurfaces, 
\bel{eq:normal}
{\partial f \over \partial t} + {\partial f \over \partial x^i} 
{ dx^i \over  dt} + {\partial f \over \partial p}{d p \over dt} 
+ {\partial f \over \partial \gamma^i}{d\gamma^i \over dt} = 0.
\ee
Notice that $d\gamma^i /dt \ne 0$ only in the presence of curvature
from $K$ or $\Phi$
because otherwise photon geodesics are straight lines. 
Since the anisotropy $\partial f /\partial \gamma^i$ is already first order
in the perturbation, it may be dropped if the background curvature $K=0$.
In the presence of negative curvature, it makes photon geodesics
deviate from each other exponentially with distance.  
Two photons which are
observed to have a given angular separation were in the past separated by 
a larger (comoving) physical distance than euclidean analysis
would imply.  We shall see that this property allows the curvature
of the universe to be essentially read off of anisotropies in the CMB.
Formal elements of this effect are discussed in \S \ref{ss-4radiation}.

On the other hand, the redshift term $dp/dt$
is important in all cases -- even in the absence of perturbations.
Since static curvature effects are unimportant in determining the redshift
contributions, we will assume in the following that the background
three-metric is flat, \ie\ $\gamma_{ij}=\delta_{ij}$ without loss of
generality.  The energy and direction of propagation are explicitly
given by
\bel{eq:pgamma}
p^2 = p^i p_i, \qquad \gamma^i = {a \over a_0} 
{p^i \over p}( 1 + \Phi),
\ee
which implies $p^0 = (1+\Psi)p$.  
The geodesic equation (\ref{eq:momentumev}) then yields to first order in
the fluctuations
\bel{eq:p0}
{1 \over p}{dp^0 \over dt} = -\left( {\partial \Psi \over \partial t} + 
{d a \over dt} {1 \over a}(1-\Psi) 
+ {\partial\Phi \over \partial t} + 2 {\partial \Psi \over \partial x^i}
{a_0 \over a} \gamma^i \right).
\ee
From this relation, we obtain
\beal{eq:energyev}
{1 \over p}{dp \over dt} 
\eal {1 \over p}{d p^0 \over dt}(1+\Psi) + {\partial \Psi
\over \partial t } + {\partial \Psi \over \partial x^i}
{d x^i \over dt} \nonumber\\ 
 \eal -\left({da \over dt} {1 \over a} + {\partial \Phi \over \partial t} 
+ {\partial \Psi \over \partial x^i} {a_0 \over a}  \gamma^i \right),
\eea
which governs the gravitational and cosmological 
redshift effects on the photons.

Now let us discuss the physical interpretation of the energy
equation (\ref{eq:energyev}).  
Consider first a small region where we can neglect the spatial variation
of $\Psi$ and $\Phi$.  
In the presence of a gravitational potential, clocks naturally ticking
at intervals $\Delta t$ run slow by
the dilation factor (see \eg\ \cite{Weinberg}),
\bel{eq:dilation}
\delta t = (-g_{00})^{-1/2} \Delta t \approx (1-\Psi) \Delta t.
\ee
For light emitted from the point 1, crests leave spaced by $\delta t_1 
= [1-\Psi(t_1)]\Delta t$.  If they arrive at the origin spaced
by $\delta t_0$, 
they should be compared with a local 
oscillator with crests spaced as $[1-\Psi(t_0)]\Delta t$, \ie\ 
the shift in frequency (energy) is
\bel{eq:compare0}
{p_1  \over p_0} = [1+\Psi(t_1)-\Psi(t_0)]{\delta t_1 \over \delta t_0}.
\ee
Now we have to calculate the in-transit 
delay factor $\delta t_1 / \delta t_0$.  
Since null 
geodesics from the origin are radial in the FRW metric, choose
angular coordinates such that along the $\chi(t)$ geodesic
\bel{eq:nullgeo}
-(1+2\Psi) dt^2 + (a/a_0)^2 (-K)^{-1} (1+2\Phi) d\chi^2 = 0.
\ee
A wave crest emitted at $(t_1,\chi_1)$ is received at $(t_0,0)$ where
the two are related by
\bel{eq:nullrelation}
\int_{t_1}^{t_0} (1+\Psi-\Phi) {a_0 \over a} dt = \int_0^{\chi_1} 
(-K)^{1/2} d\chi.
\ee
At $\chi_1$, the source emits a second crest after $\delta t_1$ which
is received at the origin at $t_0 + \delta t_0$ where
\bel{eq:equategeos}
\int_{t_1}^{t_0} (1+\Psi-\Phi) {a_0 \over a} dt = 
\int_{t_1 +\delta t_0}^{t_0+\delta t_0} (1+\Psi-\Phi) {a_0 \over a} dt.
\ee
This can be manipulated to give
\bel{eq:equategeos2}
\int_{t_1}^{t_1 + \delta t_1} (1+\Psi-\Phi) {a_0 \over a} dt = 
\int_{t_0}^{t_0 + \delta t_0} (1+\Psi-\Phi) {a_0 \over a} dt,
\ee
or
\bel{eq:delay}
{\delta t_1 \over \delta t_0} = {a(t_1) \over a(t_0)} {1 -\Psi(t_1) + \Phi(t_1)
\over 1 - \Psi(t_0) + \Phi(t_0)}.
\ee
Inserting this into equation \eqn{compare0},
 the ratio of energies becomes
\bel{eq:compare}
{p_1  \over p_0} = {a(t_1)[1+\Phi(t_1)] \over a(t_0) [1 +\Phi(t_0)]}.
\ee
Notice that the space curvature $\Phi$ but {\it not} the
Newtonian potential $\Psi$ enters this expression. 
This is easy to interpret.  Heuristically, the wavelength of the photon
itself scales with the space-space component of the metric, \ie\ 
$a (1+\Phi)$.  In the background, this leads to the universal redshift
of photons with the expansion.  The presence of a space curvature perturbation
$\Phi$ also stretches space.  We shall see that it arises from density 
fluctuations through the Einstein equations (see \S\ref{ss-4einstein}).  
Overdense
regions create positive curvature and underdense regions negative curvature. 
From equation \eqn{compare}, 
the rate of change of the energy is therefore given by
\bel{eq:energyrate}
{1 \over p }{\partial p \over \partial t} = -{da \over dt}{1 \over a} 
- {\partial \Phi \over \partial t},
\ee
which explains two of three of the terms in equation (\ref{eq:energyev}).

Now let us consider the effects of spatial variations.  Equation 
(\ref{eq:compare0}) becomes 
\bel{eq:compare2}
{p_1  \over p_0} = [1+\Psi(t_1,\chi_1)-\Psi(t_0,0)]{\delta t_1 
\over \delta t_0}.
\ee
The additional factor here is the potential difference in space.  Photons
suffer gravitational redshifts climbing in and out of potentials.  Thus 
the gradient of the potential along the direction of propagation
leads to a redshift of the photons, \ie\
\beal{eq:gravredshift}
{1 \over p}{\partial p \over \partial x^i}{d x^i \over dt} \eal -{\partial \Psi
\over \partial x^i} {d x^i \over dt} \nonumber\\
 \eal -{\partial \Psi \over \partial x^i}  {a_0 \over a} \gamma^i,
\eea
as required.  This explains why a uniform $\Psi$ does not lead to an effect on 
the photon energy and completes the physical interpretation of 
equation (\ref{eq:energyev}).  

\subsection{Collisionless Brightness Equation}
\label{ss-2collisionlesstemp}

The fractional shift in frequency from gravitational
effects is independent of frequency $p' = p(1+\delta p/p)$.  
Thus, a blackbody distribution will remain a blackbody,
\beal{eq:fbb}
f'(p') = f(p)  \eal \left\{\exp[p'/T(1+\delta p/p)]-1\right\}^{-1}  
	\nonumber\\
	\eal \left\{\exp[p'/T']-1\right\}^{-1} ,
\eea
with a temperature shift $\delta T/T = \delta p/p$.
Let us therefore integrate the collisionless
Boltzmann equation over energy, \ie\ define
\bel{eq:tempoperator}
4\Theta \equiv
{1 \over \pi^2 \rho_\gamma} \int p^3 dp f - 1 = 
{\delta \rho_\gamma \over {\rho_\gamma}},
\ee
where 
$\rho_\gamma$ is the spatially and directionally averaged 
energy density of the photons.  Since $\rho_\gamma \propto T^4$,
$\Theta(\eta,\bx,\bg)$ is 
the fractional temperature fluctuation for a blackbody. 

Employing equation \eqn{energyev} in \eqn{normal} and integrating 
over frequencies, we obtain the  collisionless Boltzmann (or brightness)
equation, 
\bel{eq:collisionlessfinal}
\dot \Theta + \gamma^i {\partial \over \partial x^i} (\Theta + \Psi)
+ \dot \gamma^i {\partial \over \partial \gamma^i} \Theta + \dot \Phi = 0,
\ee
where the overdots represent derivatives with respect to conformal
time $d\eta = dt/a$. 
Notice that since the potential $\Psi(\eta,\bx)$ is not an explicit 
function of angle $\gamma$ and $\gamma^i = \dot x^i$, we can write this
in a more compact and suggestive form,
\bel{eq:collisionless2}
{d \over d\eta}[\Theta+\Psi](\eta,\bx,\bg) = \dot \Psi - \dot \Phi,
\ee
which also shows that in a static potential
$\Theta +\Psi$ is conserved.  Thus the temperature fluctuation is just
given by the potential difference:
\bel{eq:sachswolfesimple}
\Theta(\eta_0, \bx_0,\bg_0) = 
\Theta(\eta_1, \bx_1,\bg_1) + [\Psi(\eta_1,\bx_1) - \Psi(\eta_0,\bx_0)].
\ee
This is the Sachs-Wolfe effect \cite{SW} in its simplest form.

\section{Compton Scattering}
\label{sec-2compton}

Compton scattering
$\gamma(p) + e(q)
\leftrightarrow \gamma(p') + e(q')$ dominates the interaction of CMB 
photons with electrons.  By allowing energy exchange 
between the photons and electrons, it is the primary mechanism for the
thermalization of the CMB.  It also governs the mutual evolution of photon and 
baryon inhomogeneities before last scattering.
The goal of this section is to derive its collision term in the
Boltzmann equation
to second order in the small energy transfer due to scattering.
The approach taken here provides a coherent framework
for all Compton scattering effects.
In the proper limits, the equation derived below reduces to more 
familiar forms, \eg\ the
Kompaneets equation in the homogeneous and isotropic limit and the
temperature Boltzmann equation for blackbody spectra.
Furthermore, new truly second order effects
such as the quadratic Doppler effect which mix
spectral distortions and anisotropies result \cite{HSSa}.
 
We make the following assumptions in deriving the equations:
\begin{enumerate}
\item The Thomson limit applies, \ie\ the fractional energy transfer $\delta p/p
\ll 1$ in the rest frame of the
background radiation.

\item The radiation is unpolarized and remains so.

\item
The density of electrons is low so that Pauli suppression terms may be
ignored. 

\item The electron distribution is thermal about some bulk flow
velocity determined by the baryons $\V_b$.
\end{enumerate}

Approximations (1), (3), and (4) are valid for most situations of cosmological
interest.  The approximation regarding polarization is not strictly true.
Polarization is generated at the last scattering surface by Compton
scattering of anisotropic radiation 
%because
%the Thomson cross section depends on angle
%as $|\epsilon_f\cdot\epsilon_i|^2$,
%where $\epsilon_f$ and $\epsilon_i$ are the final and initial polarization
%vectors respectively
%\cite{BE84,Kaiser83}.  
%Furthermore, polarization feeds back into anisotropies.
%Averaging over incident and summing over
%final
%polarizations leads to an angular dependence: $1+\cos^2\beta$.
%Since the scattering of linearly polarized radiation will in general have
%a different angular dependence than this, the scattering term in the Boltzmann
%equation for temperature perturbations  will be modified by polarization.
However, since anisotropies themselves tend to be small,
polarization is only generated at the $\sim 10\%$ level 
compared with temperature perturbations \cite{Kaiser83}.
The feedback effect into the temperature 
only represents a $\sim 5\%$ correction
to the temperature evolution and thus is only important for high 
precision calculations.  We will consider its effects in
greater detail in Appendix \S\ref{ss-5polarization}. 

\subsection{Collision Integral}
\label{ss-2collisionint}
\smallskip
Again employing a locally orthonormal, \ie\ Minkowski, frame we may
in general express the collision term as
\cite{Bernstein}
\beal{eq:CGen}
C[f] \eal {1 \over {2E(p)}} \int Dq Dq' Dp' (2\pi)^4 \delta^{(4)}
(p+q-p'-q') |M|^2 \nonumber\\
& & \times \Bigg\{ g(t,\X,\Q')f(t,\X,\P') \left[ 1 + f(t,\X,\P) \right] 
\nonumber\\
& & - g(t,\X,\Q)f(t,\X,\P)\left[ 1 + f(t,\X,\P') \right] \Bigg\} , 
\eea
where $|M|^2$ is the Lorentz invariant matrix element, $f(t,\X,\P)$ is the
photon distribution function, $g(t,\X,\Q)$ is the electron distribution function
and
\bel{eq:Lorentz}
Dq = { d^3 q \over {(2\pi)^3 2E(q)}},
\ee
is the Lorentz invariant phase space element.  The terms in 
equation (\ref{eq:CGen}) which
contain the distribution functions are just the contributions from
scattering into and out of the momentum state $\P$ including
stimulated emission effects.

\input chap2/geometry.tex 
We will assume that the electrons are thermally distributed about some
bulk flow velocity $\V_b$,
\bel{eq:EDist}
g(t,\X,\Q) = (2\pi)^3 x_e n_e
(2\pi m_e T_e)^{-3/2} \exp \left\{ {- [\Q-m_e \V_b]^2
\over 2m_e T_e } \right\} ,
\ee
where $x_e$ is the ionization fraction, $n_e$ is the 
electron number density,
$m_e$ is the electron mass, and we employ units with $c=\hbar=k_B=1$
here and throughout.
Expressed in the rest frame of the electron,
the matrix element for Compton scattering summed over polarization is
given by
\cite{MS}
\bel{eq:MCompton}
|M|^2 = 2 (4\pi)^2 \alpha^2 \left[ {\tilde p' \over \tilde p }
+ {\tilde p \over \tilde p'} - \sin^2 \tilde \beta \right],
\ee
where the tilde denotes quantities in the rest frame of the electron,
$\alpha$ is the fine structure constant,
and
$\cos  \tilde \beta = {\tilde \bg \cdot \tilde \bg'}$
is the scattering angle (see Fig.~\ref{fig:2geometry}).  
The Lorentz transformation gives
\bel{eq:lorentz}
{p \over \tilde p} = {\sqrt{1-q^2/m_e^2} \over 1- {\bf p}\cdot {\bf q}/pm_e },
\ee
and the identity $\tilde p_\mu \tilde p'^\mu = p_\mu p'^\mu$ relates
the scattering angles.

We now expand in the energy transfer $p-p'$ from scattering.
There are several small quantities involved in this expansion.
It is worthwhile to compare these terms.
To first order, there is only the bulk velocity of the electrons $v_b$.  
In second order, many more terms appear.
The quantity $T_e/m_e$ characterizes the kinetic energy of the electrons and 
is to be compared with $p/m_e$ or essentially $T/m_e 
\approx 5 \times 10^{-10} (1+z_*),$ where $T$ is the
temperature of the photons.  
Before a redshift  $z_{cool} \approx 8.0 (\Omega_0h^2)^{1/5}
x_e^{-2/5}$, where $x_e$ is the ionization fraction (this corresponds to
$z \simgt 500(\Omega_bh^2)^{2/5}$ for standard recombination),
the tight coupling between photons and electrons via Compton scattering
requires these two temperatures to be comparable (see \S \ref{ss-3comptonization}).
At lower redshifts, it is possible that
$T_e \gg T$, which produces distortions in the radiation via the
Sunyaev-Zel'dovich (SZ) effect as discussed in section 
\S \ref{ss-3comptonization}.
Note that the term $T_e/m_e$ may also be thought of as the
average thermal velocity squared $\langle v^2_{\rm therm} \rangle = 3T_e/m_e$.
This is to be compared with the bulk velocity squared $v_b^2$ and
will depend on the specific means of ionization.  Terms of order
$(q/m_e)^2$ contain both effects.

Let us evaluate the collision integral keeping track of the order
of the terms.
The matrix element 
expressed in terms of the corresponding quantities in the
frame of the radiation is 
\bel{eq:MComptonBack}
|M|^2 = 2 (4\pi)^2 \alpha^2 \left( {\cal M}_0 + {\cal M}_{q/m_e} 
+ {\cal M}_{(q/m_e)^2} + {\cal M}_{(qp/m_e^2)} 
+ {\cal M}_{(p/m_e)^2} \right) + h.o.,
\ee
where 
\beal{eq:Mterms}
{\cal M}_0 \eal 1 + \cos^2 \beta,  \nonumber\\
{\cal M}_{q/m_e} \eal - 2\bomb \bigg[
 	{\pq \over m_e p} + {\ppq  \over m_e p'} \bigg], \nonumber\\
{\cal M}_{(q/m_e)^2} \eal \bomb
	{q^2 \over m_e^2}, \nonumber\\
{\cal M}_{qp/m_e^2} \eal
	(1-\cosb)(1-3\cosb) \bigg[ {\pq \over m_e p}  
	+ {\ppq \over m_e p'} \bigg]^2  \nonumber\\
& &
	+ 2\bomb {(\pq)(\ppq) 
	\over {m_e^2} {pp'}} \Bigg\}, \nonumber\\
{\cal M}_{(p/m_e)^2} \eal
	(1-\cosb)^2 {p^2 \over m_e^2} .
\eea
Notice that the zeroth order term gives an angular dependence of
$1+\cos^2\beta$ which is the familiar Thomson cross section result.

Likewise, the electron energies can be expressed as
\bel{eq:Electen}
{1 \over E_q E_q'} = {1 \over m_e^2} [1 
- {\cal E}_{(q/m_e)^2}
- {\cal E}_{qp/m_e^2}
- {\cal E}_{(p/m_e)^2} ],
\ee
where
\beal{eq:Electenterms}
{\cal E}_{(q/m_e)^2} \eal {q^2 \over m_e^2} ,  \nonumber\\
{\cal E}_{qp/m_e^2} \eal {\pmp \cdot \Q \over  m_e^2} , \nonumber\\
{\cal E}_{(p/m_e)^2} \eal {\pmp^2 \over 2m_e^2}. 
\eea
The following identities are very useful for the calculation.
Expansion to second order in energy transfer can be handled in a quite
compact way by ``Taylor expanding''
the delta function for energy conservation
in $\delta p = q-q'$,
\beal{eq:deltafn}
\delta(p+q-p'-q') \eal \del + ({\cal D}_{q/m_e} + 
{\cal D}_{p/m_e})p \ddel \nonumber\\
&& + {1 \over 2} ({\cal D}_{q/m_e} + {\cal D}_{p/m_e})^2 p^2 \dddel + h.o.,
\eea
where
\beal{eq:deltaterms}
{\cal D}_{q/m_e} ={1 \over m_e p} {\pmp\cdot\Q } , \nonumber\\
{\cal D}_{p/m_e} ={1 \over m_e p} {\pmp^2}.
\eea
This is of course defined and justified by integration by parts.
Integrals over the electron distribution function are trivial,
\beal{eq:electronint}
\int {d^3 \Q \over (2\pi)^3} g(\Q) \eal x_e n_e, \nonumber\\
\int {d^3 \Q \over (2\pi)^3} q^i g(\Q) \eal m_e v_b^i x_e n_e, \nonumber\\
\int {d^3 \Q \over (2\pi)^3} q^i q^j g(\Q) \eal m_e^2 v_b^i v_b^j x_e n_e +
m_e T_e \delta^{ij} x_e n_e.
\eea
Thus while the terms of ${\cal O}(q/m_e) \rightarrow {\cal O}(v_b)$,
the ${\cal O}(q^2/m_e^2)$ terms give two contributions:
${\cal O}(v_b^2)$ due to the bulk velocity and ${\cal O}(T_e/m_e)$
from the thermal velocity. 

The result of integrating over the electron momenta can be written
\bel{eq:CImplicit}
C[f] = {d \tau \over dt} \int dp' {p' \over p}
\int {d\Omega' \over {4\pi}} {3 \over 4}
\left[ {\cal C}_0 + {\cal C}_{p/m_e} + {\cal C}_{v_b} + {\cal C}_{v_b^2} 
+ {\cal C}_{T_e/m_e} +
{\cal C}_{v_b p/m_e} + {\cal C}_{(p/m_e)^2}  \right],
\ee
where we have kept terms to second order in $\delta p/p$ and the
optical depth to Thomson scattering $\tau$ is defined through the scattering
rate
\bel{eq:opticaldepthder}
{d\tau  \over dt} \equiv x_e n_e \sigma_T,
\ee
with 
\bel{eq:Thomson}
\sigma_T = 8\pi \alpha^2 /3m_e^2,
\ee
as the Thomson cross section.
Equation \eqn{CImplicit} may be
considered as the source equation for all first and second order
Compton scattering effects.

\subsection{Individual Terms}
\label{ss-2terms}
 
In most cases of interest, only
a few of the terms in equation \eqn{CImplicit} will ever contribute.  
Let us now consider each in turn.  It will be useful
to define two combinations of distribution functions
\beal{eq:usefuldefs}
F_1(t,\X,\P,\P') \eal f(t,\X,\P') - f(t,\X,\P), \nonumber\\
F_2(t,\X,\P,\P') \eal f(t,\X,\P) + 2f(t,\X,\P)f(t,\X,\P') + f(t,\X,\P'),
\eea
which will appear in the explicit evaluation of the collision term.

\subsubsection{(a) Anisotropy Suppression: ${\cal C}_0$}

Scattering makes the photon distribution isotropic in the 
electron rest frame.  
Microphysically this is accomplished via
scattering into and out of a given direction.
Since the electron velocity is assumed to be
first order in the perturbation, to zeroth order
scattering makes the radiation isotropic $\delta\negthinspace f \equiv f - f_0 
\rightarrow 0$, where $f_0$ is the isotropic component of the 
distribution function. 

Its primary function then is the 
suppression of anisotropies as seen by the scatterers.
Since isotropic perturbations are not damped,
{\it inhomogeneities} in the distribution persist.  Inhomogeneities
at a distance are seen as anisotropies provided there are no intermediate
scattering events, \ie\ they are on the last scattering surface. 
They are the dominant
source of primary anisotropies (see \S \ref{ch-primary}) and an 
important contributor to secondary anisotropies (see \S \ref{ss-6regeneration}).

Explicitly the suppression term is
\bel{eq:C0}
{\cal C}_0 = \del \opc \fmf.
\ee
Inserting this into equation \eqn{CImplicit} for the integration
over incoming angles and noting
that $\cos\beta = \bg \cdot \bg'$, 
we obtain the contribution
\bel{eq:C0ev}  
C_0 [f] = {d \tau \over dt}[(f_0 - f) + \gamma_i \gamma_j f^{ij}],
\ee
where $f_0$ is the isotropic component of the distribution and the
 $f^{ij}$ are proportional to the quadrupole moments of the 
distribution
\bel{eq:fij}
f^{ij}(t,\bx,p) = {3 \over 4} \int {d \Omega \over 4\pi} (\gamma^i
\gamma^j - {1 \over 3} \delta^{ij}) f.
\ee
The angular dependence of Compton scattering sources a 
quadrupole anisotropy damp more slowly than the higher
moments.\footnote{This can generate viscosity in the photon-baryon
fluid and affects diffusion damping of anisotropies as 
we show in Appendix \ref{ss-5polarization}.}
Even so $C_0$ vanishes only if the distribution is isotropic $f=f_0$.
Furthermore, 
since the zeroth order effect of scattering is to isotropize the distribution,
in most cases any anisotropy is at most first order in the perturbative 
expansion.  This enormously simplifies the form of the other terms.

\subsubsection{(b) Linear and Quadratic Doppler Effect: ${\cal C}_{v_b}$ 
and ${\cal C}_{v_b^2}$} 

Aside from the small electron recoil (see c),
the kinematics of Thomson scattering
require that no energy be transferred in the rest frame of the electron
\ie\ $\tilde p' = \tilde p$.
Nevertheless, the transformation from and back into the background frame
induces a Doppler shift, 
\bel{eq:dopplershift}
{\delta p \over p} = {{1 -  \V_b \cdot \bg'} \over
{1 -  \V_b \cdot \bg}} - 1 =  {\V_b}\cdot (\bg - \bg')
+ (\V_b \cdot \bg)\V_b \cdot (\bg - \bg') + {\cal O}(v_b^3).
\ee
Notice that in addition to the usual linear term, there is also
a term quadratic in $v_b$.  Furthermore, quadratic contributions do not
disappear upon averaging over incoming and outgoing directions.  They
represent a net energy gain and/or loss by the CMB.  

Let us first consider the case that scattering is rapid, 
\eg\ before recombination, such that all
CMB photons scatter before traversing a coherence scale of the velocity 
field.   After averaging over incoming directions the net 
first order contribution is $\delta p/p = \bg \cdot \V_b$. 
As one might expect, 
this is just the Doppler shifted signal we 
expect from radiation that is isotropic in
the electron rest frame.
The spectrum therefore is a blackbody with a dipole signature $\V_\gamma$
in angle: $\delta T/T = \bg \cdot \V_\gamma = \bg \cdot \V_b$.  To
${\cal O}(v_b^2)$, there is a net energy transfer.  Scattering
brings the photons into
kinetic equilibrium with the electrons. 
This equalization amounts
to an energy gain by the photons 
if $v_\gamma < v_b$,  and a loss in the opposite case.  
The energy transfer occurs only until kinetic equilibrium
is attained.  In other words,
once the photons are isotropic in the electron rest frame $\V_\gamma 
= \V_b$, scattering has no further effect.  

On the other hand, if the mean free path of the photons due to Compton
scattering is much greater than the typical coherence scale of the 
velocity, the photons are in the diffusion limit.  This can occur in
reionized scenarios.
Scattering is not rapid enough to ever make the distribution isotropic
in the local rest frame of the electrons.  
Say some fraction $d\tau = n_e \sigma_T dt$
of the CMB scatters within a coherence scale.  Then the Doppler shift will be
reduced to $\bg \cdot \V_b d\tau$ and the energy transfer will be of
order ${\cal O}(v_b^2 d\tau)$.  As the photons continue to scatter, the
first order Doppler term vanishes since redshifts and blueshifts from
regions with different orientations of the electron velocity will mainly
cancel (see \S \ref{ss-6cancellation}).  
The second order term will however be positive definite: 
${\cal O}(\int v_b^2 d\tau)$. 

Is the resultant spectrum also a blackbody?
In averaging over angles and space above,
we have really superimposed many Doppler shifts
for individual scattering events.
Therefore the resulting
spectrum is a superposition of blackbodies with a range of temperatures
$\Delta T /T = {\cal O}(v_b)$.  Zel'dovich, Illarionov, \& Sunyaev 
\cite{ZIS}
have shown that this sort of superposition leads to spectral distortions
of the \Cy\ type with $y = {\cal O}(v_b^2)$ (see \S \ref{ss-3comptonization}).

Now let us write down the explicit form of these effects.
The linear term is given by 
\beal{eq:vterm}
{\cal C}_{v_b} \eal \Bigg\{ \ddel \opc \V_b \cdot \pmp \nonumber \\
& & - \del 2\bomb {\left[
{\pv \over p} + {\ppv \over p'} \right]} \Bigg\} \fmf. 
\eea
Assuming that the anisotropy is at most first order in the perturbation
$\delta\negthinspace f \equiv f-f_0 \simlt {\cal O}(v_b)$, 
the contribution to the collision term can be explicitly evaluated as
\bel{eq:vev}
C_{v_b} [f] = -{d \tau \over dt} \left[
 (\bg \cdot \V_b)p{\partial f_0  \over \partial p}
- {\cal O}(\delta\negthinspace f \, v_b) \right].
\ee
The ${\cal O}(\delta\negthinspace f\, v_b)$ 
term is not necessarily small compared with
other second order terms.  However, we already know its effect.  If 
scattering is sufficiently rapid, the anisotropy $\delta\negthinspace f$
will be a 
dipole corresponding to the electron velocity $\V_b$.  In this case,
its effects will cancel the ${\cal O}(v_b^2)$ quadratic term.
Notice that to first order equilibrium will be reached between the zeroth
and first order terms when
\bel{eq:equilibrium} f_0 - f
-p (\bg \cdot \V_b) 
{\partial f \over \partial p} 
= {\cal O}(v_b^2), 
\ee
assuming negligible quadrupole.  
For a blackbody, $T(\partial f / \partial T)
= - p(\partial f / \partial p)$.  Thus the equilibrium configuration
represents a temperature shift
$\delta T/T =  \bg \cdot
\V_b$.  This formally shows that the ${\cal O}(v_b)$ term makes
the photons isotropic in the baryon rest frame.

The quadratic term, given explicitly by
\beal{eq:vvterm}
{\cal C}_{v_b^2} 
\eal {1 \over 2} \dddel \opc [\V_b \cdot \pmp]^2 \fmf \nonumber \\
& & - \ddel 2\bomb \left[{\pv \over p}+ {\ppv \over p'}\right]
\V_b \cdot \pmp \fmf \nonumber \\
& &+ \del \Bigg\{ -(1-2\cosb+3\cos^2\beta)v_b^2
+ 2\bomb {(\pv)(\ppv) \over {p p'}}  \qquad \nonumber\\
& & +
(1-\cosb)(1-3\cosb)
\left[ {\pv \over p} + {\ppv \over p'} \right]^2
\Bigg\} \fmf,
\eea
can also be evaluated under the assumption of small anisotropy,
\beal{eq:vvev}
C_{v_b^2} [f] = {d \tau \over dt} \left\{ 
  \left[ (\bg \cdot \V_b)^2 + v_b^2 \right]
   p{\partial f \over \partial p} + 
  \left[ {11 \over 20} (\bg \cdot \V_b)^2 + 
   {3 \over 20} v_b^2 \right] 
   p^2 {\partial^2 f \over \partial p^2} 
		    \right\}.
\eea

\subsubsection{(c) Thermal Doppler Effect and Recoil: $C_{T_e/m_e}$ and $C_{p/m_e}$ 
}

Of course, we have artificially separated out the bulk and thermal components
of the electron velocity.  The thermal velocity leads to a quadratic
Doppler effect exactly as described above if we
make the replacement $\langle v^2_b \rangle
\rightarrow \langle v^2_{\rm therm} \rangle = 3T_e/m_e$.
For an isotropic
distribution of photons, this
leads to the familiar Sunyaev-Zel'dovich (SZ) effect \cite{SZ}.  
The SZ effect can therefore be understood as the
second order spectral distortion and energy transfer due to the superposition
of Doppler shifts from individual scattering events off electrons in
thermal motion.
It can
also be naturally interpreted macrophysically:
hot electrons transfer energy to
the photons via Compton scattering.  Since the number of photons is
conserved in the scattering, spectral distortions must result.
Low energy photons are shifted upward in frequency, leading to
the Rayleigh-Jeans depletion and the Wien tail enhancement characteristic
of \Cy\ distortions.  We will consider this process in more detail in 
\S \ref{ss-3comptonization}.

If the photons have energies comparable to the electrons (\ie\ the
electron and photon temperatures are nearly equal), there is also a
significant
correction due to the recoil of the electron.  The
scattering kinematics tell us that
\bel{eq:Recoil}
{\tilde p' \over \tilde p} =
\left[ 1 + {\tilde p \over m_e}(1 - \cos \tilde \beta)
\right]^{-1}.
\ee
Thus to lowest order, the recoil effects are ${\cal O}(p/m_e)$.  Together with
the thermal Doppler effect, these terms form the familiar Kompaneets
equation
in the limit where the radiation is isotropic
and drive the photons
toward kinetic equilibrium as
  a Bose-Einstein distribution of temperature $T_e$ (see \S\ref{ss-3be}).
A blackbody distribution cannot generally
be established since Compton scattering
requires conservation of the photon number.

Explicitly, the recoil term 
\bel{eq:recoil}
{\cal C}_{p/m_e} = -\ddel \opc {{\pmp}^2 \over {2m_e}} \ffff ,
\ee
yields
\bel{eq:recoilev}
C_{T_e/m_e}[f] = {d \tau \over dt} {p \over m_e} \left[ 4f(1+f) +  (1 + 2f)f
{\partial f \over \partial p} \right];
\ee
whereas the thermal term 
\beal{eq:thermal}
{\cal C}_{T_e/m_e} \eal \Bigg\{ \dddel \opc {{\pmp}^2 \over 2}
- \ddel 2\cos\beta \nonumber\\
& & \times (1 - \cos^2\beta) (p-p') \del [4\cos^3 \beta - 9\cos^2\beta -1]
\Bigg\}
 {T_e \over m_e} \fmf, \qquad 
\eea
gives
\bel{eq:thermalev}
C_{T_e/m_e}[f] = {d \tau \over dt} {T_e \over m_e} \left( 4p {\partial f \over \partial p}
+ p^2 {\partial^2 f \over \partial p^2} \right).
\ee
 
\subsubsection{(d) Higher Order Recoil Effects: $C_{v_b p/m_e}$ and 
$C_{(p/m_e)^2}$}

These terms represent the next order in corrections due to the recoil
effect. Explicit forms are provided in \cite{HSSa}.
In almost all cases, they are
entirely negligible.  Specifically, for most cosmological models, the baryon bulk 
flow grows by gravitational instability and is small until relatively 
recently.  On the other hand the photon energy redshifts with the expansion and
is more important early on.  Thus their cross term is never important for
cosmology.    Furthermore, since there
is no cancellation in the $C_{p/m}$ term, $C_{(p/m)^2}$ will
never produce the dominant effect.  
We will hereafter drop these terms in our consideration.

\subsection{Generalized Kompaneets Equation}
\label{ss-2isotropic}

Even for
an initially anisotropic radiation field,
multiple scattering off electrons will have the zeroth order 
effect of erasing the anisotropy.  Therefore when the optical
depth is high, we can approximate the radiation field as nearly isotropic.
Under the assumption of full isotropy, the individual
effects from equations \eqn{vev}, \eqn{vvev}, 
\eqn{recoilev} and \eqn{thermalev} combine to form
the collision term
\beal{eq:CIso}
C[f] \eal {d \tau \over dt} \Bigg\{
 - \bg \cdot \V_b
p \df + \Bigg( \left[ (\bg \cdot \V_b)^2 + v_b^2
\right]
p\df
+ \left[ {3 \over 20} v_b^2  +  {11 \over 20} (\bg \cdot \V_b)^2 
 \right] \qquad
\\
& & \times \  p^2 \ddf\Bigg)
+ {1 \over m_ep^2} {\partial \over \partial p} \left[ p^4 \left\{
T_e \df + f(1+f) \right\} \right]  \Bigg\}. 
\eea
The first and second terms represent the linear
and quadratic Doppler effects respectively.  The final term is the
usual Kompaneets equation.
Notice that in the limit of many scattering
regions, 
we can average over the direction of  the electron velocity.
The first order linear Doppler effect
primarily cancels in this case.
We can then reduce equation (\ref{eq:CIso}) to
\bel{eq:CIsoAvg}
C[f] = {d \tau \over dt}
\Bigg\{{\langle v_b^2 \rangle \over 3} {1 \over p^2}
{\partial \over \partial p} \left[p^4 \df \right] +
 {1 \over m_e p^2} {\partial \over \partial p} \left[ p^4 \left\{
T_e \df + f(1+f) \right\} \right]  \Bigg\}.
\ee
Under the replacement $\langle v^2_{\rm therm}\rangle = 3T_e/m \rightarrow v_b^2$,
the SZ (thermal Doppler) portion of the Kompaneets equation
and quadratic Doppler equation
have the same form.
Thus, spectral distortions due to bulk flow have exactly the same form
as SZ distortions and can be characterized by the \Cy\ parameter
(see \S \ref{ss-3comptonization})
given in its full form by 
\bel{eq:Y}
y = \int {d\tau \over dt} \left[ {1 \over 3} \langle v_b^2(t) \rangle
+ {T_e - T \over m_e} \right] dt  \qquad v_b \gg v_\gamma.
\ee
The appearance of the photon temperature $T$ in equation (\ref{eq:Y})
is due to the recoil terms in the Kompaneets
equation.

The quadratic Doppler effect only contributes
when the electron velocity is much greater than the photon dipole or
bulk velocity.  Just
as the thermal term vanishes when the temperatures are equal, 
the ``kinetic'' part vanishes if the bulk velocities are equal. 
The effect therefore contributes only in the diffusion limit where the photons
can be approximated a weakly anisotropic distribution 
diffusing through independently moving baryons.  
However above redshift $z_{d}
\approx 160 (\Omega_0 h^2)^{1/5}x_e^{-2/5}$ 
(see \S \ref{ss-4comptondrag}), Compton drag on the electrons
keeps the electrons coupled to the photons and requires $v_b \sim v_\gamma$.
For a fully ionized,
{\it COBE} normalized CDM model, integrating \eqn{Y} up until the drag epoch yields
a quadratic Doppler contribution of 
CDM equal to $y(z_{d})
\approx 5 \times 10^{-7}$, almost two orders of magnitude below the current
limits.  
Almost certainly the thermal effect in clusters will completely 
mask this effect.  We will henceforth ignore its contributions when 
discussing spectral distortions.

\subsection{Collisional Brightness Equation}
\label{ss-2blackbody}

We have shown that if the photons and baryons are in 
equilibrium, the effects which create spectral distortions vanish.  
In this case, we may integrate over the spectrum to form the temperature
perturbation.
Combining the collisional zeroth and ${\cal O}(v_b)$ parts,
equations \eqn{C0ev} and \eqn{vev} respectively,
with equation \eqn{collisionlessfinal} for the collisionless part, 
we obtain for the temperature perturbation evolution 
in conformal time
$\Theta(\eta,\bx,\bg)$
\bel{eq:boltzmannfinal}
\dot \Theta + \gamma^i {\partial \over \partial x^i} (\Theta + \Psi)
+ \dot \gamma^i {\partial \over \partial \gamma^i} \Theta + \dot \Phi = 
\dot \tau (\Theta_0 - \Theta - \gamma_i v_b^i + {1 \over 16} \gamma_i
\gamma_j \Pi^{ij}_\gamma),
\ee
where 
\beal{eq:Piij}
\Pi^{ij}_\gamma \eal {4 \over \pi^2 \rho_\gamma} \int p^3 dp f^{ij}(\eta,
\bx) 
\nonumber\\
\eal {1 \over \pi^2 \rho_\gamma} \int p^3 dp \int {d\Omega \over 4\pi} 
(3\gamma^i \gamma^j - \delta^{ij}) f(\eta,\bx,\bg) \nonumber \\
\eal \int {d \Omega \over 4\pi}(3 \gamma^i \gamma^j - \delta^{ij}) 
4\Theta(\eta,\bx,\bg).
\eea
The quantities $\Pi_\gamma^{ij}$ are the quadrupole moments of the
energy distribution.  Since the pressure $p_\gamma = {1 \over 3}\rho_\gamma$,
they are related to the anisotropic
stress.  To generalize this relation to open universes, merely
replace the flat space metric $\delta^{ij}$ with $\gamma^{ij}$.
Equation \eqn{boltzmannfinal} is the fundamental equation for primary anisotropy
formation (see \S\ref{ch-primary}).  
We will revisit second order effects in \S\ref{ch-secondary}
when we discuss reionized
scenarios.

%% file: chap2/geometry.tex
\begin{figure}[t]
%\vphantom{marker} \vskip 0.5truecm
\centerline{ \hskip -1truecm
\epsfxsize=4.0in \epsfbox{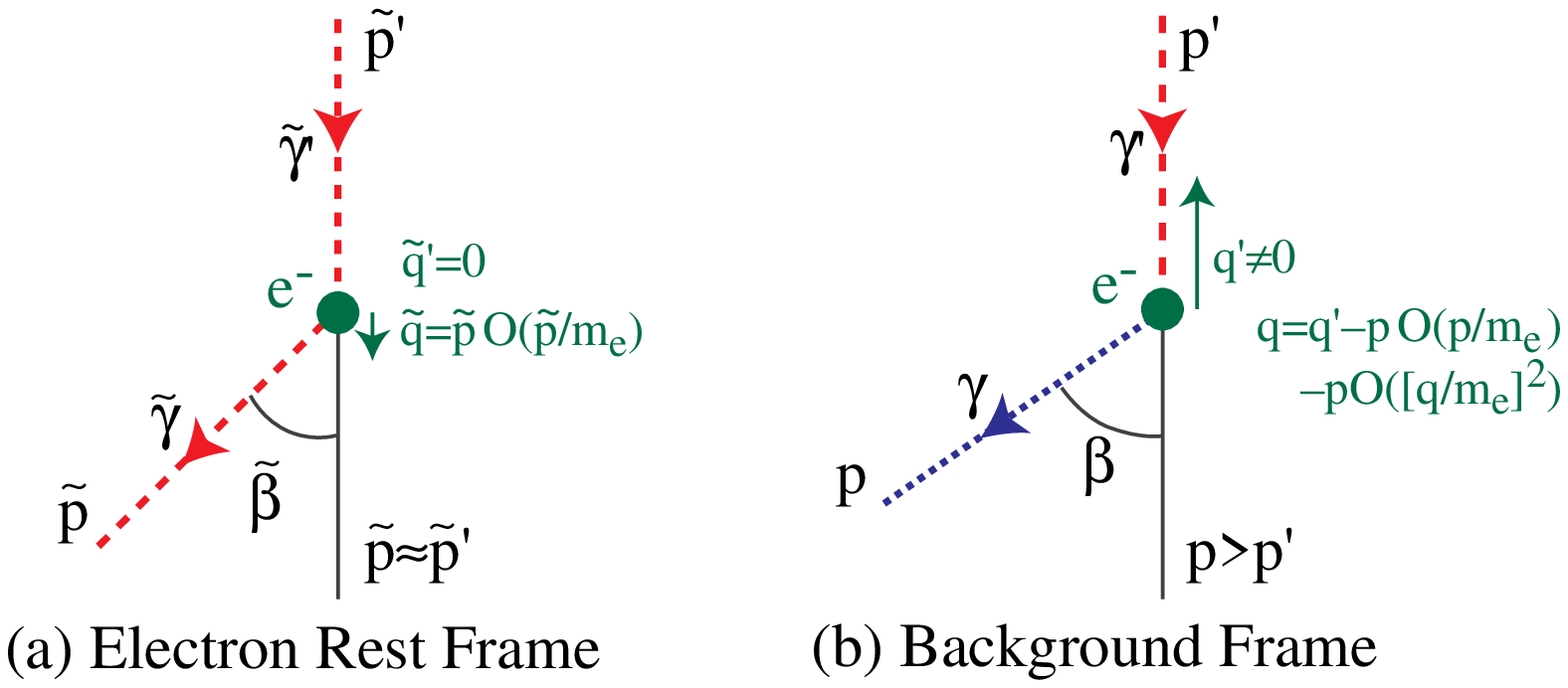}}
\vskip -0.5truecm
\caption{Scattering Geometry}
\mycaption{In the electron rest frame, scattering only transfers energy
to order ${\cal O}(\tilde p/m_e)$ due to the recoil of the electron.  
The Doppler shift into the background frame however
induces a dipole which is aligned 
with the electron velocity.  Dash length represents the photon 
wavelength. Aside from the energy shift 
due to recoil, the quadratic Doppler effect transfers energy
to the photons $\delta p/p = {\cal O}[v_e^2=(q/m_e)^2]$.
The change in scattering angle
is due to relativistic beaming effects.
}
\label{fig:2geometry}
\end{figure}

%% file: chap3.tex
\chapter{Thermalization and Spectral Distortions}
\label{ch-spectral}
\begin{quote}
\footnotesize\it
To be continuously transformed with other things is to be untransformed
once and for all.
\vskip 0.1truecm
\centerline{--Chuang-tzu, 25}
\end{quote}
The CMB exhibits a perfect blackbody form to the precision of current
measurements.  The deviations from the intensity of a blackbody
are no more than 
$3 \times 10^{-4}$ of the peak intensity \cite{Mather}. 
The question arises: how does the 
blackbody spectrum form and how is it maintained?  We have seen in
Chapter \ref{ch-Boltzmann} that spectral distortions occur when the photons
and electrons are not in equilibrium.
Many processes may thus contribute
to spectral distortions.
For example, energy may be dumped into the CMB through
out-of-equilibrium particle decays, dissipation of turbulence 
and acoustic waves in the density fluctuations,  
early phase transition relics such as unstable domain walls or
strings, and any astrophysical process that heat the electrons. 
Moreover, full thermalization of
distortions requires the creation and annihilation of photons. The 
relevant interactions for cosmology, bremsstrahlung (\eg\ \cite{DD77})
and double Compton
scattering \cite{Lightman}, are ineffective below a redshift 
of $z \simlt 10^7$.  Thus spectral distortions are the earliest
{\it direct} observational probe of cosmology.  

There is always the possibility that an experimental determination of 
distortions from a blackbody
spectrum will be confirmed: historically, there have been several false alarms,
and even at present, the low frequency measurements continue to show marginally
significant evidence of distortion. To understand the
implications of the presence or absence of spectral distortions, 
we undertake here a thorough analytic and numerical 
study \cite{HS} of thermalization processes
in the early universe.
 
\section{Collision Equations}
\label{sec-3collision}

\subsection{Compton Scattering Revisited}
\label{ss-3compton}

By far, the dominant interaction that thermally couples 
photons and electrons
before recombination is \SC.
Assuming homogeneity and isotropy, we can reduce the collisionless
Boltzmann equation \eqn{normal} and the Compton collision term
\eqn{CIso} to
\bel{eq:boltzmannagain}
{\partial f \over \partial t} - {\partial f \over \partial p}{da \over dt}
{1 \over a} =  {1 \over m p^2}\nsc {\partial \over \partial p}
\left[ p^4 \left( T_e {\partial f \over \partial p} + f(1+f) \right)\right],
\ee
where recall that $d\tau/dt = x_e n_e \sigma_T$.
For convenience, we can transform variables into a dimensionless energy
$\xe = p/T_e$, not to be confused with $x_e$ the ionization fraction.
The Boltzmann equation then becomes
\bel{eq:Ka}
\lp\dndt\rp_{K} = \lp \nsc \ktmc\rp {1 \over \xe^2} {\part  ;{{\xe}} }
\left[ \xe^4 \lp {\part f;{{\xe}} } + f + f^2 \rp \right]
+ \xe {\part f;{{\xe}} } {\part  ;t } \lp \ln
{T_e \over {T_0 (1+z)}} \rp.
\ee
As we shall see, early on the electron temperature is tightly coupled to the
photon 
temperature and thus scales with the expansion as $T_e \propto (1+z)$.  In 
the late universe, the expansion time is long enough so that during the
scattering
by say, hot electrons in clusters, the expansion may be ignored. 
Hence the last term is usually negligible.  Dropping this term,
we obtain the standard form of the Kompaneets equation.

\USC\ cannot change the number
of photons, but can only redistribute them in frequency.
This may be directly verified by integrating
the Kompaneets equation~(\ref{eq:Ka}) to form the change in the total number
density $n_{\gamma}$:
\bel{eq:Num}
\lp{d a^3 n_{\gamma} \over dt}\rp_K
\propto \int d{\xe} {\xe}^2 \lp \dndt \rp_K = 0.
\ee
The energy density evolution can likewise be obtained from
integration of equation \eqn{boltzmannagain} over frequency
\bel{eq:energyevol}
{1 \over a^4 \rho_\gamma}{\partial a^4\rho_\gamma  \over \partial t}
= 4 \nsc {1 \over m_e} 
\left( T_e - {1 \over 4\rho_\gamma \pi^2}
\int_0^\infty p^4 f(1+f)dp \right),
\ee
where the first and second terms on the right represent the
energy transfer from the thermal Doppler 
and recoil effects respectively.  

\subsection{Electron Temperature Evolution}
\label{ss-3electrontemp}

The electron distribution is correspondingly coupled to the photons by
Compton scattering.  Since Coulomb interactions with the baryons are
extremely rapid, the distribution is to good approximation Maxwellian
at all times and has the same temperature as the baryons.  We can 
determine the evolution of the electron temperature by considering 
the first law of thermodynamics for the photon-electron-baryon system
\bel{eq:firstlaw} 
d(\rho a^3) + pda^3 = dQ,
\ee
where $dQ$ is a source external to the system.  With $p_\gamma = {1 \over
3}\rho_\gamma$, $\rho_e = m_e 
+ {3 \over 2} n_e T_e$, $p_e = n_e T_e$ and similarly for the
hydrogen and helium nuclei, this reduces to
\bel{eq:firstlawexpanded}
a^3 d\rho_\gamma + {4 \over 3} \rho_\gamma da^3 + {3 \over 2}a^3 (x_e n_e+n_H + n_{He}) dT_e
+ (x_e n_e + n_H + n_{He}) T_e da^3 = dQ.
\ee
where $n_H$ and $n_{He}$ are the total number density in ionized and
neutral hydrogen and helium.
If $Y_p$ is the primordial helium mass fraction, then
\beal{eq:heliumfractions}
n_e \eal (1-Y_p/2)n_b, \nonumber\\
n_H \eal (1-Y_p)n_b, \nonumber\\
n_{He} \eal (Y_p/4)n_b. 
\eea
Thus,
with equation \eqn{energyevol}, 
the evolution equation for the electron temperature becomes
\beal{eq:electrontempev}
{d T_e \over dt} \eal  {1 \over 3n_b} [(1+x_e)/2 - (3+2x_e)Y_p/2]^{-1} 
\left( {q \over a^3 } - {1 \over a^4 }{d a^4 \rho_\gamma \over dt} \right)
- 2{da \over dt}{1 \over a} T_e \nonumber\\
\eal
{q \over 3 a^3 n_b }[(1+x_e)/2-(3+2x_e)Y_p/2]^{-1}  
- 2{da \over dt}{1 \over a} T_e  \nonumber\\ 
&& - {1 \over t_{e \gamma}} \left(
T_e - {1 \over 4\rho_\gamma \pi^2}\int_0^\infty p^4 f(1+f)dp \right),
\eea
where the rate of energy injection per comoving volume $q = a^{-3} dQ/dt$, and
\bel{eq:Teg}
t_{e\gamma} = {3  \over 4} {m_e \over \sigma_T \rho_\gamma } f_{cool},
%=
%7.66 \times 10^{19} \To^{-4} {1-Y_p/4 \over 1-Y_p/2} (1+z)^{-4} \sec.
\ee
with
\bel{eq:fx}
f_{cool} = [(1+x_e)/2 - (3+2x_e)Y_p/8] (1-Y_p/2)^{-1} x_e^{-1},
\ee
which has the limiting forms
\bel{eq:fxreduced}
x_e(1-Y_p/2) f_{cool} = \cases { (1-5Y_p/8) & $x_e=1$ \cr
				{1 \over 2}(1-3Y_p/4).& $x_e \ll 1$ \cr}
\ee
Thus the electron temperature is determined by a balance of adiabatic
cooling from the expansion, heating from external
sources $q$, and Compton cooling from the CMB.  In the early universe,
the latter wins as we shall see. 

\subsection{Bremsstrahlung and Double Compton Scattering}
\label{ss-3br+dc}
    
For cosmology, 
the most effective photon number changing processes are
bremsstrahlung, $e^- + X \rightarrow e^- + X + \gamma$ (where
$X$ is an ion), and inelastic, henceforth referred to as
\DC\ $e^- + \gamma \rightarrow e^- + \gamma +
\gamma$. 
The kinetic equation for bremsstrahlung takes the form \cite{Lightman}:
\bel{eq:BRa}
\lp\dndt\rp_{br} = Q_{br} \nsc {g({\xe}) \over \e^{{\xe}}} {1 \over {\xe}^3}
\left[{1 - (\exe)f} \right],
\ee
where
\bel{eq:Qfact}
Q_{br} = \sqrt{ 2 \over \pi} \lp \ktmc \rp^{-1/2} \alpha T_e^{-3}
{\sum n_i Z_i^2} .
\ee
Here $n_i$ is the number density of ions with atomic number $Z_i$, and
$\alpha$ is the fine structure constant.
For a H + He plasma, $\sum n_i Z_i^2 = [x_H + (x_{He} - x_H)Y_p]n_b \approx
x_e n_b$ if the hydrogen and helium
are similarly ionized.
The Gaunt factor is
given by,
\bel{eq:Gaunt}
g({\xe}) \approx {\cases {
\ln (2.25/{\xe}),  & ${\xe} \le 0.37$, \cr
\pi / \sqrt{3},  & ${\xe} \ge 0.37$.  \cr
}}
\ee
We can re-express this in a particularly suggestive form
\bel{eq:BRb}
\lp\dndt\rp_{br} = t_{br}^{-1} \left[ {1 \over \exe} - f \right],
\ee
where
\bel{eq:BRt}
t_{br} =
3.81 \times 10^{23} 
{ e^{\xe} \over g(\xe)}{\xe^3 \over \exe}
\yf^{-1} (x_e\obh)^{-2} \To^{7/2}
\lp\sc\rp^{7/2} z^{-5/2} \sec,
\ee
where $\To=T_0/2.7$K.
Apparently, this is the time scale on which bremsstrahlung 
can establish a blackbody distribution $f = (e^{\xe} -1)^{-1}$ at frequency
$x_p$.
 
Much of the early work on the 
thermalization problem \cite{SZ,ZS69,CJ,ISa,ISb}
assumed that
brems\-strahlung is the dominant photon-creating process in
the early universe.
As we can see from the scaling of equation \eqn{BRt},
in the low baryon density
universe implied by the nucleosynthesis constraint
$\obh = 0.01-0.02$,
bremsstrahlung
is rather inefficient.  \UDC\ cannot be neglected
under such conditions.  Lightman \cite{Lightman} first
derived the kinetic equation for \DC:
\beal{eq:DCa}
\lp \dndt \rp_{dc} \eal \nsc {4\alpha \over {3\pi}} \lp \ktmc \rp^2
{1 \over {\xe}^3} \left[1-(\exe)f\right] \int d{\xe} {\xe}^4(1+f)f \\
\eal t_{dc}^{-1} {I(t) \over I_P} \left[ {1 \over \exe} -f \right] ,
\eea
where
\bel{eq:DCt}
t_{dc} =
6.96 \times 10^{39} {\xe^3 \over \exe} I_P^{-1}  \yf^{-1} (x_e \obh)^{-1} \To^{-2}
\lp\sc\rp^{-2} z^{-5} \sec,
\ee
and
\bel{eq:DCI}
I(t) = \int d{\xe} {\xe}^4 (1+f)f.
\ee
Note that since $(1+f_P)f_P = -\partial f / \partial x_p$, integration
by parts yields $I(t) = I_P= 4\pi^4/15$ 
for a blackbody of temperature $T_e$.
Equation~(\ref{eq:DCa}) is only strictly valid for ${\xe} < 1$ since
its derivation assumes that the photon produced is
lower in energy than the incoming photon.
\UDC\ is, of course, inefficient at creating photons above
the mean energy of the photons in the spectrum.
However, we
will only be concerned with the effects of \DC\
in the low frequency regime where it is efficient.
Comparing equations (\ref{eq:BRt})  and (\ref{eq:DCt}) for the time-scales,
 we see that
in a low $\obh$ universe and at high redshifts,
\DC\ will dominate over bremsstrahlung.  We will quantify this 
statement in \S \ref{ss-3blackbody}.
 
The full kinetic equation to lowest order now reads
\bel{eq:Full}
\lp\dndt\rp = \lp\dndt\rp_K + \lp\dndt\rp_{dc} + \lp\dndt\rp_{br}.
\ee
%Notice that the equilibrium solution for the full kinetic
%equation includes only the Planck distribution $f = (e^{\xe}-1)^{-1}$, 
%\ie a Bose-Einstein spectrum with $\mu=0$.
Evolution of an arbitrary spectrum under this kinetic equation
must in general be solved numerically.
To do so, we employ a fully
implicit iterative modified Youngs approach \cite{Larson}.

\section{Thermalization Optical Depths and Rates}
\label{sec-3optdepth}

Although the 
Compton scattering time,
\bel{eq:Comptontime}
t_{C} = (\nsc)^{-1} = 4.47 \times 10^{18} (1+z)^{-3}
(1-Y_P/2)^{-1} (x_e \obh)^{-1}
\sec,
\ee
from equation \eqn{boltzmannagain} 
is quite short compared with most other 
time scales, its thermalization abilities are hindered by 
two properties:

\begin{enumerate}

\item There is no energy transfer in the Thompson limit.  Energy
exchange only occurs to ${\cal O}(v_e^2)$, \ie\ 
${\cal O}(T_e/m_e)$.  

\item There is no change in photon number by Compton scattering.

\end{enumerate}

We will first examine 
the effects of energy transfer and define 
an optical depth to Comptonization.  At low Comptonization
optical depth, the effect
of Compton scattering is to transfer any excess thermal energy from the
electrons to the photons.  At higher optical depth, energy exchange can
bring the whole distribution to kinetic equilibrium and create 
a Bose-Einstein distribution.  
Since Compton scattering does not 
change the photon number, a blackbody distribution cannot be attained
unless the optical depth to absorption/emission from bremsstrahlung and
double Compton scattering is high.  We quantify these arguments below. 

\subsection{Comptonization}
\label{ss-3comptonization}

The time scale for energy exchange through Compton scattering
is given by equation \eqn{energyevol} as 
\bel{eq:Kt}
t_K = \left( 4 \nsc {T_e \over m_e} \right)^{-1} =
  2.45 \times 10^{27} \yf^{-1}
(\obh)^{-1} \To^{-1} \lp\sc\rp^{-1} (1+ z)^{-4} \sec.
\ee
Notice that the rate increases with the free electron density and 
temperature.  
Conversely, the time scale associated with changes in electron energy
from Compton scattering
is controlled by the photon density.  From equation \eqn{Teg}, the Compton
cooling rate is
\beal{eq:Tegvalue}
t_{cool} \eal {3  \over 4} {m_e \over \sigma_T \rho_\gamma } {f_{cool}} 
\nonumber\\
\eal
7.66 \times 10^{19} f_{cool} \To^{-4} (1+z)^{-4} \sec,
\eea
where recall $f_{cool}$ was defined in equation
\eqn{fx}.
The difference in the time scales reflects the fact that $n_\gamma \gg 
n_e$ since a given electron scatters more frequently with photons that
a given photon with electrons.  Alternatively,
the heat capacity of the photons is much greater than
that of the electrons.  

There are two other rates associated with the evolution of the electron energy.
The expansion causes adiabatic cooling in the electrons on the Hubble
time scale
\bel{eq:TauExp}
t_{exp} \equiv  H^{-1} = {a \over da/dt} \approx 4.88 \times 10^{19}
{(z + z_{eq}+2)}^{-1/2} \To^{-2} (1+z)^{-3/2} \sec,
\ee
where recall that the redshift of equality 
$z_{eq}=4.20 \sci 4 \oh \To^{-4} (1-f_\nu)$ with $f_\nu$ as the neutrino
fraction $f_\nu = \rho_\nu/(\rho_\nu+\rho_\gamma)$.
The Compton and expansion cooling rates are equal at redshift
\bel{eq:compcoolz}
1+z_{cool} = 
	9.08 \To^{-16/5} 
	(\Omega_0 h^2)^{1/5}
	f_{cool}^{2/5}.
\ee
Thus for an ionized plasma, Compton cooling dominates until late
times. 
However, astrophysical or other processes can continuously
inject energy into the electrons
at some rate $q$ associated perhaps with structure formation.   
There are two limits of Comptonization to consider then: when the energy
injection is strong such that $T_e \gg T$ and when it is weak and the
system is dominated by Compton cooling.

\subsubsection{a. Hot Electrons and Compton-$y$ Distortions}

% put definitions somewhere -----------------------------
%for a fully ionized plasma. Here $\Theta_{2.7} = T_0/2.7K$ and
%the primordial
%mass fraction of helium $Y_p$ enters into the equation since
%$n_e = (1-Y_p/2)n_B$ for a H + He plasma with $n_B$ the
%number density of baryons.  For numerical purposes,
%we will always assume $Y_p = 0.23$.  We define $T = (1+z) 2.7$K
%to represent the effective photon temperature for the
%non-equilibrium photon distribution.  

\input chap3/y.tex

If the electrons are strongly heated,
$T_e/m_e \gg  p/m_e$ at the peak of the spectrum, and we 
can ignore the recoil term in equation \eqn{energyevol}, 
\bel{eq:energyheat}
{1 \over a^4 \rho_\gamma}{\partial a^4\rho_\gamma  \over \partial t}
= t_K^{-1}.
\ee
This suggests that we may define the ``optical depth'' to Comptonization,
as
\bel{eq:ydef}
\tau_K = \int dt/t_K = \int 4 \nsc {T_e \over m_e} dt.
\ee
The fractional energy distortion from Comptonization thus becomes 
$\delta \rho_\gamma /\rho_\gamma = \tau_K$.

With this parameter, the Kompaneets equation itself takes on a simple form
if recoil is neglected,
\bel{eq:Comptonization}
{\partial f \over \partial \tau_K} = {4 \over \xe^2}{\partial 
\over \partial \xe} \left( \xe^4 {\partial f \over \partial \xe} \right),
\ee
which is merely a diffusion equation in energy corresponding to the upscattering
in frequency from the thermal Doppler effect.  
This equation has the
exact solution \cite{ZS69}
\bel{eq:comptonspect}
f(\tau_K,\xe) = {1 \over \sqrt{\pi\tau_K}} \int_0^\infty f(0,w) \exp
\left( - {(\ln \xe - \ln w + 3\tau_K/4)^2 \over \tau_K} \right) {dw \over w}.
\ee
For an initial spectrum $f(0,\xe)$ of a blackbody, small deviations
can alternately be solved iteratively by inserting
$f(0,\xe)$ on the right hand side of \eqn{Comptonization}.  This yields
the characteristic ``Compton-$y$ distortion'' 
\cite{ZS69}
\bel{eq:comptonapprox}
{\delta f \over f} = y {\xe e^{\xe} \over e^{\xe} -1 } \left[ \xe \left(
{e^{\xe} +1 \over e^{\xe} -1} \right) -4 \right] ,
\ee
where here $y = \tau_K/4$.
This approximation breaks down in the Wien limit where fractional 
deviations from a blackbody can be quite large due to exponential suppression
in $f$ (see Fig.~\ref{fig:3y}).
In the $\xe \ll 1$ Rayleigh-Jeans limit, this becomes $\delta f/f 
= (\delta T/T)_{RJ} = -2y = - \tau_K/2$  and reflects the fact that upscattering causes
a photon deficit at low energies.  

The  Comptonization optical depth $\tau_K/4 = y  \sim (T_e/m_e)\tau$  is 
generally smaller than the Compton optical depth $\tau$.  However if the
electrons are sufficiently hot, distortions are measurable.
In clusters of galaxies, $\tau \approx 0.01-0.1$
but $T_e \approx 1-10$ keV yielding a distortion of the type given by
equation \eqn{comptonapprox} with $y \approx 10^{-5} - 10^{-3}$.  
This is the cluster Sunyaev-Zel'dovich effect \cite{SZ}.
Distortions in the upper portion of this range represent 
a significant Rayleigh-Jeans
decrement and  have been detected in several bright X-ray clusters 
\cite{Jones,BH}.  It can be cleanly separated from distortionless 
temperature shifts through the Doppler and gravitational redshift 
effects by its spectral signature.  In particular, note that independent
of the value of $y$, there is a null in the distortion at $x_p \approx 3.8$.
On the other hand, 
no isotropic or average $y$-distortion has yet been detected on the 
sky, $y < 2.5 \times 10^{-5}$ (95\% CL) \cite{Mather}.  This places
serious constraints on the amount of global reheating and ionization
allowable and consequently on some models of structure formation
(see \S \ref{ss-6PIB}). 

\subsubsection{b. Compton Cooled Limit}

Before $z_{cool}$, Compton cooling is so efficient that the
electrons are strongly thermally coupled to the photons.
In this case, the electron and photon temperature never deviates by a 
large amount, and we must retain the recoil terms in
the Kompaneets equation.   If the spectrum is initially blackbody before
some injection of energy, we may employ iterative techniques to 
solve the equation.
A blackbody spectrum of temperature $T$
satisfies $f + f^2 = -(\partial f / \partial p) T$.  Thus for small
deviations from a blackbody, the Kompaneets equation takes the form of
the diffusion equation \eqn{Comptonization} if instead of $y=\tau_K/4$
we employ
\bel{eq:ydef2}
y = \int \nsc {T_e - T \over m_e} dt.
\ee
Thus we see that small deviations from a blackbody due to heating of the
electrons can always be expressed as a Compton-$y$ distortion of the form
\eqn{comptonapprox} below the Wien tail and before the Comptonization
optical depth becomes large.  

\input chap3/ytobe.tex

If $\tau_K \gg 1$, energy exchange brings the distribution into
kinetic equilibrium (see Fig.~\ref{fig:3ytobe}).  
Since Compton scattering conserves
photon number, 
the kinetic equilibrium solution is 
a Bose-Einstein spectrum at the electron temperature,
\bel{eq:BE}
f_{BE} = {1 \over {\e^{{\xe} + \mu}-1}},
\ee
where $\mu$ is the dimensionless chemical potential.
If $T_e \approx T$, then this occurs near
\bel{eq:YtauK}
\tau_K \approx {1 \over 2} {t_{exp} \over t_K} \approx 
{1 \over 2} \left( {z \over z_K } 
\right)^2  \approx 1,
\ee
assuming radiation domination.  Here 
\bel{eq:ZK}
z_{K} \approx 7.09 \times 10^3 \yf^{-1/2} ( x_e \obh)^{-1/2} \To^{1/2}.
\ee
Notice that $z_K$ is
the redshift at which the energy exchange time scale $t_K$ equals the
expansion time scale $t_{exp}$.  Rate comparison thus serves as a
simple and useful rule of thumb for estimation purposes.
In reality, a pure Bose-Einstein distribution will form for
\cite{BDDa}
\bel{eq:critmu}
z \simgt 4\sqrt{2} z_K,
\ee
whereas only if
\bel{eq:crity}
z \simlt z_K/8 
\ee
will the spectrum be adequately described as a Compton-$y$ distortion 
of equation \eqn{comptonapprox}.  In the intermediate regime, the 
distortion appears as the Rayleigh-Jeans decrement of the $y$ distortion
but a less substantial Wien enhancement (see Fig.~\ref{fig:3ytobe}).

After external 
 electron heating stops, the electron temperature rapidly
approaches its equilibrium value
\cite{Peyraud,ZL},
\bel{eq:Te}
T_e = {1 \over 4}  {{\int p^4 f(f+1) dp} \over
{\int p^3 f dp}},
\ee
by Compton cooling off an arbitrary photon distribution.
The total energy density of the photons, except for expansion, henceforth 
does not change as it evolves,
\bel{eq:En}
\lp {d a^4 \rho_\gamma \over dt}\rp_K = 0,
\ee
as we can see from
equation~\eqn{energyevol}. 
It is easy to check that if $f=f_{BE}$, a Bose-Einstein distribution at
temperature $T$,
\bel{eq:BEenergy}
f_{BE}(1+f_{BE}) = -{\partial  f_{BE} \over \partial p} T
\ee
and equation
\eqn{Te} implies $T_e=T$, as one would expect in the
equilibrium state. 

\subsection{Chemical Potential Formation}
\label{ss-3be} 

Let us consider the Bose-Einstein distribution and its formation more 
carefully.  Spectral distortions leave the regime of Comptonization 
when the optical depth to energy transfer $\tau_K \approx 1$ or 
$z \simgt z_K$.  
In the absence of external sources,
Compton scattering does not change the number [equation \eqn{Num}] or 
energy [equation \eqn{En}] density
 of the 
photons during the era when the electrons are thermally coupled. 
Thus any external energy injection 
can be characterized by two quantities: the fractional number density of 
photons $\delta 
n_\gamma/n_\gamma$ and energy density $\delta \rho_\gamma/\rho_\gamma$
involved.  
Moreover, the equilibrium distribution is described by a single number, the
chemical potential $\mu$, and collapses 
this two dimensional parameter
space onto one.  There will therefore be some degeneracy between 
number and energy injection.  Let us quantify this.

The energy in a Bose-Einstein distribution can be expressed
as
\bel{eq:EBE}
\rho_{\gamma BE} = {1 \over \pi^2} \int f_{BE}
 p^3 dp = \rho_{\gamma P}
(T_e) \psi(\mu),
\ee
where
\bel{eq:FMU}
\psi(\mu) \approx \cases {
{6 \over I_3} \exp(-\mu), &$\mu \gg 1,$\cr
1 - 3{I_{2_{\vphantom{a}}} \over I_3^{\vphantom{a}}}\mu, &$\mu \ll 1,$\cr }
\ee
and $\rho_{\gamma P}(T_e)= I_3T_e^4/\pi^2 
= aT_e^4=4\sigma_BT_e^4$, the energy
density of blackbody
radiation, with $\sigma_B = \pi^2 k_B^2/60 \hbar^3 c^2 = \pi^2/60$ 
as the Stefan-Boltzmann
constant.
Similarly, the number density is given by
\bel{eq:NBE}
n_{\gamma BE} = {1 \over \pi^2} \int f_{BE} p^2 dp = n_{\gamma P}
(T_e) \phi(\mu),
\ee
where
\bel{eq:PHIMU}
\phi(\mu) \approx \cases {
{2 \over I_2} \exp(-\mu), &$\quad \mu \gg 1,$ \cr
1 - 2{I_{1_{\vphantom{a}}} \over I_2^{\vphantom{a}}} 
\mu, &$\quad \mu\ll 1 ,$\cr }
\ee
with
$n_{\gamma P}(T) = (I_2/I_3) aT^3 = I_2 T^3/\pi^2$.  
Here the constants $I_n$ are
defined by the Riemann Zeta function as follows:
$I_n = \int_0^\infty dx {x^n \over e^x-1} = n! \zeta (n+1)$, \eg\
$I_1 = \pi^2 / 6 \approx 1.645, I_2 = 2\zeta(3)  \approx 2.404, I_3 = 
\pi^4/15 \approx 6.494$.
 
The number of photons in a Bose-Einstein distribution decreases
with increasing chemical potential.
In particular,
a spectrum with $\mu<0$ has more photons than a blackbody, $\mu=0$;
conversely,
 a spectrum with $\mu>0$ has fewer photons.
Parenthetically, note that with equation \eqn{BEenergy},
we can express the \DC\ integral
[equation~\eqn{DCI}] as
\bel{eq:DCIBE}
I_{BE} = \int d{\xe} {\xe}^4 (1+f_{BE})f_{BE} = 4I_3 \psi(\mu),
\ee
for the case of a Bose-Einstein distribution.
 
Now if we require energy and number conservation, equations
(\ref{eq:EBE}) and (\ref{eq:NBE}) tell us:
\bel{eq:EBEP}
\rho_{\gamma BE} = {I_3 \over \pi^2} T^4_e \psi(\mu) = 
\rho_{\gamma P}(T_i)(1 + \delta\rho_\gamma/
\rho_\gamma) = {I_3 \over \pi^2} T_i^4 (1 + \delta\rho_\gamma/\rho_\gamma),
\ee
and
\bel{eq:NBEP}
n_{\gamma BE} = {I_2 \over \pi^2} {T^3_e} \phi(\mu)
= n_{\gamma P}
(1 + \delta n_\gamma /n_\gamma) = 
{I_2 \over \pi^2} T_i^3 (1 + \delta n_\gamma / n_\gamma),
\ee
where $T_i$ represents the temperature of the
radiation before injection.
For small chemical potentials, we may solve equations
\eqn{EBEP} and \eqn{NBEP} simultaneously to obtain:
\bel{eq:MU}
\mu_{pred}(z_h) \approx {1 \over 2.143}
\left[ 3 {\delta \rho_\gamma\over\rho_\gamma}
- 4 {\delta n_\gamma \over n_\gamma}  \right]
\qquad \mu \ll 1,
\ee 
to first order in the perturbations.  The numerical factor
comes from $8 I_1/I_2 - 9 I_2/I_3$.
This is the chemical potential
established near the epoch of heating $z_h$ after a time $t > t_K$
but {\it before} photon-creating processes have taken effect.

The end state Bose-Einstein spectrum
is independent of the precise form of the injection and is a 
function of the {\it total} number of photons and energy density of the photons injected.  This 
is a very powerful result.  For instance, direct heating of the electrons is 
equivalent to injecting a negligible number of high energy photons.  
Furthermore, an arbitrary distribution of injected photons can be parameterized by
the single quantity $\mu(\ddn,\dde)$ alone.
Given the independence of the evolution to the specifics of the
injection for most cases,
it is convenient to employ
injections which may be represented as
``delta functions'' (\ie\ peaked functions localized in frequency)
located at some frequency $x_h$.  
%Note that for this argument to hold, we
%must have $x_h > x_c$ so that bremsstrahlung and \DC\ can be ignored
%during the initial thermalization.

Let us examine the qualitative behavior of equation~\eqn{MU}.
Injection of energy even in the form of photons
tends to heat the electrons and cause $T_e > T_i$ [see equation \eqn{Te}].
Since the
number of photons in a blackbody is proportional to $T^3$,
this would make the spectrum underpopulated with respect to
the blackbody at $T_e$.  However, this deficit of photons can be
partially or wholly compensated by the number of photons
involved in the injection.
In fact, unlike the case of pure electron heating where
$\delta n_\gamma / n_\gamma =0$,
the chemical potential can become negative if the
energy is injected at a frequency $x_h \simlt 3.6$.
An even more curious
effect happens if energy is injected either  at, or symmetrically about,
this critical value.
In this case, the number of photons and the corresponding energy injected
is just enough so that the
electrons are heated to a temperature at which there are exactly
enough photons to create a blackbody spectrum.
This implies that an arbitrarily
large amount of energy may be injected at this critical frequency
and, given sufficient time for the photons to redistribute,
still leave $\mu = 0$, \ie\ the spectrum will remain a
 perfect blackbody.
This effect will be considered more carefully in \S \ref{ss-3lowbal}.
Presumably, however, any physically realistic
process will inject photons over a
wide range of frequencies and destroy this balance.

In the absence of number changing processes, negative chemical potential
spectra become Bose-Einstein condensates from the downscattering of
excess photons \cite{ISa}.
However, although it may be that \DC\ and bremsstrahlung
are ineffective near the frequency of injection, their effect at
low frequencies
plays a crucial
role in the evolution of the whole  spectrum.  \USC\
will move excess photons downward in frequency only until they
can be absorbed by \DC\ and/or bremsstrahlung.  
We therefore expect stability against condensation if 
$|\mu|$ is less than or
equal to the frequency at which the photon absorbing processes
are effective.  This limits the 
range of accessible negative chemical 
potentials. 
To better quantify these considerations, we must examine the role
of number changing processes in thermalization.  It is to this
subject we now turn.

\subsection{Blackbody Formation}
\label{ss-3blackbody}

Blackbody formation must involve bremsstrahlung and/or double Compton scattering
to create and destroy photons and reduce the chemical potential 
to zero.  
Let us examine the rates
of these processes.
The full kinetic equation (\ref{eq:Full}) 
shows that
at high redshifts, Compton and double Compton scattering will
dominate over bremsstrahlung.  
Thus early on, \DC\
will be responsible for creating/absorbing photons at low
frequencies, while \SC\ will redistribute them in frequency.
The net effect  will be that a blackbody distribution is
efficiently established.
Notice that \eqn{BRt} and \eqn{DCt} imply that
\DC\ and bremsstrah\-lung become increasingly efficient 
as the photon frequency decreases.
Even at low redshifts, bremsstrahlung can return the spectrum
to a blackbody form at low frequencies.  
 
Now let us examine the rates quantitatively.  It is useful to define
an optical
depth to absorption by the double Compton or bremsstrahlung processes.
\beal{eq:Ys}
\tau_{abs} \eal \int_{t_{h}}^t dt' (t_{br}^{-1} + t_{dc}^{-1}) \nonumber\\
 \eapp {1 \over 3}{t_{exp} \over t_{br}} + 2{t_{exp} \over t_{br}}, 
\eea
where the last line assumes radiation domination.  
%the expansion time is given by
%\bel{eq:TauExp}
%t_{exp} \equiv {a \over da/dt} \approx 4.88 \times 10^{19}
%{(z + z_{eq})}^{-1/2} \To^{-2} z^{-3/2} \sec,
%\ee
%with $z_{eq}=2.50 \sci 4 \oh \To^{-4}$
%as the redshift of matter-radiation equality.
For the double Compton process, we also assume the integral 
\eqn{DCI} $I(t) \approx I_P $, as is
appropriate if deviations from a Planck distribution
in the high frequency regime are small. 
Note that if there were no photons to begin with, $I(t)=0$ and
\DC\ does not occur.  This is because there must be an incoming 
photon for the scattering to take place.
\UDC\ itself
cannot create a Planck distribution {\it ex nihilo}. Bremsstrahlung
can since it only needs electrons and ions in the initial state.

Thus above the redshift at which $t_{br} = t_{dc}$, double
Compton should be the dominant
photon-creating process.  This occurs at
\bel{eq:Zdcbr}
z_{dc,br} \approx 8.69 \times 10^{5} ( x_e \obh)^{2/5}
\To^{-11/5} [g({\xe})]^{2/5}, \qquad {\xe} \ll 1,
\ee
which is roughly independent of frequency due to similar scaling of
their rates. 
For estimation purposes, we assume that $T_e \approx T$ here and below.

Ignoring Compton scattering for the moment, 
we can write down the kinetic equation 
as a trivial ordinary differential equation
\bel{eq:NY}
{{\partial f} \over {\partial \tau_{abs}}} = {1 \over \exe} - f,
\ee
where we hold the frequency $\xe$ fixed.  This has the immediate solution
\bel{eq:Yabssolution}
f(\tau_{abs},\xe) = (\exe)^{-1} \left\{ 1 - [1-f(0,\xe)] \exp(-\tau_{abs})
\right\}.
\ee
The initial spectrum $f(0,\xe)$ is exponentially damped with optical
depth leaving a blackbody in its place.  This is natural since
the fraction of photons which have not been affected by absorption 
decreases as $e^{-\tau_{abs}}$. 

When the optical depth to absorption drops below unity, thermalization
becomes inefficient.  
As equation \eqn{Ys} shows, this is approximately
when the absorption time scales $t_{br}$ and $t_{dc}$ equal the
expansion time scale $t_{exp}$. Since the absorption rate is frequency
dependent, the photon absorbing processes are effective below a
frequency
\beal{eq:Xexpbrdc}
x_{exp,br} \eapp 1.1 \times 10^{-2} \yf^{-1/2}
[g(x_{exp,br})]^{1/2} x_e \obh
\To^{-11/4} z^{1/4},\nonumber\\
x_{exp,dc} \eapp 4.3 \times 10^{-10} \yf^{-1/2}
( x_e \obh)^{1/2} \To^{-1}
 z^{3/2},
\eea
where $t_{br}(x_{exp,br})=t_{exp}$ and $t_{dc}(x_{exp,dc})
=t_{exp} $.
Combining the two, we obtain
\bel{eq:Xexp}
x_{exp}^2 = x_{exp,br}^2 + x_{exp,dc}^2,
\ee
as the frequency above which photon creation and absorption are 
ineffective.

Now let us include Compton scattering.  
The time-scale for establishing a Bose-Einstein
distribution via \SC\ $t_K$ is independent of frequency.  
We therefore expect that the number-changing
processes will dominate over \SC\ below the frequency at which the
rates are equal.  For ${\xe} \ll 1$, we may approximate this as:
\beal{eq:Xcs}
x_{c,br} \eapp 8.0 \times 10^1 [g(x_{c,br})]^{1/2} ( x_e \obh)^{1/2} \To^{-9/4}
z^{-3/4}, \nonumber\\
x_{c,dc} \eapp 3.0 \times 10^{-6} \To^{1/2} z^{1/2} ,\qquad  \qquad
\eea
where $t_{br}(x_{c,br}) = t_K(x_{c,br})$ and $t_{dc}(x_{c,dc})
= t_K(x_{c,dc})$.  Note that $g({\xe})$ is only logarithmically dependent
on frequency.  Let us define,
\bel{eq:Xc}
x_c^2 = x_{c,br}^2 + x_{c,dc}^2.
\ee
Above the frequency $x_c$, the spectrum will be Bose-Einstein
given sufficient time to establish equilibrium.
This is true even if number changing processes are effective compared with
the expansion because the created photons are rapidly carried
away by Comptonization to higher frequencies. 
Below this frequency, the spectrum returns to
a Planck distribution if either the bremsstrahlung or double Compton processes
are effective compared with the expansion.  

\input chap3/crit.tex

Figure \ref{fig:3crit} displays these critical frequencies
and redshifts for the
representative choices of
$\obh=0.025$ and $0.0125$.
Notice the transition to \DC\ dominance for $z>z_{dc,br}$
and small deviations from the simple power law approximations
%given by equations~(\ref{eq:Xcs},\ref{eq:Xexp})
for ${\xe} \approx 1$ and $z < z_{eq}$.  Here $\oh=0.25$, but
the total matter content plays only a small role in the thermalization
process, entering only through the expansion rate for $z<z_{eq}$.

\label{sec-3evolution}

\section{Low Frequency Evolution}
\label{sec-3lowev}
 
The quantitative study of thermalization involves the time evolution of
the spectrum.  Let us assume that it is distorted at a reheat redshift
$z_h$ by some non-equilibrium process that injects an arbitrary amount
of energy and/or 
photons into the CMB.  
In this general case, thermalization must be studied numerically. 
However, we shall see that for small distortions, analytic
approximations are accurate and useful in understanding the
thermalization process.

We shall see that thermalization to blackbody
is determined at low frequencies where
photons are most efficiently created and destroyed.  Moreover, the 
low frequency regime carries the largest temperature distortions
and is not yet well constrained by observation (see Fig.~\ref{fig:1firas}
and note that $x_p=1$ is $\nu = 1.9 $cm$^{-1}$).

At last scattering $z_*$, early spectral distortions are frozen in. 
However, Compton energy exchange is already ineffective 
at a higher redshift $z_K$.  
Up to $z=z_K$, Compton scattering moves 
the photons produced at low frequencies up or
excesses at high frequencies down.  It therefore plays a
crucial role in the reduction of the
chemical potential. After
$z < z_K$, the high frequency chemical potential
distortion is effectively frozen in, but the low frequency side 
can continue to evolve under bremsstrahlung.

An analytic approximation 
first employed by Zel'dovich and Sunyaev \cite{ZS69}
and extended by Danese and De Zotti \cite{DD82} to include \DC\ is
quite useful for understanding the evolution. It assumes
that one or more of the three processes
are effective enough to establish quasi-static conditions:
\bel{eq:QEq}
\lp \dndt \rp = \lp \dndt \rp_K + \lp\dndt\rp_{br} +
\lp\dndt\rp_{dc} \approx 0  ,
\ee
\ie\ the rate of change of the spectrum can
be considered slow.  Because of the frequency dependence of 
\DC\ and bremsstrahlung, equation~\eqn{QEq} is valid for the entire spectrum
only when $z \gg z_K$.

\subsection{Chemical Potential Era}
\label{ss-lowearly}

Let us first consider early evolution.  
We may always re-express the spectrum
in terms of a frequency-dependent
``chemical potential,''
\bel{eq:NBEX}
f({\xe}) = {1 \over {\exp[{\xe}+\mup]-1}},
\ee
without loss of generality.
The complete kinetic equation in the quasi-static approximation, \ie\
equation~\eqn{QEq}, then becomes
\beal{eq:KinQEq}
&& {1 \over \xe^2} {d \over d \xe} \left[ \xe^4
{\exp[\xe + \mup] \over \lp\exp[\xe + \mup] -1 \rp^2}
\ {d\mu' \over d \xe }\right]
=\nonumber\\
&& \qquad \qquad  \lp {t_{K}\over t_{br}}
+ {t_{K} \over t_{dc}}{I_{BE}\over I_P} 
\rp {\e^{\xe} \over \exe} {\exp[\mup]-1 \over
\exp[\xe + \mup] -1}. 
\eea
If we make the further approximation that $g({\xe}) \approx g(x_{c,br})$,
we may express this as
\bel{eq:KQEqb}
{1 \over \xe^2} {d \over d {\xe}} \left[ \xe^4
{\exp[{\xe} + \mup] \over \lp\exp[{\xe} + \mup] -1\rp^2} \
{d\mu' \over d {\xe}} \right]
=  {4 x_c^2 }
{\e^{{\xe}} \over \xe^3} {\exp[\mup]-1 \over
\exp[{\xe} + \mup] -1},
\ee
for ${\xe} \ll 1$.  Here we have used the relations $t_K/t_{br} = (x_{c,br}/\xe)^2$ and $t_K/t_{dc} = (x_{c,dc}/\xe)^2$.
 
For $\mup \ll {\xe}$, equation \eqn{KQEqb} has the solution
\bel{eq:MUPa}
\mup = C_1 \exp[-2x_c /{\xe}].
\ee
We have taken the solution corresponding to $\mu'(0) =0$, since
at very low frequency the spectrum is a Planck distribution.
At high frequencies ${\xe} \gg x_c$, we expect that the spectrum will
be Bose-Einstein with chemical potential $\mu$.  Thus if $\mu < x_c$ 
as is relevant for small distortions,
the two solutions must match at the junction, \ie\
$C_1 = \mu$.  

It is convenient to describe these distortions from a blackbody
spectrum as a ratio of the frequency dependent effective temperature
to the temperature of an equilibrium distribution at $T_e$ before
last scattering,  
\bel{eq:Teff}
{T \over T_e} = {{\xe} \over \ln [(1+f)/f]}.
\ee
Notice that a spectrum of the form given by equation \eqn{MUPa}
obtains its peak distortions at
\bel{eq:Xpeak}
x_{peak} = 2x_c,\qquad \mu < x_c(z)
\ee
at a value
\bel{eq:Peak}
\lp\ln {T \over T_e}\rp_{max} = \ln \lp 1 + {C_1 \over x_c \e}
\rp
= \ln \lp 1 + {\mu \over 2x_c \e} \rp.
\ee
\input chap3/evol.tex

Figure \ref{fig:3evol} shows the evolution of a spectrum, with $\mu(z_h) > 0$
($\ddn = 2.5 \times 10^{-3},\ \dde = 5.5 \times 10^{-3}$)
from the heating
epoch $z_h = 6 \times 10^{5}$ to recombination.
In Fig.~\ref{fig:3evol}a, the initial delta function injection is thermalized
by \SC\ and forms a Bose-Einstein distribution at high frequencies
on a time scale comparable to $t_K$.  Figure \ref{fig:3evol}b displays
the further quasi-static evolution of the spectrum and the
gradual freeze-out of the processes for
$z \simlt z_K \approx 5 \sci 4 $.
Notice that significant evolution of the low frequency spectrum occurs
between $z_{*} < z < z_K$, where quasi-static equilibrium cannot
be maintained across the spectrum.
 
It is instructive to consider
the evolution of this spectrum in some detail. Figure \ref{fig:3evol}a 
displays the process of chemical potential formation via \SC.
At the epoch of heating $z_h$, the energy injected 
rapidly heats the electrons by Compton heating.  Initially,
the temperature of the photons is thus lower than $T_e$ across the spectrum.
Therefore,
there is a deficit of photons in comparison with a Planck distribution
at temperature $T_e$.  Scattering
off hot electrons then comptonizes the spectrum, causing 
low frequency photons to gain energy.  
The high frequency deficit is consequently
reduced at the expense of
the low frequency until a Bose-Einstein distribution is attained
at high frequencies.  At this point, the spectrum ceases to evolve 
rapidly and comes into quasi-equilibrium. 
Bremsstrahlung and at the low redshifts considered 
here, to a lesser extent \DC,
supplies photons at low
frequencies.  Thus the low frequency spectrum returns to a blackbody
distribution at ${\xe} \ll x_c(z)$.  The overall spectrum is
described well by equation \eqn{MUPa}.  For example, at $z=4.75 \sci 5 $,
$x_{peak} \approx 6 \sci -3 $ whereas $2x_c = 5.6 \sci -3 $.
The peak value is slightly underestimated by \eqn{Peak} due
to the finite rate of \SC.  The peak amplitude of distortions
is $(\log T/T_0)= 0.184 $ whereas equation~\eqn{Peak} predicts $0.183$.
The chemical potential is accurately predicted
by equation~\eqn{MU}: at $z = 4.75 \sci 5 $ has
a high frequency tail with 
$\mu = 3.05 \sci -3 $ whereas $\mu_{pred} =
3.06 \sci -3 $.  
 
Figure \ref{fig:3evol}b displays the subsequent quasi-static evolution of
the spectrum.
At
$x_c(z) < {\xe} < x_{exp}(z)$,
photons are effectively produced {\it and} can be scattered up
to affect the high frequency spectrum (\ie\ reduce the chemical
potential). Low frequency
photons produced at ${\xe} < x_c(z)$ are absorbed by inverse bremsstrahlung
and inverse \DC\ before they can be scattered up in frequency.
Under the joint action of \SC\ and the
photon-creating processes,
the spectrum evolves under equation~\eqn{MUPa}.  The peak of the
distortion moves to higher frequencies since
photons created by bremsstrahlung and \DC\ reduce the low frequency
distortions.
 Higher frequency distortions are also affected as the newly created photons
are scattered to higher and higher frequencies.  However at these 
low redshifts, there is insufficient time to alter the chemical 
potential significantly.  We will return to consider these effects in
\S \ref{sec-3highevol}.

\subsection{Chemical Potential Freeze Out}
\label{ss-3lowlate} 

Compton upscattering
ceases to be effective when the fractional 
energy shift drops below unity.  Numerical results \cite{BDDa,HS} show that
at $\tau_K =16$ or after
\beal{eq:Comptonz}
z_{\rm freeze} \eal 4\sqrt{2} z_K \nonumber\\
 \eal 4.01 \times 10^4 \yf^{-1/2} ( x_e \obh)^{-1/2} \To^{1/2},
\eea
assuming radiation domination, the
spectrum begins to deviate from Compton quasi-equi\-librium 
equation \eqn{NBEX},
\bel{eq:initialsp}
f(z_{\rm freeze},\xe) = [\exp(\xe + \mu e^{-2x_c(z_{\rm freeze})/\xe})-1]^{-1}.
\ee
 However
number changing processes are still effective at low frequencies 
(see Fig.~\ref{fig:3evol}b) and continue to return the spectrum to blackbody
at higher and higher frequencies.

\input chap3/low.tex

Let us see how to characterize the distribution \cite{DD80,BDDa}.  
The kinetic equation in the absence of Compton upscattering can be described
by the quasistatic condition
\bel{eq:quasinocompton}
{\partial f \over \partial t} \approx 
\lp {\partial f \over \partial t} \rp_{br}
+ \lp {\partial f \over \partial t} \rp_{dc} \approx 0.
\ee
We have already shown  in
equation \eqn{Yabssolution}
that its solution 
given an initial spectrum $f(z_{\rm freeze},\xe)$ is 
\bel{eq:DDformula}
f(z,\xe) = (e^{\xe}-1)^{-1} \{ 1 - [1 - f(z_{\rm freeze},\xe)]
\exp(-\tau_{abs}(z_{abs},\xe))\},
\ee
where $z_{abs}$ is the redshift at which photon creating processes can
act independently of Compton scattering. 
If bremsstrahlung dominates over \DC\ and radiation over matter,
\bel{eq:DDconfusing}
\tau_{abs}(z_{abs},\xe) = 2 {t_{exp}(z_{abs}) \over t_{br}(z_{abs},\xe)} = 1 ,
\ee
but bremsstrahlung only returns the spectrum to a blackbody after
\bel{eq:bropdepth}
{t_K(z_{abs}) \over t_{br}(z_{abs},\xe)} = 1.
\ee
Thus the optical depth reaches unity {\it and} can create a blackbody
only after
\bel{eq:exstepz}
2{t_{exp}(z_{abs}) \over t_K(z_{abs})} = 1.
\ee
Employing equation \eqn{YtauK}, we obtain the absorption redshift for 
equation \eqn{DDformula}
\beal{eq:absorpz}
   z_{abs} \eal \sqrt{2} z_K \\
 \eapp 1.00 \times 10^4 \yf^{-1/2} (x_e \obh)^{-1/2} \To^{1/2},
\eea 
which is of course close to but not exactly equal to $z_K$.
As Fig.~\ref{fig:3low}a shows, the agreement between this approximation
and the numerical results is excellent.
Notice that the final low frequency spectrum is quite sensitive to the baryon 
content $\obh$ since it is bremsstrahlung that returns the spectrum 
to blackbody (see Fig.~\ref{fig:3low}b).

\input chap3/neg.tex

\subsection{Negative Chemical Potentials}
\label{ss-3lowneg}

The simple analysis of energy and number balance of equation \eqn{MU} shows
us that negative chemical potentials are possible if the injection involves
substantial photon number.  Unlike positive chemical potentials however,
at $\xe \le |\mu|$, the spectrum becomes unphysical and 
{\it requires} the presence of photon absorbing processes to insure 
stability.  If the predicted $\mu \simlt x_{exp}(z)$, absorption is
rapid enough to stabilize the spectrum.  If not, down scattering will continue 
until $\mu$ is reduced to this level.  
Let us therefore first consider small negative chemical potentials where
the stability criterion is satisfied. 

Figure \ref{fig:3neg}
displays the time evolution of a  small $\mu<0$ injection
($\ddn = 7.5 {\sci -3 },$ $\dde = 2.7 {\sci -3 },$ $z_h = 4 {\sci 5 }$)
for \standO.   Thermalization progresses in Fig.~\ref{fig:3neg}a as 
excess photons are downscattered until quasi-equilibrium is
established with a $\mu < 0$ high frequency tail.  
In this case, number changing processes are effective at 
the $\xe = |\mu|$ instability and
equation~\eqn{MU} gives a reasonable approximation to
the chemical potential: $\mu=-9.8 \times 10^{-3},\
\mu_{pred}= -1.0 \times 10^{-2}$.
Quasi-static evolution is shown in Fig. \ref{fig:3neg}b.
During this stage, a
small negative chemical potential behaves very much like a small positive
chemical potential and obeys the form given by equation \eqn{MUPa}.
After $z_K$, bremsstrahlung and \DC\ no
longer have to compete with \SC\ and sharply reduce the low frequency
distortions, leaving the high frequency spectrum untouched.
Again, the evolution of the spectrum between $z_K$
and $z_{rec}$  moves the peak of the distortion slightly upward in
frequency.  The analytic prediction of equation \eqn{DDformula} accurately
locates the frequency of the peak distortion but somewhat overestimates
its magnitude due to the instability at $\xe \le |\mu|$.  For larger
negative chemical potentials, this instability leads to rapid evolution as we
shall show in \S \ref{sec-3highevol}.

\input chap3/zero.tex

\subsection{Balanced Injection}
\label{ss-3lowbal}

One exceptional case is worth considering.  When energy and number
balance predicts
$\mu \approx 0$ by equation
\eqn{MU}, a more careful analysis is necessary.
For injection
at the critical frequency, $x_h \approx 3.6$, $\mu$ vanishes to
first order in the perturbations.  
However,
there is a difference between a $\mu \approx 0$ case
in which $\ddn$ and $\dde$
are balanced so as to in effect cancel, and a case in which
$\mu \approx 0$ purely due to the intrinsic
smallness of perturbations.  Given
sufficient time, the two {\it will}
evolve toward the same final spectrum.
However,
the spectrum may not reach equilibrium by recombination since in the
balanced case
we can inject an arbitrarily large amount of energy.  Large
distortions take longer to thermalize even under \SC.  Specifically,
the spectrum does not relax down to the final equilibrium configuration
implied by equation~\eqn{MU} on a time-scale $t_K$.  Instead, another type
of quasi-equilibrium spectrum
is established which in turn relaxes toward
the actual equilibrium at a slower rate.  

At injection, the electrons
are heated as in the case of a positive chemical potential.  Photons
are then scattered up from low frequencies leaving a low frequency
deficit of photons.   However, just as in
the case of the negative chemical
potential, there is also an excess of photons at high frequencies.
In fact, there is exactly the number needed to fill in the deficit
at low frequencies.  A quasi-equilibrium spectrum forms in which the
high frequency spectrum behaves like a Bose-Einstein
distribution with negative chemical potential, whereas the low
frequency spectrum mimics one of a positive chemical potential.   Given
sufficient time, redistribution in frequency will reduce both the
high frequency excess and the low frequency deficit.
However, it is quite possible that the Comptonization process 
will freeze out
before this has occurred.
 
Figure \ref{fig:3zero}a displays an example.
A large injection, $\ddn= 0.16,\ \dde=0.22$,
is introduced at $x_h = 3.7$ and
$z_h = 2.5 {\sci 5 }$ in a universe with $\oh=0.25,\ \obh=0.015$.  The
small shift in the critical frequency is due to the finite width of
our so called ``delta function'' injection and second order effects.
The high frequency spectrum fits well to
$\mu = -2.89 {\sci -3 }$,
whereas the low frequency spectrum behaves as if
$\mu \approx 10^{-2}$ -- 
almost an order of magnitude greater than the actual chemical
potential at high frequencies.
 
If the injection occurs at earlier times, we
expect that distortions will be reduced by the mechanism described
above.  Figure \ref{fig:3zero}b displays the dependence on $z_h$ for the same
initial spectrum described for Fig. \ref{fig:3zero}a.
In order of decreasing distortions,
the curves represent $z_h = 3.0 \sci 5 ,\ 4.0 \sci 5 ,\
5.0 \sci 5 ,\ 6.0 \sci 5 ,\
7.0 \sci 5 $.   The high frequency regions can be fit to a
Bose-Einstein spectrum of $\mu = -1.35 \sci -3  ,\ -3.22 \sci -4  ,\
-8.02  \sci -5  ,\ -2.87  \sci -5 $ and $\mu \approx 0$ respectively.
For a redshift of $z_h = 7.0 {\sci 5 }$,
the spectrum is fully thermalized under \SC, leaving essentially
no distortions from blackbody.
 
Notice also that even these curious spectra
retain the same structure for the peak temperature distortion.
This is because the analysis above for the location of the peak
depends only on the balance
between the number-changing processes and \SC. This balance, in turn,
depends on $\obh$ alone not the details of the positive, negative,
or ``zero'' chemical potential injection.  
Equivalently, a measurement of the peak
frequency yields information on the baryon density $\obh$ of the
universe.

\section{High Frequency Evolution}
\label{sec-3highevol}

In \S \ref{ss-3lowbal}, 
we have seen a special case in which the chemical potential
can evolve purely under \SC. However in the general case, the
chemical potential only evolves if photons can be produced or
absorbed at low frequencies.  Furthermore,
significant evolution of the high frequency spectrum, ${\xe} \gg x_c$, can
only occur at $z > z_K$ since \SC\ must be
effective to redistribute these photons.  

\subsection{Analytic Approximations}
\label{ss-3analyticmu}

The low frequency behavior
governs the rate at which photons may be produced or absorbed and thus
is critical in determining the evolution of the chemical potential.
If there is no energy release after the epoch of heating $z_h$, the
rate of change of the chemical potential can be derived
in a fashion similar
to equation~\eqn{MU} for a static chemical potential.
If we
consider the number and energy density in the spectrum to be dominated
by the high frequency Bose-Einstein form, equations~\eqn{NBE} and \eqn{EBE}
tell us
\beal{eq:NE}
{1 \over a^3 \ngbe}
{d a^3\ngbe  \over dt} \eal 
	{1 \over  n_{\gamma P}}{d n_{\gamma P} \over dT_e}{dT_e \over dt} 
	+ 3 {1 \over a}{da \over dt} + {1 \over \phi} 
	{d \phi \over d\mu}{d\mu \over dt}, \nonumber\\
{1 \over a^4 \rho_{\gamma BE}} 
{d a^4\rho_{\gamma BE} \over dt} \eal 
	{1 \over \rho_{\gamma P}} {d \rho_{\gamma P} \over dT_e} 
	{dT_e \over dt} + 4 {1 \over a}{da \over dt} + 
	{1 \over \psi} {d \psi \over d\mu}{d\mu \over dt} = 0.
\eea
We may solve these two equations simultaneously to obtain:
\bel{eq:MUT}
{d\mu \over dt} = - \lp {4 \over a^3 \ng}{d a^3\ng \over dt} \rp /B(\mu) ,
\ee
where
\bel{eq:Bfact}
B(\mu) = 3 {d \ln \psi(\mu) \over d\mu} - 4 {d \ln \phi(\mu) \over d\mu}.
\ee
Equation \eqn{MUT} was first derived by Sunyaev and Zel'dovich \cite{SZ}.
 
The rate of change of the number density is given by integrating
the kinetic equation~\eqn{Full}:
\beal{eq:dndt}
{1 \over a^3 \ng} {d a^3\ng \over dt} \eal {1 \over I_2\phi(\mu)}
\int d{\xe} {\xe}^2 {\partial f_{BE} \over \partial t} \nonumber\\
\eal
{1  \over I_2 \phi(\mu)} \lp {I_{BE}\over I_P} J_{dc}
+ J_{br}\rp, 
\eea
where $I_{BE}$ is defined in equation~\eqn{DCIBE}
 and
\beal{eq:Source}
J_{dc} \eal\int_{0}^{x_M} d\xe \xe^2 {1 \over t_{dc}}
\left[{1 \over \exe} - f \right], \nonumber\\
J_{br} \eal\int_0^\infty d\xe \xe^2 { 1 \over t_{br}}
\left[{1 \over \exe} - f \right]. 
\eea
We have introduced a cutoff $x_M \approx 1$
 in the integration for the \DC\ source
term since the kinetic equation \eqn{DCa} is not valid for high frequencies.
However, since \DC\ is extremely inefficient at high frequencies, we expect
that the error involved in truncating the integral is negligible.
 
As we can see from equation~\eqn{Source},
the change in the number of photons depends
on the integral of the low frequency spectrum.
From equation \eqn{MUPa}, we employ
\bel{eq:NMUP}
f(t,{\xe}) = { 1 \over \exp[{\xe} + \mu(t)\exp(-2x_{c} / {\xe})] -1},
\ee
which is valid for small chemical potentials, $\mu(t) < x_c$.
In the limit that only \DC\ is effective, we obtain
\bel{eq:DMUTDC}
{d\mu \over dt} = -{\mu \over t_{\mu,dc}(z)},
\ee
by employing equation~\eqn{MUT}.  Here,
\beal{eq:TMUDC}
t_{\mu, dc}(z) \eal {1 \over 2} B I_2 {t_K \over x_{c,dc}} \nonumber\\
\eal 2.09 \times 10^{33} \yf^{-1}
( x_e \obh)^{-1} \To^{-3/2} z^{-9/2} \sec .
\eea
As one might have guessed, the time scale is on order the Compton upscattering
time $t_K$ weighted by the portion of the spectrum where photons can be
created and efficiently upscattered. 
The solution at the present time is
\bel{eq:MUTDC}
\mu(z=0) = \mu(z_h) \exp [-(z_h/z_{\mu,dc})^{5/2}],
\ee
with
\bel{eq:ZMUDC}
z_{\mu, dc} = 4.09 \times 10^5 \yf^{-2/5}
\To^{1/5} ( x_e \obh)^{-2/5}.
\ee
This solution was first obtained by Danese and De Zotti \cite{DD82}.
 
For the case that bremsstrahlung dominates,
 a very similar equation holds:
\bel{eq:DMUTBR}
{d\mu \over dt} = -{\mu \over t_{\mu,br}(z)},
\ee
where
\beal{eq:TMUBR}
t_{\mu, br}(z) \eal {1 \over 2} B I_2 {t_K \over x_{c,br}}, \nonumber\\
\eapp 3.4 \times 10^{25} \yf^{-1}
( x_e \obh)^{-3/2} \To^{5/4} z^{-13/4} \sec,
\eea
and we have approximated $g(x_{c,br}) \approx 5.4$.
These equations yield the solution
\bel{eq:MUTBR}
\mu(z=0) = \mu(z_h) \exp [-(z_h/z_{\mu,br})^{5/4}],
\ee
where
\bel{eq:ZMUBR}
z_{\mu,br} \approx 5.6 \times 10^{4} \yf^{-4/5}
( x_e \obh)^{-6/5} \To^{13/5}.
\ee 
Let us call the smaller of these two redshifts $z_{\mu}$.
The characteristic redshifts for \DC\ and bremsstrahlung are equal
for a universe with
\bel{eq:OBHc}
\obh_{dc,br} \approx 0.084 \yf^{-1/2} x_e^{-1} \To^3.
\ee
For a universe with a higher baryon density, bremsstrahlung should dominate
the evolution of the chemical potential.

\input chap3/mu.tex

\subsection{Numerical Results}
\label{ss-numericalbe}
 
The analytic solutions are only valid in the case $
\mu(z) < x_c(z) \ll 1$, for
all $z$.
In many cases, $\mu(z) <
x_c(z)$ during some but not all epochs of interest $z < z_h$.
Furthermore, a small chemical potential today could have originated from
a large distortion $\mu \simgt 1$ at high redshifts.
Thus
we must examine the behavior numerically and look for deviations from
the forms of equations~\eqn{MUTDC} and \eqn{MUTBR}.
 
Let us now examine the evolution of the chemical potential in a low
$\obh$ universe as implied by nucleosynthesis.
Numerical solutions suggest that equation~\eqn{MUTDC}
 is indeed a good approximation
for sufficiently small chemical potentials.  The bottom curve of
Fig.~\ref{fig:3mu}a shows such
a case (the solid line is the numerical result, the dotted line is the
best fit) for a initial spectrum $\mu(z_h)=3.15 {\sci -3 }$
with \standO.
For comparison, $z_{\mu,pred} = 1.9 \times 10^6$ whereas
$z_{\mu,fit}= 2.0  \times 10 ^6$.
For very low redshifts,
there has been insufficient time to scatter photons upwards in
frequency to establish a perfect Bose-Einstein spectrum.
Thus the effective
chemical potential deviates toward larger distortions that equation~\eqn{MU} 
predicts.
The top curve of Fig.~\ref{fig:3mu}a  
shows an intermediate case: $\mu(z_h)=1.84 \sci -2 $
 for \standO.  We see
that equation~\eqn{MUTDC} still describes the evolution adequately
but not entirely.  The best fit value of the critical redshift has
shifted upwards however, $z_{\mu,fit}=2.2 {\sci 6 }$.  This is because
$\mu > x_c(z)$ for a significant portion
of the evolution and the spectrum evolves more slowly
than the exponential suppression given in equation~\eqn{MUTDC} suggests.  
 
We can see this effect quite clearly for larger chemical potentials.
Figure \ref{fig:3mu}b shows
such an evolution again for \standO. The top curve 
has an initial spectrum with $\mu(z_h) = 4.9 {\sci -1 }$
and
the bottom $\mu(z_h) = 1.6 {\sci -1 }$.
Dashed lines represent the predictions of equation~\eqn{MUTDC}.
For redshifts much less than $z_\mu$, the
chemical potential stays roughly constant, evolving more slowly than
predictions.
However, the fall off at high redshifts is correspondingly
much more
precipitous than equation~\eqn{MUTDC} would imply.
The effective redshift at which
a substantial suppression of the chemical potential occurs
is increased but only by a factor of order unity.  Attempts
to fit the curves to the form of equation~\eqn{MUTDC} yield
$z_{\mu,fit} = 3.5 {\sci 6 },\ 3.0 {\sci 6 }$  for (A) and (B) respectively.
Note that in these cases,
unlike Fig.~\ref{fig:3mu}a, the form of equation~\eqn{MUTDC},
even leaving $z_\mu$ arbitrary,
does not accurately trace the evolution.
In general
then, a large positive chemical potential will exhibit stability
up to a redshift $z \approx z_\mu$ and then fall dramatically.

\input chap3/negmu.tex

For negative chemical potentials, the spectrum can only establish
such a quasi-static equilibrium as required for the analytic form if
$|\mu| < x_{exp}(z_\mu)$.  For larger negative chemical potentials
and $z > z_{\mu}$, inverse \DC\ absorbs excess 
photons and returns the distribution to $|\mu| \approx
x_{exp}(z_\mu)$ nearly instantaneously.  
Thus {\it regardless of initial input of photons}
the evolution for $z > z_{\mu}$ will be approximately the same.
Figure \ref{fig:3negmu} displays this effect for \standO.
Here, we inject successively larger numbers of photons and energies
at the same frequency $x_h = 1$ (see
figure captions for details).
At high redshifts, we see that $\mu(z)$ saturates
at some maximum value regardless of the initial input.
For lower redshifts $z < z_{\mu}$, \DC\ is not sufficiently
efficient and must wait for Compton scattering to bring photons down to 
low enough frequencies to be absorbed.  
As can be seen in Fig.~\ref{fig:3negmu}, 
large negative chemical potentials are
rapidly evolved away under such a process. 
Quasi-equilibrium
is never established and deviations from equation~\eqn{MUTDC} are large.
Small
negative chemical potentials (A) exhibit the same stability as
positive chemical potentials at redshifts $z < z_\mu$.
Note also that this effect is only weakly dependent on $\obh$
(assuming \DC\ dominance):
\bel{eq:Xexpdc}
x_{exp,dc} (z_{\mu}) \approx 0.1 ( x_e \obh)^{-1/10} \To^{-7/10},
\ee
and so the critical chemical potential $\mu_c \sim -x_{exp}(z_{\mu})$
is roughly independent of both energy injection and $\obh$.
Fig. 13 (curves A, B) shows the evolution of the same
initial spectra as Fig. 12 (curves E, F) for $\obh = 0.10$.
Notice that $\mu_c$ is
roughly the same in both cases.
Of course,
we expect the estimate of the numerical constant above to be extremely
crude, since $z_{\mu}$ itself is only an order of magnitude
estimate of the epoch of effectiveness of \DC.  Figures
12 and 13 show that
the actual value
is $\mu_c \approx -0.02$ and is reasonably independent of $\obh$.
Thus, elastic and double Compton scattering conspire to eliminate
negative chemical potentials greater than a few percent.  This
result is approximately independent of the details of injection
given reasonable choices of the cosmological
parameters.
 
In summary, the analytic formulae
equations~\eqn{MUTDC} and \eqn{MUTBR} describe the
evolution adequately (to order of magnitude)
within the range $-10^{-2} \simlt \mu \simlt 1$.
The existence of a small positive chemical potential would place
tight constraints on the energy injection mechanism.  If the
injection took place at $z_K < z < z_\mu$, the energy injected would
have to be correspondingly small.  Only if it took
place in the narrow region, $z_\mu < z < {\rm few} \times z_\mu$, would
a large energy injection and a small chemical potential be consistent.
Any earlier, and an arbitrarily large
injection would be thermalized.  On the other
hand, the existence of a small negative chemical potential is not
{\it a priori}
as restrictive, since a large amount of energy can be injected and
still lead to a small value for $|\mu|$.
However, for an extremely small negative chemical potential, $\mu
\simlt -3.3 \times 10^{-4}$ as required by observation, these considerations
do not apply since we have determined numerically that
the critical chemical potential
for stability is $\mu \approx -10^{-2}$.
Extremely small negative chemical potentials are stable and
equally as restrictive
as small positive chemical potentials.

The {\it non-existence} of
$\mu$-distortions of course would rule out non-standard cosmologies
with energy injection in the range $z_K < z < z_\mu$ but say very
little about the physics for $z > z_\mu$.
The one case that escapes these consideration is  the 
balanced injection scenario.  The chemical potential
is driven to zero not by photon-creating processes but by \SC\ itself
and thus $z_\mu$ is not the critical redshift for this process.
Furthermore, an arbitrary amount of energy can be injected
and still maintain a small chemical potential even
at comparatively low redshifts.
However, even this case is likely to leave a low frequency signature
which is potentially observable (\S \ref{ss-3lowbal}).  
Thus the lack of low
frequency distortions would set tight bounds on all possible injections
in this redshift range. Let us now consider the observational 
status of spectral distortions.

\input chap3/dat.tex

\section{Comparisons and Constraints}

\subsection{Observational Data}

The {\it COBE} FIRAS experiment \cite{Mather} places tight constraints
on the presence of a Bose-Einstein
distortion in the Wien tail, $|\mu| < 3.3 \times 10^{-4}$.
However, as
we have shown in \S \ref{sec-3lowev}, 
the Rayleigh-Jeans regime is also interesting.
It is there that we expect to see the largest
temperature distortions, specifically at the frequency
$x_{peak} \approx 2x_c(z_K)$.  For a positive
chemical potential, the effective temperature of the
Rayleigh-Jeans part of the spectrum is
lower than that of the Wien tail.
 Figure \ref{fig:3dat}
plots the observational results.  
As is immediately
obvious,
the average effective temperature
of the CMB in the Rayleigh-Jeans region
{\it is} apparently lower than that of
the Wien tail.  Note we normalize the distortions
so that ``$T_e$'' is
the temperature of 
the Wien tail which is fixed by FIRAS to be 2.726K \cite{Mather}.
We have also plotted the results of our numerical integration for comparison.
This marginally significant distortion implies a quite
large chemical potential in the Wien tail (dotted line, $\mu = 0.005$)
that is inconsistent with
the FIRAS results.
If we were to require that the
Wien distortions be consistent with FIRAS (solid line,
$\mu = 3.3 \times 10^{-4}$), the predicted distortions in the Rayleigh-Jeans
region are far too small to explain the effect of the systematically low
effective temperature.

There exists one loophole: the case of balanced
injection (see \S \ref{ss-3lowbal}). 
Although, $\mu \rightarrow 0$ given sufficient Comptonization, 
the distortions will typically freeze
in before this occurs. 
Particularly interesting
is the fact that Rayleigh-Jeans distortions can be significant
while Wien distortions remain minimal (see Fig.~\ref{fig:3dat}b).
Note that the distortions on the low frequency side are consistent
with large deviations,
implied by the low effective temperature of the measurements, 
even when high frequency distortions are
consistent with the already restrictive $|\mu| <  3.3 \times 10^{-4}$.
Alternatively, we can say that the injection of a large amount
of energy even for this exceptional case in which high frequency
distortions vanish will lead to significant low frequency
distortions in many cases.

\input chap3/baryon.tex

Low frequency distortions of this type may eventually be confirmed, and it
is therefore interesting to see what information can be gained from them.
As described in \S \ref{sec-3lowev},  their behavior is governed
by the balance between bremsstrahlung
and \SC\ which is in turn sensitive to $\obh$ [see equation~\eqn{Xcs}].
At low frequencies, bremsstrahlung returns the spectrum to a Planck
distribution.
Thus, the critical frequency at which distortions peak is a {\it measure}
of $\obh$.
For illustrative purposes, Fig.~\ref{fig:3baryon} displays
the spectra obtained numerically
 for $\obh=0.0025,\ 0.015,\ 0.050,\ 0.25$ respectively,
for a fixed Bose-Einstein Wien tail with $\mu=3.3 \times 10^{-4}$.
Note that the distortions are
independent of the heating epoch, $z_h$, and the details of injection
as long as the Wien tail is fixed in this manner. 
On the other hand,
the location of the peak distortions is measurably different
for various choices of $\obh$.
Even in the balanced case, the 
dependence of the peak distortion on $\obh$ is
essentially unchanged.
Thus improved measurements in the Rayleigh-Jeans regime
are desirable for a twofold purpose.
If distortions are seen, they will give an interesting constraint
on $\obh$ in all possible cases.  If they are not seen,
it will close the last loophole in the regime $z_{rec} < z < z_\mu$
for significant injection of energy.  Let us now consider two specific
examples of energy injection constraints implied by the FIRAS measurement.

\input chap3/decay.tex

\subsection{Constraints on Decaying Particles}

If the energy injection arises from the
decay of a massive particle, we may translate the constraint on
$\mu$ into one on
the mass $m_X$, lifetime $t_X$ and branching ratio $f_X$ for decay to photons
of such a species \cite{SC,Ellis,HSl}.  
For this case, the number density of photons
injected is negligible compared with that in the background.  Therefore,
the spectral distortions are determined by the integral of the
fractional contributions to the CMB energy 
per comoving volume during the
decay.
Assuming that the comoving number density
of species
$X$ decays exponentially in time with lifetime $t_X$, we obtain
\bel{eq:XEnergy}
{\delta \epsilon_\gamma \over \epsilon_\gamma}
={m_X \over T(t_{e\! f\! f\!})}  \left( {n_X \over n_\gamma} \right) f_X ,
\ee
where $T(t)$ is the CMB temperature and $(n_X / n_\gamma)$ is the
ratio of the number densities before decay.  The functional form of
equation~\eqn{XEnergy} is identical to the case in which all particles
decayed at a time $t_{e\!f\!f} = [\Gamma(1+\beta)]^{1/\beta} t_X$
for a time
temperature relation of $T \propto t^{-\beta}$.  Here $\Gamma$ is
the usual gamma function.
 
Let us first consider
the case of a low $\obh$ universe as implied by nucleosynthesis
where \DC\ dominates the thermalization process.  For
small energy injection, the analytic considerations of 
\S \ref{ss-3analyticmu} yield
\beal{eq:CDC}
\mu_0 \eapp 4.0 \sci 2
\lp t_X \over \sec \rp^{1/2} \exp \left[-\lp t_{\mu,dc} /
t_X \rp^{5/4} \right] 
\lp m_X \over 1 {\rm GeV} \rp f_X n_X / n_\gamma  \nonumber\\
& &< 3.3 \times 10^{-4}, 
\eea
where
\bel{eq:taudc}
t_{\mu,dc} = 1.46 \sci 8  \To^{-12/5} (x_e \obh)^{4/5} (1-Y_p/2)^{4/5} \sec.
\ee
We have assumed here that we
are in the radiation-dominated epoch where $T \propto t^{-1/2}$.
 
If $\obh \simgt 0.1$,
brems\-strahlung dominates and this constraint becomes
\beal{eq:CBR}
\mu_0 \eal 4.0 \sci 2   \lp { t_X \over
\sec} \rp^{1/2}  \exp \left[ -\lp t_{\mu,br} /
t_X \rp^{5/8} \right]
\lp m_X \over 1 {\rm GeV} \rp f_X n_X / n_\gamma \nonumber\\
\elt 3.3 \times 10^{-4} 
\eea
where
\bel{eq:taubr}
t_{\mu,br} \approx 7.7 \sci 9
\To^{-36/5} (x_e \obh)^{12/5} (1-Y_p/2)^{8/5} \sec.
\ee
The {\it weaker} of the two constraints,
equations~\eqn{CDC} and \eqn{CBR}, is
the relevant one to consider for intermediate cases.
 
Since the analytic formulae are only valid
for small injections of energy $\delta \rho_\gamma / \rho_\gamma \ll 1$,
we 
expect deviations from these predictions when
particles decay near the thermalization epoch.
Large distortions are thermalized less rapidly than
the analytic approximations above would imply.
Figure \ref{fig:3decay} displays the results of numerical integration for
(a) $\obh=0.015$ and (b) $\obh=0.25$.
In both cases,
particles with a short lifetime that decay during the critical epoch
for thermalization
are more stringently constrained than analytic predictions, also
plotted, would
suggest. For late decays, \SC\
can no longer establish
a Bose-Einstein spectrum.  Instead, the spectrum can be described by
the Compton-$y$ parameter which is related to the energy release by
$\dde = 4y$.
We also plot the constraints implied by the most
current value of $y < 2.5 \sci {-5} $ \cite{Mather}.

\subsection{Dissipation of Acoustic Waves}

Energy injection into the CMB occurs even in standard models for structure
formation through the dissipation of acoustic waves by photon diffusion
(see \S \ref{ss-4dampedacoustic}).  The energy stored in the perturbations
of the spatial distribution of the photons is transferred to distortions
in the spectrum.  The lack of observable spectral distortions can be used
to limit the amount of power in acoustic waves before dissipation.
By comparing this with the amount of power measured at large scales
by the {\it COBE} DMR experiment, we can constrain the slope of the
primordial power spectrum \cite{SZb,EvilOne,Wright,HSSb}.

By employing the relation between energy injection and chemical 
potential distortions equation \eqn{MU},
we can generalize equation \eqn{DMUTDC} for the evolution of
chemical potential distortions to the case where energy is being 
continuously injected into the CMB,
\bel{eq:DMUTDCgen}
{d\mu \over dt} \approx -{\mu \over t_{\mu,dc}} + 1.4{Q \over \rho_\gamma},
\ee
where $Q/\rho_\gamma$ is the rate of fractional energy injection.
This equation can immediately be solved as
\bel{eq:Solution}
\mu \approx 1.4 \int_0^{t(z_{\rm freeze})} dt {Q(t) \over \rho_{\gamma}}
\exp[-(z/z_{\mu,dc})^{5/2}],
\ee
where $z_{\mu,dc}$ is given in \eqn{ZMUDC} and $t(z_{\rm freeze})$ is the time of 
Bose-Einstein freeze out when energy injection
can no longer be thermalized [see equation \eqn{Comptonz}]. 

The average energy density in a plane acoustic wave in the photon-baryon 
fluid is given by
$\rho_s \approx \rho_{\gamma b} c_s^2 \langle \Delta^2_{\gamma b} \rangle,$
where $\rho_{\gamma b} = \rho_{\gamma}+\rho_{b}$ and $\Delta_{
\gamma b} $ are the density
and density perturbation in the photon-baryon fluid, and the brackets
denote an average over an oscillation of the acoustic wave.
Since $\mu$ distortions arise at $z > z_{\rm freeze} > z_{eq} $,
we can take the
radiation--dominated limit, where the sound speed is  
$c^2_s = 1/3$, and
\bel{eq:SoundEn}
\langle \delta^2_{\gamma b} \rangle \approx
\langle |\Delta_\gamma(t,k)|^2 \rangle  = {1 \over 2} |\Delta_\gamma(\eta,k)|^2.
\ee
Therefore, the rate of fractional energy injection
\bel{eq:SoundSolution}
{Q(t) \over \rho_{\gamma}} 
= - \sum_k {1 \over 3} {d \langle |\Delta_\gamma(k,t)|^2
\rangle \over dt}.
\ee
The energy density perturbation in the photons $\Delta_\gamma$
in the acoustic phase
is discussed in \S \ref{ss-4dampedacoustic} and found to be related
to the initial potential perturbation $\Phi(0,k)$ by
\bel{eq:deltagammaacoustic}
\Delta_\gamma(t,k) = 6\Phi(0,k)\exp[-(k/k_D)^2],
\ee
for adiabatic perturbations,
where the diffusion scale is 
$k_D(z) = 2.34 \times 10^{-5} \Theta_{2.7} (1-Y_p/2)^{1/2}
(\Omega_B h^2)^{1/2} z^{3/2}$ Mpc$^{-1}$.

To perform the sum over $k$ modes, we must make an assumption about the
form of the initial power spectrum.
The simplest and most often employed assumption is a pure power law
$k^3 |\Phi(0,k)|^2 = B k^{n-1}$, where $n=1$ is the scale-invariant
Harrison-Zel'dovich spectrum.   
Inserting these expressions into equation \eqn{Solution}, 
both the sum over $k$ and the integral over time can be performed
analytically for leading to,
\bel{eq:Final}
\mu = 1.4 F(n) {V_x \over 2\pi^2} 36 k_D^3 |\Phi(0,k_D)|^2 \bigg|_{z=z_{\mu}} ,
\ee
where
\bel{eq:Constant}
F(n) = {1 \over 10} \Gamma[(n+1)/2]\,\Gamma[3(n-1)/5,(z_{\rm freeze}/z_\mu)^{5/2}].
\ee
with $\Gamma(m,x)$ as the incomplete gamma function.
If $n$ is significantly greater than unity,
the incomplete gamma function
$\Gamma(m,x) \rightarrow \Gamma(m)$
since $z_{\rm freeze}/z_\mu \ll 1$
and $F(n)$ is roughly of order unity.
 
It is easy to interpret this result.  If $n>1$, the smallest waves
carry the most energy, and the distortion comes almost entirely
from the waves that damped at the thermalization epoch.  Prior
to thermalization,
no distortion survives due
to the rapidity of the double Compton process.
On the other hand if $n<1$, the fractional energy injection from dissipation 
will be a
maximum at the latest relevant time, \ie\ recombination.
This implies that the constraint from spectral distortions
will come from the upper limit on Compton-$y$ distortions.
However if the spectrum is normalized at large scales, for $n<1$ the
power decreases at small scales leaving no useful constraint.

Let us see how perturbations on the damping scale are related to 
the large scale temperature fluctuations seen by the {\it COBE} satellite.
In 
adiabatic models, these arise mainly from the Sachs-Wolfe effect,
\bel{eq:SWform}
C_\ell 
\approx { 9 \over 200 \sqrt{\pi}} BV \eta_0^{1-n}
        { \Gamma[(3-n)/2]\Gamma[\ell+(n-1)/2] \over
          \Gamma[(4-n)/2]\Gamma[\ell+(5-n)/2] }
\ee
from equation \eqn{SWformula}, where the observed rms anisotropy 
is
\bel{eqn:DTOT}
\left( { \Delta T \over T} \right)^2_{rms} = {1 \over 4\pi}
\sum_{\ell=2}^\infty (2\ell+1)W_\ell C_\ell ,
\ee
and is measured to be $(\Delta T /T)_{rms} = 1.12 \pm 0.10 \times 10^{-5}$
\cite{Bennett94}.
The COBE window function is approximately
$W_\ell = \exp[-\ell(\ell+1) \sigma^2]$,
with $\sigma = 0.0742$ being the gaussian width
of the $10^\circ$ FWHM beam.  This relation sets the normalization $B$ for
the initial conditions $k^3|\Phi(0,k) |^2
=Bk^{n-1}$ as a function of the spectral index $n$.  Substitution
back into 
equation \eqn{Final} yields the amplitude of the chemical potential 
distortion. 

Note that the dependence on the cosmological parameters $\Omega_0$,
$\Omega_b$, $h$
is quite weak:
approximately $\mu \propto (\Omega_b^{1/10} h_{\vphantom{b}}^{6/5})^
{1-n}_{\vphantom{b}}
\Omega_0^{(2-n)/2}$.
Hence for $n \approx 1$, $\mu$ is  completely
independent of $h$ and $\Omega_b$.  Moreover, $\mu$ is nearly
independent of $\Omega_b$ for {\it all} $n$, since
raising $\Omega_b$ makes both the damping length shorter and
the thermalization redshift smaller.

\input chap3/diffuse.tex

It is also useful to provide an approximate inversion of
equation \eqn{Final}:
\bel{eq:N}
n \approx 1 + {\ln [C_1 \Omega_0^{-0.46} \mu / (\Delta T/T)^2_{10^\circ}]
        \over \ln[ C_2 (\Omega_b h^2)^{-1/10} (\Omega_0 h^2)^{-1/2}
        I(\Omega_0)]}
\ee
where we find the constants $C_1 = 5.6 \times 10^{-3}$ and $C_2 =
8.9 \times 10^{5}$ and the small logarithmic correction
$I(\Omega_0) \approx 1 - 0.085 \ln \Omega_0$.
One can verify that this is an excellent approximation within
the range $1.0 < n < 2.0$ and the allowable cosmological
parameters.
Note that the dependence of $n-1$ on $\mu$ and the normalization
is only logarithmic, and its dependence on the cosmological
parameters is almost entirely negligible.
Even
relatively large changes in $\mu$ or the normalization
 will not greatly affect the
constraint on $n$. 

The best fit value of $\mu$ to the spectral data from the FIRAS
experiment is $\mu = -1.2 \pm 1.1 \times 10^{-4}$ (68\% CL)
\cite{Mather}. 
Naively speaking, this provides an
upper limit on {\it positive} $\mu < 0.6 \times 10^{-4}$ (95\% CL).
However since $\mu \ge 0$ for damping distortions, a more
conservative bound is obtained by renormalizing the quoted
probability distribution, assumed to be Gaussian, under
the condition that $\mu$ is positive.  This is clearly the most
reasonable approach if $\mu < 0$ were unphysical, which is not
necessarily the case.  Nonetheless, since this
method provides a {\it conservative} limit, we employ it for the
main result of our analysis.
Taking into account the COBE DMR measurement errors and adopting
a $4 \mu$K
cosmic variance, $(\Delta T / T)_{rms}
(10^\circ)=1.12 \pm 0.18 \times 10^{-5}$ \cite{Bennett94},
we find
\bel{eq:Mulimit}
{\mu \over (\Delta T /T)^2_{10^\circ}} < 1.4 \times 10^4 \quad
(95\% {\rm CL}).
\ee
This would be equivalent to an upper limit of $\mu < 1.76
\times 10^{-4}$ for
a fixed normalization at the mean value of the DMR detection.
Using equation \eqn{Mulimit}, we set a limit on the slope
$n < 1.60$ for $h=0.5$ and
$n < 1.63$ for $h=1.0$ (see Fig.~\ref{fig:3diffuse}) for $\Omega_0 = 1$
with similar but slightly less stringent limits for $\Lambda$-dominated 
universes ($\Omega_0 < 1$).
If we were even more conservative,
using $|\mu|<3.3
\times 10^{-4}$ to imply $\mu < 3.3
\times 10^{-4}$, the constraints shift negligibly:
$n < 1.65$ $ (h=0.5)$ and $n < 1.68$ $(h=1.0)$.
On the other hand, employing
the more optimistic bound and taking into account the
COBE DMR measurement, we obtain $\mu < 0.63 \times 10^{-4}$
which implies
$n < 1.54$ ($h=0.5$) and $n<1.56$ ($h=1.0$).
These limits are
nearly independent of $\Omega_B$ and do not change within
the nucleosynthesis bounds of $0.011 < \Omega_B h^2 < 0.016$ \cite{Walker,Smith}.

%% file: chap3/y.tex
\begin{figure}[t]
\centerline{ \hskip-0.5truecm
\epsfxsize=3.5in \epsfbox{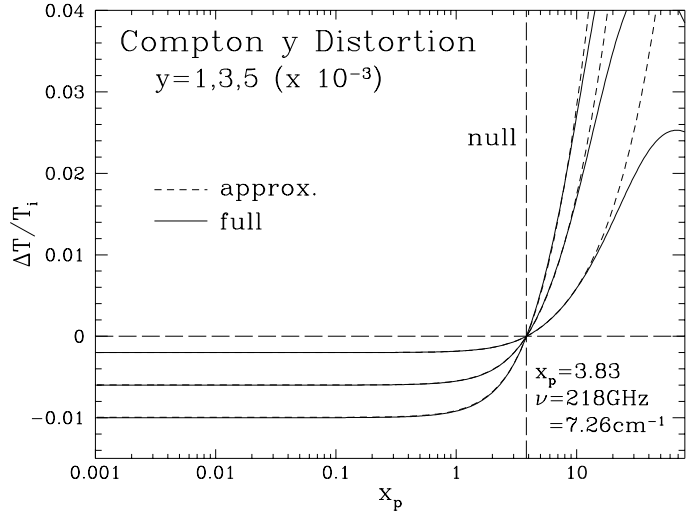}}
\vskip -0.5truecm
\caption{Compton-$y$ distortion}
\mycaption{Compton upscattering by hot electrons leaves a constant 
Rayleigh-Jeans decrement of $y \equiv (\Delta T/T)_{RJ}$ and a Wien
excess that is overestimated by equation \eqn{comptonapprox} as
compared with the diffusion integral \eqn{comptonspect}.  The crossover
is at $x_p = 3.83$ and is independent of $y$ and allows a clean separation
between $y$ distorted and temperature shifted spectra.  
}
\label{fig:3y}
\end{figure}

%% file: chap3/ytobe.tex
\begin{figure}[t]
\centerline{ \hskip-0.5truecm
\epsfxsize=3.5in \epsfbox{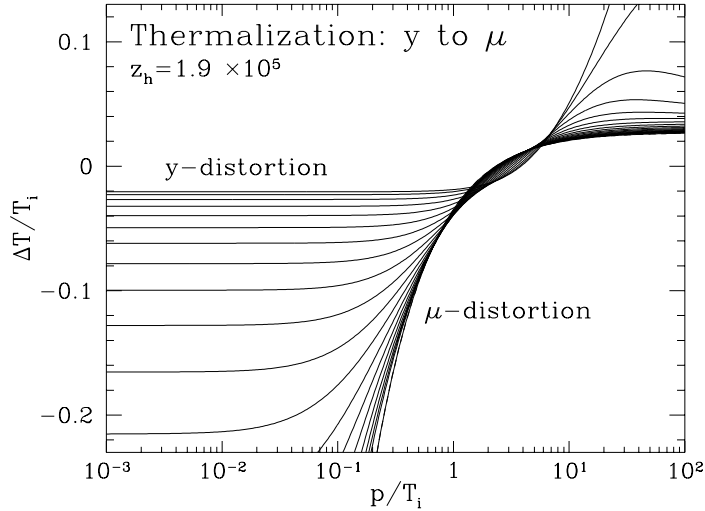}}
\vskip -0.5truecm
\caption{Thermalization from $y$ to $\mu$}
\mycaption{An initial $y$-distortion at $z=1.9 \times 10^5$ with the 
characteristic Rayleigh-Jeans suppression thermalizes to a Bose-Einstein
distribution as both low-frequency and high-frequency photons are shifted
to $\xe \sim 1$ by Compton scattering.  Curves are equally spaced in 
redshift between $z=1.9 \times 10^5-0.1 \times 10^5$ from highest to lowest
$(\Delta T/T)_{RJ}$. 
Bremsstrahlung 
and double Compton scattering have been artificially turned off.
}
\label{fig:3ytobe}
\end{figure}

%% file: chap3/crit.tex
\begin{figure}[t]
\centerline{ \hskip -0.5truecm
\epsfxsize=3.5in \epsfbox{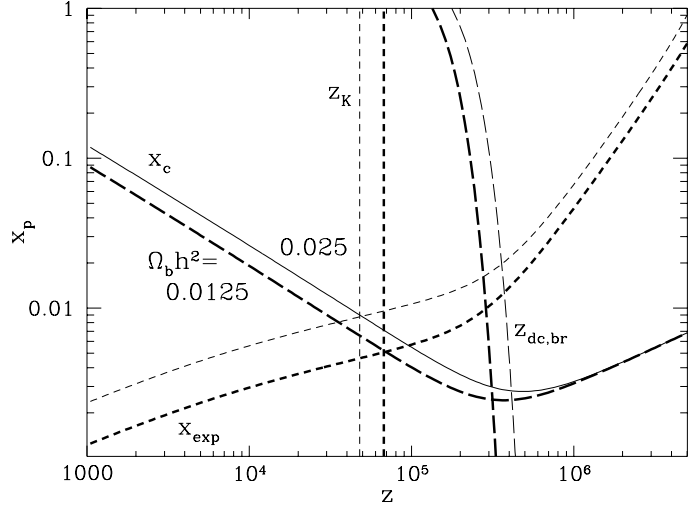}}
\vskip -0.5truecm
\caption{Critical Frequencies and Redshifts}
\mycaption{Comparative rates for an
$\Omega_0 h^2 = 0.25$ universe and $\Omega_b h^2 = 0.0125$ (heavy)
and $0.025$ (light)
Solid line is $x_c$, dashed line is $x_{exp}$, long dashed
lines represent critical redshifts as labeled. 
}
\label{fig:3crit}
\end{figure}

%% file: chap3/evol.tex
\begin{figure}[t]
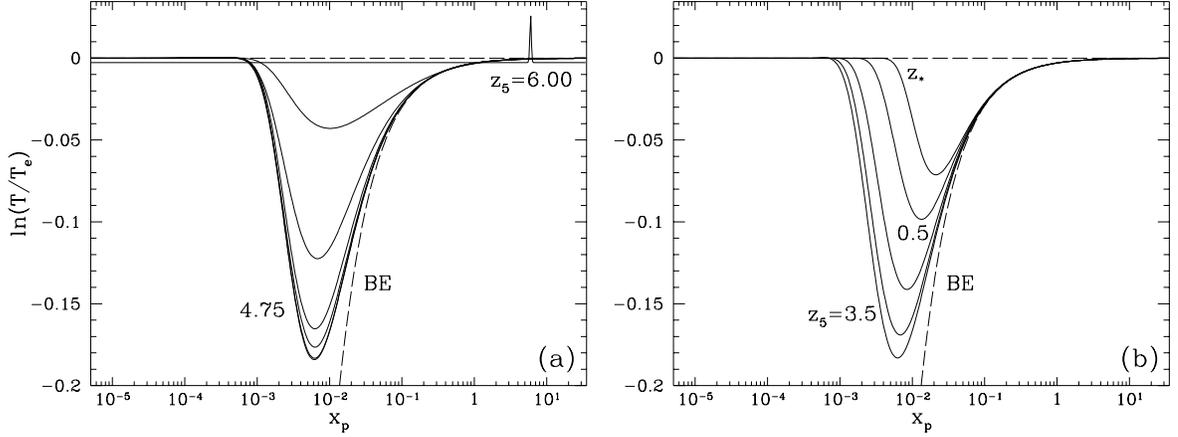

\centerline{
\epsfxsize=3.0in \epsfbox{chap3/evola.epsf} 
\epsfxsize=3.0in \epsfbox{chap3/evolb.epsf}}
\vskip -0.5truecm
\caption{Low Frequency Evolution}
\mycaption{Positive chemical potentials.
Initial spectrum: injection
at $x_h=6,\ z_h = 6 \times 10^{5}$ with
$\ddn=2.5 \times 10^{-3},\ \dde=5.5 \times 10^{-3},$
for $\oh=0.25,\ \obh=0.025$.
(a)  Establishment of the Bose-Einstein spectrum,
$4.75 \sci 5 < z < 6.00 \sci 5 $ where $z_5 = z / 10^5$ and
curves are equally spaced in redshift.
(b)  Quasi-static evolution, $z_{*} < z < 3.5 \sci 5 $ and 
Bose-Einstein freeze out $z_K < z < z_*$.
Long dashes represents best
fit Bose-Einstein spectrum and the undistorted Planck distribution.
}
\label{fig:3evol}
\end{figure}

%% file: chap3/low.tex
\begin{figure}[t]
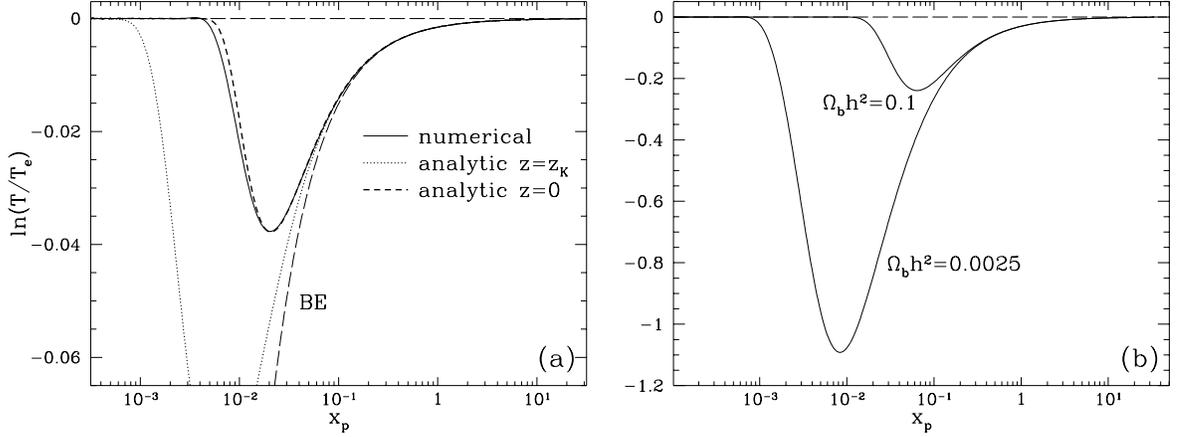

\centerline{
\epsfxsize=3.0in \epsfbox{chap3/lowa.epsf}
\epsfxsize=3.0in \epsfbox{chap3/lowb.epsf}}
\vskip -0.5truecm
\caption{Low Frequency Spectrum}
\mycaption{Positive chemical potentials
(a) Comparison with analytic results. Initial spectrum: injection
with $\ddn=1.2 \times 10^{-3},\ \dde=
2.7 \times 10^{-3}$ at
$x_h=6,\ z_h = 4 \times 10^5$ for $\oh=0.25,\
\obh=0.025$. The spectrum evolves significantly from Bose-Einstein freeze
out at $z_c = 4\sqrt{2} z_K$ due to bremsstrahlung at low frequencies.
The analytic estimation of the absorption optical depth
provides an accurate description of the spectrum.
(b) Baryon dependence of bremsstrahlung absorption.
Initial spectrum:
injection with $\ddn=1.2 \sci -2 ,\ \dde =2.7 \sci -2 $ at
$x_h=6,$ for $\oh=0.25.$,
$\obh=0.0025,\ z_h=1.2 {\sci 6 } $ and $\obh=0.10,\ z_h=1.5 {\sci 5 }$.
The peak distortion measures the baryon content.
}
\label{fig:3low}
\end{figure}

%% file: chap3/neg.tex
\begin{figure}[t]
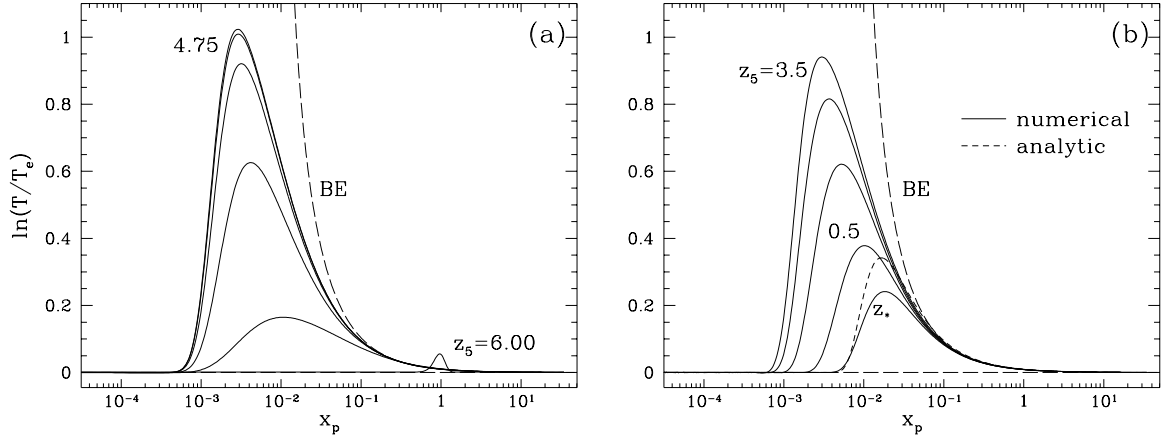

\centerline{
\epsfxsize=3.0in \epsfbox{chap3/nega.epsf}
\epsfxsize=3.0in \epsfbox{chap3/negb.epsf}}
\vskip -0.5truecm
\caption{Negative Chemical Potentials}
\mycaption{Initial spectrum: injection
at $x_h=1,\ z_h = 6 \times 10^{5}$ with
$\ddn=7.5 \times 10^{-3},\ \dde=2.7 \times 10^{-3},$
for $\oh=0.25,\ \obh=0.025$.
(a)  Establishment of the Bose-Einstein spectrum
 $4.75 \sci 5 < z < 6.00 \sci 5 $ where $z_{5} = z / 10^5$ (equally
spaced in redshift).
(b)  Quasi-static evolution and freeze out $z_{*} < z < 3.5 \sci 5 $.
The analytic approximation for bremsstrahlung absorption is adequate but
less accurate than for positive chemical potentials.  
}
\label{fig:3neg}
\end{figure}

%% file: chap3/zero.tex
\begin{figure}[t]
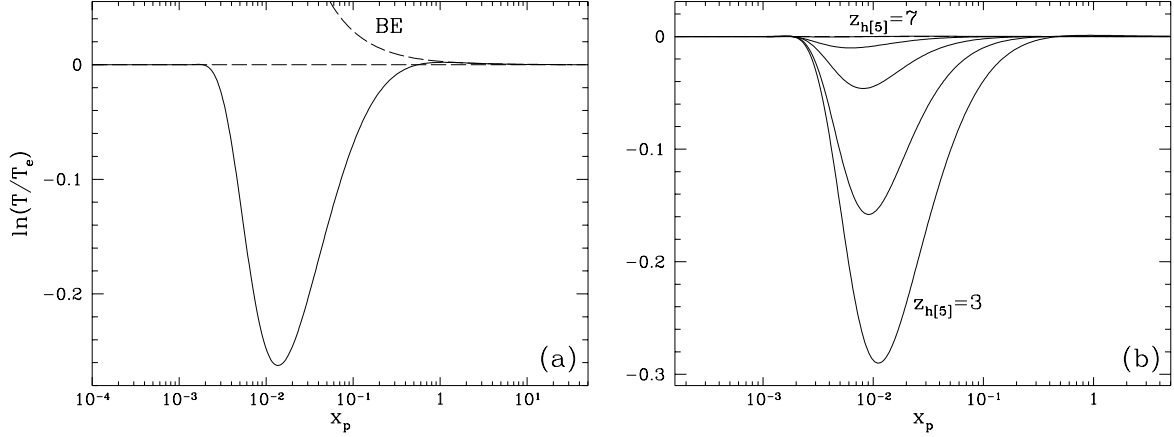

\centerline{
\epsfxsize=3.0in \epsfbox{chap3/zeroa.epsf}
\epsfxsize=3.0in \epsfbox{chap3/zerob.epsf}}
\vskip -0.5truecm
\caption{Balanced Injection}
\mycaption{The number and energy injection are balanced to give
$\mu =0$ after reshuffling by Compton scattering. (a) 
Initial spectrum: injection
at $x_h=3.7,\ z_h = 2.5 \sci 5 $ with $\ddn=0.16,\ \dde=0.22$ for
\standO.  Dashed lines are best analytic
fit for high frequencies (Bose-Einstein
with negative
chemical potential) and low frequencies (exponentially suppressed
positive chemical potential).  Note that low frequency distortions
can be much larger than high frequency distortions would imply. (b) 
Time evolution of the spectrum for the same parameters as (a) 
save that 
$z_h = 3.0 \sci 5 ,\
4.0 \sci 5 ,\ 5.0 \sci 5 , \ 6.0 \sci 5 ,\  7.0 \sci 5 $ in order
of decreasing distortions. 
}
\label{fig:3zero}
\end{figure}

%% file: chap3/mu.tex
\begin{figure}[t]
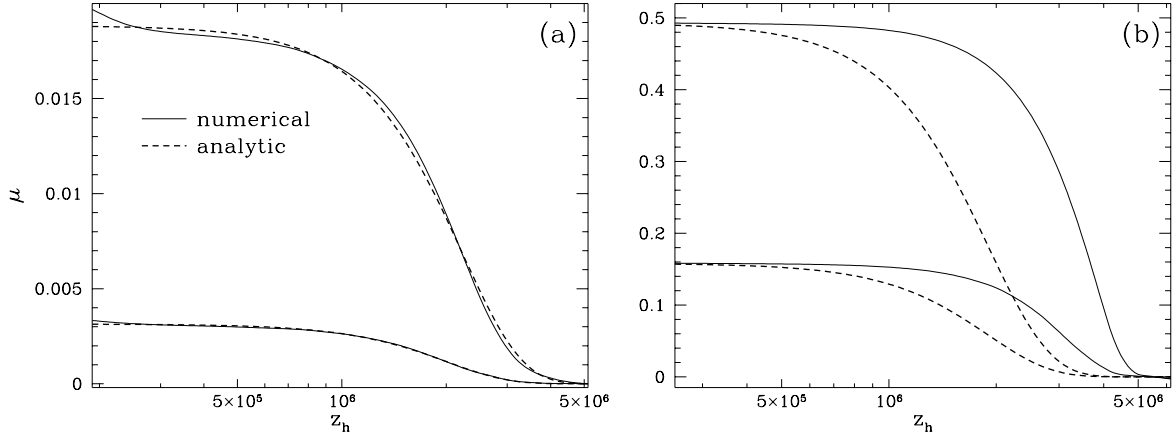

\centerline{
\epsfxsize=3.0in \epsfbox{chap3/mua.epsf}
\epsfxsize=3.0in \epsfbox{chap3/mub.epsf}}
\vskip -0.5truecm
\caption{Positive Chemical Potential Evolution}
\mycaption{(a) Small $\mu$.
Initial spectrum: injection
of $\ddn=5.4 \times 10^{-3},\ \dde=
2.0 \times 10^{-2}$ (top) and
of $\ddn=2.5 \times 10^{-3},\ \dde=
5.5 \times 10^{-3}$ (bottom) 
(b) Large $\mu$.
Initial spectrum: injection
of $\ddn=1.5 \times 10^{-1},\ \dde=
5.5 \times 10^{-1}$ (top) and
of $\ddn=4.4 \times 10^{-2},\ \dde=
1.6 \times 10^{-1}$ (bottom).
All injections at $x_h=6$ with $\oh=0.25,\ \obh=0.025$.
}
\label{fig:3mu}
\end{figure}

%% file: chap3/negmu.tex
\begin{figure}[t]
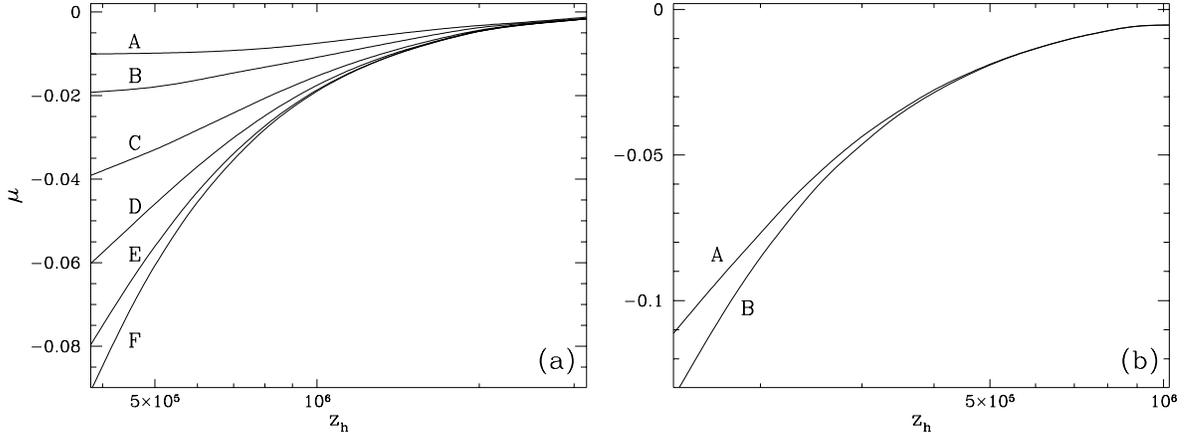

\centerline{
\epsfxsize=3.0in \epsfbox{chap3/negmua.epsf}
\epsfxsize=3.0in \epsfbox{chap3/negmub.epsf}}
\vskip -0.5truecm
\caption{Negative Chemical Potential Evolution}
\mycaption{(a) Time evolution and instability:  
(A)
$\ddn=7.5 \times 10^{-3},\ \dde=
2.7 \times 10^{-3}$;
(B)
$\ddn=1.5 \times 10^{-2},\ \dde=
5.5 \times 10^{-3}$;
(C)
$\ddn=3.8 \times 10^{-2},\ \dde=
1.4 \times 10^{-2}$;
(D)
$\ddn=7.5 \times 10^{-2},\ \dde=
2.7 \times 10^{-2}$;
(E)
$\ddn=1.5 \times 10^{-1},\ \dde=
5.5 \times 10^{-2}$;
(F)
$\ddn=3.0 \times 10^{-1},\ \dde=
1.1 \times 10^{-1}$.
All for injection at $x_h=1$ with $\oh=0.25,\ \obh=0.025$.
(b) High baryon case. (A)
$\ddn=1.5 \times 10^{-1},\ \dde=
5.5 \times 10^{-2}$;
(B)
$\ddn=3.0 \times 10^{-1},\ \dde=
1.1 \times 10^{-1}$,
for injections at $x_h=1$ with $\oh=0.25,\ \obh=0.10$.
}
\label{fig:3negmu}
\end{figure}

%% file: chap3/dat.tex
\begin{figure}[t]
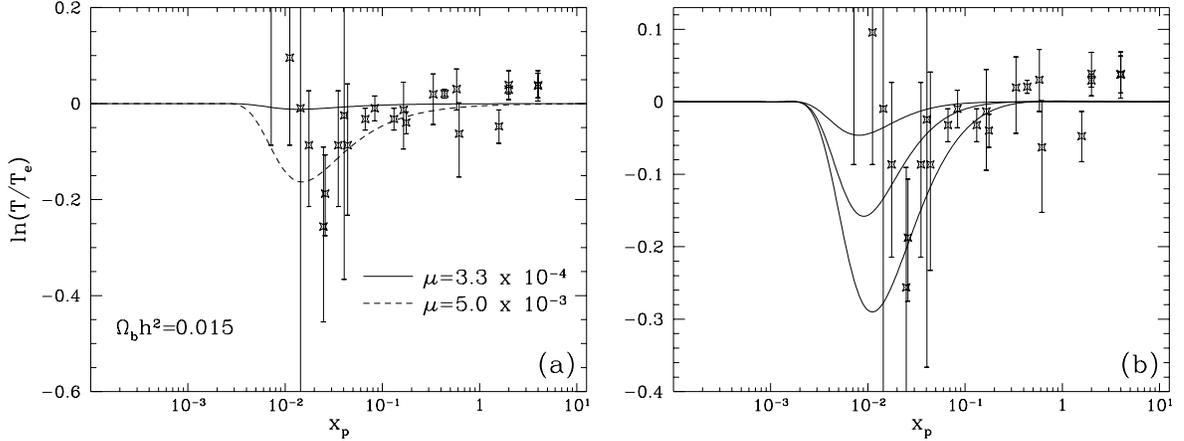

\centerline{
\epsfxsize=3.0in \epsfbox{chap3/data.epsf}
\epsfxsize=3.0in \epsfbox{chap3/datb.epsf}}
\vskip -0.5truecm
\caption{Comparison with Observational Data}
\mycaption{Observational low frequency data compared with (a) numerical
results for $\mu = 0.005$ (dotted) and $\mu = 3.3 \times 10^{-4}$ (solid
line) with $\obh = 0.015$ and $\Omega_0 h^2 = 0.25$. Only the latter
satisfies the high frequency FIRAS data \cite{Mather}. (b) Balanced
injection $\mu = -1.3 \times 10^{-3}$, $-3.2 \times 10^{-4}$, $-8.0
\times 10^{-5}$ in the Wien tail.  Notice that this special case has large
low frequency and small high frequency distortions.
}
\label{fig:3dat}
\end{figure}

%% file: chap3/baryon.tex
\begin{figure}[t]
\centerline{ \hskip -0.75truecm
\epsfxsize=3.5in \epsfbox{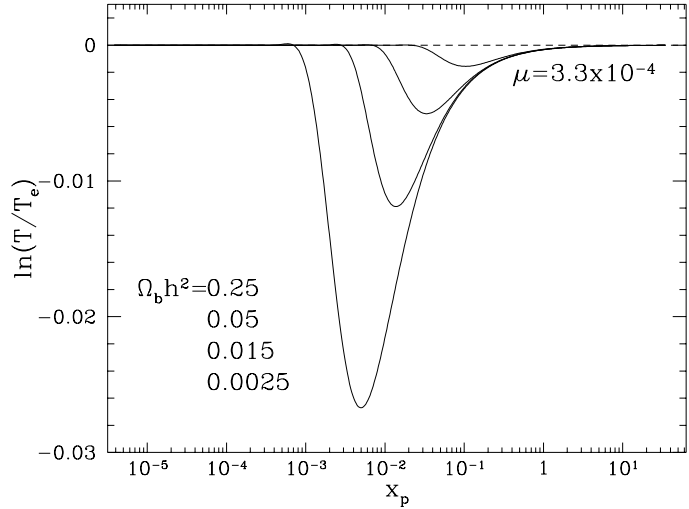}}
\vskip -0.5truecm
\caption{Rayleigh-Jeans Baryon Dependence}
\mycaption{Predicted spectral distortions with $\mu = 3.3 \times 10^{-4}$
for $\obh =$ 0.0025, 0.015, 0.050, 0.25 in order of decreasing
distortions for $\oh = 0.25$.
}
\label{fig:3baryon}
\end{figure}

%% file: chap3/decay.tex
\begin{figure}[t]
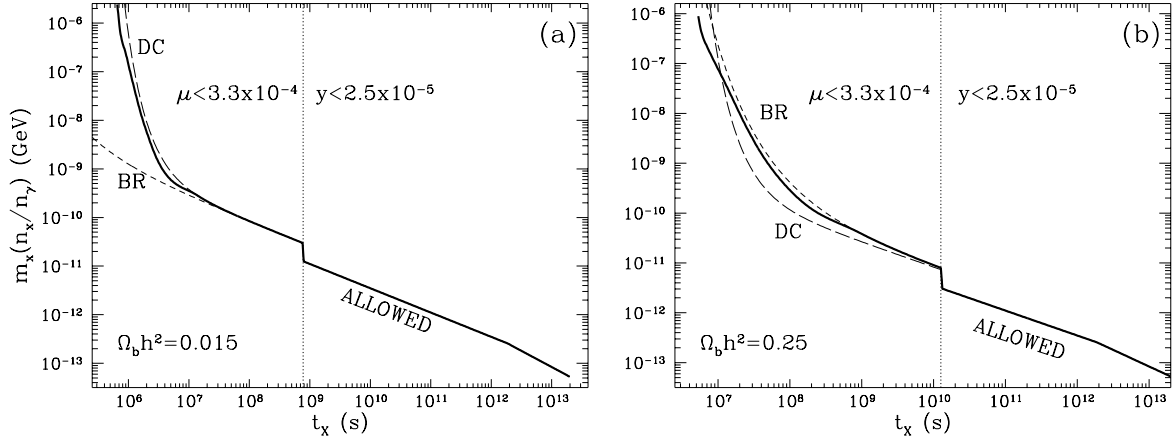

\centerline{
\epsfxsize=3.0in \epsfbox{chap3/decaya.epsf}
\epsfxsize=3.0in \epsfbox{chap3/decayb.epsf}}
\vskip -0.5truecm
\caption{Particle Decay Constraints}
\mycaption{FIRAS constraints on $\mu$ and $y$ limit the energy injection from
massive unstable particles.  Dashed lines are the analytic approximation.
Dotted vertical lines mark the approximate transition between $\mu$ 
and $y$ distortions.  In reality the constraint curve makes a smooth
transition between the two.
For low $\obh$, double Compton dominates the thermalization,
whereas for high $\obh$ bremsstrahlung is most efficient. 
}
\label{fig:3decay}
\end{figure}

%% file: chap3/diffuse.tex
\begin{figure}[t]
\centerline{ \hskip-0.5truecm
\epsfxsize=3.5in \epsfbox{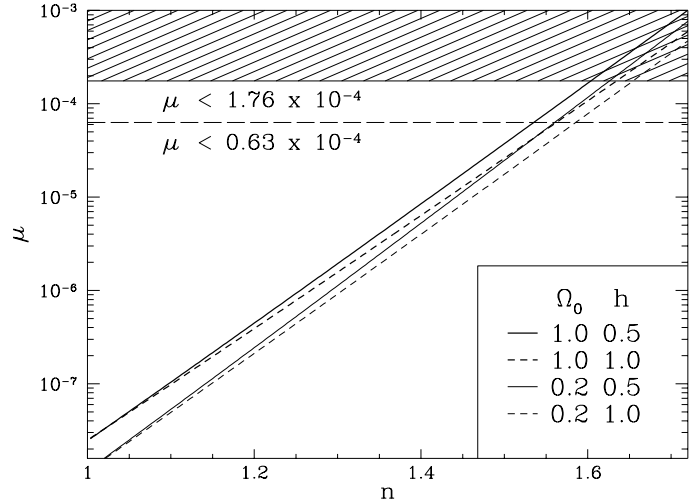}}
\vskip -0.5truecm
\caption{Diffusion Dissipation and Limits on $n$}
\mycaption{Spectral distortions from the dissipation of acoustic waves
for an initial adiabatic density perturbation spectrum of $k^3 |\Psi(0,k)|^2
= Bk^{n-1}$ in an $\Omega_0+\Omega_\Lambda=1$ flat universe, 
normalized to give the {\it COBE} DMR rms of 
$(\Delta T/T)_{10^\circ} = 1.12 \times 10^{-5}$.  With the uncertainties on both
the DMR and FIRAS measurements, the conservative $95\%$ upper limit is
effectively $\mu < 1.76 \times 10^{-4}$.  The constraint on $n$ is weakly
dependent on cosmological parameters.  We have also plotted the optimistic
limit of $\mu < 0.63 \times 10^{-4}$ discussed in the text. 
}
\label{fig:3diffuse}
\end{figure}

%% file: chap4.tex
\chapter{Multifluid Perturbation Theory}
\label{ch-perturbation}

\begin{quote}
\footnotesize\it
It is the nature of things that they {\rm are} ties to each other.
\vskip 0.1truecm
\centerline{--Chuang-tzu, 20}
\end{quote}
In the standard scenario, small perturbations in the early universe
grow by gravitational instability to form the wealth of structure observable 
today.  At the early stages of this process, relevant for CMB work, fluctuations
are still small and can be described in linear perturbation theory. 
What makes the problem non-trivial is the fact that  
different components such as the photons, baryons,
neutrinos, and collisionless dark matter, have different equations of 
state and interactions.  It is therefore necessary to employ a
fully relativistic 
multifluid treatment to describe the coupled evolution of the individual
particle species.

In this chapter, we discuss the framework for the evolution of 
fluctuations.  Since in linear theory, each normal mode evolves independently
we undertake a mode by mode analysis.   
In open universes, this decomposition implies a lack
of structure above the curvature scale for random-field perturbations.
We show why this arises and how it might be avoided by generalizing
the random field condition \cite{LW}.  The evolution itself is governed
by the energy momentum conservation equations in the perturbed space-time 
and feeds back into the metric fluctuations through the
Einstein equations.  In Newtonian gauge, they generalize the Poisson
equation familiar from the non-relativistic theory.

It is often useful to express
the evolution in other gauges, \eg\ the popular synchronous gauge and the
total matter gauge.  We discuss the general 
issue of gauge transformations and their effect on the 
interpretation of perturbations.  
Various aspects of the evolution appear simplest for different choices
of gauge.  
Those that involve the photons are most straightforward to analyze in
Newtonian form where redshift and infall correspond to classical 
intuition.  On the other hand, the evolution of the matter and consequently
the metric perturbations themselves becomes simpler on its own
rest frame.
We therefore advocate a hybrid representation for perturbations
based on the so-called ``gauge invariant'' formalism.

\section{Normal Mode Decomposition}
\label{sec-4normal}

\subsection{Laplacian Eigenfunctions}
\label{ss-4laplacian}

Any scalar fluctuation may be decomposed in eigenmodes
of the Laplacian
\beal{eq:eigenmodes}
\nabla^2 Q \equiv \gamma^{ij} Q_{|ij} = -k^2 Q,
\eea
where `$|$' represents a covariant derivative with respect to the
three metric $\gamma_{ij}$ of constant curvature $K = -H_0^2 (1-\Omega_0
-\Omega_\Lambda)$. In flat space $\gamma_{ij}=\delta_{ij}$,
and $Q$ is a plane wave $\exp(i\bk\cdot\bx)$.  
As we shall discuss further in \S \ref{ss-4completeness}, the
eigenfunctions are complete for $k \ge \sqrt{-K}$.  Therefore
we define the 
transform of an arbitrary square integrable function $F(\bx)$ as 
\cite{LS,LW}
\bel{eq:decompconvention}
F(\bx) = \sum_{|\bk| \ge \sqrt{-K}} F(\bk) Q(\bx,\bk) 
	= {V \over (2\pi)^3} \int_{|\bk|\ge\sqrt{-K}}^{\infty} d^3 k
F(\bk)Q(\bx,\bk).
\ee
In the literature, an alternate convention is often employed in order
to make the form appear more like the flat space convention 
\cite{Wilson,HSb},
\bel{eq:decompconventionalt}
F(\bx) = \sum_{\tilde k} \tilde F(\tilde\bk) Q(\bx,\tilde\bk) 
	= {V \over (2\pi)^3} \int_0^\infty d^3 \tilde k
\tilde F(\bk)Q(\bx,\tilde \bk),
\ee
where the auxiliary variable $\tilde k^2 = k^2 + K$. 
The relation between the two conventions is
\beal{eq:conversion}
\tilde k |\tilde F(\tilde k)|^2 \eal k |F(k)|^2 \nonumber\\
	\eal (\tilde k^2 - K)^{1/2} |F([\tilde k^2 - K]^{1/2})|^2
\eea
and should be kept in mind when comparing predictions. In particular, note
that power law conditions in $\tilde k$ for $\tilde F$ are not the
same as in $k$ for $F$.  

Vectors and tensors needed in the description of the velocity and stress
perturbation can be constructed from the covariant derivatives of $Q$ and the
metric tensor,
\beal{eq:vector}
Q_i \!\!\!&\equiv&\!\!\! -k^{-1} Q_{|i}, \nonumber\\
Q_{ij} \!\!\!&\equiv&\!\!\! k^{-2} Q_{|ij} + {1 \over 3} \gamma_{ij}Q ,
\eea
where the indices are to be raised and lowered by the three metric
$\gamma_{ij}$ and $\gamma^{ij}$.
The following identities can be derived from 
these definitions 
and 
the communtation relation for covariant derivatives (see \eg\ \cite{Weinberg}
eqn.~8.5.1)
\cite{KS84}
\beal{eq:Qidentities}
Q_i^{\,|i} \eal kQ, \nonumber\\
\nabla^2 Q_i \eal -(k^2 -3K)Q_i, \nonumber\\
Q_{i|j} \eal -k(Q_{ij} - {1 \over 3} \gamma_{ij} Q), \nonumber\\
Q^i_{\ i} \eal 0, \nonumber\\
Q_{ij}^{\ \ |j} \eal {2 \over 3} k^{-1} (k^2 - 3K)Q_i ,
\eea
and will be useful in simplifying the evolution equations.

\subsection{Radial Representation}
\label{ss-4radial}

To gain intuition about these
functions, let us examine an explicit representation.
In radial coordinates,
the 3-metric becomes
\bel{AeqnMetric}
\gamma_{ij} dx^{i} dx^{j} = -K^{-1} [d\chi^2 + \sinh^2\chi
(d\theta^2 + \sin^2\theta d\phi^2)],
\ee
where the distance is scaled to the curvature radius $\chi = \sqrt{-K}\eta$.  
Notice that the (comoving) 
angular diameter distance is $\sinh\chi$, leading to 
an exponential increase in the surface area of a shell with radial distance
$\chi \gg 1$. 
The Laplacian can now be written as
\bel{eq:AeqnLaplacian}
\gamma^{ij} Q_{|ij} = -K \sinh^{-2} \chi
\left[
      {\partial \over \partial \chi}
\left(
      \sinh^2\chi {\partial Q \over \partial \chi }
\right)
+ \sin^{-1}\theta {\partial \over \partial \theta}
\left(
      \sin\theta  {\partial Q \over \partial \theta}
\right)
+ \sin^{-2}\theta {\partial^2 Q \over \partial \phi^2}
\right].
\ee
Since the angular part is independent of curvature, we
may separate variables such that
$Q = X_\nu^\ell(\chi)Y_\ell^m(\theta,\phi)$,
where
$\nu^2 = \tilde k^2/(-K)= -(k^2/K+1)$.
From equation \eqn{AeqnLaplacian}, it is obvious that
the spherically
symmetric $\ell =0$ function is
\bel{eq:AeqnXIso}
X_\nu^0(\chi) = {\sin(\nu\chi) \over \nu \sinh\chi}.
%= \sqrt{-K} {\sin(\tilde k \Delta \eta) \over \tilde k
%\sinh(\Delta\eta\sqrt{-K})}.
\ee
As expected, the change in the area element from a flat to
curved geometry causes $\chi
\rightarrow \sinh\chi$ in the denominator.
The higher modes are explicitly given by
\cite{Liftshitz,Harrison}
\bel{eq:AeqnHorribleFunc}
X_\nu^\ell(\chi) = (-1)^{\ell+1} M_\ell^{-1/2}
\nu^{-2} (\nu^2+1)^{-\ell/2}
\sinh^\ell \chi
{d^{\ell+1} (\cos \nu \chi) \over d(\cosh \chi)^{\ell+1}},
\ee
and become $j_\ell(k\eta)$ in the flat space limit, where
\beal{eq:AeqnMFact}
M_\ell \!\!\!&\equiv&\!\!\! \prod_{\ell'=0}^{\ell} K_{\ell'}, \nonumber\\
K_0 \eal  1, \nonumber\\
K_\ell \eal 1-(\ell^2-1)K/k^2,  \qquad \ell \ge 1,
\eea 
which all reduce to unity as $K \rightarrow 0$.  This factor represents our
convention for the normalization of the open universe functions,
\bel{eq:Xnormalization}
\int X_\nu^\ell(\chi) X_{\nu'}^{\ell'}(\chi) \sinh^2 \chi d\chi 
= {\pi \over 2\nu^2} \delta(\nu-\nu') \delta(\ell-\ell'),
\ee  
and is chosen to be similar to the flat space case.  In the literature,
the normalization is often chosen such that $\tilde X_\nu^\ell = 
X_\nu^\ell M_\ell^{-1/2}$ is employed as 
the radial eigenfunction \cite{Wilson,HSb}.
 
It is often more convenient to generate these functions
from their recursion relations.
One particularly useful relation is \cite{AS}
\bel{eq:AeqnRecursion}
{d \over d \eta} X_\nu^\ell = {\ell \over 2\ell+1} k K_\ell^{1/2} X_\nu^{\ell-1}
-{{\ell+1}\over{2\ell+1}}k K_{\ell+1}^{1/2} X_\nu^{\ell+1}.
\ee
Since radiation free streams on radial null geodesics, we shall see that 
the collisionless Boltzmann equation takes on the same form as equation
\eqn{AeqnRecursion}.

\subsection{Completeness and Super Curvature Modes}
\label{ss-4completeness}

Open universe eigenfunctions possess the curious property that they
are complete for $k \ge \sqrt{-K}$.  Mathematically, this is easier to see
with a choice of three metric such that $\gamma_{ij} = \delta_{ij}/(-Kz^2)$,
the so-called flat-surface representation \cite{Wilson,LW}. In this
system $-\infty < x < \infty$, $-\infty < y < \infty$, $0 \le z < 
\infty$ and surfaces of constant $z$ are flat.  The Laplacian
\bel{eq:flatsurflap}
\nabla^2 Q = -Kz^2 \left( 
{ \partial^2 Q \over \partial x^2}
+{ \partial^2 Q \over \partial y^2}
+{ \partial^2 Q \over \partial z^2} \right)
+ Kz {\partial Q \over \partial z},
\ee
has eigenfunctions
\bel{eq:flatsurfeigen}
Q = z\, \exp(ik_1 x + ik_2 y) K_{i\nu}(k_\perp z),
\ee
where $K_{i\nu}$ 
is the modified Bessel function and $k^2_\perp = k^2_1 + k^2_2$.
Since the $x$ and $y$ dependences are just those of plane waves, which
we know are complete, we need only concern ourselves with the $z$
coordinate.  As pointed out by Wilson \cite{Wilson}, it reduces to 
a Kontorovich-Lebedev transform,
\beal{eq:kontorovich}
g(y) \eal \int_0^\infty f(x) K_{ix}(y)dx, \nonumber\\
f(x) \eal 2\pi^{-2} x \sinh(x\pi) \int_0^\infty g(y) K_{ix}(y) y^{-1} dy,
\eea
\ie\ there exists a completeness relation,
\bel{eq:completeness}
\int_0^\infty d\nu \nu \sinh(\pi \nu) K_{i\nu}(k_\perp z) K_{i\nu}
(k_\perp z') = {\pi^2 \over 2} z \delta(z-z').
\ee
Therefore an arbitrary square integrable function $F(\bx)$ can be
decomposed into a sum of eigenmodes of $\nu \ge 0$,
\beal{eq:flatsurfdecomp}
F(\bx) \eal \int_0^\infty \nu \sinh(\pi\nu)d\nu \int_{-\infty}^{\infty} dk_1
\int_{-\infty}^{\infty} dk_2 F(\bk) Q(\bx,\bk), \nonumber\\
F(\bk) \eal {1 \over 2\pi^4} \int_0^\infty {dz \over z^3} \int_0^\infty dx
	\int_0^\infty dy F(\bx) Q(\bx,\bk),
\eea
where $Q$ is given by equation \eqn{flatsurfeigen} and $\bx=(x,y,z)$
and $\bk=(k_1,k_2,\nu)$.  Since $\nu \ge 0 $ implies $k \ge \sqrt{-K}$,
this establishes the claimed completeness.

\input chap4/X.tex

This completeness property leads to a seemingly bizarre consequence if
we consider random fields, \ie\ randomly phased superpositions of these
eigenfunctions.  To see this, return to the radial representation.
In Fig.~\ref{fig:4X}, we plot the spherically
symmetric $\ell=0$ mode given by equation~\eqn{AeqnXIso}.
Notice that its first zero is at $\chi = \pi/\nu$. This is
related to the completeness property: as $\nu \rightarrow 0$, we can obtain
arbitrarily large structures. 
For this reason, $\nu$ or more specifically $\tilde k = \nu \sqrt{-K}$ 
is often thought of as the wavenumber \cite{Wilson,KamSper}.  
 However, the {\it amplitude} of the structure above the curvature scale is
suppressed as $e^{-\chi}$.  Prominent structure lies only at the curvature 
scale as $\nu \rightarrow 0$. 
In this sense, $k$ should be regarded as the effective wavelength.  
This is important to bear in mind when considering the meaning of
``scale invariant'' fluctuations.
In fact, the $e^{-\chi}$ behavior is {\it independent}
of the wavenumber and $\ell$,
if $\chi \gg 1$.
 
This peculiarity in the eigenmodes has significant consequences.
Any random phase superposition of the eigenmodes $X_\nu^\ell$
will have exponentially suppressed structure larger than
the curvature radius.  Even though completeness tells us that
arbitrarily large structure can be built out of the $X_\nu^\ell$
functions,  it {\it cannot} be done without correlating the modes.
This is true even if the function is square integrable,
\ie\ has support only to a finite radius
possibly above the curvature scale.

Is the lack of structure
above the curvature scale reasonable?
The fundamental difference between open and flat universes is
that the volume increases exponentially with the radial
coordinate above the curvature scale
$V(\chi_c) \sim [\sinh(2\chi_c) -2\chi_c]$. 
Structure above the curvature scale implies correlations
over vast volumes \cite{KamSper}.
It is in fact difficult to
conceive of a model where correlations do not die exponentially
above the curvature radius.  The random phase hypothesis has
been proven to be valid for 
 inflationary perturbations in a pre-existing open geometry
\cite{LS} and only mildly violated for bubble nucleated open
universes \cite{YST}.

Lyth and Woszcynza \cite{LW} show that the simplest way to generalize random
fields to include supercurvature scale structure is to employ an
overcomplete set of eigenfunctions extended by analytic continuation
of the modes to $k \rightarrow 0$.  Of course,
random phase conditions in the overcomplete set can alternatively be 
expressed as initially phase correlated modes of the complete set. 
In linear theory, the evolution of each mode is independent and thus
there is no distinction between the two.   
Including supercurvature perturbations merely amounts to extending
the treatment to the full range of k: $0 \le k < \infty$.  All of
the equations presented here may be extended in this manner with the
understanding that $\nu \rightarrow |\nu|$.  

\subsection{Higher Angular Functions}
\label{ss-4angular}

We will often need to represent a general function
of position $\bx$ {\it and} angular direction $\bg$, \eg\ for the
radiation distribution.  As we have seen, vector and tensors
constructed from $Q$ and its covariant derivatives
 can be used to represent dipoles and 
quadrupoles, $G_1 = \gamma^i Q_i$ and $G_2 = {3 \over 2}\gamma^i\gamma^j
Q_{ij}$.   
We can generalize these considerations and form the full multipole
decomposition \cite{Wilson}
\bel{eq:AeqnLDecomposition}
F(\bx,\bg) = \sum_{\tilde \bk}
\sum_{\ell=0}^{\infty} \tilde F_\ell(\bk)G_\ell (\bx,\bg,\bk),
\ee
where
\bel{eq:AeqnGl}
G_\ell(\bx,\bg,\bk) =
(-k)^{-\ell} Q_{|i_1...i_\ell}(\bx,\bk)P_\ell^{i_1...i_\ell}(\bx,\bg),
\ee
and
\beal{eq:AeqnP}
P_{0} \eal 1,  \qquad
P^i_{1} = \gamma^i, \nonumber\\
P^{ij}_{2} \eal {1 \over 2}(3\gamma^i \gamma^j - \gamma^{ij}), \nonumber\\
P^{i_1...i_{\ell+1}}_{\ell+1} \eal
{2\ell+1 \over \ell+1} \gamma_{\vphantom{\ell}}^{(i_1} P_\ell^{i_2...i_{\ell+1})
}
 - {\ell \over \ell+1} \gamma_{\vphantom{\ell}}^{(i_1 i_2}P_{\ell-1}^
{i_3..i_{\ell+1})}, 
\eea
with parentheses denoting symmetrization about the indices.
For flat space, this becomes
$G_\ell = (-i)^\ell \exp(i\bk \cdot \bx)P_\ell(\bk \cdot \bg)$,
where $P_\ell$ is an ordinary Legendre polynomial.
Notice that along a path defined by fixed $\bg$, the flat
$G_\ell$ becomes
$j_\ell(k\eta)$ after averaging over $k$-directions.
Traveling on a fixed direction away from a point is
the same as following a radial path outwards.  Thus fluctuations {\it along}
this path can be decomposed in the radial eigenfunction.  It is therefore
no surprise that $G_\ell$ obeys a recursion relation similar to $X_\nu^\ell$, 
\beal{eq:AeqnLLPM} 
\gamma^i G_{\ell|i}
\eal {d \over d\eta}G[\bx(\eta),\bg(\eta)] =
\dot x^i {\partial \over \partial x^i} G_\ell + {\dot \gamma^i}{\partial \over 
\partial \gamma^i}
G_\ell \nonumber\\
\eal k \left\{ {\ell \over 2\ell+1} 
K_\ell G_{\ell-1} - {\ell+1 \over 2\ell+1} G_{\ell+1} \right\}, 
\eea
which follows from equation~\eqn{AeqnGl} and \eqn{AeqnP} via an 
exercise in combinatorics involving commutations of covariant
derivatives \cite{GSS91}.
Here we take $\bx(\eta)$ to be the integral path along $\bg$.
By comparing equations~\eqn{AeqnRecursion} and \eqn{AeqnLLPM}, the open
universe generalization of the relation between $G_\ell$ and
the radial eigenfunction is now apparent:
\bel{eq:Gradialrelation}
G_\ell[\bx(\eta),\bg(\eta)] = M_\ell^{1/2} X_\nu^\ell(\eta).
\ee
The only conceptual difference is that for the radial path that we decompose
fluctuations on, $\bg$ is not constant.
The normalization also suggests that to maintain close
similarity to the flat space case, the multipole moments should be
redefined as
\bel{eq:Decompositionconv}
F(\bx,\bg) = \sum_{|\bk|\ge \sqrt{-K}} \,
\sum_{\ell=0}^{\infty}  F_\ell(\bk) M_\ell^{-1/2} G_\ell (\bx,\bg,\bk),
\ee
which again differ from the conventions of \cite{Wilson,HSb} by
a factor $M_\ell^{1/2}$.

\section{Newtonian Gauge Evolution}
\label{sec-4newtonian}

\subsection{Metric Fluctuations}
\label{sec-4metric}

In linear theory, the evolution of each $k$ mode is independent.  We
can therefore assume without loss of generality that
the equation of motion for the $k$th mode can be obtained by
taking a metric of the form, 
\beal{eq:metriclower}
g_{00} \eal -(a/a_0)^2 (1+2\Psi Q), \nonumber\\
g_{0i} \eal 0, \nonumber\\
g_{ij} \eal (a/a_0)^2 (1+2\Phi Q)\gamma_{ij},
\eea
assuming the Newtonian representation, and correspondingly
\beal{eq:metricupper}
g^{00} \eal -(a_0/a)^2 (1-2\Psi Q), \nonumber\\
g^{0i} \eal 0, \nonumber\\
g^{ij} \eal (a_0/a)^2 (1-2\Phi Q)\gamma^{ij},
\eea
where employ the notation $\Psi(\eta,\bx) = \Psi(\eta)Q(\bx)$, 
{\it etc.}~and drop the $k$ index
where no confusion will arise.  Note that we have
switched from time to conformal time as the zero component.
The Christoffel symbols can now be written as
\def\hp{\hphantom{i}}
\def\hpm{\hphantom{i}\kern -1pt}
\beal{eq:Christawful}
\Gamma^{0}_{\hpm 00} \eal {\dot a \over a} + \dot \Psi Q, \nonumber\\
\Gamma^0_{\hp 0i} \eal -k\Psi Q_i, \nonumber\\
\Gamma^i_{\hpm 00} \eal -k\Psi Q^i, \nonumber\\
\Gamma^i_{\hpm 0j} \eal ({\dot a \over a} + \dot \Phi Q) \delta^i_{\hpm j}, 
				\nonumber\\
\Gamma^0_{\hp ij} \eal [{\dot a \over a} +
	 (-2{\dot a \over a}\Psi + 2{\dot a \over a}\Phi 
 	+ \dot \Phi)Q]\gamma_{ij},  \nonumber\\
\Gamma^i_{\hpm jk} \eal ^{(s)}\Gamma^i_{\hp jk} - k\Phi(\delta^i_{\hpm j} Q_k 
	+ \delta^i_{\hpm k}Q_j - \gamma_{jk}Q^i),
\eea
where $^{(s)}\Gamma^i_{\hp jk}$ is the Christoffel symbol on the unperturbed
3-surface $\gamma_{ij}$.

Finally we can write the Einstein tensor as $G_{\mu\nu} = \bar G_{\mu\nu}
+ \delta G_{\mu\nu}$,  where
\beal{eq:Gmunubackground}
\bar G^0_{\ 0} \eal -3 \left({a_0 \over a}\right)^2
 \left[ \left({ \dot a \over a }\right)^2 
	+ K \right], \nonumber\\
\bar G^i_{\hpm j} \eal - \left({a_0 \over a^2}\right)^2
	 \left[ 2{\ddot a \over a} -
	\left( { \dot a \over a }\right)^2 + K \right] \delta^i_{\hpm j}, 
			\nonumber\\
\bar G^0_{\ i} \eal \bar G^i_{\hp 0} = 0 
\eea
are the background contributions and 
\beal{eq:Gmunuperturbation}
\delta G^0_{\ 0} \eal 2 \left( {a_0 \over a} \right)^2 
	\left[ 3 \left( {\dot a \over a} \right)^2 \Psi
	      -3 {\dot a \over a} \dot \Phi
	      - (k^2 -3K)\Phi \right]Q,  \nonumber\\
\delta G^0_{\ i} \eal 2 \left( {a_0 \over a} \right)^2 
	\left[ {\dot a \over a} k\Psi - k\dot\Phi \right] Q_i, \nonumber\\
\delta G^i_{\hpm 0} \eal -2 \left( {a_0 \over a} \right)^2 
	\left[ {\dot a \over a} k\Psi - k\dot\Phi \right] Q^i, \nonumber\\
\delta G^i_{\hpm j} \eal 2 \left( {a_0 \over a} \right)^2  \Bigg\{
	\left[2 {\ddot a \over a} - \left( {\dot a \over a} \right)^2 \right]
	\Psi + {\dot a \over a}[\dot \Psi-\dot \Phi] \nonumber\\
& & - {k^2 \over 3}\Psi -\ddot\Phi - {\dot a \over a}\dot\Phi 
	- {1 \over 3}(k^2 - 3K)\Phi \Bigg\} \delta^i_{\hpm j} Q \nonumber
							\nonumber\\
& & - \left( {a_0 \over a} \right)^2  k^2 (\Psi + \Phi) Q^i_{\hpm j},
\eea
are the first order contributions from the metric fluctuations.

\subsection{Conservation Equations}
\label{ss-4conservation}

The equations of motion under gravitational interactions 
are most easily obtained by 
employing the conservation equations.  The stress-energy tensor of
a non-interacting fluid is covariantly conserved
$T^{\mu\nu}_{\ \ ; \mu} = 0$.  The $\nu=0$ equation gives energy 
density conservation, \ie\ the continuity equation; 
the $\nu = i$ equations
give momentum conservation, \ie\ the Euler equation.  
To first order, the stress energy tensor of a fluid $x$, possibly 
itself a composite of different particle species,
is
\beal{eq:stressenergy}
T^0_{\ 0} \eal -(1+{\delta_x} Q)\rho_x, \nonumber\\
T^0_{\ i} \eal (\rho_x + p_x)V_x Q_i, \nonumber\\
T^j_{\ 0} \eal -(\rho_x + p_x)V_x Q^j, \nonumber\\
T^i_{\ j} \eal p_x(\delta^i_{\hpm j} + 
	{\delta p_x \over p_x} \delta^i_{\hpm j} Q
+ \Pi_x Q^i_{\hpm j} ),
\eea
where $\rho_x$ is the energy density, $p_x$ is the pressure, $\delta_x 
= \delta \rho_x /\rho_x$ and
$\Pi_x$ is the anisotropic stress of the fluid.

\subsubsection{Continuity Equation}

The zeroth component of the conservation equation becomes
\beal{eq:zeroconserv}
-\partial_0 T^{00} \eal \partial_i T^{i0} + \Gamma^0_{\ \alpha\beta} T^{\alpha \beta}
+ \Gamma^\alpha_{\ \alpha\beta} T^{0\beta} \nonumber\\
 \eal T^{i0}_{\ \ |i} + 2\Gamma^0_{\ 00}T^{00} +
\Gamma^0_{\ ij} T^{ij} + \Gamma^i_{\ i0}T^{00} ,
\eea
where we have dropped second order terms.  
For pedagogical reasons, let us evaluate each term explicitly
\beal{eq:zeroterms}
T^{00} \eal (1 + {\delta_x} Q - 2\Psi Q ) \, (a_0/a)^2 \rho_x ,\nonumber\\
\partial_0 T^{00} \eal [(1+\delta_x Q-2\Psi Q) 
  ({\dot\rho_x \over \rho_x} -2 {\dot a \over a})  + (\dot \delta_x
  - 2\dot \Psi Q)] \, (a_0/a)^2 \rho_x ,\nonumber\\
T^{i0}_{\ \ |i} \eal (1+w_x) kV_xQ \, (a_0/a)^2 \rho_x ,\nonumber\\
\Gamma^0_{\ 00}T^{00}
  \eal [{\dot a \over a} (1 + \delta_x Q - 2\Psi Q)
  + \dot\Psi Q] \, (a_0/a)^2 \rho_x ,\nonumber\\
\Gamma^0_{\ ij}T^{ij} \eal 3w_x [{\dot  a \over a} (1 
  + {\delta p_x \over p_x} Q - 2\Psi Q ) + \dot \Phi Q] \, (a_0/a)^2 \rho_x ,
	\nonumber\\    
\Gamma^i_{\ i0}T^{00} \eal 3[{\dot a \over a} (1 + \delta_x Q 
-2\Psi Q) + \dot\Phi Q] \, (a_0/a)^2 \rho_x,
\eea
where $w_x \equiv p_x/\rho_x$ gives the equation of state of the fluid.

The zeroth order equation becomes
\bel{eq:zerozero}
{\dot \rho_x \over \rho_x} = -3(1+w_x) {\dot a \over a}.
\ee
For a constant $w_x$, $\rho_x \propto a^{-3(1+w_x)}$, \ie\
$w_r = {1 \over 3}$ and $\rho_r \propto a^{-4}$ for the radiation,
$w_m \approx 0$ and $\rho_m \propto a^{-3}$ for the matter, and
$w_v = -1$ and $\rho_v = $ constant for the vacuum or cosmological constant
contribution. 
The first order equation is the continuity equation for perturbations,
\bel{eq:mattercontinuity}
\dot \delta_x = -(1+w_x)(kV_x + 3\dot\Phi) - 3 {\dot a \over a} 
\delta w_x,
%\left( {\delta p_x \over \delta \rho_x} -w_x  \right) \delta_x,
\ee
where the fluctuation in the equation of state
\beal{eq:deltaw}
\delta w_x \eal {p_x + \delta p_x \over \rho_x + \delta\rho_x} -  w_x
	\nonumber\\
\eal \left( {\delta p_x \over \delta \rho_x} -w_x  \right) \delta_x.
\eea
This may occur for example if the temperature of a non-relativistic
fluid is spatially varying and can be important at late times when 
astrophysical processes can inject energy in local regions.

We can recast equation \eqn{mattercontinuity} into the form
\bel{eq:numbercontinuity}
{d \over d\eta}\left({\delta_x \over 1+w_x}\right) = -(kV_x + 3\dot\Phi) - 3
{\dot a \over a} {w_x \over 1+w_x} \Gamma_x,
\ee	
where the entropy fluctuation is
\bel{eq:entropygamma}
w_x \Gamma_x = (\delta p_x/\delta\rho_x - c_{x}^2)\delta_x,
\ee
with the sound speed $c_x^2 \equiv  \dot p_x / \dot \rho_x$.
Here we have used the relation 
\beal{eq:dotw}
\dot w_x \eal {\dot \rho_x \over \rho_x}({\dot p_x \over \dot \rho_x} - w_x) 
\nonumber\\
\eal -3(1+w_x)(c_{x}^2 - w_x){\dot a \over a},
\eea
which follows from equation \eqn{zerozero}.  Entropy fluctuations are
generated if the fluid is composed of species for which both the equation
of state and the number density fluctuations differ. For a single particle
fluid, this term vanishes.  

Let us interpret  equation \eqn{numbercontinuity}.  
In the limit of an ultra-relativistic
or non-relativistic single particle fluid, the quantity
\bel{eq:deltaoverw}
{\delta_x \over 1+w_x} =
{\delta n_x \over n_x}  
\ee
is the number density fluctuation in the fluid.
Equation \eqn{numbercontinuity} thus 
reduces to the ordinary continuity equation for the number density of
particles in the absence of creation and annihilation processes.
Aside from the usual $kV_x$ term, there is a $3\dot\Phi$ term.
We have shown in \S \ref{ss-2redshift} 
that this term represents the stretching of space due
to the presence of space curvature, \ie\ the spatial metric
has a factor $a(1+\Phi)$.  Just as the expansion term $a$ causes an
$a^{-3}$ dilution of number density,
there is a corresponding perturbative effect of $3\Phi$ from 
the fluctuation.  For the radiation energy density, there is also
an effect on the wavelength which brings the total to $4\Phi$ 
as equation \eqn{mattercontinuity} requires.

\subsubsection{Euler Equation}

Similarly, the conservation of momentum equation is obtained from
the space component of the conservation equation,
\beal{eq:conservmomentum}
-\partial_0 T^{0i} \eal \partial_j T^{ji} + \Gamma^i_{\ \alpha\beta}T^{\alpha\beta}
+ \Gamma^{\alpha}_{\ \alpha\beta}T^{i\beta}   \nonumber\\ 
\eal T^{ji}_{\ \ |j} + \Gamma^i_{\ 00}T^{00} + 2\Gamma^i_{\ 0j}T^{0j} + 
\Gamma^0_{\ 00}T^{i0} + \Gamma^0_{\ 0j} T^{ij} + \Gamma^j_{\ j0}T^{i0}. 
\eea
Explicitly, the contributions are
\beal{eq:termsmomentum}
\partial_0 T^{0i} \eal \{ [(1+w_x)({\dot \rho_x 
\over \rho_x} -2 {\dot a \over a}) + \dot w_x] V_x + (1+w_x)\dot V_x\}Q^i \, (a_0/a)^2 \rho_x ,  
	\nonumber\\
T^{ij}_{\ \ |j} \eal [ - {\delta p_x \over p_x}
	+ {2 \over 3} (1-3K/k^2) \Pi_x] kw_x Q^i \, (a_0/a)^2 \rho_x , 
	\nonumber\\
\Gamma^i_{\ 00} T^{00} \eal -k\Psi Q^i \, (a_0/a)^2 \rho_x , 
 	\nonumber\\
\Gamma^0_{\ 0j} T^{ij} \eal -k\Psi Q^i \, (a_0/a)^2 p_x , 
 	\nonumber\\
\Gamma^i_{\ 0j} T^{0j} 
	\eal {\dot a \over a}(1+w_x)V_x Q^i \, (a_0/a)^2 \rho_x ,
	\nonumber\\
\eal \Gamma^0_{\ 00}T^{i0} 
	\nonumber\\
\eal {1 \over 3} \Gamma^j_{\ j0} T^{i0}.
\eea
These terms are all first order in the perturbation and
form the Euler equation
\bel{eq:eulermomentum}
\dot V_x = -{\dot a \over a} (1-3w_x)V_x 
- {\dot w_x \over 1+w_x} V_x + {\delta p_x/\delta \rho_x \over 1+w_x} k\delta_x 
- {2 \over 3} {w_x \over 1+w_x} (1-3K/k^2)k\Pi_x + k\Psi.
\ee
Employing equation \eqn{dotw}  for the time variation of the equation
of state and equation \eqn{entropygamma} for the entropy, we can rewrite
this as
\bel{eq:eulermomentum2}
\dot V_x + {\dot a \over a} (1-3c_{x}^2) V_x  =
%+ {\delta p_x/\delta \rho_x \over 1+w_x} k\delta_x 
{c^2_x \over 1+w_x} k\delta_x + {w_x \over 1+w_x}k\Gamma_x
- {2 \over 3} {w_x \over 1+w_x} (1-3K/k^2)k\Pi_x + k\Psi. \quad
\ee
The gradient of the gravitational potential provides a source to velocities
from infall.  The expansion causes a drag term on the matter but not
the radiation.  This is because the expansion redshifts particle
 momenta as $a^{-1}$.
For massive particles, the velocity consequently decays as $V_m \propto
a^{-1}$.  For radiation, the particle energy or equivalently the
temperature of the distribution redshifts.
The bulk velocity $V_r$ represents a {\it fractional} temperature fluctuation
with a dipole signature.  Therefore, the decay scales out. 
Stress in the fluid, both isotropic (pressure) and anisotropic,
prevents gravitational infall.
The pressure contribution is separated into an acoustic part 
proportional to the sound speed $c_x^2$ and an entropy part which
contributes if the fluid is composed of more than one particle species.
 
\subsection{Total Matter and Its Components}
\label{ss-4total}

If the fluid $x$ in the last section is taken to be the total matter
$T$, equations \eqn{mattercontinuity} and \eqn{eulermomentum2} 
describe the evolution of the whole system.  However, even considering
the metric fluctuations $\Psi$ and $\Phi$ as external fields, the system
of equations is not closed since the anisotropic stress $\Pi_T$ and
the entropy $\Gamma_T$ remain to be defined.  The fluid must
therefore be broken into particle components for which these quantities
are known.  

We can reconstruct the total matter variables from the components via
the relations,
\beal{eq:totalconstruct}
\rho_T\delta_T \eal \sum_i \rho_i \delta_i, \\
\delta p_T \eal \sum_i  \delta p_i, \\
\label{eq:velocitytotal}
(\rho_T + p_T)V_T \eal \sum_i (\rho_i + p_i)V_i, \\
p_T \Pi_T \eal \sum_i p_i \Pi_i , \\
\dot \rho_T c_{T}^2 \eal \sum_i \dot \rho_i c^2_{i},
\eea
which follow from the form of the stress-energy tensor.  Vacuum
contributions are usually not included in the total matter.
Similarly, the
entropy fluctuation can be written
\beal{eq:totalentropy}
p_T \Gamma_T \eal 
\delta p_T - {\dot p_T \over \dot\rho_T}
	\delta\rho_T \nonumber\\
	     \eal \sum_i \delta p_i - {\dot p_i \over \dot \rho_i}
		\delta \rho_i + \left({\dot p_i \over \dot \rho_i}
		- {\dot p_T \over \dot \rho_T} \right) \delta \rho_i 
		       \nonumber\\
	     \eal \sum_i p_i \Gamma_i + (c_{i}^2 - c_{T}^2)\delta\rho_i.
\eea
Even supposing the entropy of the individual fluids vanishes, there
can be a non-zero $\Gamma_T$ due to differing density contrasts between
the components which have different equations of state $w_i$.  
If the universe consists of non-relativistic matter
and fully-relativistic radiation only, there are only two relevant equations
of state $w_r = 1/3$ for the radiation and $w_m \approx 0 $ for the matter.
The relative entropy contribution
then becomes, 
\bel{eq:entropymr}
\Gamma_T = -{4 \over 3}{1-3w_T \over 1+w_T} S,
\ee
where the $S$ is the fluctuation in the matter to radiation number density
\bel{eq:Sdefinition}
S = \delta(n_m/n_r) = \delta_m - {3 \over 4} \delta_r,
\ee
and is itself commonly referred to as the entropy fluctuation for obvious
reasons. 

Although covariant conservation applies equally well to 
particle constituents as to the total fluid, we have assumed in the last
section that the species were non-interacting.  
To generalize the conservation equations, we must consider momentum
transfer between components.  Let us see how this is done. 
\vskip 0.5truecm
\subsection{Radiation}
\label{ss-4radiation}

In the standard model for particle physics, the universe contains 
photons and three flavors of massless neutrinos as its radiation 
components.  For the photons, we must consider the momentum
transfer with the baryons through Compton 
scattering. We have in fact already obtained the full
evolution equation for the photon
component through the derivation of the Boltzmann equation in 
Chapter \ref{ch-Boltzmann}.   In real space, the temperature fluctuation
is 
given by [see equation \eqn{boltzmannfinal}]
\beal{eq:Gen}
 {d \over d\eta}(\Theta+\Psi) \!\!\! &\equiv& \!\!\!
 {\dot \Theta + \dot \Psi}  + \dot x^i {\partial \over {\partial x^i}}
   (\Theta + \Psi) + {\dot \gamma^i} {\partial \over {\partial \gamma^i}}
   (\Theta + \Psi) \nonumber\\
\eal \dot\Psi - \dot \Phi + {\dot \tau}(\Theta_0 - \Theta +
\gamma_i v^i_b + {1 \over 16}\gamma_i \gamma_j\Pi^{ij}_{\gamma}), 
\eea
recall that 
$\tau$ is the Compton
optical depth, $\Theta_0 = \delta_\gamma/4$ is the
isotropic component of $\Theta$,
and $\Pi^{ij}_{\gamma}$ the quadrupole moments of the photon energy density 
are given by 
equation \eqn{Piij}.

The angular fluctuations in a given spatial mode $Q$ can be 
expressed by the multipole decomposition  of equation \eqn{Decompositionconv}
\bel{eq:LDecomposition}
\Theta(\eta,\bx,\bg) = \sum_{\ell=0}^{\infty} \Theta_\ell(\eta)
M_\ell^{-1/2} G_\ell (\bx,\bg).
\ee
Be employing the recursion relations
\eqn{AeqnLLPM}, we can break equation \eqn{Gen} into the standard
hierarchy of 
coupled equations for the $\ell$-modes:
\beal{eq:Hierarchy}
{\dot \Theta_0} \eal -{k \over 3}\Theta_1 -{\dot \Phi},  \nonumber\\
{\dot \Theta_1} \eal k\left[ \Theta_0 + \Psi -{2 \over 5}
	K_2^{1/2} \Theta_2 \right] 
	- {\dot \tau}( {\Theta_1 - V_b}), \nonumber\\
{\dot \Theta_2} \eal k \left[ {2 \over 3} K_2^{1/2} \Theta_1
- {3 \over 7} K_3^{1/2} \Theta_3 \right]
- {9 \over 10} {\dot \tau \Theta_2}, \nonumber\\
{\dot \Theta_\ell} \eal k \left[ {\ell \over 2\ell-1}K_\ell^{1/2} 
\Theta_{\ell-1}
- {\ell+1 \over 2\ell+3} K_{\ell+1}^{1/2}
\Theta_{\ell+1} \right] - {\dot \tau}\Theta_\ell, \quad (\ell > 2)
\eea
where $\gamma_i v^i_b(\bx) =  V_b G_1(\bx,\bg)$ and
recall $K_\ell = 1-(\ell^2-1)K/k^2$. 
Since $V_\gamma=\Theta_1$, comparison with equation \eqn{eulermomentum2}
gives the relation between the anisotropic stress perturbation of the
photons and the quadrupole moment
\bel{eq:anisopi}
\Pi_\gamma = {12 \over 5} (1-3K/k^2)^{-1/2} \Theta_2.
\ee
Thus anisotropic stress is generated by the streaming of radiation from
equation \eqn{Hierarchy} once the mode enters the horizon $k\eta \simgt 1$.
The appearance of the curvature term is simply an artifact of our convention
for the multipole moment normalization.  For supercurvature modes,
it is also a convenient rescaling of the anisotropic stress since
in the Euler equation \eqn{eulermomentum2}, the term
$(1-3K/k^2)k\Pi_\gamma = 12(k^2-3K)^{1/2}\Theta_2/5$ is manifestly
finite as $k\rightarrow 0$.

By analogy to equation \eqn{Hierarchy},
we can immediately write down the corresponding  Boltzmann equation
for (massless) neutrino temperature perturbations $N(\eta,\bx,\bg)$
with the replacements
\bel{eq:replacements}
\Theta_\ell  \rightarrow N_\ell, \dot \tau \rightarrow 0,
\ee
in equation~\eqn{Hierarchy}.
This is sufficient since neutrino decoupling occurs before any scale of interest
enters the horizon.

\subsection{Matter}
\label{ss-4matter}

There are two non-relativistic components of dynamical 
importance to consider: the baryons and 
collisionless cold dark matter.
The collisionless evolution equations for the baryons are given by
\eqn{mattercontinuity} and \eqn{eulermomentum2} with $w_b \approx 0$
if $T_e /m_e \ll 1$.  However, before recombination, Compton scattering
transfers momentum between the photons and baryons.  It is unnecessary 
to
derive the baryon transport equation from first principles since the momentum
of the total photon-baryon fluid is still conserved.  
Conservation of momentum yields
\bel{eq:momentumtransfer}
(\rho_\gamma+p_\gamma)\delta V_\gamma = {4 \over 3} \rho_\gamma \delta V_\gamma
= \rho_b \delta V_b.
\ee 
Thus equations \eqn{mattercontinuity}, \eqn{eulermomentum2} and 
\eqn{Hierarchy} imply
\beal{eq:Baryon}
\dot \delta_b \eal -kV_b - 3 \dot\Phi, \nonumber\\
\dot V_b \eal -\dotaa V_b + k\Psi + \dot \tau (V_\gamma - V_b)/R,
\eea
where $R= 3\rho_b / 4\rho_\gamma$.  
The baryon continuity equation can also be
combined with the photon continuity equation [$\ell=0$ in \eqn{Hierarchy}]
to obtain
\bel{eq:Baryon2}
\dot \delta_b = -k(V_b - V_\gamma) + {3 \over 4} \dot \delta_\gamma.
\ee
As we shall see, this is useful since it has a gauge invariant 
interpretation: it represents
the evolution of the number density or entropy fluctuation [see
equation~\eqn{Sdefinition}].  
%Before recombination, scattering
%makes $V_b=V_\gamma$, which shows that the photons and baryons evolve
%adiabatically $\dot\delta_\gamma = {4 \over 3} \dot\delta_b$.  
Finally,
any collisionless non-relativistic component can described with
equation \eqn{Baryon} by dropping the interaction term $\dot\tau$.  
The equations can also be obtained 
from \eqn{mattercontinuity} and \eqn{eulermomentum2}
by noting that for a collisionless massive particle, the pressure, sound
speed and entropy fluctuation may be ignored.

\subsection{Einstein Equations}
\label{ss-4einstein}

The Einstein equations close the system by expressing the time evolution of
the metric in terms of the matter sources,
\bel{eq:einstein}
G_{\mu\nu} = 8\pi G T_{\mu\nu},
\ee
where $T_{\mu\nu}$ is now the total stress-energy tensor (including
any vacuum contributions).
The background equations give matter conservation for the space-space equation.
This is already contained in
equation \eqn{zerozero}. The time-space component vanishes leaving only
the time-time component
\bel{eq:einsteinhubble}
\left( {\dot a \over a} \right)^2 + K = {8 \pi G \over 3} 
\left( {a \over a_0} \right)^2( \rho_T + \rho_v),
\ee
where $\rho_v$ is the vacuum contribution and we have
used equation \eqn{Gmunubackground}.  This evolution equation 
for the scale factor is often written in terms of the Hubble parameter,
\beal{eq:Hubbledefinition}
H^2 \eadef \left( {1 \over a} {d a \over dt} \right)^2 
 = \left( {\dot a \over a} {a_0 \over a} \right)^2 \nonumber\\
\eal \left( {a_0 \over a} \right)^4 { a_{eq} + a \over a_{eq} + a_0 }
\Omega_0 H_0^2 
- \left({a_0 \over a}\right)^2 K
+ \Omega_\Lambda H_0^2 ,
\eea
where recall
$\Omega_0 = \rho_T/\rho_{crit}$ 
and $\Omega_\Lambda = \rho_v/\rho_{crit}$ 
with $\rho_{crit} = 3 H_0^2/8\pi G$.
Here $a_{eq}$ is the epoch of matter-radiation equality.  
Notice that as a function of $a$, the expansion will be 
dominated successively by radiation, matter, curvature, and $\Lambda$.
Of course, either or both of the latter terms may be absent in the
real universe.

The first order equations govern the evolution of $\Psi$ and $\Phi$.
They are the
time-time term,
\bel{eq:timetimeeinstein}
3 \left( {\dot a \over a} \right)^2 \Psi 
- 3 {\dot a \over a} \dot \Phi  - (k^2-3K)\Phi  =
 -{4\pi G } \left( {a \over a_0} \right)^2 \rho_T\delta_T,
\ee
the time-space term,
\bel{eq:timespaceeinstein}
{\dot a \over a}\Psi -\dot\Phi = 4\pi G 
\left( {a \over a_0} \right)^2 (1+w_T) \rho_T V_T/k,
\ee
and the traceless space-space term
\bel{eq:spacespaceeinstein}
k^2(\Psi + \Phi) = -8\pi G  
\left( {a \over a_0} \right)^2 p_T \Pi_T.
\ee
The other equations express the conservation laws which we have
already found.
Equations \eqn{timetimeeinstein} and \eqn{timespaceeinstein} can be
combined to form the generalized Poisson equation
\bel{eq:poissoneinstein}
(k^2 - 3K)\Phi = 
 {4\pi G} \left( {a \over a_0} \right)^2 \rho_T [\delta_T + 3{\dot a \over a}
(1+w_T)V_T/k]. 
\ee
Equations \eqn{spacespaceeinstein} and \eqn{poissoneinstein} form the
two fundamental evolution equations for metric perturbations in
Newtonian gauge.

Notice that the form of \eqn{poissoneinstein} reduces to the ordinary 
Poisson equation of Newtonian mechanics if the last term in 
the brackets is negligible.
Employing the matter continuity equation \eqn{mattercontinuity}, 
this occurs when  $k\eta \gg 1$, \ie\ when the fluctuation is well
inside the horizon as one would expect.  This extra piece represents
a relativistic effect and depends on the frame of reference in which the     
perturbation is defined.  This suggests that we can simplify the form
and interpretation of the evolution equations by a clever choice of
gauge. 

\section{Gauge}
\label{sec-4gauge}

\begin{quote}

\footnotesize\it
Sayings from a perspective work nine times out of ten, wise sayings
work seven times out of ten.  Adaptive sayings are new every day,
smooth them out on the whetstone of Heaven.
\vskip 0.1truecm
\centerline{--Chuang-tzu, 27}

\end{quote}

Fluctuations are defined on hypersurfaces of constant time.  Since in
general relativity, we can choose the coordinate system arbitrarily,
this leads
to an ambiguity in the definition of fluctuations referred to as {\it gauge
freedom}.  
There is {\it no} gauge invariant meaning to density fluctuations. 
For example, even a completely homogeneous and isotropic 
Friedmann-Robertson-Walker space can be expressed with an inhomogeneous
metric by choosing an alternate time slicing that is warped 
(see Fig.~\ref{fig:4gauge}). 
Conversely, a fluctuation can be thought of as existing in a homogeneous
and isotropic universe where the initial time slicing is altered 
(see \S \ref{ss-5gensol}).
Two principles are worthwhile to keep in mind when considering the gauge:

\begin{enumerate}

\item Choose a gauge whose coordinates are completely fixed.

\item Choose a gauge where the physical interpretation and/or form of
the evolution is simplest.

\end{enumerate}
\noindent
The first condition is the most important.  Historically, much confusion
has arisen from the use of a particular gauge choice, the synchronous 
gauge, which {\it alone} does not fix the coordinates entirely \cite{PV}.  
An ambiguity in the mapping onto this gauge appears, for example, 
at the initial conditions.  Usually this problem is solved by completely
specifying the initial hypersurface. 
Improper mapping can lead to artificial ``gauge modes'' 
in the solution.  
The second point is that given gauge freedom exists, 
we may as well exploit it
by choosing one which simplifies
either the calculation or the interpretation.   It turns out that
the two often conflict.  For this reason, we advocate
a hybrid choice of representation for fluctuations.  

\input chap4/gauge.tex
How is a hybrid choice implemented?  This is the realm of the
so-called ``gauge invariant'' formalism.  
Let us consider for a 
moment the meaning of the term gauge invariant.
If the coordinates are completely specified, the fluctuations are 
real geometric objects and may be represented in any coordinate system.  
They are therefore manifestly gauge invariant.  However, in the new frame 
they 
may take on a different  {\it interpretation}, \eg\ density fluctuations in 
general will not remain density fluctuations.  The ``gauge invariant''
program reduces to the task of 
writing down fluctuations in a given gauge in terms of quantities in an
arbitrary gauge.  It is therefore a problem in mapping.  The only
quantities that are not ``gauge invariant'' in this sense are 
those that are ill defined, \ie\ represent fluctuations in a gauge
whose coordinates have not been completely fixed.  This should be
distinguished from objects that actually have a gauge invariant 
interpretation.
As we shall see, quantities such as anisotropies of $\ell \ge 2$ are the
same in any frame.  This is because the coordinate system is defined
by a scalar function in space to describe the ``warping'' of the
time slicing and a 
vector to define the ``boost,'' leaving higher order quantities invariant.

\subsection{Gauge Transformations}

The most general form of a metric perturbed by scalar
fluctuations is \cite{KS84}
\beal{eq:generalmetric}
g_{00} \eal -(a/a_0)^2[1+2A^GQ], \nonumber\\
g_{0j} \eal -(a/a_0)^2 B^GQ_j,  \nonumber\\
g_{ij} \eal (a/a_0)^2 [\gamma_{ij} + 2H_L^G Q \gamma_{ij} + 2H_T^G Q_{ij}] ,
\eea
where the superscript $G$ is employed to remind the reader that the
actual values vary from gauge to gauge.
A gauge transformation is a change in the correspondence between the
perturbation and the background  represented by the coordinate shift
\beal{eq:gaugetransformation}
\tilde \eta \eal \eta + TQ, \nonumber\\
\tilde x^i \eal x^i + LQ^i.
\eea
$T$ corresponds to a choice in time slicing and $L$ the choice of the
spatial coordinate grid.  They transform the metric as
\beal{eq:gaugemetric}
\tilde g_{\mu\nu}(\eta,x^i) \eal {\partial x^\alpha \over \partial
\tilde x^\mu}{\partial x^\beta \over \partial \tilde x^\nu}
g_{\alpha\beta}(\eta-TQ,x^i-LQ^i) \nonumber\\
\eapp g_{\mu\nu}(\eta, x^i) 
	+ g_{\alpha\nu}\delta x^\alpha_{\ ,\mu} 
	+ g_{\alpha\mu}\delta x^\alpha_{\ ,\nu} 
	- g_{\mu\nu,\lambda}\delta x^\lambda.
\eea
From this, we obtain the relations for the metric fluctuations
\beal{eq:gaugeperturbations}
A^{\tilde G} \eal A^G - \dot T - {\dot a \over a}T, \nonumber\\
B^{\tilde G} \eal B^G + \dot L + kT, \nonumber\\
H_L^{\tilde G} \eal H_L^G - {k \over 3}L - {\dot a \over a}T, \nonumber\\
H_T^{\tilde G} \eal  H_T^G + kL.
\eea
An analogous treatment of the stress energy tensor shows that 
\beal{eq:gaugedensities}
v_x^{\tilde G} \eal v_x^G + \dot L, \nonumber\\
\delta_x^{\tilde G} \eal \delta_x^G + 3(1+w_x){\dot a \over a} T, \nonumber\\
\delta p_x^{\tilde G}  \eal \delta  p_x^G + 3c_{x}^2 \rho_x 
(1+w_x){\dot a \over a} T, \nonumber\\
\Pi_x^{\tilde G} \eal \Pi_x^G.
\eea
Therefore any ambiguity in the time slicing $T$ leads to freedom in
defining the density contrast $\delta_x$.  Notice that the anisotropic 
stress $\Pi_x$ has a truly gauge invariant meaning as does any
higher order tensor contribution.  Furthermore, relative quantities
do as well, \eg\
\beal{eq:relative}
{\delta_x^{\tilde G} \over 1+w_x} - {\delta_y^{\tilde G} \over 1+w_y}
\eal { \delta_x^{G} \over 1+w_x} - { \delta_y^G  \over 1 + w_y}, \nonumber\\
v_x^{\tilde G} -  v_y^{\tilde G} \eal v_x^G - v_y^G, \nonumber\\
\Gamma_x^{\tilde G} \eal \Gamma_x^G,
\eea
the relative number density, velocity, and entropy fluctuation. We hereafter
drop the superscript from such quantities. 

\subsection{Newtonian Gauge}

In the Newtonian gauge, $B^N=H_T^N=0$.  
Physically, it is a time slicing in which 
the expansion
is isotropic.
This considerably simplifies the interpretation of effects
such as gravitational infall and redshift.  
From an arbitrary coordinate system $G$,
the Newtonian gauge is reached by employing [see equation 
\eqn{gaugeperturbations}]
\beal{eq:newtoniantrans}
T \eal -B^G/k + \dot H_T^G/k^2,  \nonumber\\
L \eal - H_T^G/k. 
\eea
From equations \eqn{gaugeperturbations} and  \eqn{gaugedensities}, the 
fundamental perturbations on this choice of hypersurface slicing
are 
\beal{eq:psiphigauge}
\Psi \equiv A^N 
%\eal A - {1 \over a} {d \over d\eta}(aT) \nonumber\\
\eal A^G + {1 \over a} {d \over d\eta}[aB^G/k - a\dot H_T^G/k^2], \nonumber\\
\Phi \equiv  H_L^N \eal 
H_L^G + {1 \over 3} H_T^G + {\dot a \over a}(B^G/k-\dot H_T^G/k^2), \nonumber\\
\delta_x^N \eal \delta_x^G + 3(1+w_x){\dot a \over a} 
(-B^G/k + \dot H_T^G/k^2), \nonumber\\
\delta p_x^N \eal \delta p_x^G + 3 c_{x}^2 \rho_x(1+w_x)
{\dot a \over a}(-B^G/k + \dot H_T^G/k^2), \nonumber\\
V_x \equiv v_x^N \eal v_x^G - \dot H_T^G/k.
\eea
This is commonly referred to as the ``gauge invariant'' definition
of Newtonian perturbations.  Note that the general form of the
Poisson equation becomes
\bel{eq:generalpoisson}
\Phi = 4\pi G \left( {a \over a_0} \right)^2 \rho_T
\left({\delta_T^G + 3 {\dot a \over a} (1+w_T) (v_T^G-B^G)/k} \right).
\ee
As we have seen, density perturbations in this gauge grow due to infall
into the potential $\Psi$ and metric stretching effects from $\Phi$.
In the absence of changes in $\Phi$, they will therefore not
grow outside the horizon since causality prevents infall growth.

\subsection{Synchronous Gauge}

The synchronous gauge, defined by $A^S=B^S=0$ is a popular and in many cases
computationally useful choice.  The condition $A^S=0$ implies that proper 
time corresponds with coordinate time, and $B^S=0$ that constant space
coordinates are orthogonal to constant time hypersurfaces.  
This is the natural coordinate system for freely falling observers.

From an arbitrary coordinate choice, the synchronous condition is
satisfied by the transformation
\beal{eq:synchtrans}
T \eal a^{-1} \int d\eta a A^G + c_1 a^{-1}, \nonumber\\
L \eal -\int d\eta (B^G + kT) + c_2 ,  
\eea
where $c_1$ and $c_2$ are integration constants.  There is therefore
residual gauge freedom in synchronous gauge.  
It manifests itself as a degeneracy in the mapping of fluctuations
onto the synchronous gauge and appears, for example as an ambiguity
in $\delta_x^S$ of $3(1+w_x)c_1 \dot a/a^2$.  This represents
an unphysical gauge mode.  To eliminate it, one 
must carefully define the initial conditions.

It is a simple exercise in algebra to transform the evolution equations
from Newtonian to synchronous representation.  The metric perturbations
are commonly written as 
\beal{eq:synchmetric}
h_L \eadef 6 H_L^S, \nonumber\\
\eta_T \eadef - H_L^S - {1 \over 3} H_T^S .
\eea
Equation \eqn{synchtrans} tells us that
\beal{eq:synchtimeslice}
T \eal -\dot L/k = (v_x^N - v_x^S)/k \nonumber\\
  \eal {1 \over 2} (\dot h_L + 6\dot\eta_T)/k^2,
\eea
from which it follows
\bel{eq:phiform1}
\dot \Phi = {1 \over 6} \dot h_L - k(v_x^N-v_x^S)/3 + {d \over d\eta}
[ {\dot a \over a} \vsn/k].
\ee
Furthermore, the density and pressure relations
\beal{eq:synchdensitypressure}
\delta_x^N \eal \delta_x^S - 3(1+w_x)\dotaa \vsn/k, \nonumber\\
\delta p_x^N \eal \delta p_x^S - 3(1+w_x)c_{x}^2 \rho_x \dotaa \vsn/k ,
\eea
and equation \eqn{dotw} yields
\bel{eq:synchdotd}
\dot \delta_x^N = \dot \delta_x^S - (1+w_x) \left\{
 3(\dot \Phi - {1 \over 6} \dot h_L) + \left[k^2 - 9(c_{x}^2-w_x)
\left(\dotaa\right)^2 \right] \vsn/k \right\},
\ee
and
\bel{eq:synchvphi}
	 3 \dotaa \left( { \delta p_x^N \over \delta \rho_x^N}
			- w_x\right) \delta_x^N =  
	 3 \dotaa \left( { \delta p_x^S \over \delta \rho_x^S}
			- w_x\right) \delta_x^S + 
9(1+w_x)(c_{x}^2-w_x)\left(\dotaa\right)^2 \vsn/k.
\ee
Thus the continuity equation of \eqn{mattercontinuity} becomes
\bel{eq:synchcontinuity}
\dot \delta_x^S = -(1+w_x) (kv_x^S + \dot h_L/2) - 3 \dotaa 
	\left( { \delta p_x^S \over \delta \rho_x^S} - w_x \right) 
	\delta_x^S.
\ee
Likewise with the relation
\bel{eq:synchvelinfall}
\dot v_x^S + \dotaa v_x^S = \dot v_x^N + \dotaa v_x^N - k\Psi,
\ee
and equation \eqn{dotw},
the transformed Euler equation immediately follows:
\bel{eq:syncheuler}
\dot v_x^S = -\dotaa (1-3w_x)v_x^S - {\dot w_x \over 1+w_x}
v_x^S + {\delta p_x^S /\delta \rho_x^S \over 1+w_x} k\delta_x^S 
- {2 \over 3} {w_x \over 1+w_x}(1-3K/k^2)k\Pi_x.
\ee
Finally, one can also work in the reverse direction and obtain the
Newtonian variables in terms of the synchronous gauge perturbations.
Given the residual gauge freedom, this is a many to one mapping.
The Newtonian metric perturbation follows 
from equation \eqn{generalpoisson},  $B^S=0$, and the
gauge invariance of $\Pi_T$:
\beal{eq:synchpoisson}
(k^2-3K)\Phi \eal 4\pi G \left( {a \over a_0} \right)^2 \rho_T [\delta_T^S 
+ 3 \dotaa (1+w_T) v_T^S/k], \nonumber\\
k^2(\Psi + \Phi) \eal -8\pi G \left( {a \over a_0} \right)^2 p_T \Pi_T.
\eea
They can also be written in terms of
the synchronous gauge metric perturbations as
\beal{eq:synchpot}
\Psi \eal {1 \over 2k^2} \left[ \ddot h_L + 6\ddot\eta_T 
	+ {\dot a \over a}(\dot h_L + 6\dot\eta_T) \right], \nonumber\\
\Phi \eal -\eta_T + {1 \over 2k^2} {\dot a \over a}(\dot h_L + 6\dot \eta_T).
\eea
In fact, equations \eqn{synchpoisson} and \eqn{synchpot} close the system
by expressing the time evolution of the metric variables $\eta_T$ and 
$h_L$ in terms of the matter sources.

Now let us return to the gauge mode problem.  The time slicing freedom can be
fixed by a choice of the initial hypersurface.  The natural choice is one
in which the velocity vanishes $v_x^S(\eta_i)=0$ for some set of ``observer''
particle species $x$.  This condition fixes
$c_1$ and removes the gauge ambiguity in the density perturbations.  Notice
also that the synchronous gauge has an elegant property.  Since it is
the coordinate system of freely falling observers, if the velocity of a 
{\it non-interacting} pressureless species 
is set to zero initially it will 
remain so.  In the Euler equation \eqn{syncheuler}, the infall
term that sources velocities has been transformed away by equation
\eqn{synchvelinfall}.  Thus in the 
absence of pressure and entropy terms, there are no sources to the velocity.

The synchronous gauge therefore represents a ``Lagrangian'' 
coordinate system as opposed to the more ``Eulerian'' choice of
a Newtonian coordinate system.  In this gauge, the coordinate 
grid follows freely falling particles so that density growth due to 
infall is transformed into dilation effects from the stretching of the
grid.  
Although the coordinate grid must be redefined when particle
trajectories cross, this does not occur in linear perturbation
theory if the defining particles are non-relativistic.  
Thus in synchronous gauge, the dynamics
are simpler since we employ the rest frame of the collisionless
matter.  The only drawback to this gauge choice is that physical intuition
is more difficult to obtain since we have swept dynamical effects
into the behavior of the coordinate grid.

\subsection{Total Matter Gauge}

As an obvious extension of the ideas which make the synchronous gauge
appealing, it is convenient to employ the rest frame of the total rather than
collisionless matter.  The total matter velocity is thus
set to be orthogonal to the constant time hypersurfaces
 $v_T^T=B^T$. With the additional constraint
$H_T^T=0$, the transformation is obtained by
\beal{eq:totalmattertrans}
T \eal (v_T^G-B^G)/k , \nonumber\\
L \eal -H_T^G/k,
\eea
which fixes the coordinates completely.  The matter perturbation 
quantities become
\beal{eq:totalmatterpert}
\Delta_x \equiv \delta_x^T \eal \delta_x^G + 3(1+w_x) \dotaa (v_T^G - B^G)/k, 
	\nonumber\\
\delta p_x^T \eal \delta p_x^G + 3(1+w_x) c_{x}^2 \rho_x \dotaa (v_T^G-B^G)/k, 
	\nonumber\\
V_x \equiv V_x^T \eal v_x^G - \dot H_T^G/k.
\eea
Notice that the Newtonian gauge $B^N= H^T_N = 0$ 
and $v_x^T= v_x^N = V_x$.  In 
synchronous gauge, $B^S=0$ as well.  If the rest frame of the total matter
is the same as the collisionless non-relativistic matter, as is
the case for adiabatic conditions, $\delta^S_x \approx \Delta_x^T$ 
if $v_x^S(0) = 0$.

The evolution equations are easily obtained from Newtonian gauge with
the help of the following relations,
\bel{eq:derivhub}
{d \over d\eta}\left(\dotaa\right) = -{1\over2} \left[ \left(\dotaa\right)^2 + K \right]
(1+3w_T) + {3 \over 2}(1+w_T) \left({ a \over a_0}\right)^2 
\Omega_\Lambda H_0^2,
\ee
which follows from equation \eqn{einsteinhubble} and
\bel{eq:timespace2}
\dotaa \Psi - \dot \Phi = {3 \over 2} \left[ \left(\dotaa\right)^2 + K
- \left({ a \over a_0}\right)^2\Omega_\Lambda H_0^2  \right]
(1+w_T)V_T/k 
\ee
from equation \eqn{timespaceeinstein}.  The Newtonian Euler equation can 
also be rewritten as
\beal{eq:newtonianeulerform}
{d \over d\eta}\left( \dotaa (1+w_T)V_T \right) \eal - \left( \dotaa \right)^2
(1-3w_T)(1+w_T)V_T + \dotaa {\delta p_T^N \over \delta \rho_T^N} k\delta_T^N 
	\nonumber\\
& & -{2 \over 3} \dotaa w_T (1-3K/k^2)k\Pi_T + (1+w_T)\dotaa k\Psi 
	\nonumber\\
& & -{1 \over 2}(1+3w_T)(1+w_T) \left[\left( \dotaa\right)^2 +K \right] V_T 
	\nonumber\\
& & + {3 \over 2} (1+w_T)^2 \left( {a \over a_0} \right)^2 \Omega_\Lambda
H_0^2 V_T.
\eea
With this relation, the total matter continuity and Euler equations
readily follow,
\beal{eq:totalmatter}
\dot \Delta_T - 3w_T \dotaa \Delta_T \eal -(1-3K/k^2)(1+w_T)kV_T 
- 2 (1-3K/k^2 ) \dotaa w_T \Pi_T, \\
\dot V_T + \dotaa V_T \eal {c_T^2 \over 1+w_T}k\Delta_T + k\Psi +
 {w_T \over 1+w_T}k\Gamma_T  
 - {2 \over 3}(1-3K/k^2){w_T \over 1+w_T} k\Pi_T. \qquad
\eea
The virtue of this representation is that the evolution of the total matter
is simple.  This is reflected by the form of the Poisson equation,    
\beal{eq:totalmatterPoisson}
(k^2-3K)\Phi \eal 4\pi G \left({a \over a_0}\right)^2 \rho_T \Delta_T, \\
k^2(\Psi +\Phi) \eal -8\pi G \left({a \over a_0} \right)^2 p_T \Pi_T.
\eea
In the total matter rest frame, there are no relativistic effects from the 
velocity and hence the Poisson equation takes its non-relativistic form. 
Again the drawback is that the interpretation is muddled.

\subsection{Hybrid Formulation}

We have seen that the Newtonian gauge equations correspond closely 
with classical intuition and thus provide a simple representation for
relativistic perturbation theory.  However, since density 
perturbations grow by the causal mechanism of potential infall, we have
build a fundamental scale, the particle horizon, into the evolution.  
Frames that comove with the matter, \ie\ in which the particle velocity
vanishes, have no fundamental scale.  This simplifies 
the perturbation equations and in many cases admit scale invariant, \ie\ 
power law solutions (see \S \ref{ch-evolution}).   Two such frames are 
commonly employed: the rest frame of the collisionless non-relativistic
mater and that of the total matter. 
The former is implemented under a special choice of the synchronous gauge 
condition and the latter by the total matter gauge.  
For the case of adiabatic fluctuations, where non-relativistic and
relativistic matter behave similarly, they are essentially identical.
For entropy fluctuations, the total matter gauge is more ideal. 

Since we can express fluctuations on any given frame
by combination of variables on any other,  we can mix and match 
quantities to suit the purpose at hand.  
To be
explicit, we will hereafter employ total matter gauge 
density fluctuations $\Delta_x 
\equiv \delta^T_x$, but Newtonian temperature  $\Theta \equiv
\delta_\gamma^N/4$ and 
metric perturbations $\Psi$ and $\Phi$.  The velocity perturbation is
the same in both these frames, which we denote $V_x = v_x^N = v_x^T$.  
To avoid confusion, we will hereafter employ {\it only} this choice.  
We now turn to the solution of
these equations and their implications for the CMB.

%% file: chap4/X.tex
\begin{figure}[t]
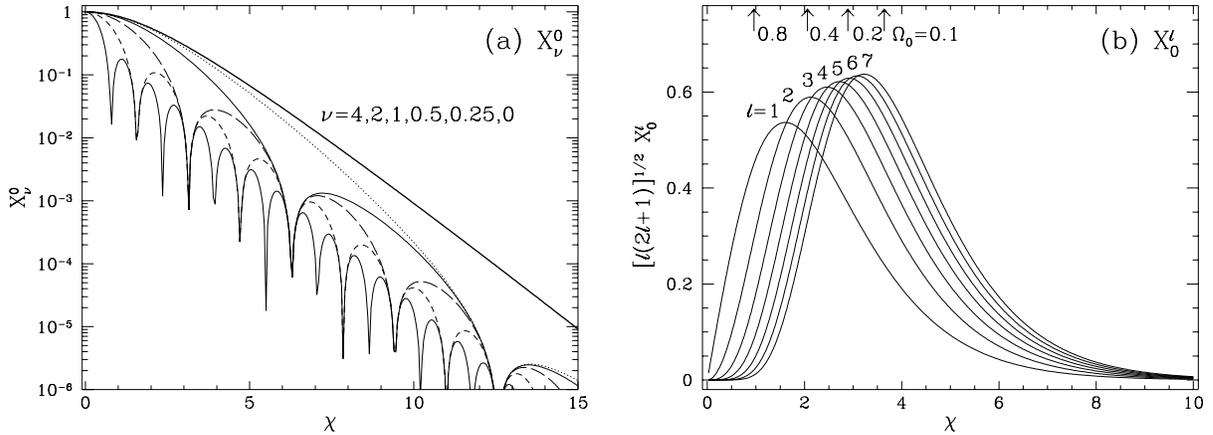

\centerline{ \hskip -0.5 truecm
\epsfxsize=3.0in \epsfbox{chap4/Xa.epsf} \hskip 0.5 truecm
\epsfxsize=3.0in \epsfbox{chap4/Xb.epsf}}
\vskip -0.5truecm
\caption{Open Radial Eigenfunctions}
\mycaption{(a) 
The isotropic $\ell = 0$ function for several
values of the wavenumber $\nu$.
The zero crossing moves out to arbitrarily large scales as
$\nu \rightarrow 0$, reflecting completeness.
However, the function retains prominent structure only
near the curvature scale $\chi \approx 1$.  A random superposition of
these low $\nu$ modes cannot produce more than exponentially
decaying structure larger than the curvature scale.
(b) Low order multipoles in the asymptotic limit $\nu \rightarrow 0$.
If most power lies on the curvature scale, the $\ell$-mode
corresponding to the angle that the curvature radius subtends will
dominate the anisotropy.  The normalization is appropriate for comparing
contributions to the anisotropy $\ell(2\ell+1)C_\ell/4\pi$.
Also shown is the location of the horizon $\chi = \eta_0 \sqrt{-K}$
for several values of $\Omega_0$.  If contributions to the anisotropy
come from a sufficiently early epoch, the dominant $\ell$-mode for 
the curvature scale will peak
at this value (see \eg\ Fig.~\ref{fig:5iso}).
}
\label{fig:4X}
\end{figure}

%% file: chap4/gauge.tex
\begin{figure}[t]
%\vphantom{marker} \vskip 0.5truecm
\centerline{ \hskip -1truecm
\epsfxsize=4.0in \epsfbox{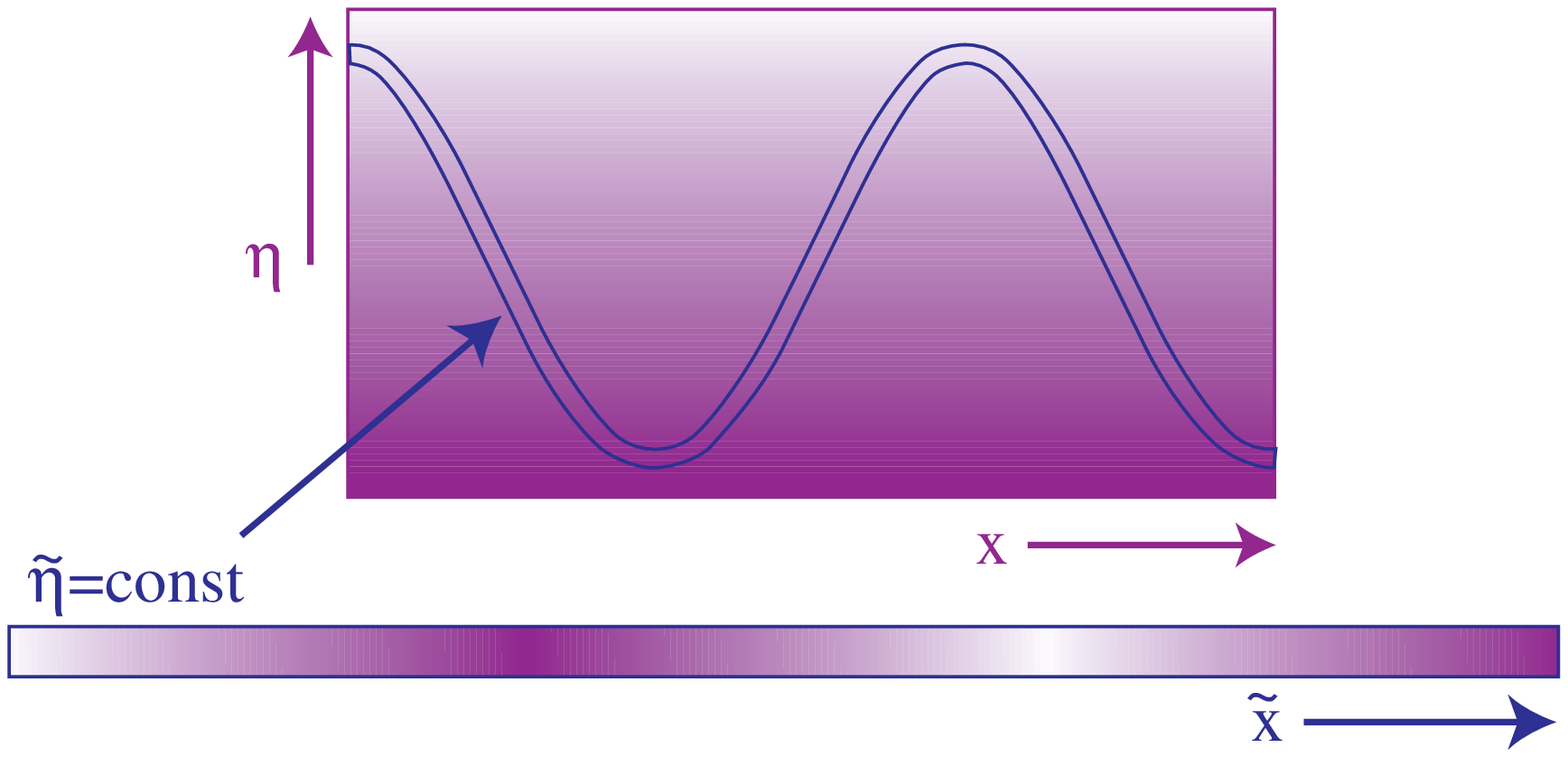}}
\vskip -0.5truecm
\caption{Gauge Ambiguity}
\mycaption{Gauge ambiguity refers to the freedom to choose the time
slicing on which perturbations are defined.  In this simple example,
a homogeneous FRW universe appears to have density perturbations for
a warped choice of time slicing.   One usually employs a set of
standard ``observers'' to define the coordinate slicing.  
The Newtonian gauge boosts observers into a frame where the expansion rate
looks isotropic (shear free).  
The synchronous gauge can be implemented to follow the 
collisionless non-relativistic particles.  The total matter gauge employs the
rest frame of the total energy density fluctuations.
}
\label{fig:4gauge}
\end{figure}

%% file: chap4b.tex
\chapter{Perturbation Evolution}
\label{ch-evolution}

\begin{quote}
\footnotesize\it
Although heaven and earth are great, their evolution is uniform. \\
Although the myriad things are numerous, their governance is unitary.
\vskip 0.1truecm
\centerline{--Chuang-tzu, 12}
\end{quote}

Superhorizon and subhorizon perturbation 
evolution take on simple asymptotic forms and interpretations
under the hybrid gauge representation developed in \S \ref{sec-4gauge}.
All component fluctuations {\it evolve} similarly above the horizon
and assume differing forms only due to the initial conditions.
We discuss the general solution to the perturbation equations valid
for an arbitrary mixture of initial curvature and entropy fluctuations
in a universe that passes from radiation to matter to curvature and/or
cosmological constant domination.
These two initial conditions distinguish the adiabatic and isocurvature
growing modes. 
Evolution during and after horizon crossing exhibits  more complicated
behavior. 
Well under the horizon but before recombination, photon
pressure in the Compton coupled photon-baryon fluid resists gravitational
compression and sets up acoustic waves.  In the intermediate case, 
gravity {\it  drives} the acoustic oscillations.  
The presence of baryons and radiation feedback
on the potentials alter the simple oscillatory form of the acoustic
wave.  These effects leave distinct signatures on CMB anisotropies 
in the degree to arcminute range.
After recombination, the baryons are released from Compton drag and their
density fluctuations can again grow by gravitational instability.
The discussion here of the evolutionary properties
of perturbations sets the stage for the analysis of anisotropy formation
in \S \ref{ch-primary} and \S \ref{ch-secondary}.

\section{Superhorizon Evolution}
\label{sec-4superhorizon}

\subsection{Total Matter Equation}
\label{ss-4totalmattereqn}

Only gravity affects the evolution of the matter and radiation above
the horizon scale in the total matter representation.  This greatly
simplifies
the evolution equations since we can treat all the particle species as
a combined total matter fluid without loss of information.  Let us
prove this assertion.  Specifically, we need to show that all particle
velocities are equal \cite{HSsmall}.  
Ignoring particle interactions which play no role
above the horizon, the Euler equations for pressureless
matter and radiation components
are given by
\beal{eq:eulermatterradiation}
\dot V_m \eal -{\dot a \over a}V_m + k\Psi, \\
\label{eq:eulerradiation}
\dot V_r \eal -{\dot a \over a}V_T + k\Psi + {1 \over 4}k\Delta_r,
\eea
where we have transformed the Newtonian Euler equation \eqn{eulermomentum2}
into the total matter representation with equation \eqn{totalmatterpert}.  We
have also neglected the small contribution from anisotropic stress.

Infall into potential wells 
sources the matter and radiation velocities alike.  
Although it attains its maximum effect near horizon crossing $k\eta \simgt 1$
due to causality,
the fact that a given eigenmode $k$ does not represent one physical
scale alone allows infall to generate a small velocity contribution of
${\cal O}(k\eta)$ when
$k\eta \simlt 1$. 
Expansion drag on the matter causes $V_m$ to decay as
$a^{-1}$.  However, the Euler equation for the radiation contains not only
a different expansion drag term but also pressure contributions
which prevent infall. Let us determine when pressure is important. 
The Poisson equation \eqn{totalmatterPoisson}
requires
\bel{eq:PoissonTM2}
(k^2-3K)\Phi = {3 \over 2}\left[ \left( \dotaa \right)^2 + K -   
 \left( {a \over a_0} \right)^2 \Omega_\Lambda H_0^2 \right] \Delta_T,
\ee
where we have employed the Hubble equation \eqn{einsteinhubble}.  
Since 
\bel{eq:dotaarmd}
{\dot a \over a} = {\cases {  1/\eta & RD \cr
		              2/\eta, & MD \cr }}
\ee
in the radiation-dominated (RD) and matter-dominated (MD) epochs,
to order of magnitude
\bel{eq:deltaTphirel}
\Delta_T \sim (k\eta)^2 \Phi,
\ee
before curvature or $\Lambda$ domination.
Since $\Psi \approx -\Phi$, pressure may be neglected compared with infall
outside the horizon where $k\eta \ll 1$.
This seemingly obvious statement 
is actually not true for the Newtonian gauge density perturbation 
since $\delta_T^N = {\cal O}(\Psi)$ if $k\eta \ll 1$.  
The appearance of the expansion drag term $V_T$
in equation \eqn{eulerradiation} is in fact due to the pressure
contributions in the Newtonian frame.  
Starting from arbitrary initial 
conditions and in the absence of infall, 
the expansion will damp away velocities until $V_T=V_m=V_r=0$.  The infall
source gives rise to equal velocities for all components. 

We can thus describe the coupled multi-component
system as a single fluid, defined by the total matter variables whose
behavior does not depend on the microphysics of the components.
Assuming the various species are all either fully  
relativistic or non-relativistic, \ie\ employing equations \eqn{Hierarchy}
and \eqn{Baryon} with their decoupled variants, we obtain
\beal{eq:Total}
\dot \Delta_T - 3w_T {\dot a \over a} \Delta_T \eal
-{\left( 1 - {3K \over k^2} \right)}(1+w_T)kV_T 
 - 2 \left( 1 -{3K \over k^2}\right) {\dot a \over a} w \Pi_T, \\
\label{eq:TotalEuler}
\dot V_T + {\dot a \over a}V_T \eal {4 \over 3} {w_T \over (1+w_T)^2} k
[\Delta_T - (1-3w_T)S] +k\Psi  \nonumber\\
& & - {2 \over 3}k
\left( 1 - {3K \over k^2}\right){w_T \over 1+w_T} \Pi_T,
\eea
where we have used the entropy relation \eqn{entropymr} and
recall $S \equiv \Delta_m - {3 \over 4}\Delta_r$.  The difference
in how $V_m$ and $V_r$ is damped by the expansion appears as an
entropy term in the total Euler equation. 
Again for superhorizon scales, we can ignore the pressure term $\propto
\Delta_T$ in the total Euler equation above.

The evolution of the entropy is given by
the continuity equation for the number density \eqn{Baryon2},  \ie\
$\dot S = k(V_r - V_m)$, where the matter and radiation velocities
are defined in a manner analogous to $V_T$
[see equation \eqn{velocitytotal}].  
Since all components have the same velocity,
$S$ is a constant before the mode enters the horizon and, if it is present,
must have been established at the initial conditions.

\subsection{General Solution}
\label{ss-5gensol}

\subsubsection{From Radiation to Matter Domination}

Before horizon crossing, radiation pressure may be neglected.  
Specifically this occurs at  $\dot a /a = k$ or
\bel{eq:horizoncr}
a_H = { 1 + \sqrt{1+ 8(k/k_{eq})^2 } \over 4(k/k_{eq})^2 }, \qquad {\rm RD/MD}
\ee
where $k_{eq} = (2\Omega_0 H_0^2 a_0 )^{1/2}$ is the scale that
passes the horizon at equality and $a_{eq} = 1$. 
Dropping the curvature and $\Lambda$ contribution to the expansion and
combining
the total continuity \eqn{Total} and Euler equations \eqn{TotalEuler} 
yields the second
order evolution equation
\bel{eq:BeqnCTotal}
\left\{ {d^2 \over da^2} - {f \over a}{d \over da}
+ {1 \over a^2}\left[\left({k \over k_{eq}}\right)^2
\left( {1 - {3K \over k^2}} \right)h - g
\right] \right\} \Delta_T = \left({k \over k_{eq}}\right)^2
\left( {1 - {3K \over k^2}} \right)j S,
\ee
where
\beal{BeqnFunctions}
f \eal {3a \over 4+3a}
        -{5 \over 2} {a \over 1+a}, \nonumber\\
g \eal 2 +{9a \over 4+3a} - {a \over 2}{6+7a \over (1+a)^2}, \nonumber\\
h \eal {8 \over 3} {a^2 \over (4+3a)(1+a)}, \nonumber\\
j \eal {8 \over 3} {a \over (4+3a)(1+a)^2}. 
\eea
Here we  have used $3w_T = (1+a)^{-1}$ and have 
dropped the anisotropic stress correction $\Pi_T$ 
(see Appendix \ref{ss-5largepotential}). 
The solutions to the homogeneous equation with $S=0$ are given by
\beal{eq:BeqnUAD}
U_G \eal \left[
      a^3
     + {2 \over 9} a^2
     - {8 \over 9}a
     -{16 \over 9}
     + {16 \over 9} \sqrt{a+1}   \right]
   {1 \over a(a+ 1)}\, , \nonumber\\
U_D \eal {1 \over a \sqrt{a+1}}\, , 
\eea
and represent the growing
and decaying mode of adiabatic perturbations respectively.
Using Green's method,
the particular solution in the presence of a {\it constant}
entropy fluctuation $S$ becomes $\Delta_T = C_G U_G + C_D U_D + S U_I$,
where $U_I$ is given by \cite{KS86}
\bel{eq:BeqnUS}
U_I = {4 \over 15} \left( k \over k_{eq} \right)^2
           \left( {1 - {3K \over k^2}} \right)
           { 3a^2 + 22a + 24 + 4(4+3a)(1+a)^{1/2}
           \over (1+a)( 3a+4)[1+(1+a)^{1/2}]^4 }a^3.
\ee

\subsubsection{From Matter to Curvature or $\Lambda$ domination}
 
After radiation becomes negligible,  both the isocurvature and adiabatic
modes evolve in the same manner
\bel{eq:BeqnDeltaT}
\ddot \Delta_T + {\dot a \over a} \dot \Delta_T = 4\pi G\rho_T \left({
a \over a_0 }\right)^2 \Delta_T.
\ee
For pressureless perturbations, each mass shell evolves as a separate
homogeneous universe.  Since a density perturbation can be viewed
as merely a different choice of the initial time surface, the evolution of the
fractional shift in the scale factor $a^{-1} \delta a/\delta t$, 
\ie\ the Hubble parameter $H$,
must coincide with $\Delta_T$.  This is an example of how a clever 
choice of gauge simplifies the analysis.  It is straightforward 
to check that the Friedman equations \eqn{einsteinhubble} and \eqn{zerozero}
do indeed imply
\bel{eq:BeqnH}
\ddot H + {\dot a \over a} \dot H = 4\pi G\rho_T \left({
a \over a_0 }\right)^2 H,
\ee
so that one solution, the decaying mode, of equation \eqn{BeqnDeltaT} is 
$\Delta_T \propto
H$ \cite{PeeblesLSS}.
The growing mode $\Delta_T \propto D$
can easily be determined by writing its form as
$D \propto H G$ and by substitution into equation \eqn{BeqnDeltaT}
\bel{eq:BeqnG}
\ddot G + \left( {\dot a \over a} + 2 {\dot H \over H} \right) \dot G = 0.
\ee
This can be immediately solved as \cite{PeeblesLSS}
\bel{eq:BeqnD}
D(a) \propto H \int {d a \over (a H)^3}.
\ee
Note that we ignore pressure contributions in $H$.
If the cosmological constant $\Lambda=0$, this integral can be performed
analytically
\bel{eq:BeqnGrowth}
D(a) \propto 1 + {3 \over x} + {3 (1+x)^{1/2} \over x^{3/2} }
        \ln [(1+x)^{1/2}-x^{1/2}],
\ee
where $x=(\Omega_0^{-1}-1)(a/a_0)$.
In the more general case, a numerical solution to this integral
must be employed.   Notice that $D \propto a$ in the matter-dominated
epoch and goes to a constant in the curvature or $\Lambda$-dominated
epoch.

\subsubsection{General Solution} 

Before curvature or $\Lambda$ domination,
$D \propto a$.
The full solution for $\Delta_T$,
where the universe is allowed to pass through radiation,
matter and curvature or $\Lambda$ domination,
can be simply obtained from equation \eqn{BeqnUAD} and \eqn{BeqnUS},
by replacing $a$ with $D$ normalized so
that $D=a$ early on, \ie\
\bel{eq:replacementD}
a \rightarrow D = {5 \over 2} a_0 \Omega_0 g(a)  \int {da \over a}{1 \over
g^3(a)} \left({a_0 \over a}\right)^2,  
\ee
where
\bel{eq:Dgfact}
g^2(a) = \left( { a_0 \over a } \right)^3 \Omega_0
         + \left( { a_0 \over a } \right)^2 (1-\Omega_0-\Omega_\Lambda)
	 + \Omega_\Lambda.
\ee
For convenience, we parameterize the initial amplitude of the homogeneous
growing mode with the initial curvature fluctuation $\Phi(0)$.
The general growing solution then becomes
\bel{eq:GenSolution}
\Delta_T = \Phi(0) U_A + S(0) U_I,
\ee
The evolutionary factors
$U_A$ and $U_I$ are given by equations \eqn{replacementD}, \eqn{BeqnUAD}, 
\eqn{BeqnUS} to be
\beal{eq:DeltaLS}
U_A \eal {6 \over 5} \left({k \over k_{eq} }\right)^2 
           \left( {1 - {3K \over k^2}} \right)
	\left[ 
      D^3
     + {2 \over 9} D^2
     - {8 \over 9}D
     -{16 \over 9}
     + {16 \over 9} \sqrt{D+1}   \right]
   {1 \over D(D+ 1)}\, , \quad\nonumber\\
U_I \eal {4 \over 15} \left( k \over k_{eq} \right)^2
           \left( {1 - {3K \over k^2}} \right)
           { 3D^2 + 22D + 24 + 4(4+3D)(1+D)^{1/2}
           \over (1+D)(4+ 3D)[1+(1+D)^{1/2}]^4 }D^3 
\eea
respectively. 
We have implicitly assumed that
curvature and $\Lambda$ dynamical contributions are
only important well after equality $a\gg1$.
Curvature dominates over matter at $a/a_0 > \Omega_0/
(1-\Omega_0 -\Omega_\Lambda)$,
whereas $\Lambda$ dominates over matter at $a/a_0  >
(\Omega_0 / \Omega_\Lambda)^{1/3}$ and over curvature at
$a/a_0 > [(1-\Omega_0-\Omega_\Lambda)/\Omega_\Lambda]^{1/2}$.
Although we will usually only consider $\Lambda$ models which are flat,
these solutions are applicable to the general case.

\subsection{Initial Conditions}

Two quantities, the initial curvature perturbation and 
entropy fluctuation serve to entirely specify the
growing solution.  Adiabatic models begin with no entropy fluctuations, \ie\
$S(0)=0$.  Isocurvature models on the other hand have no curvature
perturbations initially, \ie\ $\Phi(0) =0$.  
Note that any arbitrary mixture of adiabatic and
isocurvature modes is also covered by equation~\eqn{GenSolution}.

For a universe with photons, 3 families of massless neutrinos, baryons and cold collisionless 
matter, the entropy becomes,
\beal{eq:Scomponents}
S \eal \Delta_m - {3 \over 4} \Delta_r \nonumber\\
  \eal (1 -{\Omega_b \over \Omega_0})\Delta_c + {\Omega_b \over \Omega_0}
\Delta_b - {3 \over 4} (1-f_\nu) \Delta_\gamma - {3 \over 4} f_\nu\Delta_\nu
\nonumber\\
  \eal (1-{\Omega_b \over \Omega_0})[(1-f_\nu)S_{c\gamma} + f_\nu S_{c\nu}]
        + {\Omega_b \over \Omega_0}[(1-f_\nu)S_{b\gamma} + f_\nu S_{b\nu}],
\eea
where 
$c$ represents the cold collisionless component.
The neutrino fraction
$f_\nu = \rho_\nu/(\rho_\nu+\rho_\gamma)$ is time independent after
electron-positron annihilation, implying $f_\nu = 0.405$ for three
massless neutrinos and the standard
thermal history.  
$S_{ab}$ is the entropy or number density fluctuation between
the $a$ and $b$ components, 
\bel{eq:Sab}
S_{ab} = \delta (n_a/n_b) = {\Delta_a \over 1+w_a} - {\Delta_b \over 1+w_b}.
\ee
Entropy conservation $\dot S_{ab} =0 = \dot S$ then has an obvious
interpretation: since the components cannot separate
above the horizon, the particle number ratios must remain constant.

The axion isocurvature model introduces density perturbations $\Delta_c$
in the cold
collisionless axions in the radiation-dominated epoch without generating
curvature.  This implies that $S_{c\gamma}=S_{c\nu}= \Delta_c(0) =$constant and
$S_{b\gamma} = S_{b\nu} = 0$.  However, the scale invariant model does
not succeed in forming large scale structure and tilted models overproduce
CMB anisotropies.
The most promising isocurvature model is the baryon-dominated model of
Peebles \cite{PeeblesPIBa,PeeblesPIBb} where $\rho_c$ is assumed absent.  
By the same argument as above,
$S_{b\gamma}=S_{b\nu}=\Delta_b(0)$ initially.  
Of course, since there is no cold 
collisionless component $S_{c\gamma}=S_{c\nu}=0$ and $S_{b\gamma}=S_{b\nu}=S$.
We shall see that some versions of this model can succeed since baryon       
fluctuations can lead to early structure formation and reionization damping
of CMB anisotropies (see \S \ref{ss-6PIB}).
When displaying isocurvature models, we implicitly assume the baryonic
case.

\input chap4/cdmlarge.tex

\subsection{Component Evolution}
\label{ss-4componentevolution}

With the definition of $S$ [equation~\eqn{Scomponents}], all component
perturbations can be written in terms of
$\Delta_T$.
The velocity and potentials are constructed as
\beal{eq:BeqnContPois}
V_T \eal -{3 \over k}{\dot a \over a}\left( {1-{3K \over k^2}} \right)^{-1}
        {{1+a} \over {4+3a}}\left[ a {d\Delta_T \over da} -
        {1 \over {1+a}} \Delta_T \right], \nonumber\\
\Psi \eal - {3 \over 4} \left( {k_{eq} \over k} \right)^2
           \left( {1 - {3K \over k^2}} \right)^{-1}
           {1 + a \over a^2}
\Delta_T,
\eea
where note that constant entropy assumption
requires that all the velocities $V_i = V_T$.
The relation for the velocity may be simplified by noting that
\beal{eq:BeqnEta}
\eta \eapp {2\sqrt{2} \over k_{eq}} \left[ \sqrt{1+a} -1  \right] \qquad {
\rm RD/MD} \nonumber\\
\eapp {1 \over \sqrt{-K}} \cosh^{-1}\left[ 1 + {2 (1-\Omega_0) \over
\Omega_0} {a \over a_0} \right], \qquad {\rm MD/CD} 
\eea
where CD denotes curvature domination with $\Lambda=0$.  For $\Lambda \ne
0$, it must be evaluated by numerical integration.   Before curvature or
$\Lambda$ domination
\bel{eq:BeqnDota}
{\dot a \over a} = {(1+a)^{1/2} \over \sqrt{2} a} k_{eq},
\ee
which can be used to explicitly evaluate \eqn{BeqnContPois}.

Now let us consider the implications of
the general solution~\eqn{GenSolution}.  The results for the
adiabatic mode are extremely simple.
When the universe is dominated by radiation (RD), matter (MD), curvature
(CD) or the cosmological constant
 ($\Lambda$D), the total density fluctuation takes the form
\bel{eq:DeltaA}
\Delta_T/\Phi(0) = {\cases { \vphantom{\big( }
                        {4 \over 3} (k/k_{eq})^2(1-3K/k^2) a^2 & RD \cr
                        \vphantom{\big( }
                        {6 \over 5} (k/k_{eq})^2(1-3K/k^2) a & MD \cr
                        \vphantom{\big( }
                        {6 \over 5} (k/k_{eq})^2(1-3K/k^2) D. & CD/$\Lambda$D \cr }}
\ee
Moreover since $S=0$, the components evolve together
$\Delta_b = \Delta_c = {3 \over 4} \Delta_\gamma = {3 \over 4} \Delta_\nu$
where $\Delta_c$ is any decoupled non-relativistic component (\eg\ CDM).
The velocity
and potential are given by
\beal{eq:VPsiA}
V_T/\Phi(0) \eal {\cases {\vphantom {\big( }
                        -{\sqrt{2} \over 2} (k/k_{eq}) a & RD
\cr
                        \vphantom{\big( }
                        -{3\sqrt{2} \over 5}
                         (k/k_{eq}) a^{1/2} & MD \cr
                        \vphantom{\big( }
                        - {6 \over 5} (k/k_{eq})
			 \dot D/k_{eq}, & CD/$\Lambda$D \cr }} \\
\label{eq:PsiAdi}
\Psi/\Phi(0) = -\Phi/\Phi(0) \eal {\cases {\vphantom {\big( }
                        - 1 & RD \cr
                        \vphantom{\big( }
                        - {9 \over 10} & MD \cr
                        \vphantom{\big( }
                        - {9 \over 10} D/a. &
                        CD/$\Lambda$D  \cr }} 
\eea
An example of the evolution is plotted in Fig.~\ref{fig:4cdmlarge}.

We can also generate the Newtonian temperature perturbation from the
gauge transformation 
\bel{eq:GaugeTMN}
\Theta_0 = {\Delta_\gamma \over 4} - {\dot a \over a} {V_T \over k},
\ee
which yields
\bel{eq:ThetaA}
\Theta_0/\Phi(0) = {\cases {\vphantom {\big( }
                        {1 \over 2} & RD \cr
                        \vphantom{\big( }
                        {3 \over 5} & MD \cr
                        \vphantom{\big( }
                        {3 \over 2} -{9 \over 10} D/a. &
                        CD/$\Lambda$D  \cr }} 
\ee
In fact, these relations are far easier to derive in the Newtonian gauge
itself where $\dot \Theta_0 = \dot \Phi$.  Note that in the matter-dominated
epoch, $\Theta_0 = -{2 \over 3} \Psi$ which will be important for the
Sachs-Wolfe effect (see \S \ref{sec-5sachswolfe}).

Contrast this with the isocurvature evolution,
\bel{eq:DeltaLSCase}
\Delta_T/S(0) = {\cases {
                        \vphantom{\big( }
                \,{1 \over 6}\,( k /k_{eq} )^2 (1 - 3K/k^2)
                        a^3
                        \vphantom{\big( }
                        & RD  \cr
                {4 \over 15} ( k / k_{eq} )^2 (1- 3K/k^2)
                        a
                        \vphantom{\big( }
                        & MD  \cr
                {4 \over 15} ( k / k_{eq} )^2 (1- 3K/k^2) D.
                        \vphantom{\big( }
                        & CD/$\Lambda$D
                }}
\ee
In baryonic models
\bel{eq:BeqnDeltab}
\Delta_b  = {1 \over 4+3a }[4S + 3(1+a)\Delta_T],
\ee
and
\beal{eq:BeqnDeltar}
\Delta_\nu \eal {4 \over 3} (\Delta_b - S_{b\nu}), \nonumber\\
\Delta_\gamma \eal {4 \over 3} (\Delta_b - S_{b\gamma}). 
\eea
Recall that since the curvature perturbation vanishes initially 
$S_{b\nu}=S_{b\gamma}=S$.
From these relations, we obtain
\bel{eq:DbLS}
\Delta_b/S(0) = {\cases {
                        \vphantom{\big( }
                1 - {3 \over 4}a & RD \cr
                        \vphantom{\big( }
                {4 \over 3} \left[ a^{-1} + {1 \over 5}
                 ( k/k_{eq} )^2 (1 - 3K/k^2) a \right]
                   & MD \cr
                        \vphantom{\big( }
                {4 \over 3} \left[ a^{-1} + {1 \over 5}
                 ( k/k_{eq} )^2 (1 - 3K/k^2) D \right],
                   & CD/$\Lambda$D \cr
                }}
\ee
and
\bel{eq:DrLS}
\Delta_\gamma/S(0) = \Delta_\nu/S(0) = {\cases {
                        \vphantom{\big( }
                -a & RD \cr
                        \vphantom{\big( }
                {4 \over 3} \left[ -1 + {4 \over 15}
                 ( k/k_{eq} )^2 (1- 3K/k^2) a \right]
                   & MD \cr
                        \vphantom{\big( }
                {4 \over 3} \left[ -1 + {4 \over 15}
                 ( k/k_{eq} )^2 (1- 3K/k^2) D \right],
                   & CD/$\Lambda$D \cr }}
\ee
for the baryon and radiation components. Lastly, the velocity, potential,
and photon temperature
also have simple asymptotic forms,
\beal{eq:VPsiLS}
V_T/S(0) \eal {\cases {
                        \vphantom{\big( }
                - {\sqrt 2 \over 8} (k/k_{eq}) a^2  & RD \cr
                        \vphantom{\big( }
                - {2\sqrt 2 \over 15} (k/k_{eq}) a^{1/2}  & MD \cr
                        \vphantom{\big( }
                - {4 \over 15} (k/k_{eq}) \dot D/k_{eq}  , & CD/$\Lambda$D \cr
}} \\
\label{eq:PsiLS}
\Psi/S(0) = -\Phi/S(0) \eal {\cases {
                        \vphantom{\big( }
                - {1 \over 8} a & RD \cr
                        \vphantom{\big( }
                - {1 \over 5}  & MD \cr
                        \vphantom{\big( }
                - {1 \over 5} D/a, & CD/$\Lambda$D \cr
}} \\
\label{eq:ThetaI}
\Theta_0/S(0) \eal {\cases {
                        \vphantom{\big( }
                - {1 \over 8} a & RD \cr
                        \vphantom{\big( }
                - {1 \over 5}  & MD \cr
                        \vphantom{\big( }
                - {1 \over 5} D/a. & CD/$\Lambda$D \cr
}}
\eea
The equality of $\Theta$ and $\Psi$ is easy to understand 
in the Newtonian gauge where $\dot \Theta_0 = -\dot \Phi$.
In Fig.~\ref{fig:4largeiso}, we display an example of the
isocurvature component evolution. 
\input chap4/largeiso.tex

\subsection{Discussion}
 
Let us try to interpret these results physically.
The isocurvature condition is satisfied by initially placing
the fluctuations in
the baryons $\Delta_b = S(0)$ with
$\Delta_\gamma = 0$, so that $\Delta_T = 0$.
As the universe evolves however,
the relative significance of the baryon fluctuation
$\Delta_b \rho_b/\rho_T$ for the total density fluctuation
$\Delta_T$
 grows as $a$.  To compensate, the photon
and neutrino fluctuations grow to be equal and opposite
$\Delta_\gamma = \Delta_\nu = -a S(0)$.
The tight-coupling condition $\dot \Delta_b = {3 \over 4}\dot
\Delta_\gamma$ implies then that
the baryon fluctuation must also decrease so that $\Delta_b
= (1-3a/4)S(0)$.  The presence of $\Delta_\gamma$ means that there
is a gradient in the photon energy density.  This gradient gives rise
to a dipole $V_\gamma$ as the regions come into causal contact [see
equation~\eqn{Hierarchy}],
\ie~$V_\gamma \propto k\eta \Delta_\gamma \propto -ka^2 S(0)$.
The same argument holds for the neutrinos.
Constant entropy requires that the total fluid move with the photons
and neutrinos $V_T = V_\gamma$, and thus infall,
produced by the gradient in the velocity, yields a
total density perturbation $\Delta_T \propto -k\eta(1-3K/k^2) V_T
\propto k^2(1-3K/k^2) a^3 S(0)$ [see equation~\eqn{Total}].
This is one way of interpreting equation~\eqn{DeltaLS} 
and the fact that the entropy provides a source of total
density fluctuations in the radiation-dominated epoch \cite{Hu}
 
A similar analysis applies for adiabatic fluctuations, which begin instead
with finite potential $\Psi$.  Infall implies $V_T \propto k\eta\Psi(0)
\approx
-k\eta \Phi(0)$, which then yields $\Delta_T \propto - k\eta V_T
\propto k^2(1 - 3K/k^2) a^2 \Phi(0)$,
thereby also keeping the potential constant.
Compared to the adiabatic case, the isocurvature scenario predicts
total density perturbations which are smaller by one factor of $a$
in the radiation-dominated epoch
as might be expected from cancellation.
 
After radiation domination, both modes
grow in pressureless linear theory $\Delta_T \propto D$
[{\it c.f.} equations~\eqn{DeltaA} and \eqn{DeltaLSCase}].
Whereas in the radiation-dominated limit,
the entropy term $S$ and the gravitational infall term $\Psi$ are comparable
in equation~\eqn{Total}, the entropy source is thereafter suppressed by
$w_T = p_T/\rho_T$, making the isocurvature and adiabatic evolutions identical.
Furthermore, since the growth of $\Delta_T$ is
suppressed in open and $\Lambda$-dominated universes, the potential $\Psi$
decays which has interesting consequences for anisotropies
as we shall see in \S \ref{sec-5sachswolfe}.

\section{Subhorizon Evolution before Recombination}
\label{sec-4subhorizon}

As the perturbation enters the horizon,
we can no longer view the
system as a single fluid.  Decoupled components
such as the neutrinos free stream and change the number density, \ie\ entropy,
fluctuation.
However, above the photon diffusion scale, the photons and baryons are still
tightly coupled by Compton scattering
until recombination.  Since even then the diffusion length
is much
smaller than the horizon $\eta_*$, it is appropriate to combine
the photon and baryon fluids for study \cite{Jorgensen,Seljak}.
In this section, we show that
photon pressure resists the gravitational
compression of the photon-baryon
fluid,
leading to {\it driven} acoustic oscillations
\cite{HSa} which are then damped by photon diffusion.

\subsection{Analytic Acoustic Solutions}
\label{ss-4analyticacoustic}

At intermediate scales, neither radiation pressure nor gravity
can be ignored. Fortunately, their effects can be analytically
separated and analyzed \cite{HSa}. Since photon-baryon tight coupling still
holds,
it is appropriate to expand the Boltzmann equation~\eqn{Hierarchy}
and the Euler equation~\eqn{Baryon} for the baryons
in the Compton scattering time $\dot \tau^{-1}$ \cite{PY}.
To zeroth order, we regain the tight-coupling identities,
\beal{eq:BeqnZeroOrder}
\dot\Delta_\gamma \eal {4 \over 3}\dot\Delta_b, \qquad
{(\rm or}\ \dot\Theta_0 = {1 \over 3} \dot \delta_b^N ) \nonumber \\
\Theta_1 \eadef V_\gamma = V_b, \nonumber\\
\Theta_\ell \eal 0. \qquad \ell \ge 2 
\eea
These equations merely
express the fact that the radiation is isotropic in the baryon
rest frame and the density fluctuations in the photons grow adiabatically
with the baryons.  Substituting the zeroth order solutions back into
equations~\eqn{Hierarchy} and \eqn{Baryon}, we obtain the iterative
first order solution, 
\beal{eq:BeqnFirstOrder}
\dot \Theta_0 \eal -{k \over 3} \Theta_1 - \dot \Phi, \nonumber\\
\dot \Theta_1 \eal -{\dot R \over 1+R}\Theta_1
+ {1 \over 1+R}k\Theta_0 + k\Psi,
\eea
where we have used the relation $\dot R = (\dot a/a)R$.
The tight-coupling approximation eliminates the
multiple time scales and the infinite hierarchy of
coupled equations of the full problem.
In fact, this simple set of equations can readily be solved numerically
\cite{Seljak}.
To solve them analytically, 
let us rewrite it as a single second order equation,
\bel{eq:BeqnTightRepeat}
\ddot \Theta_0 + {\dot R \over 1+R}\dot \Theta_0 +
k^2 c_s^2 \Theta_0 = F,
\ee
where
the photon-baryon sound speed is
\bel{eq:SoundSpeed}
c^2_s \equiv { \dot p_\gamma \over \dot \rho_\gamma + \dot\rho_b}= 
{1 \over 3} {1 \over 1+R},
\ee
assuming $p_b \approx 0$ and 
\bel{eq:BeqnForceRepeat}
F = - \ddot \Phi - {\dot R \over 1+R}\dot
\Phi - {k^2 \over 3}\Psi,
\ee
is the forcing function. Here $\ddot \Phi$ represents the dilation effect,
$\dot \Phi$ the modification to expansion damping, and $\Psi$
the gravitational infall.  The homogeneous $F=0$ equation yields
the two fundamental solutions
under the adiabatic approximation,
\beal{eq:Homogeneoussoln}
\theta_a \eal (1+R)^{-1/4} \cos k r_s, \nonumber\\
\theta_b \eal (1+R)^{-1/4} \sin k r_s, 
\eea
where the sound horizon is 
\bel{eq:SoundHorizon}
r_s = \int_0^\eta c_s  d\eta'
= {2 \over 3} {1 \over k_{eq}} \sqrt{6 \over R_{eq}}
  \ln { \sqrt{1+R} + \sqrt{ R + R_{eq} }
    \over
        1 + \sqrt{ R_{eq}}}.
\ee
The phase relation $\phi=kr_s$ just reflects 
the nature of acoustic oscillations.
If the sound speed were constant, it would yield the
expected dispersion relation
$\omega = kc_s$.

The adiabatic or WKB approximation assumes that the time scale for the
variation in the sound speed is much longer than the period of the oscillation.
More specifically, the mixed $\dot R \dot \Theta_0$ is included in this
first order treatment, 
but second order terms are dropped under the assumption
that
\bel{eq:WKBassumption}
(kc_s)^2 \gg (1+R)^{1/4} {d^2 \over d\eta^2} (1+R)^{-1/4} ,
\ee
or 
\beal{eq:WKBassumption2}
(kc_s)^2 \!\!\!&\gg&\!\!\! {\dot R^2 \over (1+R)^2} \nonumber\\
 \!\!\!&\gg&\!\!\! {\ddot R \over 1+R}.
\eea
It is therefore satisfied at early times and on small scales.  Even at last   
scattering the approximation holds well for $k > 0.08 h^3$ Mpc$^{-1}$ if
$R <  1$ and $a > 1$, as is the case for the standard CDM model.  

\input chap4/wkb.tex

Now we need to take into account the forcing function $F(\eta)$ due to
the gravitational potentials $\Psi$ and $\Phi$.  Employing the Green's
method, we construct the particular solution,
\bel{eq:BeqnGreen}
\hat \Theta_0(\eta) = C_1 \theta_a(\eta) + C_2 \theta_b(\eta)
+ \int_0^\eta  {
\theta_a(\eta')\theta_b(\eta) - \theta_a(\eta)\theta_b(\eta')
               \over
\theta_a(\eta')\dot \theta_b(\eta') - \dot\theta_a(\eta') \theta_b(\eta')
}
F(\eta') d\eta'.
\ee
Equation \eqn{Homogeneoussoln} implies
\bel{eq:BeqnNumerator}
\theta_a(\eta')\theta_b(\eta) - \theta_a(\eta)\theta_b(\eta') =
[1+R(\eta)]^{-1/4}
[1+R(\eta')]^{-1/4} \sin[kr_s(\eta)-kr_s(\eta')] \, ,
\ee
and
\bel{eq:BeqnDenominator}
\theta_a(\eta')\dot \theta_b(\eta') - \dot \theta_a(\eta') \theta_b(\eta')
={k \over \sqrt{3}} [1+R(\eta')]^{-1} \, .
\ee
With $C_1$ and $C_2$ fixed by
the initial conditions, the 
solution in the presence of the source $F$ then becomes \cite{HSa}
\beal{eq:PartSoln}
[1+R(\eta)]^{1/4} \hat \Theta_0(\eta) \eal \Theta_0(0) \cos k r_s(\eta)
 + {\sqrt{3} \over k}
[\dot \Theta_0(0)+{1 \over 4}\dot R(0) \Theta_0(0)]\sin k r_s(\eta) \nonumber\\
& & +
{\sqrt{3} \over k}
\int_0^\eta d\eta' [1+R(\eta')]^{3/4}
{\sin[k r_s(\eta)-k r_s(\eta')]} F(\eta')\, ,\quad 
\eea
and $
k\Theta_1 = -3 (\dot \Theta_0 + \dot \Phi).
$
The potentials in $F$ can be approximated from their
large (\S \ref{ss-4componentevolution}) 
and small (\S \ref{ss-4dampedacoustic}) 
scale solutions.  As we shall show in
Appendix \ref{ss-5recombination}, this 
can lead to extremely accurate solutions.
To show the true power of this
technique here, we instead
employ their numerical values in Fig.~\ref{fig:4wkb}.
The excellent agreement
with the full solution
indicates that our
technique is limited only by our knowledge of the potentials.

\subsection{Driven Acoustic Oscillations}
\label{ss-4acousticdriven}

\subsubsection{Baryon Drag}
 
Some basic features of the acoustic oscillations 
are worthwhile to note.
Let us start with a toy model in which the potential is constant 
$\dot \Psi = 0 = \dot \Phi$.  This corresponds to
a universe which was always matter dominated.   Let us also
assume that the baryon-photon ratio $R$ is constant.  
Of course neither of these assumptions are valid for the real universe,
but as we shall see the generalization to realistic cases is 
qualitatively simple.   Under these assumption, the solution of
equation \eqn{BeqnTightRepeat} is obvious,
\bel{eq:simplemodel}
\hat \Theta_0(\eta) = [\Theta_0(0)+(1+R)\Psi]\cos(kr_s) + 
{ 1 \over k c_s}
\dot\Theta_0(0)\sin(kr_s) - (1+R)\Psi,
\ee
where the sound horizon reduces to $r_s = c_s \eta$.
Several basic features are worth noting:

\begin{enumerate}

\item The zero point of the oscillation 
$\Theta_0 = -(1+R)\Psi$ is increasingly shifted with the baryon content.

\item The amplitude of the oscillation increases with the baryon content $R$.

\item The redshift $\Psi$ from climbing out of potential wells cancels
the $R=0$ zero point shift.

\item Adiabatic initial conditions where $\Theta_0(0)=$ constant and
isocurvature initial conditions where $\dot \Theta_0(0) =$ constant
stimulate the cosine and sine harmonic respectively.

\end{enumerate}

\noindent 
Of course here $\dot\Theta_0(0)$ does not really describe the
isocurvature case since here $\Phi \ne 0$ in the initial conditions.  We
will see in the next section what difference this makes.

The zero point of the oscillation is the state at which the forces of
gravity and pressure are in balance.  
If the photons dominate, $R \rightarrow 0$ and this balance occurs
at
$\Theta_0=-\Psi$ reflecting the fact that in equilibrium, the
photons are compressed and hotter inside the 
potential well.  Infall not only increases the number density of photons 
but also their energy through gravitational blueshifts. 
It is evident however that when the photons
climb back out of the well, they suffer an equal and opposite effect.  
Thus the effective temperature is $\Theta_0+\Psi$.  It is this 
quantity that oscillates around zero if the baryons can be neglected.  

Baryons add gravitational and inertial mass to the fluid without raising
the pressure. 
We can rewrite the oscillator equation \eqn{BeqnTightRepeat} as
\bel{eq:EffectiveMass} 
(1+R) \ddot \Theta_0 + {k^2 \over 3} \Theta_0 = -(1+R){k^2 \over 3}\Psi,
\ee
neglecting changes in $R$ and $\Phi$.  Note that 
$m_{\rm eff} = 1+R$ represents the effective mass of the oscillator.
Baryonic infall drags the photons into potential wells and 
consequently leads to greater compression 
shifting the effective temperature to 
$-R\Psi$.  
All compressional phases will be enhanced over
rarefaction phases. 
This explains the alternating 
series of peak amplitudes in Fig.~\ref{fig:4wkb}b 
where the ratio $R$ is significant at late times.  In the lower $R$ case of
Fig.~\ref{fig:4wkb}a, the effect is less apparent.
Furthermore, a shift in the zero point 
implies larger amplitude oscillations since the initial displacement
from the zero point becomes larger.

Adiabatic and isocurvature conditions also
have different phase relations.  Peak fluctuations occur for $kr_s = m\pi$
and $kr_s = (m-1/2)\pi$ for adiabatic and isocurvature modes respectively.
Unlike their adiabatic counterpart, isocurvature
conditions are set up to resist gravitational
attraction.  
Thus the compression phase is reached for odd $m$ adiabatic peaks and even $m$
isocurvature peaks.  

\subsubsection{Doppler Effect}

The bulk velocity of the fluid along the line of sight  $V_\gamma/\sqrt{3}$
causes the observed temperature to be Doppler shifted.
From the continuity equation \eqn{BeqnFirstOrder}, the acoustic velocity
becomes
\bel{eq:simplevelocity}
{V_\gamma(\eta) \over \sqrt{3}} = -{\sqrt{3} \over k} \dot \Theta_0 
= \sqrt{3} [\Theta_0(0) + (1+R)\Psi] c_s \sin(kr_s) 
- {\sqrt{3} \over k} \dot\Theta_0(0) \cos(kr_s) ,
\ee
assuming $\dot \Phi =0$, which yields the following interesting facts:
\begin{enumerate}

\item The velocity is $\pi/2$ out of phase with the temperature.

\item The zero point of the oscillation is not displaced.

\item The amplitude of the oscillation is reduced by a factor of 
	$\sqrt{3} c_s = (1+R)^{-1/2}$ 
	compared with the temperature. 

\end{enumerate}
\noindent
Because of its phase relation, the velocity contribution will fill in 
the zeros of the temperature oscillation.  Velocity oscillations,
unlike their temperature counterparts are symmetric around zero.  
The {\it relative} amplitude of the velocity compared with 
the temperature oscillations also decreases with the baryon content $R$. 
For the same initial displacement, conservation of
energy requires a smaller velocity as the mass increases.  Together
the zero point shift and the increased amplitude of temperature perturbations
is sufficient to make compressional temperature peaks significantly more 
prominent than velocity or rarefaction peaks (see Fig.~\ref{fig:4wkb}b).

\subsubsection{Effective Mass Evolution}

In the real universe however, $R$ must grow from zero at the 
initial conditions and adiabatically changes the effective mass
of the oscillator $m_{\rm eff} = (1+R)$.
While the statements above for constant $R$ are qualitatively correct,
they overestimate the effect.  The exact solution given in equation 
\eqn{PartSoln} must 
be used for quantitative work.  

Notice that the full first order relation \eqn{BeqnFirstOrder}
is exactly an oscillator 
with time-varying mass:
\bel{eq:TimeVaryingMass} 
{d \over d\eta} (1+R) \dot \Theta_0 + {k^2 \over 3} \Theta_0 = -(1+R){k^2 \over 3}\Psi - {d \over d\eta}(1+R)\dot\Phi,
\ee
where the last term on the rhs is the dilation effect from $\dot\Theta_0 
= -\dot\Phi$.
This form exposes a new feature due to a time varying effective mass. 
Treating the effective mass as an oscillator parameter, we can solve
the homogeneous part of equation 
\eqn{TimeVaryingMass} under the adiabatic
approximation.
In classical 
mechanics, the ratio of the energy $E={1 \over 2}m_{\rm eff} \omega^2 A^2$
to the frequency $\omega$ of an oscillator
is an adiabatic invariant.  Thus the amplitude scales as
$A \propto \omega^{1/2} \propto (1+R)^{-1/4}$,
which explains the appearance of this factor in equation \eqn{Homogeneoussoln}.

\subsubsection{Driving Force and Radiation Feedback}

Now let us consider a time varying potential.  In any situation where
the matter does not fully describe the dynamics, 
feedback from the radiation into
the potential through the Poisson equation can cause time variation.  
For isocurvature conditions, we have seen that radiation feedback causes
potentials to grow from zero outside the horizon
(see \S \ref{sec-4superhorizon} and Fig.~\ref{fig:4largeiso}).
The net effect for the isocurvature mode is that outside the sound horizon,
fluctuations behave as $\Theta = -\Phi \approx \Psi$ 
[see equation \eqn{ThetaI}].  
After sound horizon crossing, radiation density perturbations cease to
grow, leading to a decay in the gravitational potential in the radiation-dominated epoch.  Thus scales that cross during matter domination
experience more growth and are enhanced over their small scale counterparts.
Furthermore,
morphologically 
$-\ddot \Phi - k^2\Psi/3 \propto \sin(kr_s)$ leading to near resonant
driving of the sine mode of the oscillation until sound horizon crossing.
This supports our claim above that $\sin(kr_s)$ represents the isocurvature
mode. 

The adiabatic mode exhibits contrasting behavior.  Here the potential is
constant outside the sound horizon and then decays like the isocurvature
case.  However, it is the decay itself that drives the oscillation
since the form of the forcing function becomes approximately  
$-\ddot \Phi -k^2\Psi/3 \propto \cos(kc_s)$ until $kc_s \eta \sim 1$ 
and then dies away.  
In other words, the gravitational force drives the first compression without
a counterbalancing effect on the subsequent rarefaction phase. 
Therefore, for the adiabatic mode, the oscillation amplitude is boosted
at sound horizon crossing in the radiation-dominated universe 
which explains the prominence of the oscillations with respect to the
superhorizon tail in 
Fig.~\ref{fig:4wkb}a. One might expect from the dilation effect 
$\dot \Theta = -\dot \Phi$ that
the temperature is boosted up to 
$\Theta(\eta) \approx \Theta(0) -
\Phi(\eta) + \Phi(0) \approx {3 \over 2} \Phi(0)$.   We shall see in the
next
section that a more detailed analysis supports this conclusion.  
Therefore, unlike the isocurvature case, adiabatic modes experience an
enhancement for scales smaller than the horizon at equality. 

\subsection{Damped Acoustic Oscillations}
\label{ss-4dampedacoustic}

Well below the sound horizon in the radiation-dominated epoch, the 
gravitational
potentials have decayed to insignificance and the 
photon-baryon fluctuations
behave as simple oscillatory functions. However photon-baryon
tight coupling breaks down at the photon diffusion
scale.  At this point, photon fluctuations
are exponentially damped due to diffusive mixing and rescattering.
We can account for this by expanding the Boltzmann and Euler equations
for the photons and baryons respectively to second order in $\dot \tau^{-1}$
(see \cite{PeeblesLSS} and Appendix \ref{ss-5polarization}).
This gives the dispersion relation an imaginary part, making
the general solution
\bel{eq:WKBSolution}
\Theta_0 = C_A (1+R)^{-1/4}{\cal D}(\eta,k) \cos kr_s + C_I (1+R)^{-1/4}
{\cal D}(\eta, k) \sin kr_s ,
\ee
where $C_A$ and $C_I$ are constants and the damping factor is
\bel{eq:Damp}
{\cal D}(\eta,k) = e^{-(k/k_D)^2} ,
\ee
with the damping scale
\bel{eq:DampLength}
k_D^{-2} = { 1 \over 6} \int d\eta {1 \over \dot \tau}
{R^2 + 4(1+R)/5 \over (1+R)^2} .
\ee
For small corrections to this relation due to the angular
dependence of Compton scattering and polarization, see \cite{Kaiser83} and
Appendix \ref{ss-5polarization}.
Since the $R$ factors in equation \eqn{DampLength}
 go to ${4 \over 5}$ for $R \ll 1$ and $1$ for
$R \gg 1$, the damping length is approximately $\lambda_D^2 \sim k_D^{-2} \sim 
\int d\eta/\dot\tau$.  This relation is easy to understand qualitatively.
The Compton
mean free path of the photons is $\lambda_C 
= \dot \tau^{-1}$.  The scale on which a photon can diffuse is given
by a random walk process $\sqrt{N} \lambda_C$ where the number of steps
is $N = \eta/\lambda_C$.  Therefore the diffusion scale is approximately
$\lambda_D \approx \sqrt{\lambda_C \eta} = \sqrt{\eta/\dot\tau}$.

The amplitudes of these oscillations, \ie\
the constants $C_A$ and $C_I$, are determined by the total effect of
the gravitational driving force in equation~\eqn{BeqnForceRepeat}.
However, a simpler argument suffices for showing its general behavior.
As shown in \S \ref{ss-4componentevolution}, isocurvature fluctuations grow like
$\Delta_\gamma \approx -a S(0)$ until sound horizon crossing.
Since the sound horizon crossing is near $a_H \sim k_{eq}/k$ (see 
equation \eqn{horizoncr}),
the isocurvature amplitude will be  suppressed by
$ k_{eq}/k $.
On the other hand, adiabatic fluctuations which grow as $a^2$
will have a $ (k_{eq}/k)^2$ suppression factor which just cancels
the factor $(k/k_{eq})^2$ from the Poisson equation [see \eqn{DeltaA}]
when expressed in terms of the initial potential.
This simple argument fixes the amplitude up to a factor of order unity.
 
\input chap4/smalliso.tex

We obtain the specific amplitude by solving equation~\eqn{Total}
under
the constant entropy assumption $\dot S = 0$.
The latter approximation is not strictly
valid since free streaming of the neutrinos will change the entropy
fluctuation.  However, since the amplitude is fixed after
sound horizon crossing,
which is only slightly after horizon
crossing, it suffices.  Under this assumption,
the equation can again be solved in the small scale limit.
Kodama \& Sasaki \cite{KS86} find that 
for adiabatic perturbations,
\bel{eq:AmpAdi}
C_A={3 \over 2} \Phi(0),
\quad C_I=0,  \qquad ({\rm adi})
\ee
from which the isocurvature solution follows via Greens method, 
\bel{eq:AmpIso}
C_A=0, \quad C_I= -{\sqrt{6} \over 4} {k_{eq} \over k} S(0), \qquad ({\rm iso})
\ee
if $k\gg k_{eq}$, $k\eta \gg 1$ and $k \gg \sqrt{-K}$.
As expected, the isocurvature mode stimulates the $\sin kr_s$ harmonic,
as opposed to $\cos kr_s$ for the adiabatic mode.
 
We can also construct the evolution of density perturbations
at small scales.  Well inside the horizon $\Delta_\gamma = 4\Theta_0$, since
total matter and Newtonian fluctuations are equivalent.
The isocurvature mode solution therefore
satisfies (RD/MD)
\bel{eq:DeltagSS}
\Delta_\gamma/ S(0) =  -\sqrt{6} \left( {k_{eq} \over k} \right) (1+R)^{-1/4}
{\cal D}(a,k) \sin kr_s.
\ee
The tight-coupling limit implies $\dot \Delta_b
= {3 \over 4} \dot \Delta_\gamma$ which requires (RD/MD),
\bel{eq:DeltabSS}
\Delta_b/ S(0) = 1  -{3\sqrt{6} \over 4} \left( {k_{eq} \over k} \right) (1+R)^{-
1/4}
{\cal D}(a,k) \sin kr_s .
\ee
This diffusive suppression of the adiabatic component for the baryon fluctuation
is known
as Silk damping \cite{Silk}.  After damping, the baryons are left with the
original entropy perturbation S(0).  Since they are surrounded by a
homogeneous and isotropic sea of photons, the baryons are unaffected by
further photon diffusion.
From the photon or baryon continuity equations at small scales,
 we obtain (RD/MD)
\bel{eq:VSS}
V_b/S(0) = V_\gamma/C_I \approx {3 \sqrt{2} \over 4} \left( {k_{eq} \over k} 
\right)
(1+R)^{-3/4}
{\cal D}(a,k) \cos k r_s .
\ee
As one would expect, the velocity oscillates $\pi /2 $
out of phase with, and increasingly suppressed compared to,
 the density perturbations.
Employing equations \eqn{DeltagSS} and \eqn{DeltabSS}, we construct
the total density perturbation by assuming that free streaming
has damped out the neutrino contribution
(RD/MD),
\bel{eq:DeltaSS}
\Delta_T/S(0) = {a \over 1+a} \left[ 1 - {3\sqrt{6} \over 4}
        {k_{eq} \over k}
         R^{-1}(1+R)^{3/4}
        {\cal D}(a,k) \sin kr_s
        \right] .
\ee
From this equation, we may derive the potential
(RD/MD),
\bel{eq:VPsiSS}
\Psi/S(0) = - {3 \over 4} \left( {k_{eq} \over k} \right)^2{1\over a}
\left[ 1 - {3\sqrt{6} \over 4}
{k_{eq} \over k}R^{-1}(1+R)^{3/4} {\cal D}(a,k) \sin kr_s
\right]  ,
\ee
which decays with the expansion since $\Delta_T$ goes to a
constant.
In Fig.~\ref{fig:4smalliso}, we compare these analytic approximations with the
numerical results. After damping eliminates the adiabatic
oscillations, the evolution of perturbations is governed by
diffusive processes.
A similar analysis for adiabatic perturbations shows that diffusion damping
almost completely eliminates small scale baryonic fluctuations.\footnote{Residual
fluctuations are on the order $R\Psi$ as discussed in Appendix \ref{ss-5polarization}.}  
Unlike
the isocurvature case, unless CDM wells are present
to reseed fluctuations,
 adiabatic models consequently fail to form
galaxies. 

\section{Matter Evolution after Recombination} 
\label{sec-4postrecombination}

At $z_* \approx 1000$, the CMB can no longer keep hydrogen ionized and the
free electron density drops precipitously.  The photons thereafter free stream
until a possible epoch of reionization.  The subsequent evolution of the
photon fluctuations will be intensely studied in \S \ref{ch-primary} and
\S \ref{ch-secondary}.  Essentially, they preserve the fluctuations they
possess at last scattering in the form of anisotropies.  Here we will
concentrate on the evolution of the matter as it is important for
structure formation and feeds back into the CMB through reionization.

\input chap4/drag.tex

\subsection{Compton Drag}
\label{ss-4comptondrag}

Baryon fluctuations 
in diffusion damped scales can be regenerated 
after Compton scattering has become ineffective. 
The critical epoch 
is that at which the photon pressure or ``Compton drag'' can no longer
prevent gravitational instability in the baryons.   
The drag on an individual baryon does not depend on the total number
of baryons but rather the number of photons and its ionization state.  
From the baryon Euler equation \eqn{Baryon} and the
Poisson equation \eqn{BeqnContPois}, 
the drag term $\propto V_b$ comes to dominate over
the gravitational infall term $\propto k\Psi$ at redshifts above $z \sim 200
(\Omega_0 h^2)_{\vphantom{e}}^{1/5} x_e^{-2/5}$.
Thus all modes are released from Compton drag at the same time,
which we take to be
\bel{eq:Zdrag}
z_d = 160 (\Omega_0 h^2)_{\vphantom{e}}^{1/5} x_e^{-2/5},
\ee
defined as the epoch when fluctuations effectively join the
growing mode of pressureless linear theory.
 
It is important to realize that the drag and the last scattering redshift
are generally not equal.  
Following the drag epoch, baryons can be treated as freely falling.  
If cold dark matter exists in the model, potential wells 
though suppressed at small scales will still exist.  In adiabatic
CDM models, the Silk damped baryon fluctuations under the photon diffusion
scale can be regenerated as the baryons fall into the dark matter
potentials (see Fig.~\ref{fig:4drag}).  
For isocurvature models, the entropy fluctuations
remaining after Silk damping are released at rest to grow in 
linear theory.  

One complication arises though.  The collapse of baryon fluctuations after
recombination can lead to small scale non-linearities.  Astrophysical
processes associated with compact object formation can inject enough
energy to reionize the universe (see \S \ref{ch-secondary} and \eg\ \cite{GO}).
Ionization again couples the baryons and photons.     
Yet even in a reionized universe, the Compton drag 
epoch eventually ends due to the decreasing number density of electrons.
In CDM-dominated adiabatic models, the baryons subsequently fall into the 
dark matter wells leaving no trace of this extra epoch of Compton coupling.
The CMB also retains no memory since last scattering occurs after
the drag epoch in reionized scenarios.   This is not the case for
baryon isocurvature models since there are no dark matter wells 
into which baryons might fall.  
Evolution in the intermediate regime therefore has
a direct effect on the amplitude of fluctuations in the matter and CMB 
today.

Reionization is also more 
likely in models where the initial power spectrum is tilted
toward small scales.  In the 
baryon isocurvature case, entropy fluctuations at small scales
can be made 
quite large since they are essentially unprocessed by the pre-recombination
evolution.  For these reasons, we will concentrate on baryon isocurvature
models in discussing Compton drag in reionized models.

\input chap4/comptondrag.tex

\subsection{Reionization in Isocurvature Models}

Let us first consider the case where the universe was reionized immediately
following standard recombination.
Well before the end of the drag epoch $z_d$, the initial entropy 
fluctuations are frozen into the baryons.  
Well afterwards, the baryon fluctuations
grow as in pressureless linear theory.   An excellent empirical 
approximation 
to the behavior at intermediate times is given by
\bel{eq:baryongrowth}
\Delta_b/S(0) = {\cal G}(a,a_d),
\ee
with the interpolation function
\bel{eq:DragGrow}
{\cal G}(a_1,a_2) = 1 + {D(a_1)\over D(a_2)}\exp(-a_2/a_1),
\ee
where if $a_1 \gg a_2$, ${\cal G}(a_1,a_2) \rightarrow D(a_1)/D(a_2)$.
The velocity $V_T$ is given by the continuity equation~\eqn{Total}.
Notice that growth in
an open and/or $\Lambda$ universe is properly accounted for.  
This approximation is depicted
in Fig.~\ref{fig:4comptondrag}a.

Now let us consider more complicated thermal histories.
Standard recombination may be followed by a significant
transparent period before reionization at $z_i$, due to some later
round of structure formation.  There are two
effects to consider here: fluctuation behavior in the transparent
regime and after reionization.  Let us begin
with the first question.  Near recombination, the baryons
are released from drag essentially at rest and thereafter
can grow in pressureless linear theory.  The joining conditions
then imply that
${3 \over 5}$ of the perturbation enters 
the growing mode $D$ \cite{PeeblesLSS},
yielding present fluctuations of
$\sim {3 \over 5} C_I D(z=0)/D(z_d)$.  This expression overestimates
the effect for low $\Omega_0 h^2$ models due to the slower growth rate
in a radiation-dominated universe.   We introduce a phenomenological
correction\footnote{A deeper and more complete 
analysis of this case is given in \cite{HSsmall}.}  
by taking the effective
drag epoch to be $z_d \approx 750$ for $\Omega_0 h^2 \approx 0.05$.
The evolution is
again well described by the interpolation function \eqn{DragGrow}
so that $\Delta_b(a) = {\cal G}(a,a_t)C_I$.
By this argument, the effective redshift to employ is $z_t \sim
{3 \over  5} z_d$. 
 
Now let us consider the effects of reionization at $z_i$. After $z_i$, Compton
drag again prevents the baryon perturbations from growing.  Therefore the final
perturbations will be $\Delta_b(a_0) \approx \Delta_b(a_i) D(a_0)/D(a_d)$.
Joining the transparent and ionized solutions, we obtain
\bel{eq:GrowLate}
\Delta_b/C_I = {\cases { {\cal G}(a,a_t)
                        & $a<a_g$  \cr
                {\cal G}(a_i,a_t){\cal G}(a,a_d) ,
                        & $a>a_g$ \cr} }
\ee
which is plotted in Fig.~\ref{fig:4comptondrag}b.
Since perturbations do not stop growing
immediately after reionization and ionization after
the drag epoch does not affect the perturbations, we take
$a_g = {\rm min}(1.1a_i,a_d)$.

For the photons, the continued ionization causes the diffusion length to
grow ever larger.  As the electron density decreases due to the expansion,
the diffusion length reaches the horizon scale and the photons effectively
last scatter.  As we have seen, diffusion destroys the intrinsic
fluctuations in the CMB.  Any residual fluctuations below the horizon
must therefore be
due to the coupling with the electrons.  Since last scattering follows the
drag epoch, the electrons can have a significant velocity at last scattering.
Thus we expect the Compton coupling to imprint a Doppler effect on the
photons at last scattering.  We will discuss this process in greater
detail in \S \ref{ch-secondary}.

%% file: chap4/cdmlarge.tex
\begin{figure}[t]
\centerline{\hskip -0.5truecm
\epsfxsize=3.5in \epsfbox{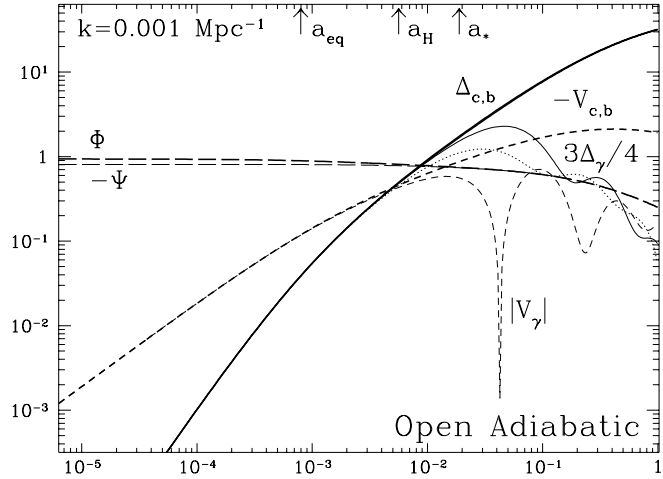}}
\vskip -0.5truecm
\caption{Large Scale Adiabatic Evolution}
\mycaption{All density fluctuations evolve adiabatically $\Delta_c =
\Delta_b = {3 \over 4} \Delta_\gamma = {3 \over 4} \Delta_\nu$ for
the cold dark matter, baryons, photons and neutrinos respectively above
the horizon $a < a_H$. 
Unlabeled dotted line is ${3 \over 4} \Delta_\nu$.
The potentials remain nearly constant until curvature
domination with a 10\% change at equality.  The small difference between
$\Phi$ and $-\Psi$ is due to the neutrino anisotropic stress (see Appendix
\ref{ss-5largepotential}).  After horizon crossing, the neutrinos free
stream as do the photons after last scattering $a_*$.  The model here
is a fully ionized adiabatic $\Omega_0=0.2$, $h=0.5$, $\Omega_b=0.06$ universe.
}
\label{fig:4cdmlarge}
\end{figure}

%% file: chap4/largeiso.tex
\begin{figure}[t]
\centerline{\hskip -0.5truecm
\epsfxsize=3.5in \epsfbox{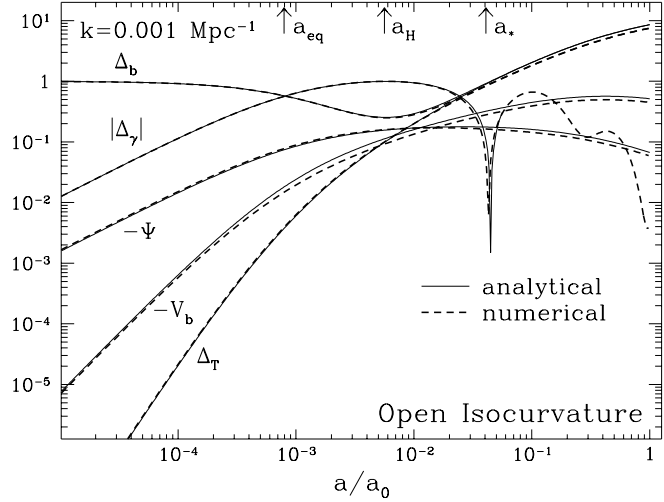}}
\vskip -0.5truecm
\caption{Large Scale Isocurvature Evolution}
\mycaption{Perturbations, which originate
in the baryons, are transferred to the radiation as the universe
becomes more matter dominated to avoid a significant curvature
perturbation.  Nonetheless, radiation fluctuations create
total density fluctuations from feedback.  These adiabatic
fluctuations in $\Delta_T$ dominate over the original entropy
perturbation near horizon crossing $a_H$
in the matter dominated
epoch.  The single fluid approximation cannot extend after last
scattering for the photons
$a_*$, since free streaming will damp $\Delta_\gamma$
away.  After curvature domination, the total density is prevented from
growing and thus leads to decay in the gravitational potential
$\Psi$.
}
\label{fig:4largeiso}
\end{figure}

%% file: chap4/wkb.tex
\begin{figure}[t]
\centerline{ \hskip -0.25 truecm
\epsfxsize=3.0in \epsfbox{chap4/wkba.epsf} \hskip 0.25 truecm
\epsfxsize=3.0in \epsfbox{chap4/wkbb.epsf}}
\vskip -0.5truecm
\caption{Acoustic Oscillations}
\mycaption{Pressure resists the gravitational forces of compressional 
(adiabatic) and rarefaction (isocurvature) leading to 
acoustic oscillations.  Baryons increase the gravitating mass
leading to higher compressional peaks, which dominate over the
rarefaction peaks and the Doppler line of sight velocity contribution
$V_\gamma/\sqrt{3}$ as $R$ is increased. 
Whereas
the isocurvature case has $\Omega_0 = \Omega_b$, the
adiabatic model has $\Omega_b=0.06$ and a consequently smaller
$R$.
Also displayed here is the semianalytic
approximation described in the text, which is essentially
exact.
The small difference in the numerical amplitudes of $\Phi$ and $\Psi$
is due to the anisotropic stress of the neutrinos (see \S \ref{ss-5largepotential}).  Here $\Omega_0=0.2$ 
and $h=0.5$.
}
\label{fig:4wkb}
\end{figure}

%% file: chap4/smalliso.tex
\begin{figure}[t]
\centerline{ \hskip -0.5truecm
\epsfxsize=3.5in \epsfbox{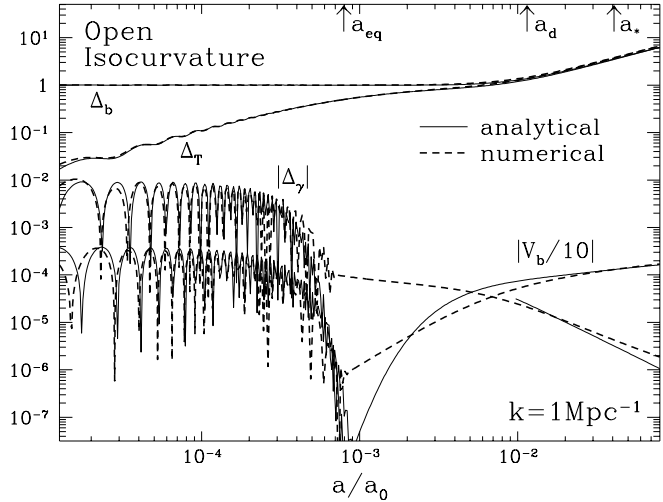}}
\vskip -0.5truecm
\caption{Small Scale Isocurvature Evolution}
\mycaption{At small scales gravity may be ignored, yielding pure adiabatic
oscillations.
Perturbations in the photons damp once the diffusion length grows
larger than the wavelength $k_D < k$.  Likewise the
adiabatic component of the baryon fluctuations also
damps leaving
them with the original entropy perturbation.  After diffusion,
the photons and baryons behave as separate fluids, allowing
the baryons to grow once Compton drag becomes negligible $a > a_d$.
Photon fluctuations are then
regenerated by the Doppler effect as
they diffuse across infalling baryons. 
The analytic approach
for the photons in this limit apply between the drag
epoch and last scattering $a_d < a < a_*$
(see \S \ref{ss-6regeneration}).
The model here is  $\Omega_0=0.2,h=0.5$, and no recombination.
}
\label{fig:4smalliso}
\end{figure}

%% file: chap4/drag.tex
\begin{figure}[t]
\centerline{\hskip -0.5truecm
\epsfxsize=3.5in \epsfbox{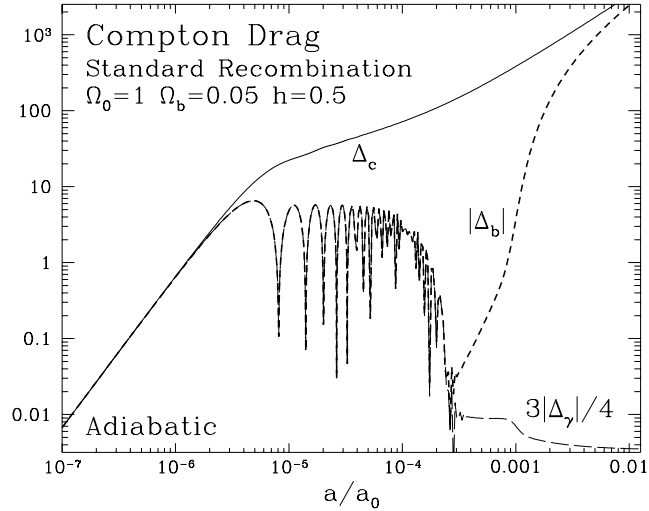}}
\vskip -0.5truecm
\caption{Compton Drag and Standard Recombination}
\mycaption{After recombination, the Compton drag on the baryons decreases
sharply.  The residual ionization after recombination however is sufficient
to slow baryon infall into dark matter wells.  The baryon 
and cold dark matter fluctuation $\Delta_c$ only converge at $z \simlt 100$.
}
\label{fig:4drag}
\end{figure}

%% file: chap4/comptondrag.tex
\begin{figure}[t]
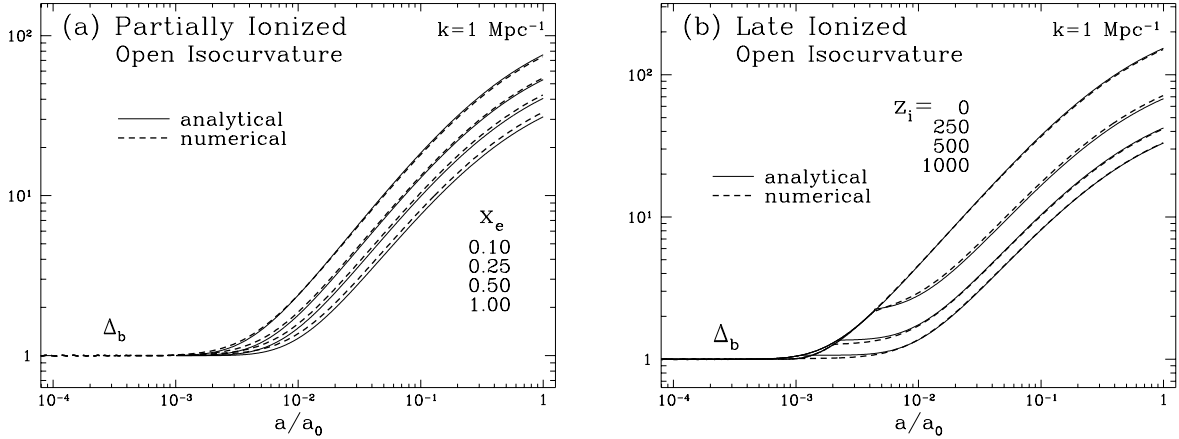

\centerline{ \hskip -0.5 truecm
\epsfxsize=3.0in \epsfbox{chap4/comptondraga.epsf} \hskip 0.5 truecm
\epsfxsize=3.0in \epsfbox{chap4/comptondragb.epsf}}
\vskip -0.5truecm
\caption{Compton Drag and Reionization}
\mycaption{(a) The baryons are released to grow in pressureless linear
theory after Compton drag becomes negligible.  Raising the ionization
fraction $x_e$ delays the end of the drag epoch and makes fluctuations
larger. (b) A transparent period between recombination and reionization
at $z_i$ leads to enhanced growth.  After reionization, fluctuations
are again suppressed until the end of the drag epoch.  The model 
here an open baryon isocurvature one with $\Omega_0=\Omega_b=0.2$ $h=0.5$.
}
\label{fig:4comptondrag}
\end{figure}

%% file: chap5.tex
\chapter{Primary Anisotropies}
\label{ch-primary}
\begin{verse}
\footnotesize\it\baselineskip=7pt
Trees in the mountains plunder themselves,

Grease in the flame sizzles itself,

Cinnamon has a taste, so they hack it down,

Lacquer has a use, so they strip it off.

All men know the uses of the useful,

No man knows the uses of the useless.

\vskip 0.1truecm
\hskip 2truecm {--Chuang-tzu, 4}
\end{verse}

\section{Overview}
\label{sec-5overview}

What can the study of anisotropy tell us about cosmology in 
general?  When the {\it COBE} DMR 
team first reported the detection of anisotropies on the $10^\circ$ 
scale and larger \cite{Smoot} at the $10^{-5}$ level, 
they were widely hailed as the panacea for all 
cosmological ills and uncertainties.  It was quickly realized however 
that that which makes the discovery so important also makes it less than 
ideal for pinning down cosmological models: anisotropies at this 
level are a generic prediction
of the gravitational instability picture for structure formation.  
The {\it COBE} DMR data {\it alone} lacks the dynamic range to distinguish
between closely related instability scenarios.  
However, combined with the smaller scale measurements of large scale
structure and the CMB itself, its true potential can be tapped. 

The CMB suffers from fewer problems of interpretation than large
scale structure since fluctuations are likely to be 
still in the linear regime
at last scattering.  It therefore has the potential to offer clean
tests of the underlying cosmology.
The current generation of anisotropy
experiments extends the angular scale coverage an order of magnitude
down to the degree scale.   The next generation of space based experiments
will probe yet another order of magnitude down to the several arcminute
scale.  It is important to realistically assess what cosmological 
information lies in the spectrum of anisotropies from arcminute scales
on up.  
The general study of anisotropy formation will be the focus of the 
remaining portion of this work. 
In this chapter, we consider primary anisotropy formation.  Specifically,
we assume that the photons free stream since recombination 
suffering only gravitational interactions
between $z_* \approx 1000$ and today.  
In the next chapter, we will consider processes in the intermediate
regime which may alter the anisotropy.

\subsection{Anisotropy Sources}
\label{ss-5sources}

At the most abstract level,
there are only two factors relevant to the formation of anisotropies:
gravitational interactions and Compton scattering.  Nevertheless,
their influence encodes a great wealth of cosmological information in
the CMB.  At the next level of detail,  
primary anisotropies are characterized by four 
quantities:

\begin{enumerate}

\item $\Theta_0(\eta_*,\bx)$: the intrinsic temperature at last scattering.

\item $\bg \cdot [\V_\gamma(\eta_*,\bx) - v_{obs}]$: the line of sight 
velocity at $\eta_*$ relative to the observer.

\item $\partial_x \Psi(\eta,\bx(\eta))$: 
the gradient of the gravitational potential along
the line of sight.

\item $\partial_\eta \Phi(\eta,\bx(\eta))$: 
the time derivative of the space curvature
along the line of sight.

\end{enumerate}

\noindent
Obviously intrinsic hot and cold spots on the last scattering surface
appear as anisotropies today.  The observed temperature of the background
is also  Doppler shifted due to the line of sight bulk motion
(dipole moment) 
of the photons at last scattering relative to 
the observer.  Our own peculiar velocity will just yield a
dipole anisotropy pattern on the sky.  The measured dipole in the CMB
is almost certainly due to this effect and implies that the local group
is moving at $627\pm22$ km/s with respect to the CMB frame \cite{Smoot91}.  
A spatial variation in the photon bulk velocity at last scattering
can result in an anisotropy at smaller angles.
Gradients in the gravitational potential cause redshifts and blueshifts
as the photons climb in and out of potential wells.  Changes in the
space curvature cause dilation effects due to the implied stretching
of space. This effect therefore has the the same origin as the 
cosmological redshift (see \S \ref{ss-2redshift}).

Even this description 
is not very 
useful unless we specify how the four quantities $\Theta_0$, $v_\gamma$,
$\Psi$ and $\Phi$ arise.   Linear perturbation theory,
developed in the last two chapters, supplies these quantities.
Let us
summarize those results.  The initial conditions and the subsequent
evolution of the total matter determines
the metric fluctuations $\Psi$ and $\Phi$ by the Poisson equation
\eqn{totalmatterPoisson}.  
These in turn feedback
on the matter and radiation through  gravitational infall 
and dilation.  For scales outside the horizon, gravitational interactions
alone determine the fluctuations and make all particle components in
the universe evolve similarly.

Inside the horizon, physical interactions must be taken into account.
Before recombination, Compton scattering couples the 
photons to the baryons.  From the Boltzmann 
equation  \eqn{Hierarchy} for the multipole moments,
this has two significant effects:

\begin{enumerate}

\item  $\V_b = \V_\gamma$: photons and baryons track each other during
their evolution.

\item $\Theta_\ell \propto e^{-\tau}, \ell \ge 2$: except for
the dipole, anisotropies are strongly  damped.

\end{enumerate}

\noindent Together they imply that the photons are isotropic in the baryon 
rest frame.  This also explains why the photons may be characterized by their
temperature and bulk velocity alone.

Since the two velocities are equal,
photons and baryons cannot stream away from each other.  This means
that number density fluctuations are frozen in, \ie\ the entropy 
fluctuation is constant [see
equation \eqn{Sdefinition}].
The photons and  baryons therefore evolve adiabatically and may be thought
of as a single photon-baryon fluid.
Photon pressure resists the gravitational compression of the
fluid and sets up acoustic waves.  The oscillations are frozen in
at last scattering leading to intrinsic temperature fluctuations $\Theta_0$ from
compression and rarefaction as well as bulk motion of the fluid $v_\gamma$.
At the smallest scales, photon diffusion amongst the baryons and subsequent
rescattering collisionally
damps fluctuations as $e^{-\tau}$ leading to a small scale
cut off in the spectrum at last scattering.  

\subsection{Projection and Free Streaming}
\label{ss-5projection}

After last scattering,
photons free stream toward the observer on radial null geodesics and
suffer only the gravitational interactions of redshift and dilation.  
Spatial fluctuations on the last scattering surface are observed as
anisotropies in the sky.  Free streaming thus transfers $\ell=0$
inhomogeneities and $\ell=1$ bulk velocities to
high multipoles as the $\ell$-mode coupling of the Boltzmann 
equation \eqn{Hierarchy}
suggests.   
Microphysically, this occurs because the paths of photons 
coming from hot and cold regions
on the last scattering surface cross.  Isotropic $\ell=0$ density
perturbations are thus averaged away collisionlessly. 
It is also evident that this conversion 
does not occur for superhorizon scales $k\eta \ll 1$ 
since the photons can travel only a small fraction of a wavelength. 

The background curvature also affects the photons in the free streaming
limit.  Due to the more rapid deviation of geodesics, a given length
scale will correspond to a smaller angle in an open universe than a flat
one.  Thus the {\it only} effect of negative
spatial curvature during free streaming is to speed the transfer of
power to higher multipoles [see equation \eqn{Hierarchy}].  
Its effect is noticeable if the angular
scale $\theta \sim \ell^{-1}$ is less than the ratio of the physical
scale to the curvature radius $ \sqrt{-K}/k$.
Notice that even the lowest eigenmode,
$k = \sqrt{-K}$ possesses $\ell$-mode coupling and hence 
free streaming damping of low-order multipoles,
once the horizon becomes larger than the curvature radius
$\eta\sqrt{-K} \simgt 1$.  
As discussed in \S \ref{ss-4completeness}, this is simply because
the $k = \sqrt{-K}$ eigenmode has structure only as large as the curvature
scale.  After the curvature scale passes inside the horizon, structure
at this scale is seen as an anisotropy on the sky as opposed to the 
featureless lowest flat eigenmode $k=0$.
If a truly scale invariant spectrum is desired, the modes must
be ``over-completed'' by taking $k \rightarrow 0$ in the open case
as well.

\subsection{Mathematical Description}
\label{ss-5mathematical}

A full description of the photon temperature must be two dimensional
to account for both the spatial and angular distribution
$\Theta(\eta,\bx,\bg)$. 
However, we can only observe the CMB from one location and hence have
information on the angular distribution alone.   The ensemble average
of the angular temperature correlation function can be decomposed
into the moments of the normal modes as 
\bel{eq:temperaturecorrelation}
\left< \Theta^*(\eta,\bx,\bg)\Theta(\eta,\bx,\bg') \right>
= {V \over 2\pi^2} \int_{k \ge \sqrt{-K}}^\infty \sum_\ell
{1 \over 2\ell+1} k^3 |\Theta_\ell(\eta,k)|^2 P_\ell (\bg'\cdot \bg),
\ee
where $P_\ell$ is a Legendre polynomial.  
Orthogonality of the $P_\ell$'s insures that $\ell$ modes do not
couple and the random phase assumption does the same for the $k$ modes. 
For models which predict 
supercurvature fluctuations, the lower limit of the integral should
be taken to zero.  The power in the $\ell$th 
multipole is usually denoted $C_\ell$, where
\bel{eq:Cl}
{2 \ell + 1 \over 4\pi} C_\ell = {V \over 2\pi^2} \int_{k \ge\sqrt{-K}}^\infty {d k \over k}
 k^3 {|\Theta_\ell(\eta,k)|^2 \over 2\ell+1}. \\
\ee
Note that the ensemble average anisotropy predicted for an experiment with
window function $W_\ell$ is $(\Delta T/T)^2 = \sum (2\ell+1)W_\ell
C_\ell/4\pi$ with $C_\ell$ 
evaluated at the present.\footnote{We only observe one
realization of the ensemble and thus $C_\ell$ must be estimated with
$2\ell+1$ measurements. Reversing this statement, there is a 
``cosmic variance,'' associated with a $\chi^2$ distribution of $2\ell+1$
degrees of freedom,  in the theoretical predictions for even an ideal
measurement.}
We can also sum in $\ell$
to obtain
\bel{eq:RMSL}
|\Theta + \Psi|_{rms}^2 \equiv |\Theta_0 + \Psi|^2 +
\sum_{\ell=1}^{\infty} 
{|\Theta_\ell|^2 \over 2\ell +1},
\ee
which measures the total power in a single $k$-mode.
Since fluctuations are merely transferred to high multipoles by
free streaming, the rms is conserved if $\dot \Phi = \dot \Psi =
\dot \tau = 0$,
as is evident from equation~\eqn{Gen}.  This merely indicates that the 
blueshift from falling into a static potential is exactly cancelled
by the redshift climbing out.

Up until this point, the initial spectrum in $k$ has been left arbitrary
since $k$ modes evolve independently.   
It is always possible to encorporate the evolution as a transfer function 
in $k$.  
However, each multipole $\ell$ of $C_\ell$
contains a sum over modes and does not evolve independently.
We will often employ as
examples simple power law initial spectra for which 
${\cal P}(k) = k^3 |\Phi(0,k)|^2 \propto k^{n-1}$ and $k^3 |S(0,k)|^2 \propto
k^{m+3}$ for adiabatic and isocurvature modes respectively.  
Thus $n=1$ and $m=-3$ are the scale invariant choices for
the spectrum.  Here scale invariance represents equal power per
logarithmic $k$ interval and is not equivalent to the commonly
employed choice of equal power per logarithmic $\tilde k = (k^2+K)^{1/2}$
interval (see \S \ref{ss-4laplacian} and Appendix \ref{sec-Bnormalization}). 

It is often instructive to consider 
the full angular and spatial information contained in
the two dimensional transfer function
\bel{eq:Weight}
T_\ell^2(k)  {\cal P}(k) \equiv
{V \over 2\pi^2} { 1 \over 2\ell+1} k^3 |\Theta_\ell|^2.
\ee
which satisfies 
$(2\ell+1)C_\ell/4\pi = \int T_\ell^2(k) {\cal P}(k) d\ln k$ for any
initial spectra.  Note that $\ell T_\ell^2$ also represents the power per   
logarithmic interval in $k$ and $\ell$ of anisotropies in the 
scale invariant model. 
 
\section{Sachs-Wolfe Effect}
\label{sec-5sachswolfe}

\input chap5/swbasic.tex

On large scales, gravity dominates the anisotropy through redshift
and dilation \cite{SW}.  Its
effects are usually broken up into two parts.  Contributions at or
before last scattering combine to form 
the ordinary Sachs-Wolfe (SW) effect.
Those occuring after last scattering are referred to as 
the integrated Sachs-Wolfe
(ISW) effect. After first describing their general nature, we will 
examine in detail their manifestation in a critical, open and $\Lambda$-dominated, adiabatic or isocurvature model. 

\subsection{Ordinary Sachs-Wolfe Effect}
\label{ss-5SW}
As the photons climb out of potential wells at last scattering,
gravity redshifts the temperature to $\Theta_0 \rightarrow \Theta_0 + \Psi$,
where $|\Psi| < 0$ in a potential well.  The effective
perturbation at last scattering is thus $[\Theta+\Psi](\eta_*)$.  
The combination of intrinsic temperature
fluctuations and gravitational redshift is called the ordinary 
Sachs-Wolfe (SW) effect \cite{SW}.  For a gauge choice other than
Newtonian, the two
may be divided up in different ways.

The intrinsic fluctuations at $\eta_*$ are in turn determined
by gravitational effects {\it before} last scattering.  If $k\eta \ll 1$,
the Boltzmann equation~\eqn{Hierarchy} reduces to the dilation effect
\bel{eq:Reduced}
\dot \Theta_0 = -\dot \Phi \approx \dot \Psi,
\ee
or 
\bel{eq:SWdifference}
[\Theta_0+\Psi](\eta) \approx \Theta_0(0) + 2\Psi(\eta_*) - \Psi(0).
\ee
Here we have again assumed $\Pi_T=0$, which causes a $\sim 10\%$
error (see \S \ref{ss-5largepotential}).  

\subsubsection{Isocurvature and Adiabatic Cases}
 
Since the isocurvature
initial conditions satisfy
$\Psi(0) = 0 = \Theta_0(0)$, equation \eqn{Reduced} 
implies $\Theta_0(\eta) = \Psi(\eta)$.
The effective
superhorizon scale temperature
perturbation for isocurvature fluctuations
is therefore
\bel{eq:SWIso}
\Theta_0 + \Psi = 2\Psi. \qquad ({\rm iso})
\ee
The growing potential stretches space so as to dilute the
photon density in the well.  
Gravitational redshift out of
the well subsequently doubles the
effect. Note however that in a low $\Omega_0h^2$ model
with standard
recombination, the potential may not reach its full
matter-dominated value of $\Psi = -{1 \over 5} S(0)$ from
equation~\eqn{PsiLS} by last scattering (see Fig.~\ref{fig:5swbasic}).
 
For adiabatic perturbations, 
the initial conditions require
$
\Theta_0(0) = -{1 \over 2} \Psi(0)
$
[see equation~\eqn{DeltaLS}], reflecting the fact that the 
photons are overdense inside the
potential well [see equation \eqn{ThetaA}].  
Although the potential is constant in both
the matter- and radiation-dominated epoch, it changes to
$
\Psi(a) = {9 \over 10}\Psi(0)
$
through equality.
The dilation effect then brings the photon temperature perturbation in
the matter-dominated epoch
to $\Theta(\eta) = - { 2 \over 3} \Psi(\eta)$.  Thus the 
effective perturbation is
\bel{eq:SWAdi}
[\Theta_0 + \Psi] = \cases{ 	{1 \over 2} \Psi & RD \cr
				{1 \over 3} \Psi, & MD  \cr } \qquad {\rm (adi)}
\ee
where the latter is the familiar Sachs-Wolfe result.
Again since last scattering may occur before full matter
domination, one should employ the full form of equation \eqn{SWdifference}
instead of the asymptotic form from equations \eqn{PsiLS} and \eqn{SWAdi}.
After $a_*$, the photons climb out of the potential wells,
leaving the quantity $[\Theta_0 + \Psi](\eta_*)$ to be viewed 
as temperature fluctuations on the sky today.  

\subsubsection{Free Streaming Solution}

To determine the exact nature of the resultant anisotropies, one must follow
the photons from last scattering to the present.
%Since photon geodesics
%are radial in the absence of scattering, we may use
%the radial eigenfunctions of the Laplacian $X_\nu^\ell$ 
%to solve for the free
%streaming behavior (see \S \ref{ss-4radial}). 
The collisionless Boltzmann equation for $(\Theta+\Psi)/(2\ell+1)$
takes the
same form as the recursion relation for the radial eigenfunctions of 
the Laplacian
[{\it c.f.} equations \eqn{AeqnRecursion} and \eqn{Hierarchy}].  This
is natural since the radiation free streams on null geodesics.
Thus the spatial fluctuation represented by $[\Theta+\Psi](\eta_*,k)$ is seen 
by the distant observer as
an anisotropy of
\bel{eq:SWstream}
{\Theta_\ell(\eta,k) \over 2\ell+1} =
[\Theta_0+\Psi](\eta_*,k) X_\nu^\ell (\chi - \chi_*),
\ee
where recall that $\chi=\sqrt{-K}\eta$. 
In the flat case, $X_\nu^\ell \rightarrow j_\ell$ which peaks at $\ell
\approx k(\eta-\eta_*)$.  If the distance traveled by the photon is
under a wavelength, \ie\ $k(\eta-\eta_*) \ll 1$, then only $j_0$ has
weight and fluctuations remain in the monopole.  As time progresses,
power is transferred from the monopole to high $\ell$ as one would
expect from the projection effect (see Fig.~\ref{fig:5swbasic}).  

In the adiabatic flat case, power law models for the initial conditions 
$k^3 |\Phi(0)|^2 = B k^{n-1}$ have a simple form for the
Sachs-Wolfe  contribution to $C_\ell$.  If we assume that
the universe was matter dominated at last scattering, $ \Theta_0 + \Psi
= {1 \over 3} \Psi$.  From equation \eqn{Cl},
\beal{eq:SWformula}
C_\ell^{SW} \eapp \left({1 \over 3}{\Psi(\eta_*) \over \Phi(0)}\right)^2
{2 \over \pi} BV \int{dk \over k} k^{n-1} j_\ell^2(k\eta_0) \nonumber\\
\eapp { 9 \over 200 \sqrt{\pi}} BV \eta_0^{1-n} 
	{ \Gamma[(3-n)/2]\Gamma[\ell+(n-1)/2] \over
	  \Gamma[(4-n)/2]\Gamma[\ell+(5-n)/2] },
\eea
where we have employed the relation $\Psi(\eta_*)/\Phi(0) = -9/10$
of equation \eqn{PsiAdi}.  
In this flat model,
\bel{eq:lambdaeta0}
\eta_0 \approx 2 (\Omega_0 H_0^2)^{-1/2} (1 + \ln\Omega_0^{0.085}),
\ee
where the small logarithmic correction is from the rapid expansion
at the present in a $\Lambda$ universe.
Notice that for scale invariant spectra,
the projection factor $\eta_0^{n-1}$ vanishes.  With equal power at 
all scales, it does not matter which physical scale gets mapped onto
a given angular scale.  

Equation \eqn{SWformula} is more commonly expressed in terms of the
amplitude of the
matter power spectrum today $|\Delta_T(\eta_0,k)|^2 = Ak^n$.  
From equation \eqn{DeltaA}, the relation between the two normalizations
is 
\beal{eq:ABnormalization}
B \eal {25 \over 36} k_{eq}^4 D^{-2} A \nonumber\\
  \eal {25 \over 9} (\Omega_0 H_0^2)^2 (a_0/D_0)^2 A,
\eea
where $D_0=D(\eta_0)$ and 
recall that $D$ is the pressureless growth factor normalized at
equality.  Since in a $\Lambda$ universe, growth is suppressed and
$a_0/D_0 < 1$, the same matter power spectrum normalization $A$ 
implies a greater Sachs-Wolfe anisotropy since it was generated
when the potentials were larger.  The final expression becomes
\bel{eq:GammaFormula}
C_\ell^{SW} \approx
{ 1 \over 8\sqrt{\pi}} AV H_0^4 \Omega_0^2 (a_0/D_0)^2 \eta_0^{1-n} 
	{ \Gamma[(3-n)/2]\Gamma[\ell+(n-1)/2] \over
	  \Gamma[(4-n)/2]\Gamma[\ell+(5-n)/2] }.
\ee
The factor $\Omega_0^2 (a_0/D_0)^2 \approx \Omega_0^{1.54}$ 
for $\Lambda$ models
\cite{EBW}.
Since $\Gamma(\ell+2)/\Gamma(\ell) = \ell(\ell+1)$, the Sachs-Wolfe
contribution for a scale invariant $n=1$ spectrum
is flat in $\ell(\ell+1)C_\ell$.  We will therefore
occasionally plot $\ell(\ell+1)C_\ell/2\pi$ instead of the logarithmic power 
$\ell(2\ell+1)C_\ell/4\pi$ as has become standard convention.  
For $\ell \gg 1$, the two conventions yield identical
results.  Note that this formula describes only 
the Sachs-Wolfe contributions 
and does not account for the early ISW and acoustic contributions, which
push the high $\ell$ tail up, and the late ISW effect, which enhances the low 
$\ell$ multipoles.

\input chap5/earlyisw.tex

\subsection{Integrated Sachs-Wolfe Effect}
\label{ss-5ISW}

If the potentials vary with time, the photon will experience differential
redshifts due to the gradient of $\Psi$, which no longer yield equal and
opposite contributions as the photons enter and exit the potential well,
and time dilation from $\Phi$.
They act like an impulse $(\dot \Psi - \dot
\Phi) \delta \eta$ at some intermediate time $\eta$ which then free
streams to the present.  
The sum 
of these contributions along the line of sight is called the integrated
Sachs-Wolfe (ISW) effect.  
By the same reasoning that lead to the
solution for the Sachs-Wolfe effect, one can immediately write
down the solution for the combined effect:
\bel{eq:SachsWolfe}
{\Theta_\ell(\eta,k) \over 2\ell+1} =
[\Theta_0+\Psi](\eta_*,k) X_\nu^\ell (\chi -\chi_*)
+ \int_{\eta_*}^{\eta} [\dot \Psi - \dot \Phi](\eta',k)
X_\nu^\ell (\chi-\chi') d\eta'.
\ee

\noindent
Since the potentials for both the adiabatic and isocurvature modes
are constant in the matter-dominated epoch, the ISW contribution is
separated into two parts:

\begin{enumerate}

\item The early ISW effect from radiation domination: (a) 
isocurvature growth before horizon 
crossing and (b) pressure growth suppression after
horizon crossing for either mode.

\item The late ISW effect due to expansion growth
suppression
in the $\Lambda$- or curvature-dominated epoch.

\end{enumerate}

\noindent In adiabatic models, scales which cross the sound horizon in 
the radiation-dominated epoch experience a boost from the
decay of the potential (see Fig.~\ref{fig:5earlyisw}a).  
Since the effect is due to radiation pressure 
and depends only on the epoch of equality, open
and $\Lambda$ models predict identical contributions.  These scales
will furthermore not experience significant late ISW effects since
the potential has already decayed by $\Lambda$ or curvature domination.
On the other hand, larger scales are 
unaffected by the early ISW effect and suffer only
the consequences of the late ISW effect.  Because $\Lambda$ domination
occurs only recently if $\Omega_0 \simgt 0.1$, the potential will not
have had a chance to fully decay and the net effect is smaller than 
in the corresponding open case.  

For isocurvature models, potential growth outside the sound horizon
in the radiation-dominated epoch forces the temperature fluctuation to
grow with it through the dilation effect (see Fig.~\ref{fig:5earlyisw}b).  
Modes which cross
only after matter domination experience the full effect of growth.  
For scales that cross during radiation domination, radiation 
pressure suppresses further growth.  Thus large scale modes are 
enhanced over small scale modes.  Since isocurvature models are
dominated by this early ISW effect, the difference 
between open and $\Lambda$ models is smaller than in adiabatic models.

The total Sachs-Wolfe effect predicts rich structure in the anisotropy 
spectra.
To understand the full Sachs-Wolfe spectrum, it is necessary to examine
simultaneously the spatial and angular information in the radiation.
It will therefore be instructive to consider the radiation transfer 
function $T_\ell(k)$,  rather than $C_\ell$ for any one model. 
Note that $\ell T^2_\ell(k)$ is equivalent to the logarithmic
contribution in 
$k$ and $\ell$ to the anisotropy of a scale invariant model [see
equation~\eqn{Weight}].  Summing in $k$ produces $\ell (2\ell+1)C_\ell/4\pi$
and in $\ell$ yields $k^3|\Theta+\Psi|^2_{rms}$ for this model.

%\subsection{Sachs-Wolfe Spectrum}
%\label{ss-5sachswolfe2}
%\smallskip

\subsection{Adiabatic $\Omega_0=1$ models}
\label{ss-5omega1}

\input chap5/cflat.tex

To build intuition for equation~\eqn{SachsWolfe},
let us first
consider the familiar adiabatic $\Omega_0=1$ model in which the
ISW term represents only a small correction.
A given $k$-mode contributes maximally to the angle that scale subtends
on the sky at last scattering.  The transfer function therefore
displays a sharp ridge corresponding 
to this correlation (see Fig.~\ref{fig:5cflat}a),
\bel{eq:angledistance}
\ell_{main} + {1 \over 2} \approx k r_\theta(\eta_*) ,
\ee
where the comoving angular diameter distance is
\bel{eq:rtheta}
r_\theta(\eta) = (-K)^{-1/2} \sinh(\chi_0-\chi) ,
\ee
and reduces to $r_\theta= \eta_0-\eta_*$ as $K \rightarrow 0$.
It is evident from Fig.~\ref{fig:5cflat}a that the full result contains more 
than just this main correlation ridge.
The conversion of fluctuations in a spatial eigenmode $k$
on the last
scattering surface into anisotropies on the sky
is basically a projection of the eigenmode in the spherical geometry.
For example, a plane wave $\exp(ik\Delta\eta)$ can be written as a 
sum over $j_\ell(k\Delta\eta)Y_\ell^m$.  
Since the projection is not precisely one-to-one, a given mode will
project onto a range of angles.  In fact, it will alias angles equal to
and larger than what the main
face on $\bk \perp \bg$ projection of equation \eqn{angledistance} predicts,
\ie\ $\ell \le \ell_{\rm main}$, as is clear from Fig.~\ref{fig:1projection}.  
This
is expressed by the oscillatory structure of the radial
eigenfunction.  Comparing panels in Fig.~\ref{fig:5cflat},  we see that
the structure in the transfer function is indeed due to this effect. 

\input chap5/flatisw.tex

Even with $\Omega_0=1$, a low $h \approx 0.5$ model has 
additional contributions 
after last scattering. The early ISW effect affects modes that 
cross the sound 
horizon between last scattering and full matter domination. 
Since these contributions come from near last 
scattering, the ISW integral \eqn{SachsWolfe} may be approximated as
\beal{eq:ISWapprox1}
\int_{\eta_*}^{\eta_0} [\dot \Psi - \dot \Phi] j_\ell(k(\eta_0-\eta)) d\eta
\eapp 
\int_{\eta_*}^{\eta_0} [\dot \Psi - \dot \Phi] j_\ell(k\eta_0) d\eta 
\nonumber\\
\eal [\Delta\Psi-\Delta\Phi] j_\ell(k\eta_0),
\eea
which is strictly only valid for contributions from $k\eta \ll 1$.  Contributions to the $k$th mode in fact occur near horizon crossing where $k\eta \approx 1$.
Nevertheless this approximation is instructive.

The early ISW effect adds nearly coherently with the SW effect and in fact cancels 
it by removing the redshift that the photon would otherwise suffer.  
At large scales, this brings the total effect down to the matter-dominated
${1 \over 3}\Psi(\eta_0)$ value and thus changes the large scale 
normalization.  At scales approaching the sound horizon
at last scattering, it increases the effective temperature from the 
acoustic compression again by removing the cancelling redshift.  In
Fig.~\ref{fig:5flatisw}, 
we compare the approximation of equation \eqn{ISWapprox1} to the the full
integral and the effect of dropping the contribution entirely.  
Notice that, aside from its affect on the normalization,
the early ISW contribution fills in scales somewhat larger
than the sound horizon at last scattering.  The approximation underestimates
the angular scale somewhat by assuming that the contribution comes from
the further distance $\eta_0$ as opposed to the true distance $\eta_0-\eta$. 

\input chap5/lamb.tex

\subsection{Adiabatic $\Lambda$ Models}
\label{ss-5lambda}

Now let us move onto the more complicated $\Lambda$ case.
For $\Lambda$ models, the ISW term in equation \eqn{SachsWolfe}
yields both early and late type contributions.  The boost on intermediate
scales from the early ISW effect is much more dramatic than for the
high $\Omega_0 h^2$ models. 
In the transfer function,
this appears as a high ridge crossing larger angles for the same $k$
as the SW effect, due to its origin closer to 
the observer. 
The maximum contribution the early ISW effect can make is if the
potential decayed to zero between last scattering and the present.
From the relation $\dot \Theta_0 = -\dot \Phi$, this  
would yield $\Theta_0(\eta) = \Theta_0(0) + \Phi(0) = {3 \over 2}
\Phi(0)$.  Compared with the matter-dominated SW tail of ${1 \over 3} \Psi
= - {3 \over 10} \Phi(0)$, the early ISW effect can approach a height
5 times greater than the SW tail.  Note that the same decay drives
the acoustic oscillation to a similar height so that this effect
will join smoothly onto the acoustic peaks as we shall see below.
The lack of potential decay for scales that enter the horizon during matter 
domination makes the early ISW ridge drop off at large scales 
(see Fig.~\ref{fig:5lamb}).
 
\input chap5/lateisw.tex

After $\Lambda$ domination $a_\Lambda/a_0 =
 (\Omega_0/\Omega_\Lambda)^{1/3}$,
the potential once again decays. 
For typical values of $\Omega_0 \simgt 0.1$, this occurs only 
recently. Furthermore,  the potential
at all scales decays at the same rate.  The expansion time scale at 
$\Lambda$ domination $\eta_\Lambda=\eta(a_\Lambda)$ sets 
a critical wavelength corresponding to $k\eta_\Lambda = 1$.  The ISW
integral takes on different form in the two regimes separated by this 
division 
\bel{eq:ISWapprox}
\int_{\eta_*}^{\eta_0} [\dot \Psi - \dot \Phi] 
j_\ell[k(\eta_0-\eta)] d\eta 
\approx 
\cases { 
[\Delta\Psi-\Delta\Phi] j_\ell[(k\eta_0 
- k\eta_\Lambda)] & $k\eta_\Lambda \ll 1$ \cr
[\dot\Psi-\dot\Phi](\eta_k) I_\ell/k, & $k\eta_\Lambda \gg 1$ \cr}
\ee
where $\Delta\Phi$ and $\Delta\Psi$ are the changes in the potential
from the matter-dominated form of \eqn{PsiAdi} to the present.  
We have used the angle-distance relation \eqn{angledistance} to
find the peak of $j_\ell$ at $\eta_k = \eta_0 -(\ell+1/2)/k$.  
The integral $I_\ell$ is given by 
\bel{eq:jlintegral}
I_\ell \equiv \int_0^\infty dx j_\ell(x) = {\sqrt{\pi} \over 2} 
	{\Gamma[{1 \over 2}(\ell+1)] 
	\over \Gamma[{1 \over 2}(\ell+2)]}.
\ee
The limits correspond physically to two cases:

\begin{enumerate}

\item If the wavelength is much longer than distance a photon can 
travel during the decay, photons 
essentially receive an instantaneous kick.  The result is similar to
the SW and early ISW effects.

\item In the opposite limit, the photon traverses many wavelengths during
the decay and suffers alternating red and blueshifts from crests and 
troughs.  The result is a cancellation of contributions.

\end{enumerate} 

\noindent Since $\Lambda$ domination occurs near the present, the critical
wavelength is approximately the horizon size at present and 
yields $\ell=0$
monopole contributions along the projection ridge.  Thus most contributions
will come from the cancellation regime if the $k$ modes are weighted 
equally.  
We can verify this by comparing the cancellation approximation
with the full integral for the scale invariant model.
Fig.~\ref{fig:5lateisw} shows
that in this case the cancellation approximation is excellent.  
Compared with the SW effect which
predicts a flat spectrum, the late ISW $\Lambda$ contributions fall with
$\ell$ due to cancellation.
For the more general case of 
power law initial spectra $k^3 |\Phi(0,k)|^2 = Bk^{n-1}$, the 
total contribution becomes
\bel{eq:ClISW}
C_\ell^{ISW} \approx 2 \left( { 9 \over 10} \right)^2 BV 
\left( {\Gamma[(\ell+1)/2] \over \Gamma[(\ell+2)/2]} \right)^2 \int_0^\infty
{d k \over k} k^{n-3} \left[ {D \over a} \left( {\dot D \over D} - 
{\dot a \over a} \right) \right]^2_{\eta=\eta_k},
\ee
where we have employed equation \eqn{PsiAdi} for the potentials
and recall that the growth factor $D$ is normalized such that
$D(a_{eq})=a_{eq}=1$.

\input chap5/sepadi.tex

Let us take a closer look at the transfer function in Fig.~\ref{fig:5lamb}.
For $k\eta_\Lambda \ll 1$, cancellation is ineffective 
and like its early counterpart, 
the late ISW effect opposes the SW effect.
In Fig.~\ref{fig:5sepadi}a, we plot the analytic
decomposition of contributions to a $k$-mode slice corresponding
to these large scales.  As one can see from equation \eqn{ISWapprox},
these modes contribute little to $\ell \ge 2$, since $k(\eta_0-\eta_\Lambda)
\ll 1$.  For intermediate scales, the late ISW effect itself is partially 
cancelled. 
The ridge structure of
Fig.~\ref{fig:5lamb} at the low multipoles 
is due to the late ISW effect adding with every other ridge
in the SW free streaming oscillation (see Fig.~\ref{fig:5sepadi}b).
At the smallest scales, those which
would ordinarily contribute to higher order multipoles, the late ISW
effect is entirely cancelled.  Again this implies that 
typical adiabatic $\Lambda$ spectra have 
a small boost in anisotropies from the late ISW effect only at the
lowest multipoles (see Fig.~\ref{fig:5lateisw}b).

\input chap5/open.tex

\subsection{Adiabatic Open Models}
\label{ss-5open}

Open adiabatic models follow similar physical principles. 
The early ISW effect depends only on the matter-radiation ratio near 
last scattering from $\Omega_0h^2$ and thus is identical to the
$\Lambda$ case.  However, photons curve on their geodesics so that
the projection takes the same physical scale to a significantly
smaller angular 
scale.   This is quantified by
the angle to distance relation \eqn{rtheta}.  In the transfer
function, one sees that the early ISW ridge is pushed to significantly
higher $\ell$ (see Fig.~\ref{fig:5open}). 

Curvature dominates at $a/a_0 = \Omega_0 /(1-\Omega_0)$ leaving
the potential more time to decay than in the $\Lambda$ model.  
The late ISW effect will therefore
be more significant in this model (see Fig.~\ref{fig:5earlyisw}a). 
Moreover, the cancellation scale
is smaller leading to a less sharp decline with $k$ (or $\ell$) of the effect.
The net result is that the late ISW cancellation tail merges smoothly
onto the early ISW rise for sufficiently low $\Omega_0$.  For 
$\Omega_0 \approx 0.1-0.3$, they overwhelm the SW effect on 
all scales.

\input chap5/ratra.tex

Unlike the flat case, there is a lowest eigenmode corresponding to
the curvature scale $k=\sqrt{-K}$.  Supercurvature scales that would
ordinarily contribute to low order multipoles are absent unless the
modes are ``overcompleted'' (see \S \ref{ss-4completeness}).  For the
low $\Omega_0=0.1$ example displayed in Fig.~\ref{fig:5open}, 
this cutoff at $\log(k*$Mpc$)\approx 3.8$ chops off 
some of the main projection ridge of the late ISW effect 
for the lowest multipoles.  Thus the 
absence of supercurvature modes in the sum over $k$ can lead to
a slight suppression of the lowest multipoles.  With scale 
invariant weighting of the $k$-modes, the spectrum has the form 
shown in Fig.~\ref{fig:5ratra}.  
Note that this is the typical 
\cite{LS,RP} but not unique \cite{Bucher} prediction 
of open inflationary models.

\input chap5/iso.tex

Due to its more recent origin, the 
late ISW effect projects onto a significantly larger angle
than the SW effect for a given $k$.
Examining the individual contributions in 
Fig.~\ref{fig:5sepadi}c, we 
see that indeed at the curvature scale, the late ISW effect
affects 
the lowest multipoles, whereas the SW effect peaks around $\ell \approx 10$.
Thus the presence or lack of supercurvature modes is not as significant
as one might naively expect from the fact that the curvature scale subtends
$\ell \approx 10$ at the horizon distance in an $\Omega_0=0.1$ universe.
For a smaller scale chosen to intersect the main late ISW projection ridge
in Fig.~\ref{fig:5sepadi}d, 
we see that the late ISW effect completely dominates the
SW effect as claimed.

\input chap5/sepiso.tex

\subsection{Isocurvature $\Lambda$ and Open Models}
\label{ss-5iso}
 
Isocurvature models differ significantly
 in that the potentials {\it grow} until
full matter domination.  Strong early ISW contributions which
are qualitatively similar to the SW term will
occur {\it directly} after recombination and continue until full matter
domination (see Fig.~\ref{fig:5swbasic}).  Thus the projection
of scales onto angles will follow a continuous
sequence which merges the SW and
early ISW ridges (see Fig.~\ref{fig:5iso}).

\input chap5/alias.tex

For the $\Lambda$ case, the early ISW effect
completely dominates that of the late
ISW effect.  Thus the analytic separation shows that the 
ISW and SW effects
make morphologically similar contributions and the boost in low order multipoles
is not manifest.  Moreover, the two add coherently creating a
greater total effect unlike the adiabatic case (see Fig.~5a,b).
 Open isocurvature models
behave similarly except that the late ISW contributions
near its maximum (late ISW ridge) is not entirely negligible.
It is thus
similar to the adiabatic case ({\it c.f.} Fig.~\ref{fig:5sepadi}d and 
\ref{fig:5sepiso}d) except that it
does not usually
dominate the {\it total} anisotropy.  Note that the curvature cutoff 
can strongly affect the anisotropy spectrum since the curvature scale
projects onto $\ell \approx 10$ for the SW and early ISW contributions
in the $\Omega_0=0.1$ model. 
There will be a deficit of power at $\ell \simlt 10$ if no supercurvature
contributions are considered.

\input chap5/neff.tex

On the other hand, the scale invariant model represented here
does not present a viable model for structure formation.
As discussed in \S \ref{ss-5ISW}, potential growth leads to an enhancement
of large over small scale power.
The initially 
scale invariant isocurvature $m=-3$ model has insufficient small 
scale power to form galaxies.  The problem
can be alleviated by increasing the spectral index to $m \approx -1$.  
This has significant effects on the anisotropy.  By heavily weighting
the small physical scales, we enhance the projection aliasing contribution
from the higher ridges of Fig.~\ref{fig:5iso}.  
 This aliasing or power bleeding from small scales makes
the anisotropy spectrum less steep (blue) than the spatial power spectrum 
(see Fig.~\ref{fig:5alias}).

In fact, there is an upper limit as to how fast anisotropies can rise with
$\ell$.  Suppose that the spectrum is so blue as to have all contributions
come from the smallest physical scale in the problem $k_{cut}$, 
\eg\ the photon
diffusion scale at last scattering.  In this case, $j_\ell(k_{cut}\Delta\eta)$
becomes independent of $\ell$ and thus $\Theta_\ell \propto 2\ell+1$ 
from equation \eqn{Cl} or $C_\ell \approx$ constant.  This corresponds
to an effective large scale slope of $n_{\rm eff} = 3$ as compared
with the adiabatic SW prediction of equation \eqn{GammaFormula}.  
Isocurvature
$m=-1$ models are an intermediate case with $n_{\rm eff} \approx 2$.  
Since the effect is from small scale power aliasing
for $m \simgt -2$,
the effective
anisotropy slope will only weakly depend on the initial power spectrum
slope $m$.  In Fig.~\ref{fig:5neff}, 
we plot the dependence of isocurvature large
scale anisotropies with $m$.  Note that because the power comes from
small scales, large scale anisotropies are not sensitive to the initial
spectrum at large spatial scales.
In particular, possible curvature scale ambiguities, such as the absence
(or presence) of supercurvature modes which can suppress (enhance) the low
order multipoles,  have little effect on the result for $m \simgt -2$.

\input chap5/power.tex

\section{Acoustic Peaks}
\label{sec-5acoustic}

On scales below the sound horizon,
acoustic oscillations imprint hot
and cold spots from regions caught in compression and rarefaction at 
last scattering.
Viewed today, these become peaks in the anisotropy
power spectrum.   Since acoustic oscillations are generic in the
gravitational
instability scenario for structure formation, these peaks contain valuable
model-independent cosmological information.

\subsection{Mathematical Description}

Acoustic contributions are described by the phase and the amplitude of the
sound waves at last scattering.  
Since different $k$ modes are frozen at different phases of their oscillation,
there will in general be a series of peaks in the temperature and 
velocity spectra at last scattering.  The bulk velocity of the
photon fluid contributes as a Doppler shift in the observed temperature. 
The fluctuations captured at last
scattering for a scale invariant adiabatic model is displayed in Fig.~\ref{fig:5power}.  

These fluctuations are projected onto anisotropies as
\beal{eq:FreeStream}
{\Theta_\ell(\eta) \over 2\ell+1} \eal [\Theta_0+\Psi](\eta_*,k)
        X_\nu^\ell(\chi-\chi_*) + \Theta_1(\eta_*,k)
        {1\over k}{d \over d \eta}X_\nu^\ell(\chi-\chi_*) \nonumber\\
        & & + \int_{\eta_*}^{\eta} (\dot \Psi -\dot\Phi)
        X_\nu^\ell(\chi-\chi')
        d\eta' \, , 
\eea
(see Appendix \ref{ss-5analytic} for a derivation).  The 
dipole projects in a different manner than the monopole because
of its angular dependence.  
The face on $\bk \perp \bg$ mode of the ``main projection'' (see
Fig.~\ref{fig:5cflat} and \ref{fig:1projection}) vanishes for the 
Doppler effect which arises because of the line of sight velocity.  
This causes velocity contributions to be out of phase with the temperature as 
the derivative structure suggests and indicates that the two effects add 
in quadrature.

Due to the finite duration of last scattering, the effective 
fluctuations $[\Theta_0+\Psi](\eta_*)$
and $\Theta_1(\eta_*)$ are more severely diffusion damped than one might
naively expect.  As the 
ionization fraction drops due to recombination, the mean free path and hence
the diffusion length increases.  We will see how this affects the amplitude
of oscillations in \S \ref{ss-5diffusion}.  Once this is accounted for, the
tight coupling description of the acoustic oscillations from \S \ref{sec-4subhorizon}
leads to
an excellent description of the resultant anisotropy (see \cite{HSa}
and Appendix \ref{ss-5analytic}).  
It is useful however to extract a few simple 
model-independent results.

\input chap5/peak.tex

\subsection{Location of the Peaks}
\label{ss-5location}

The most robust feature of the acoustic oscillations is the angular
location of the peaks.
Consider first, the
spatial power spectrum at last scattering.   Peaks will occur at extrema
of the oscillations, \ie\
\bel{eq:physicalpeaks}
k_p r_s(\eta_*) = \cases{ p\pi & adi \cr
			  (p-1/2)\pi, & iso \cr }
\ee
where the sound horizon at last scattering is
\bel{eq:soundls}
r_s(\eta_*) = \int_0^{\eta_*} c_s  d\eta'
= {2 \over 3} {1 \over k_{eq}} \sqrt{6 \over R_{eq}}
  \ln { \sqrt{1+R_*} + \sqrt{ R_* + R_{eq} }
    \over
        1 + \sqrt{ R_{eq}}},
\ee
with $k_{eq} = (2\Omega_0 H_0^2 a_0/a_{eq})^{1/2},$ 
$a_{eq}/a_0 = 2.38 \times 10^{-5} \Theta_{2.7}^4 (\Omega_0 h^2)^{-1} 
(1-f_\nu)^{-1}$ and recall $R=3\rho_b/4\rho_\gamma$, \ie\
\beal{eq:rfactors}
R_{eq} \eal {1 \over 1-f_\nu}{3 \over 4}{\Omega_b \over \Omega_0},  \nonumber\\
R_* \eal 31.5 \Omega_b h^2 \Theta_{2.7}^{-4} (z_*/10^3)^{-1},
\eea
$\Theta_{2.7}=T_0/2.7$K and $(1-f_\nu)^{-1} = 1.68$ for
three massless neutrinos.

From equation \eqn{angledistance}, the scale $k_p$ subtends an angle
\bel{eq:ellpeak}
\ell_p \approx k_p r_\theta(\eta_*),
\ee
where 
\bel{eq:rthetaeval}
r_\theta(\eta_*) \approx \cases {2 (\Omega_0 H_0)^{-1} & $\Omega_\Lambda=0$ \cr
				 2 (\Omega_0 H_0^2)^{-1/2} (1+
				\ln\Omega_0^{0.085}). &  
					$\Omega_\Lambda+\Omega_0=1$ \cr}
\ee		
For low $\Omega_b h^2$, $R_* \ll 1$ and the sound horizon at last 
scattering reduces to
\bel{eq:soundhorizonapp}
r_s(\eta_*) \approx {1 \over \sqrt{3}}\eta_* 
 \approx {2 \over \sqrt{3}} (\Omega_0 H_0^2)^{-1/2}[(1+x_R)^{1/2}-x_R^{1/2}]
z_*^{-1/2},
\ee
where the radiation contribution at last scattering produces the modification
factor in square brackets with
$x_R = {a_{eq}/a_*}$.
Note that the correction factor in equation \eqn{soundhorizonapp} goes
asymptotically to $1$ and ${1 \over 2} x_R^{-1/2} \propto (\Omega_0 h^2)^{1/2}
(1-f_\nu)^{1/2}$ in the high and low $\Omega_0 h^2$ limits respectively. 

Let us summarize these results.  Adiabatic models will possess peaks in
$\ell$ that follow a series $(1:2:3:4...)$, whereas isocurvature models
obey the relation
$(1:3:5:7...)$ due to their phase difference (see \S\ref{ss-4acousticdriven}). 
The fundamental angular scale on which these series are based is 
that which is subtended by the sound horizon at last scattering.
It is purely dependent on the background
dynamics, matter content, and geometry
and thus can be used as a robust probe of
these fundamental cosmological parameters.  The scale is 
only weakly sensitive
to the baryon content if it is near the value required by nucleosynthesis
$\Omega_b h^2 \approx 10^{-2}$ but becomes increasingly sensitive as 
$\Omega_b h^2$ 
increases beyond the point at which 
the photon-baryon fluid is baryon dominated at last scattering
$\Omega_b h^2 \simgt 0.03$.  
The radiation content at last scattering increases the expansion rate
and thus decreases the horizon scale at last scattering.  
If $\Omega_0 h^2$ is sufficiently
low, the location of the peaks can provide an interesting 
constraint on the matter-radiation ratio, including perhaps the
number of relativistic (massless) neutrino species.   Otherwise, changes in the
age of the universe through $\Omega_0 h^2$ and $\Omega_\Lambda$ largely
scale out of the ratio between the two scales but may provide some
constraint on large $\Lambda$ models.  

The location of the peaks is by far the most sensitive to the presence of
curvature in the universe.  Curvature makes the sound horizon at last 
scattering subtend a much smaller angle in the sky than a flat universe.
In Fig.~\ref{fig:5peak}, we compare open and $\Lambda$ geometric 
effects.
The corresponding spectra are plotted in Figs.~\ref{fig:5lateisw}b and
\ref{fig:5ratra}.  
Notice that aside from the first peak, the numerical results agree quite well
with the simple projection scaling.  This
is because the first peak also obtains contributions from the early ISW
effect.  Because of its later generation, those contributions subtend
a larger angle on the sky.   They also are generated when radiation
is less important.  Thus for example, in an open universe, the angular
location scales close to $\Omega_0^{1/2}$ even in a low $\Omega_0 h^2$
model. 

\input chap5/heights.tex

\subsection{Heights of the Peaks}
\label{ss-5heights}

The heights of the peaks are somewhat more model dependent than their locations
since they will be controlled by the initial spectrum of fluctuations.
However, for initial conditions that are featureless (\eg\ the commonly
assumed power law models) in the decade or so of scales that yield observable
peaks, the {\it relative} heights again contain nearly model independent
information.

Aside from the initial spectrum,
essentially two quantities control the heights of the peaks: the baryon-photon
ratio $\Omega_b h^2$ and the matter-radiation ratio $\Omega_0 h^2 (1-f_\nu)$
(see Fig.~\ref{fig:5heights}).
The presence of baryons increases the gravitating mass of the fluid leading
to more gravitational compression of the fluid from baryon drag.  
Thus every other peak
will be enhanced by gravitational effects on the baryons.  As discussed in
\S \ref{ss-4acousticdriven}, 
these are the odd peaks for the adiabatic mode and the even for
the isocurvature.  Enhancement only occurs if the gravitational potential is
still significant.  In the radiation-dominated epoch, the gravitational
potential decays after sound horizon crossing.  Thus the alternating
series of peaks only occurs for scales that cross after radiation domination
leading to a pattern that is dependent on the matter-radiation ratio.

\input chap5/diffusion.tex

In adiabatic models, the decay of the potentials $\Psi$ and $\Phi$ lead to 
driving effects from infall and dilation.  This boosts oscillations by
a factor of $\sim 5$ in amplitude for modes that cross in radiation domination.
By delaying equality through lowering $\Omega_0 h^2 (1-f_\nu)$, we can 
bring this effect to larger scales and thus boost more of the
peaks.   For isocurvature models, the opposite occurs.  By delaying equality,
we take away potential growth from larger and larger scales.  This
lowers the radiation fluctuation.  

\input chap5/damping.tex

\subsection{Diffusion Damping at Recombination}
\label{ss-5diffusion}

At small scales, the features described above for the heights of the
peaks can be hidden by diffusion damping. 
We obtain the diffusion damped fluctuation at last scattering
from the acoustic solutions of
equation~\eqn{PartSoln}, denoted by an overhat, with 
the relations (see Appendix \ref{ss-5polarization} \cite{HSa})
\beal{eq:DampCorr}
[\Theta_0 + \Psi](\eta_*) \eal [\hat \Theta_0+\Psi](\eta_*)
{\cal D}(\eta_*,k), \nonumber\\
\Theta_1(\eta_*) \eal  \hat \Theta_1(\eta_*)
{\cal D}(\eta_*,k), 
\eea
where we assume $R\Psi(\eta_*) \ll \Theta_0$ and
the damping factor is weighted by the visibility function 
\bel{eq:DampLS}
{\cal D}(\eta_*,k) = \int_0^{\eta_*} d\eta \dot\tau e^{-\tau}
                        e^{-(k/k_D)^2}.
\ee
with the damping scale $k_D(\eta)$ calculated from equation \eqn{DampLength}.
Since the visibility function $\dot\tau e^{-\tau}$ goes to a delta function
for large $\tau$, this definition also coincides with its 
tight-coupling definition from equation~\eqn{Damp}.  Note that the ionization
history enters in two places: the increase in the diffusion length $k_D^{-1}$
and the visibility function weighting.  Since the visibility function
peaks at $z \approx 10^3$ nearly independent of cosmological parameters
and is by definition normalized to have unit area, much of the qualitative
behavior of the damping can be determined by examining $k_D^{-1}$.

Recall from \S \ref{ss-4dampedacoustic} 
that the diffusion length is approximately the distance 
a photon can random walk by $\eta_*$, $k_D^{-1} \propto 
\sqrt{\eta_* \lambda_C}$, where the Compton mean
free path is $\lambda_C \propto (x_e n_b)^{-1}$.  The behavior of the
diffusion length through last scattering will be determined by the evolution
of the ionization fraction.  In Appendix \ref{ss-5analytic}, 
we will show how to
construct the diffusion length from a realistic treatment of recombination.
However, to obtain simple scaling results, the Saha approximation for 
the equilibrium ionization suffices.

The Saha equation assumes that photoionization and recombination of hydrogen
$e + p \leftrightarrow H + \gamma$ are in  equilibrium.  If 
the photon chemical potential is vanishingly small as required by 
the FIRAS observation \cite{Mather}, the chemical potentials of the
other species must satisfy $\mu_e + \mu_p = \mu_H$.  The
number density of a non-relativistic species $x$ is given by
\bel{eq:boltzmannfactor}
n_x = g_x \left( { m_x T_x \over 2\pi} \right)^{3/2} e^{(\mu_x -m_x)/T_x}  
\ee
where $g_x$ is the spin multiplicity.  This chemical potential 
relation then  implies the Saha 
equation
\bel{eq:Saha}
{n_e n_p \over n_H n_b} = {x^2_e \over 1-x_e} = {1 \over n_b}
	\left( {  m_e T \over 2\pi } \right)^{3/2} e^{-(m_e+m_p-m_H)/T}
\ee
where we neglect the helium fraction, $n_b = n_p + n_H$ and
 the strong thermal coupling between photons, electrons, and baryons
at last scattering has allowed us to set all the temperatures equal
(see \S \ref{ss-3electrontemp}).  Note that $m_e + m_p - m_H = 13.6$eV,
the electron binding energy. 

The interesting result here is that as the ionization drops to zero, its
parameter dependence goes to $x_e \propto (\Omega_b h^2)^{-1/2}$ at
fixed redshift (or temperature).  
The final damping length
approximately scales as $k_D^{-1}(\eta_*) \propto \eta_*^{1/2} 
(\Omega_b h^2)^{-1/4}$.  The damping angular scale therefore becomes
\bel{eq:dampingangle}
\ell_D \propto \eta_*^{-1/2} (\Omega_b h^2)^{1/4} r_\theta(\eta_*)
\ee
At asymptotically 
high and low $\Omega_0 h^2$, this goes to $\Omega_0^{-3/4} \Omega_b^{1/4}$
and $\Omega_0^{-5/4} \Omega_b^{1/4} h^{-1/2}$ in an open universe
and $\Omega_0^{-1/4}\Omega_b^{1/4}$ and $\Omega_0^{-1/2}\Omega_b^{1/4}
h^{-1/2}$ in a $\Lambda$ universe.
The damping scale is thus somewhat more strongly dependent on $\Omega_b$
than the acoustic scale but even more weakly dependent on $h$ alone (see
Fig.~\ref{fig:5damping}).  The Saha prediction requires modification 
for high $\Omega_b h^2$ models due to the increasing importance of the
Lyman-$\alpha$ opacity at last scattering \cite{HSsmall}.

%% file: chap5/swbasic.tex
\begin{figure}[p]
\centerline{\hskip -0.5truecm
\epsfxsize=5.5in \epsfbox{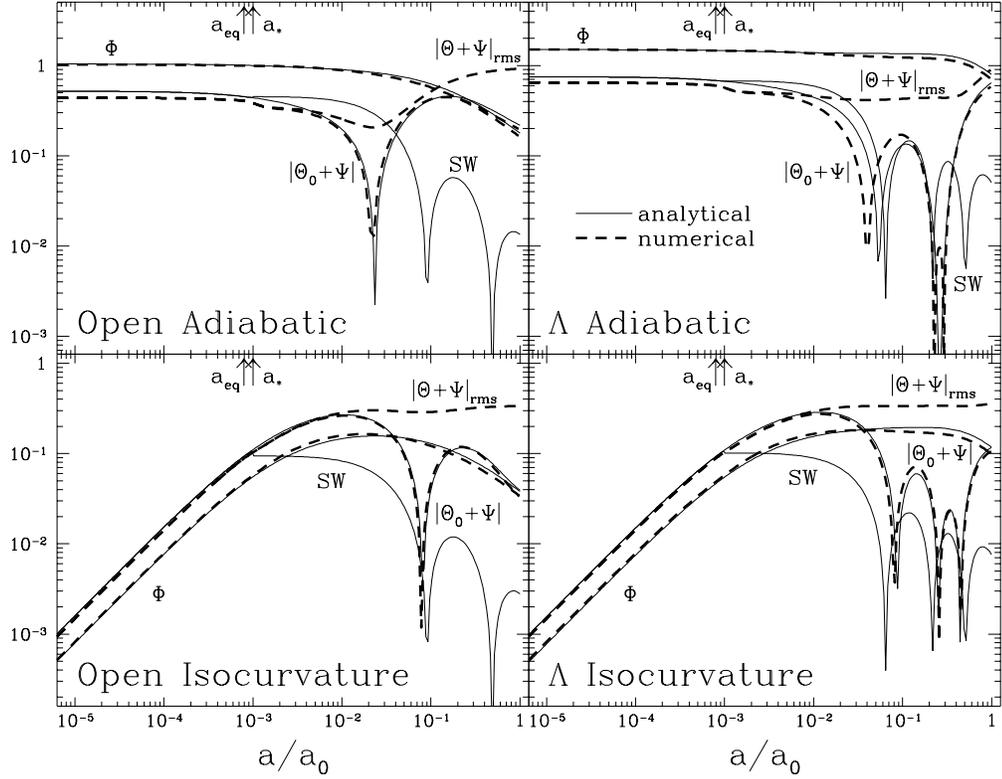}}
 \vskip -0.5truecm
\caption{Sachs-Wolfe Evolution}
\mycaption{In the adiabatic case, temperature fluctuations are
enhanced in gravitational wells such that $\Theta_0$ and $\Psi$
cancel, yielding $\Theta_0+\Psi=\Psi/3$ in the
matter dominated epoch.  For the isocurvature case,
the dilation effect creates a net total of
$\Theta_0+\Psi = 2\Psi$ reflecting the anticorrelated nature
of radiation and total density fluctuations.
After last scattering at $a_*$, this SW contribution
(analytic only) collisionlessly damps from the monopole and
transfers power to anisotropies. The rms
temperature fluctuations (numerical only) acquires contributions
after $a_*$ from the ISW effect due to the radiation (early) {\it and}
curvature or $\Lambda$ (late) contributions.  The scale here is
chosen to be $k=4 \times 10^{-4} $Mpc$^{-1}$ in an $\Omega_0=0.1$
$h=0.5$ universe.}
\label{fig:5swbasic}
\end{figure}

%% file: chap5/earlyisw.tex
\begin{figure}[t]
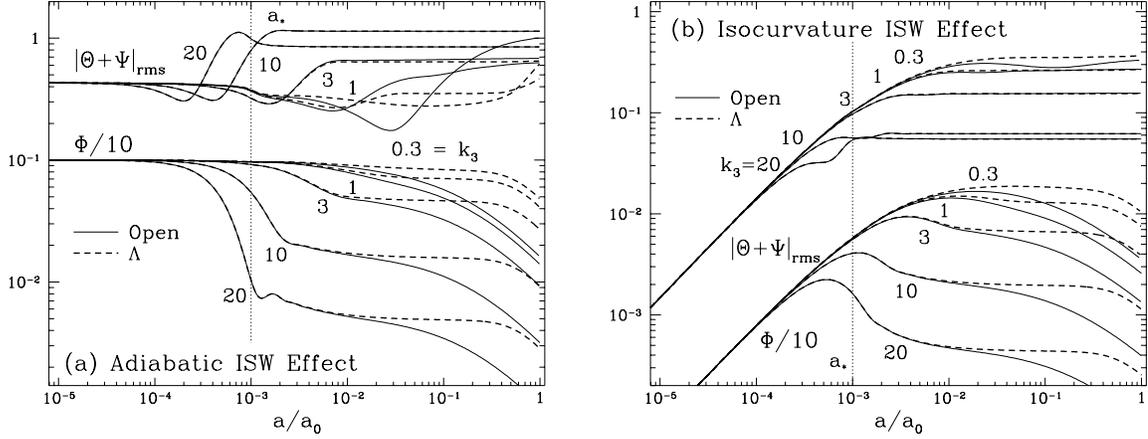

\centerline{ \hskip -0.25 truecm
\epsfxsize=3.0in \epsfbox{chap5/earlyiswa.epsf} \hskip 0.25 truecm
\epsfxsize=3.0in \epsfbox{chap5/earlyiswb.epsf}}
 \vskip -0.5truecm
\caption{ISW Effect}
\mycaption{(a) Adiabatic models.  Potential decay 
at horizon crossing during radiation domination boosts 
scales approaching
the first acoustic oscillation through the early ISW effect.  
Larger scales suffer only the late ISW
effects due to the rapid expansion in open and $\Lambda$ models,
leaving a deficit at intermediate scales.
(b) Isocurvature models.  For small scales, potential growth halts
after horizon crossing in the radiation dominated epoch leading
to a relative boost for large scale fluctuations.  Since this early ISW
effect dominates, there is little distinction between open and
$\Lambda$ models.   
All models have $\Omega_0=0.1,h=0.5$ with
standard recombination and $k = k_3 \times 10^{-3}$Mpc$^{-1}$.
}
\label{fig:5earlyisw}
\end{figure}

%% file: chap5/cflat.tex
\begin{figure}[t]
\vphantom{marker} \vskip 0.5truecm
\centerline{ \hskip -0.25 truecm \hskip 0.1in
\epsfxsize=2.95in \epsfbox{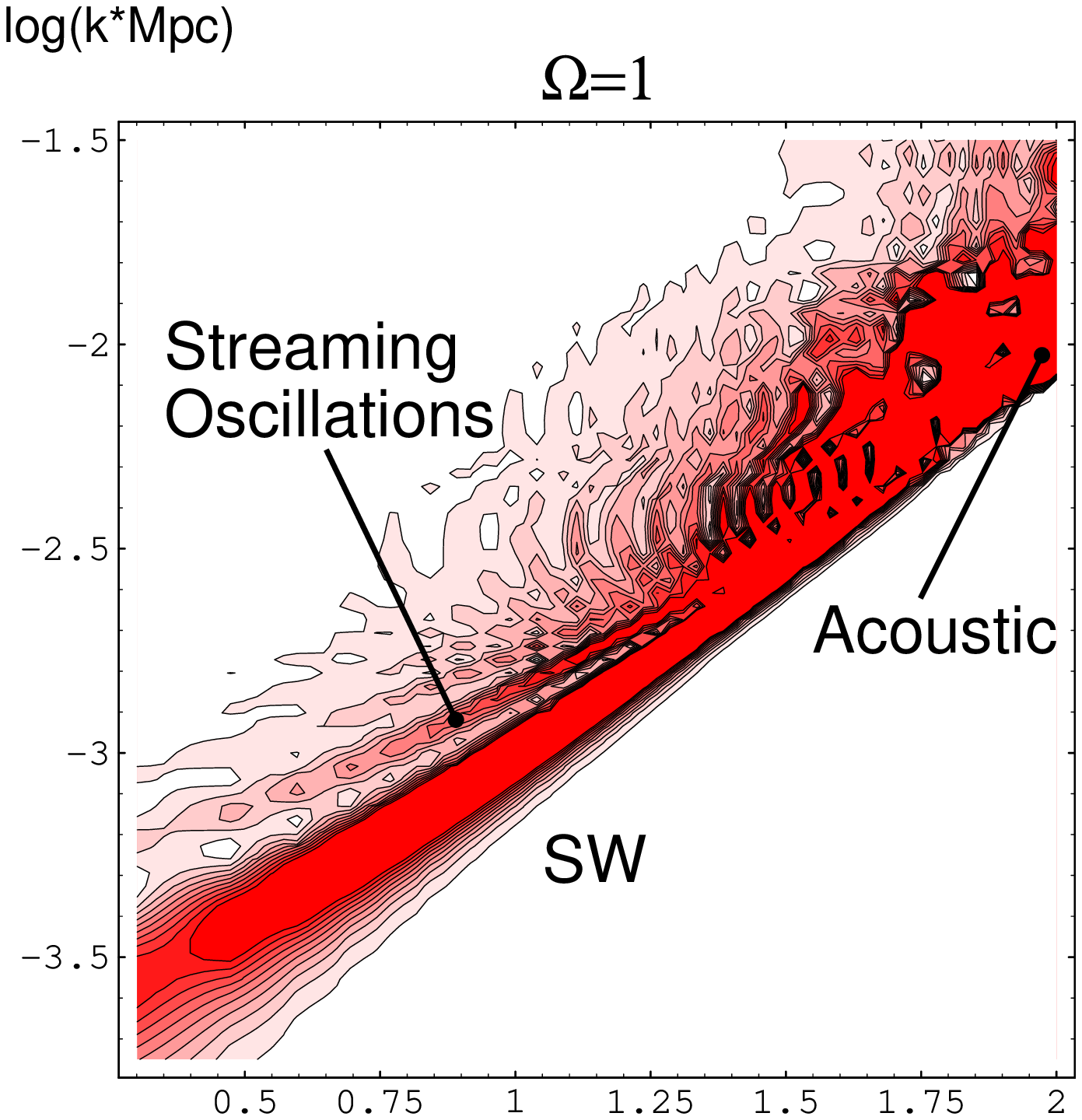} \hskip -0.25truecm
\epsfxsize=2.95in \epsfbox{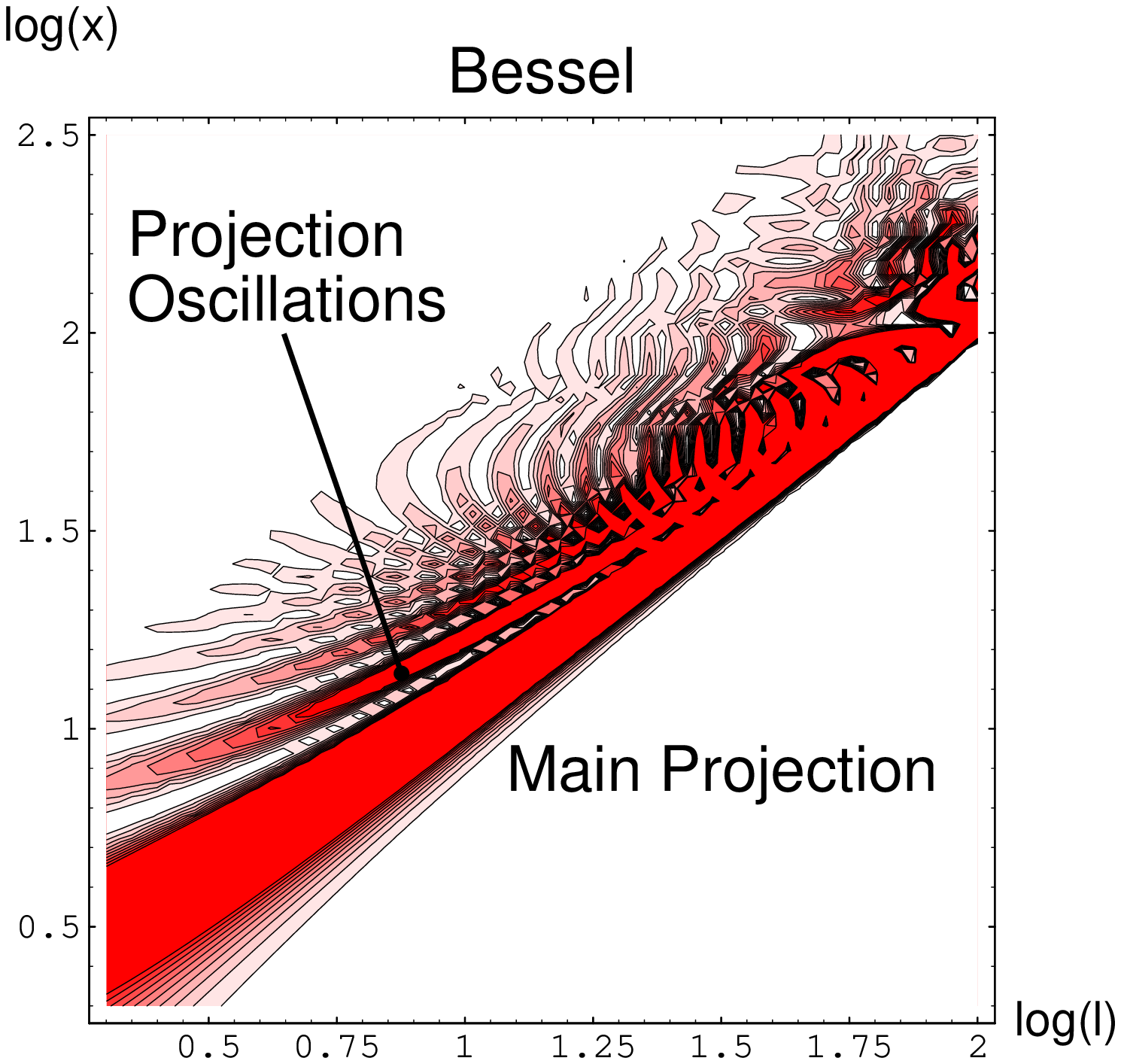}}
 \vskip -0.75truecm
\caption{$\Omega_0=1$ Radiation Transfer Function}
\mycaption{Shown here and
in Figs.~\ref{fig:5lamb}, \ref{fig:5open}, and \ref{fig:5iso} is
the weighted squared transfer function 
$\ell T^2_\ell(k)$ which also represents the
anisotropy contribution per logarithmic $k$ and $\ell$ interval 
in a scale invariant model. 
Contours are equally spaced up to a cut off set to best display the features.
The strong correlation between $\ell$ and $k$ merely reflects 
the projection of a scale on
the last scattering surface to an angle on the sky.
At $\log\ell \simgt 2$, SW contributions fall off and
are replaced by the acoustic peaks
(saturated here).
The detailed structure can
be traced to the radial eigenfunction $X_\nu^\ell(\chi)=j_\ell(x)$
which governs
the projection and free streaming oscillations.
}
\label{fig:5cflat}
\end{figure}

%% file: chap5/flatisw.tex
\begin{figure}[t]
\centerline{\hskip -0.5truecm
\epsfxsize=3.5in \epsfbox{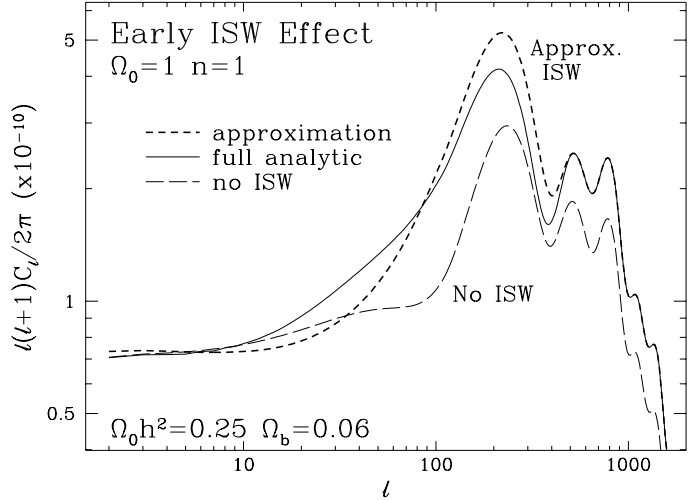}}
 \vskip -0.5truecm
\caption{$\Omega_0=1$ Early ISW Spectrum}
\mycaption{Even in an $\Omega_0=1$ $\Omega_0 h^2=0.25$ high matter
content universe, early ISW contributions from radiation pressure are
non-negligible. 
Ignoring the ISW effect entirely leads
to a significant error in both the large scale normalization 
and shape of the
anisotropies.  
Approximating {\it all} of the ISW contribution
to occur near recombination through equation \eqn{ISWapprox1}
leads to $10-15\%$
errors in temperature since it
comes from  more recent times where the fluctuation subtends a larger angle 
angle on
the sky.  The full integration therefore has more power at larger
angular scales and makes the rise to the first Doppler peak more
gradual. These are analytic results from Appendix  \ref{ss-5recombination}.
}
\label{fig:5flatisw}
\end{figure}

%% file: chap5/lamb.tex
\begin{figure}[t]
\vphantom{marker} \vskip 0.5 truecm
\centerline{ \hskip -0.25 truecm \hskip 0.1in 
\epsfxsize=2.95in \epsfbox{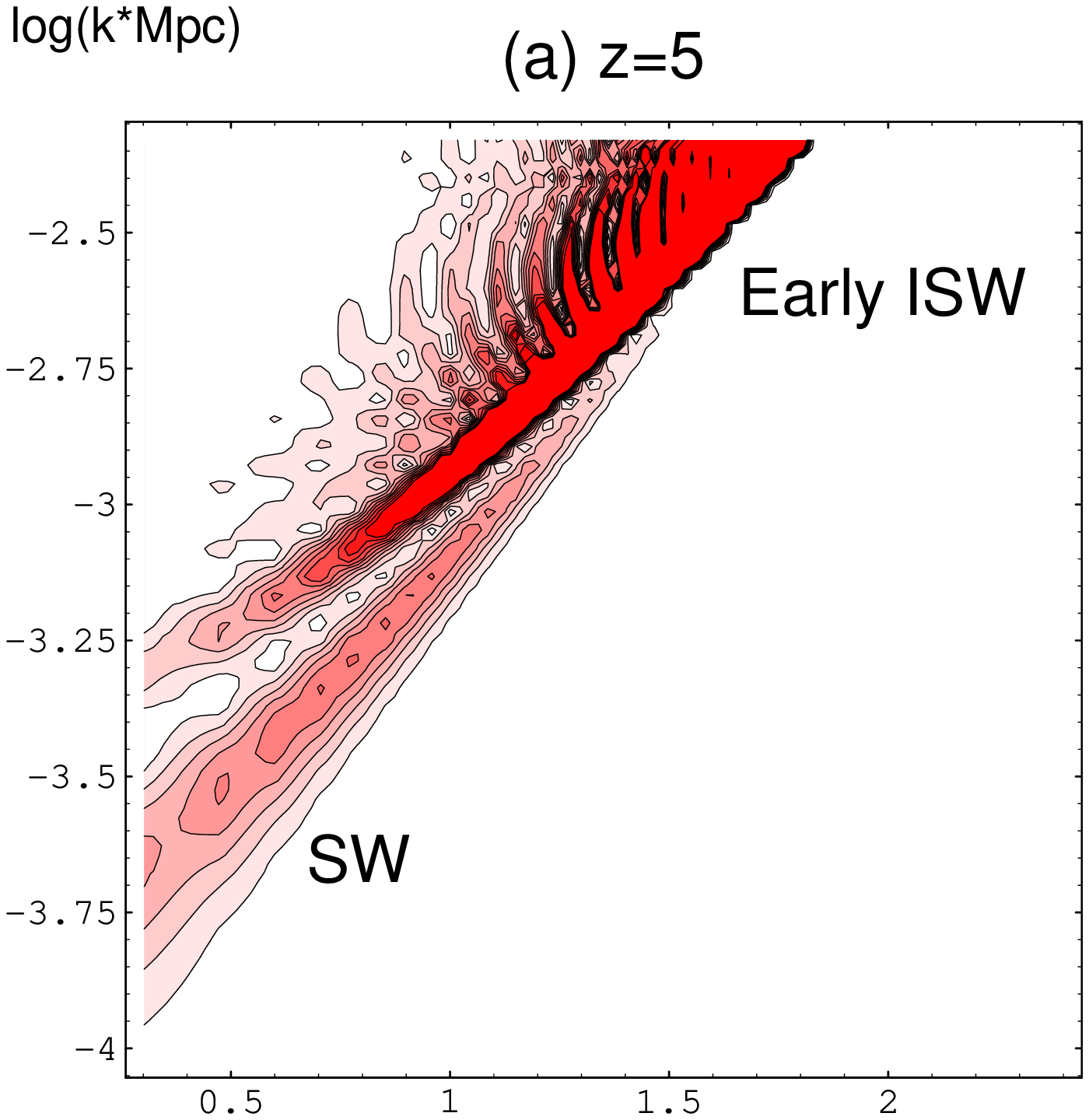} \hskip -0.truecm
\epsfxsize=2.95in \epsfbox{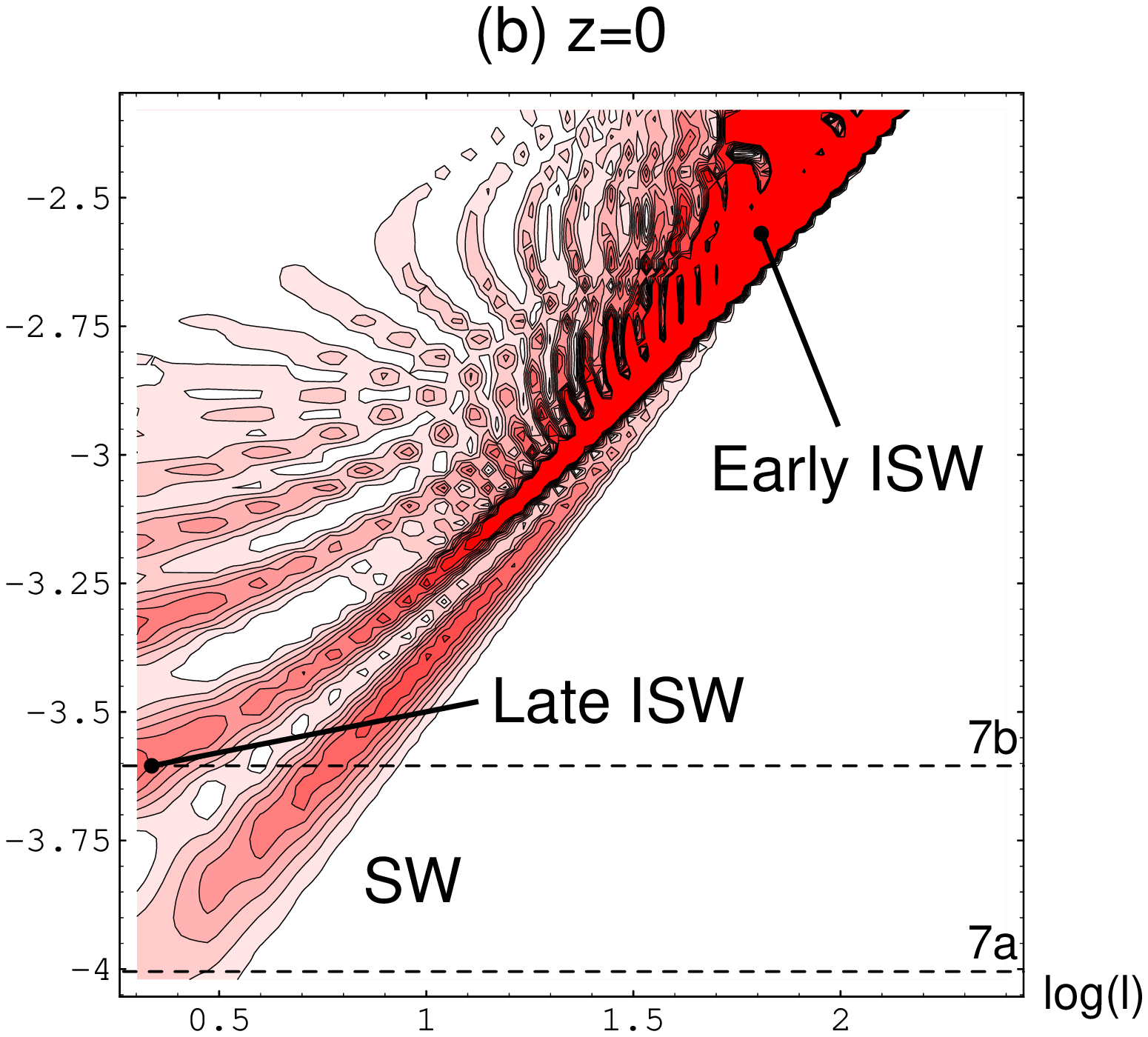}}
 \vskip -0.75truecm
\caption{$\Lambda$ Adiabatic Radiation Transfer Function}
\mycaption{Unlike the $\Omega_0=1$
case, this scenario has strong contributions after
last scattering from the early and late
ISW effect.  (a) The early ISW effect projects
onto a second ridge
which is more prominent than the SW
ridge at intermediate but not large angles.
(b) After $\Lambda$ domination, the late ISW contributions
come free streaming in from the monopole yielding a boost in
the low order multipoles for a small range in $k$, due to cancellation
with SW contributions at the largest scales and crest-trough cancellation
at smaller scales.  Scales depicted in Fig.~\ref{fig:5sepadi} 
are marked here by
dashed lines. The model here is $\Omega_0=0.1,h=0.5$ with
standard recombination.
}
\label{fig:5lamb}
\end{figure}

%% file: chap5/lateisw.tex
\begin{figure}[t]
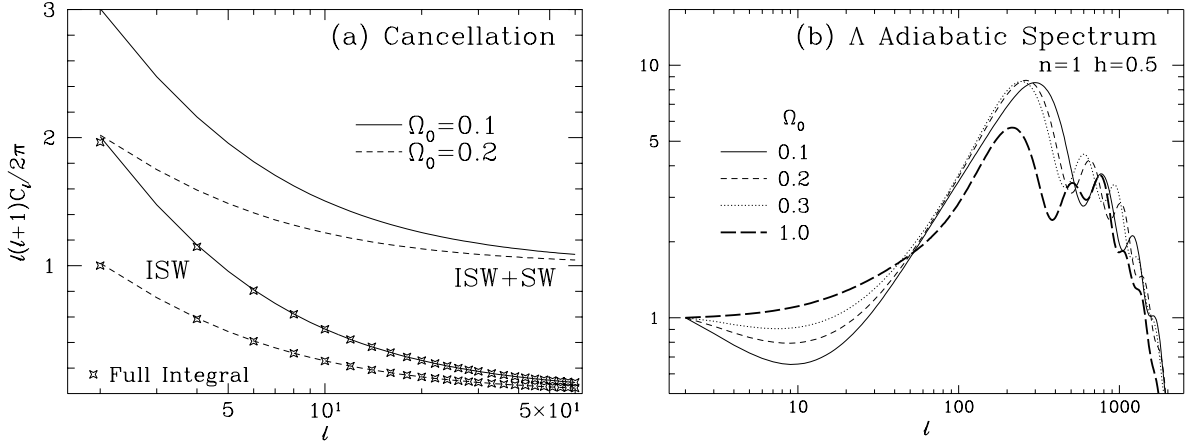

\centerline{ \hskip -0.25 truecm
\epsfxsize=3.0in \epsfbox{chap5/iswanat.epsf} \hskip 0.25 truecm
\epsfxsize=3.0in \epsfbox{chap5/lateiswb.epsf}}
 \vskip -0.5truecm
\caption{$\Lambda$ Late ISW Spectrum}
\mycaption{(a) Analytic Separation.
The late ISW effect is cancelled as
photons stream through many wavelengths of the perturbation during
the decay.  The comparison here of the full late ISW integral to
the cancellation approximation shows that even at the largest angles,
the late ISW contributions are well inside the cancellation regime. 
The SW effect on the other hand is flat in this representation.
As $\Lambda$ increases, the contribution of the late ISW effect relative
to the SW effect increases at low multipoles and appears as a boost.
(b) Numerical results.  The early ISW effect contributes significantly
at scales not much smaller than the cancellation tail of the late ISW 
effect bending the spectrum back up. 
}
\label{fig:5lateisw}
\end{figure}

%% file: chap5/sepadi.tex
\begin{figure}[t]
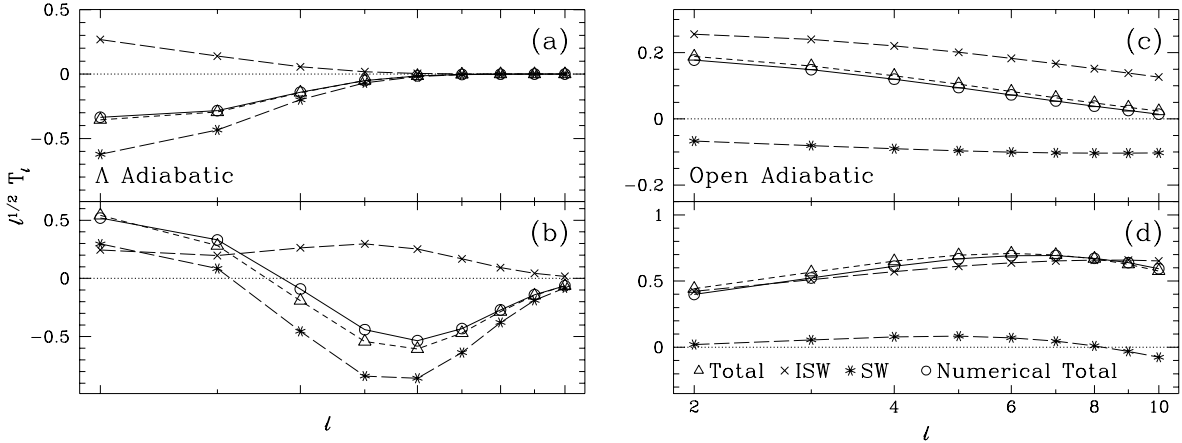

\centerline{ \hskip -0.15 truecm
\epsfxsize=3.0in \epsfbox{chap5/sepadia.epsf} \hskip 0.15 truecm
\epsfxsize=3.0in \epsfbox{chap5/sepadib.epsf}}
 \vskip -0.5truecm
\caption{Analytic Decomposition: Adiabatic Models}
\mycaption{Scales are chosen to match the features in
Fig.~\ref{fig:5lamb} and \ref{fig:5open}.
$\Lambda$ models:  (a) At the largest scales, \eg\ here $k=10^{-4}$Mpc$^{-1}$, 
the SW effect dominates over, but is partially cancelled by,
 the late ISW effect.
(b) Intermediate scale peaks in Fig.~\ref{fig:5lamb} 
are due to the late ISW boost of the higher SW projection ridges.
Open models: (c)
The maximum scale
corresponds to the curvature radius $k = \sqrt{-K}$.
For the SW effect, this scale projects broadly in $\ell$
peaking near
$\ell \sim 10$.  For the late ISW effect,
this scale projects onto the monopole and
dipole near curvature domination
thus leaving the ISW contributions to decrease smoothly with $\ell$.
(d) At smaller scales,  corresponding to the large ridge in Fig.~6,
the late ISW effect projects onto $\ell \approx
2-10$ and completely dominates leading to a rising spectrum of
anisotropies.
The models are for $\Omega_0=0.1$ $h=0.5$ with standard recombination and 
arbitrary normalization. 
}
\label{fig:5sepadi}
\end{figure}

%% file: chap5/open.tex
\begin{figure}[t]
\vphantom{marker} \vskip 0.5truecm
\centerline{ \hskip -0.25 truecm \hskip 0.1in
\epsfxsize=2.95in \epsfbox{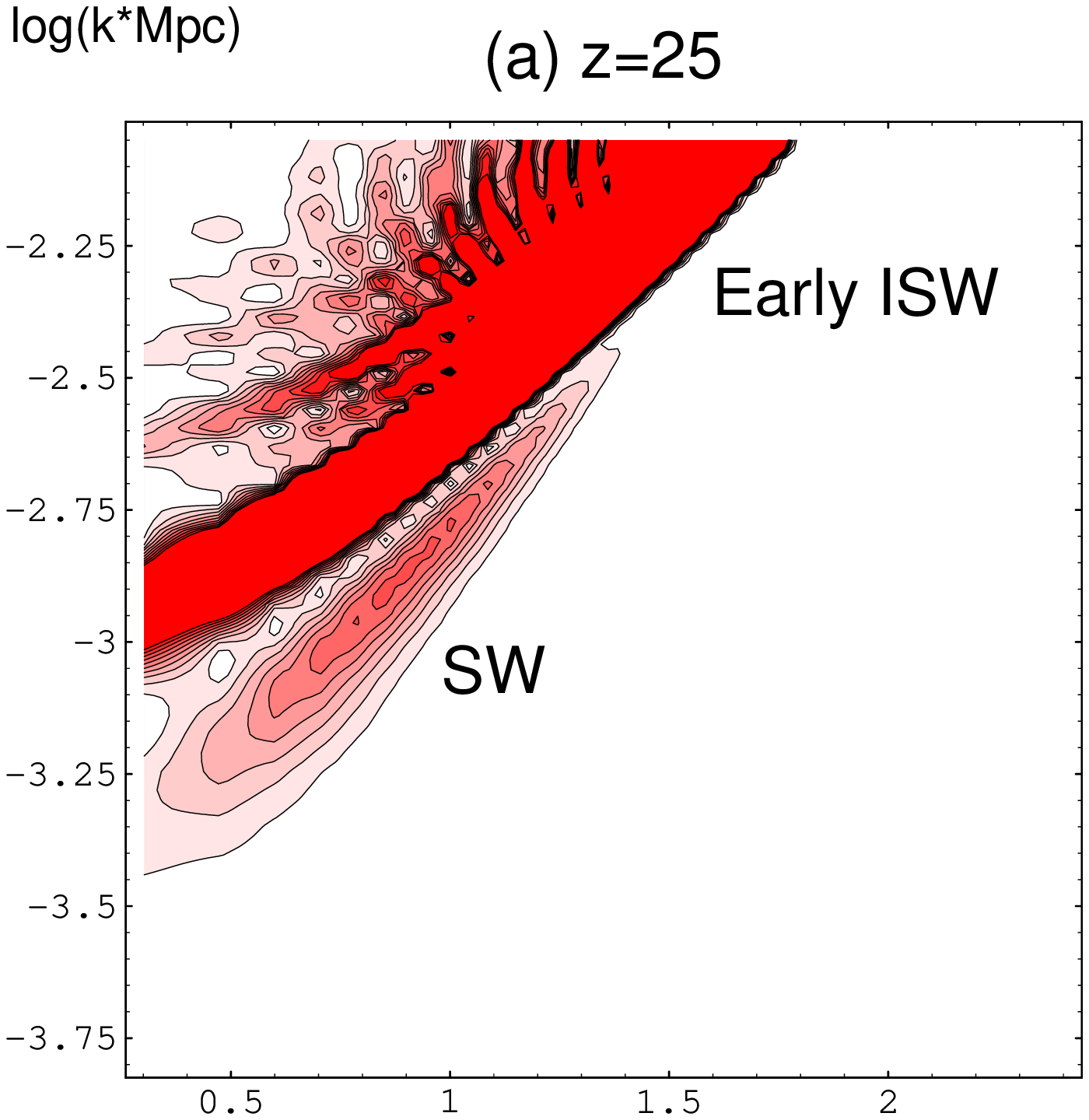} \hskip -0.truecm
\epsfxsize=2.95in \epsfbox{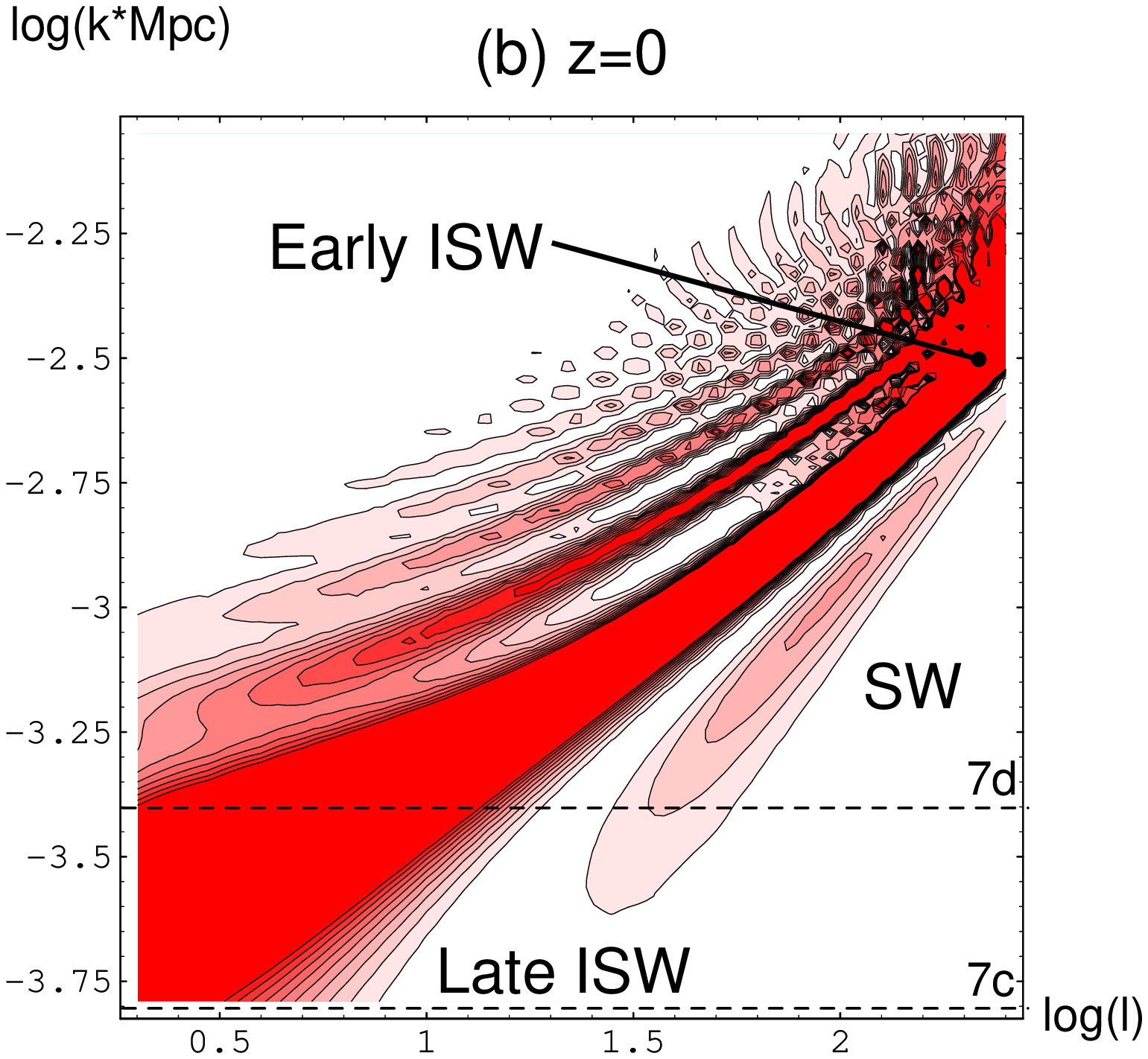}}
 \vskip -0.75truecm
\caption{Open Adiabatic Radiation Transfer Function}
\mycaption{(a) Like the $\Lambda$ case, the radiation ISW effect
contributes significantly to intermediate angle anisotropies.
(b) The late ISW effect
appearing at the left
is much more significant than the corresponding $\Lambda$ effect.
Thus on all angular scales,
the total ISW contribution
dominates the SW effect. 
The curvature scale $\log(k*$Mpc$)=-3.8$ intersects the late ISW ridge
near the lowest multipoles.  Absence of supercurvature contributions 
can suppress these multipoles.  Dashed lines represent scales
in Fig.~\ref{fig:5sepadi}.
The model is $\Omega_0=0.1,
h=0.5$, with standard recombination.
}
\label{fig:5open}
\end{figure}

%% file: chap5/ratra.tex
\begin{figure}[t]
\centerline{ \hskip -0.5truecm
\epsfxsize=3.5in \epsfbox{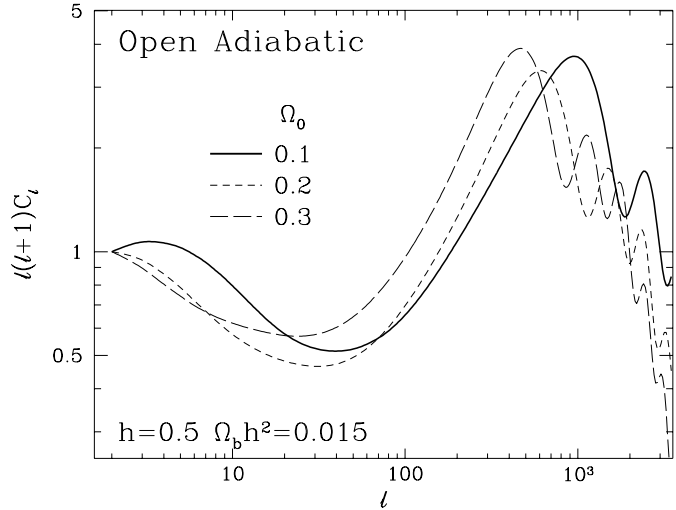}}
 \vskip -0.5truecm
\caption{Open Adiabatic Spectrum}
\mycaption{Scale invariant $n=1$ model: equal power per logarithmic $k$ 
interval to the curvature scale $k = \sqrt{-K}$.  
The early ISW effect merges with the 
cancellation tail of the late ISW effect completely dominating the SW 
contributions.  The lack of supercurvature modes can lead to a suppression
of low order multipoles as the curvature scale becomes significantly smaller
than the horizon (see also Fig.~\ref{fig:5sepadi}d).  
Notice also that geodesic deviation shifts the 
acoustic contributions more than the early ISW contributions and broaden
out the first peak.  
}
\label{fig:5ratra}
\end{figure}

%% file: chap5/iso.tex
\begin{figure}[t]
\vphantom{marker} \vskip 0.5truecm
\centerline{ \hskip -0.25 truecm \hskip 0.1in
\epsfxsize=2.95in \epsfbox{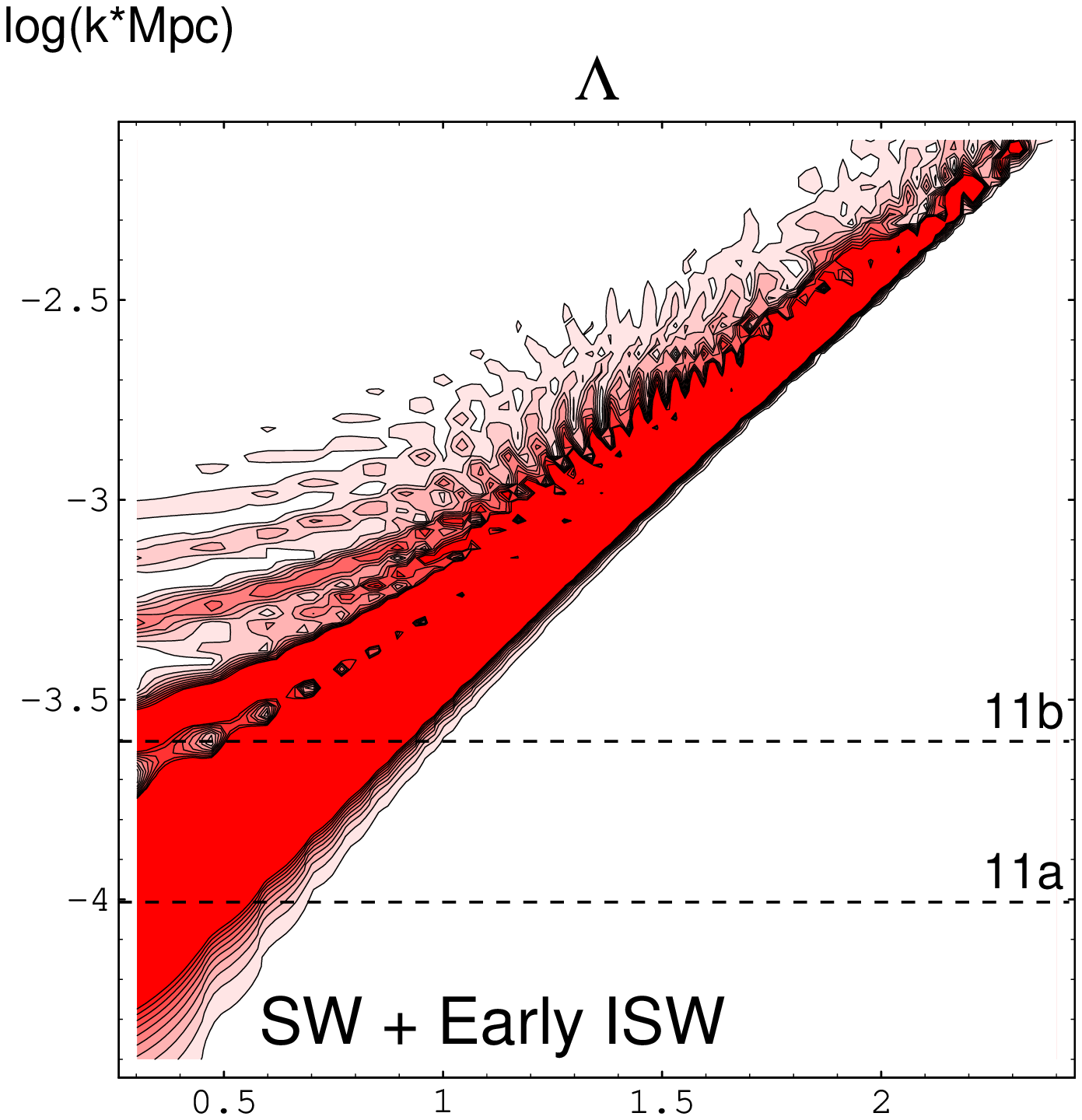} \hskip -0.25truecm
\epsfxsize=2.95in \epsfbox{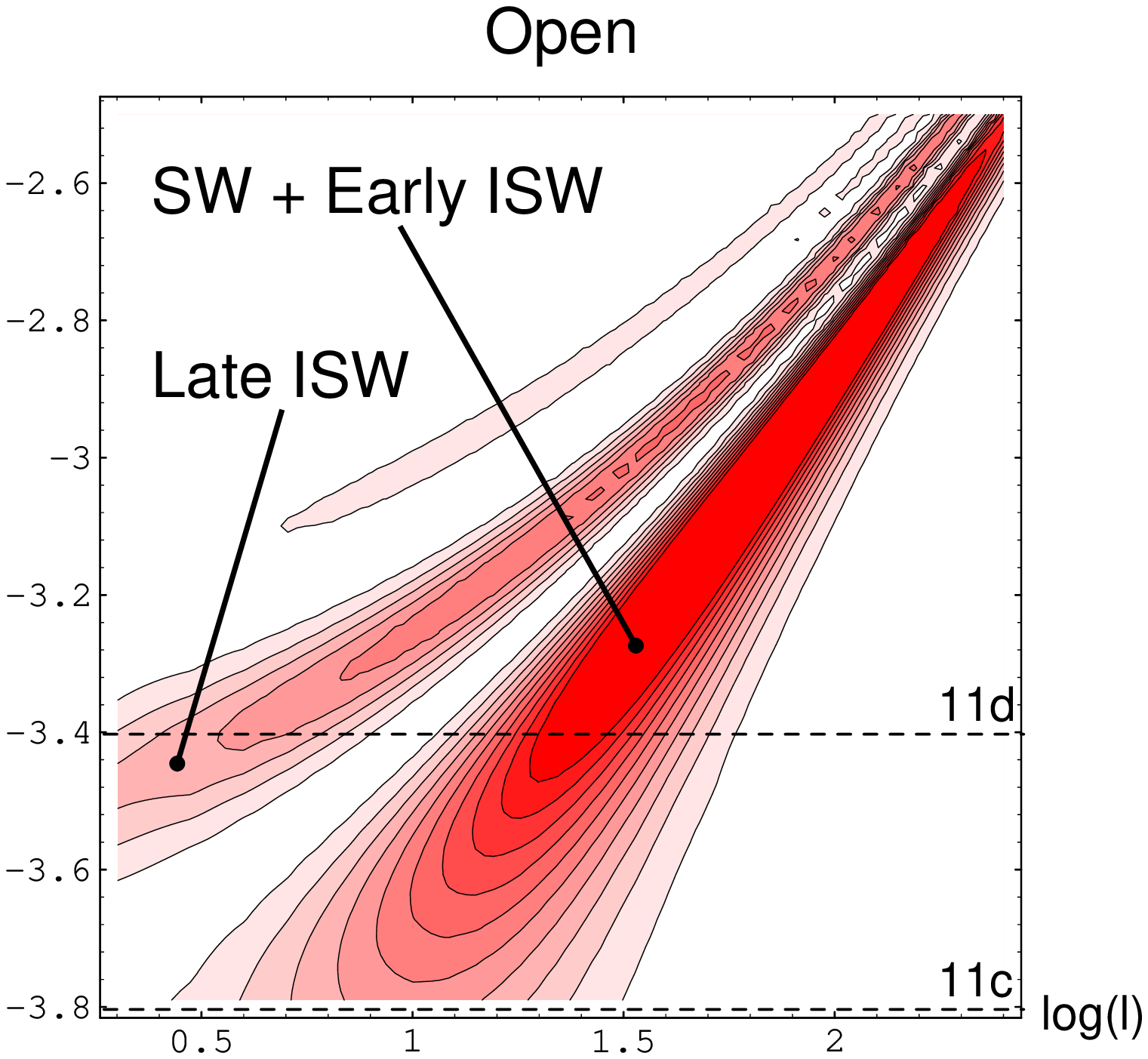}}
 \vskip -0.75truecm
\caption{Isocurvature Radiation Transfer Function}
\mycaption{Unlike their
adiabatic counterparts, the potential {\it grows} in the
radiation domination era only to turn over and decay in the
curvature and $\Lambda$ dominated era.
The ISW contribution will thus smoothly match onto the SW contribution.
This has the effect of merging the SW and ISW ridges to make a
wide feature that contributes broadly in $\ell$.
For $\Lambda$ models, the early ISW effect completely dominates
over the late ISW effect. Scales depicted in Fig.~\ref{fig:5sepiso} are
marked here in dashed lines. The model here is $\Omega_0=\Omega_b=0.1$, 
$h=0.5$ with 
standard recombination.
}
\label{fig:5iso}
\end{figure}

%% file: chap5/sepiso.tex
\begin{figure}[t]
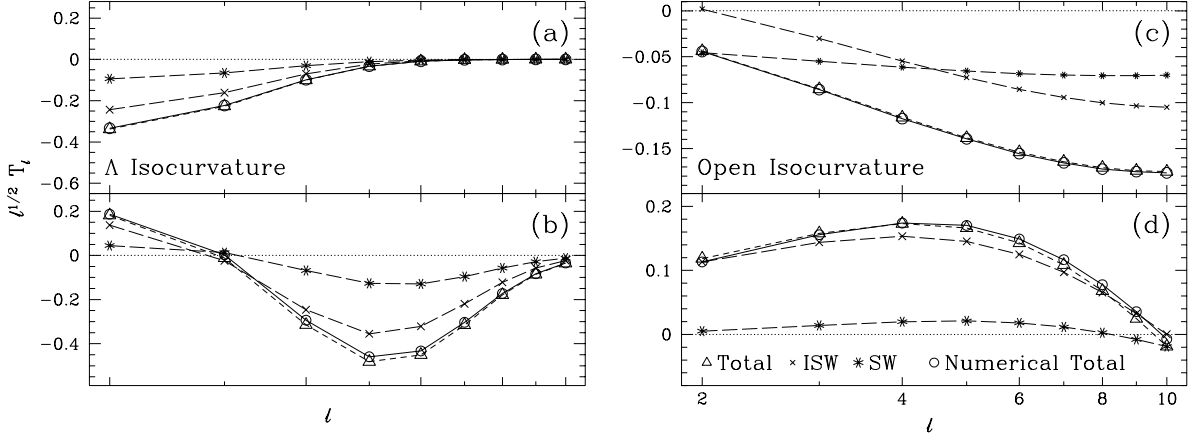

\centerline{ \hskip -0.25 truecm
\epsfxsize=3.0in \epsfbox{chap5/sepisoa.epsf} \hskip 0.25 truecm
\epsfxsize=3.0in \epsfbox{chap5/sepisob.epsf}}
 \vskip -0.5truecm
\caption{Analytic Decomposition: Isocurvature Models}
\mycaption{In general, isocurvature models have strong early ISW contributions
which mimic and coherently boost the SW effect.
Scales are chosen to match
the features in Fig.~\ref{fig:5iso}.
$\Lambda$
models:
(a) Notice that the shape of the SW and ISW effects are identical at
large scales.
(b) Even at the late ISW peak, the early ISW contributions are so strong
that the late contributions are never apparent unlike the
adiabatic model.
Open models:  (c) As with
$\Lambda$ models, early ISW and SW contributions are similar
in form
at large scales.
(d) Near the peak of the late ISW contribution however,
the relative contributions are similar to the adiabatic case.
The model here is $\Omega_0=\Omega_b =0.1$, $h=0.5$ with standard 
recombination.
}
\label{fig:5sepiso}
\end{figure}

%% file: chap5/alias.tex
\begin{figure}[t]
\vphantom{marker} \vskip 0.5truecm
\centerline{\hskip 0.2truein \hskip 0.1in
\epsfxsize=2.95in \epsfbox{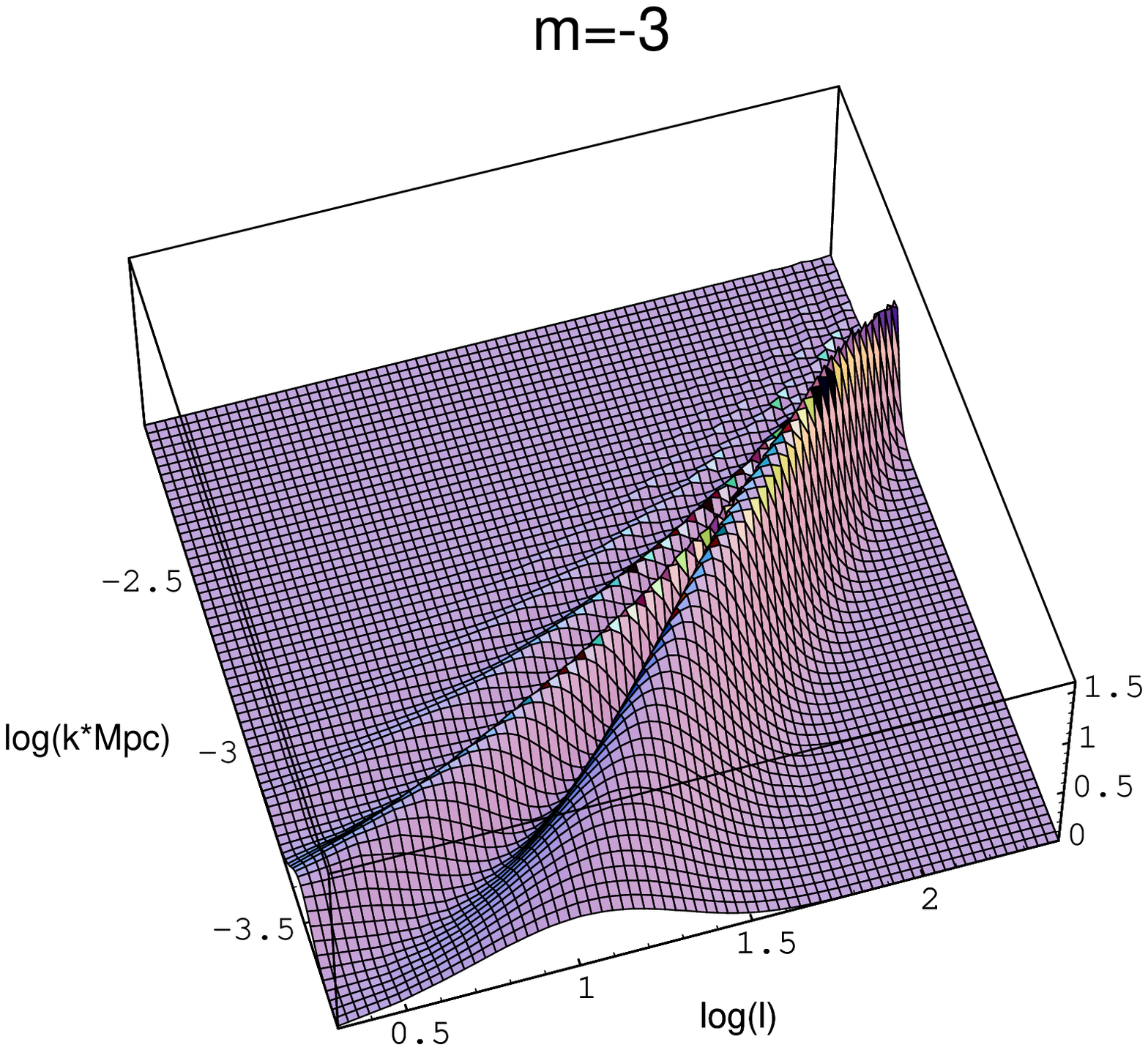} \hskip -0.2truein
\epsfxsize=2.95in \epsfbox{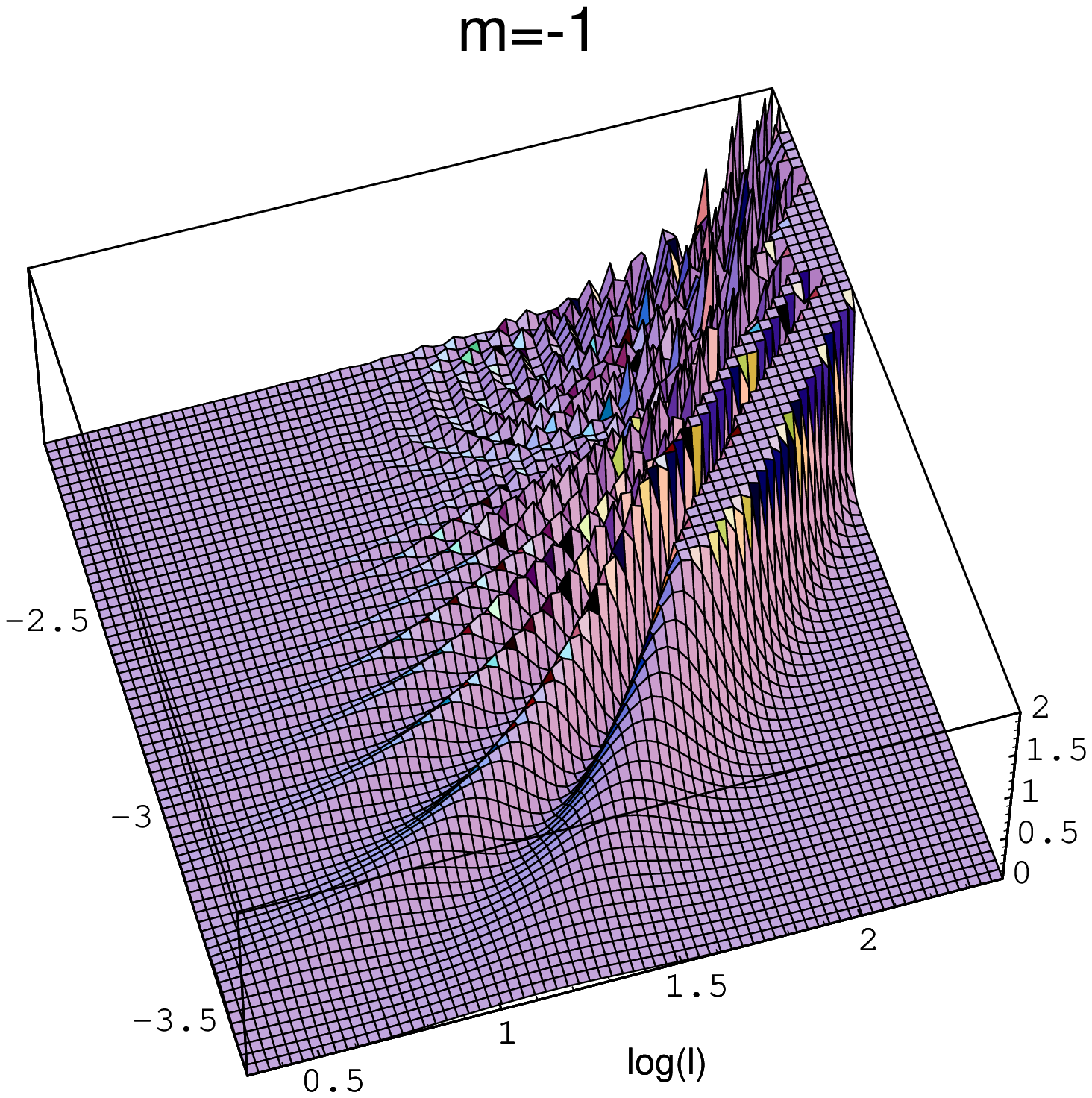}}
\vskip -0.75truecm
\caption{Aliasing Effect}
\mycaption{The full open isocurvature photon power spectrum 
 for
$k^3|S(0)|^2 \propto k^{m+3}$.
(a) Scale invariant $m = -3$.  (b) Blue $m=-1$.  The steeply blue
spectrum required by large scale structure constraints suffers
projection aliasing.  Large scale anisotropies are dominated
by small scale power leaking through the projection.  The anisotropy
spectrum is thus less blue than the spatial power spectrum and
insensitive to the large scale power spectrum. }
\label{fig:5alias}
\end{figure}

%% file: chap5/neff.tex
\begin{figure}[t]
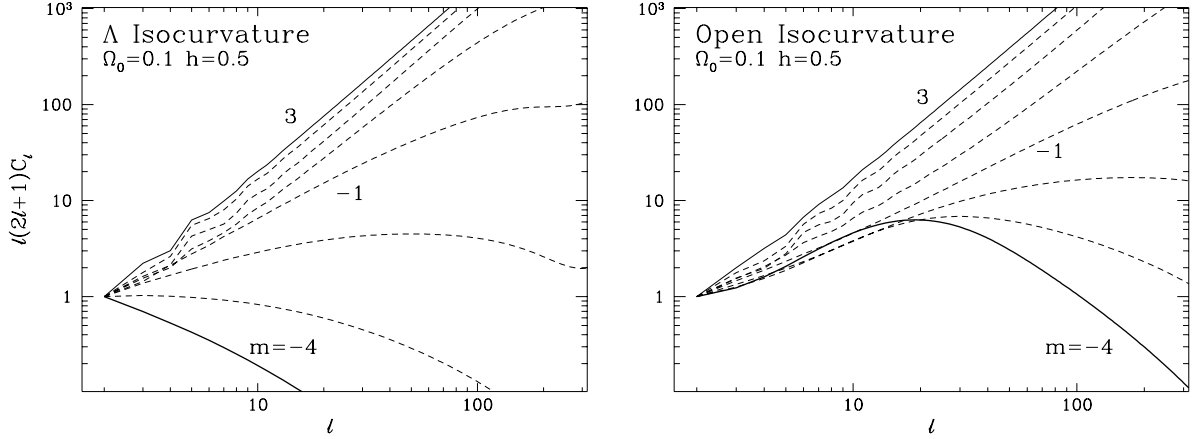

\centerline{ \hskip -0.25 truecm
\epsfxsize=3.0in \epsfbox{chap5/neffa.epsf} \hskip 0.25 truecm
\epsfxsize=3.0in \epsfbox{chap5/neffb.epsf}}
 \vskip -0.5truecm
\caption{The $m$ Dependence of Isocurvature Spectra}
\mycaption{Blue spectra $m \simgt -2$ are dominated by small scale power
aliased onto large angle anisotropies.  The effective slope never exceeds
$n_{\rm eff}=3$.  In the $m \approx 1$ regime $n_{\rm eff} \approx 2$ for
both open and $\Lambda$ models.  Red spectra show different open and 
$\Lambda$ models due to the lack of supercurvature modes in the open case
which cuts off anisotropies.  This is more severe in isocurvature models
since the curvature scale at early ISW formation scales projects onto smaller
angles than for their adiabatic late ISW counterparts.
}
\label{fig:5neff}
\end{figure}

%% file: chap5/power.tex
\begin{figure}[t]
\centerline{
\epsfxsize=3.5in \epsfbox{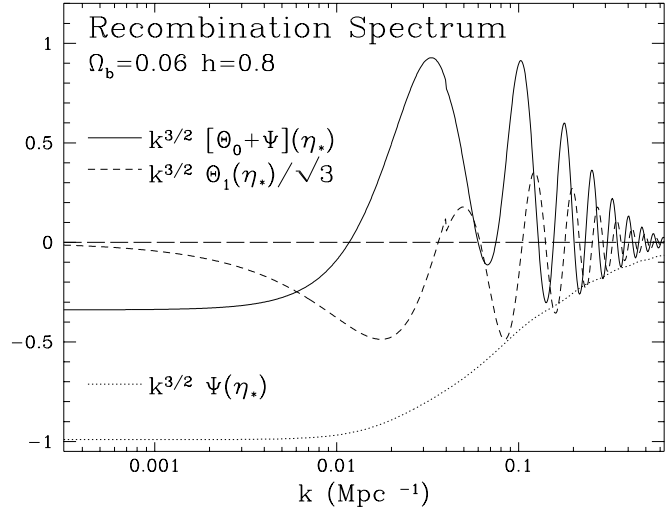}}
 \vskip -0.5truecm
\caption{Fluctuations at Last Scattering}
\mycaption{Analytic separation from Appendix \ref{ss-5analytic}.
Notice that the dipole is significantly
smaller than the monopole as expected but is not negligible,
especially near the zeros of the monopole oscillations.  In particular, along
with the early ISW effect, it fills in fluctuations {\it before} the first
acoustic peak. Due to baryon contributions, gravity is able
to shift the equilibrium position of the fluctuations, leading to
a modulation of the monopole peaks (see \S 3.2).  We
have drawn in the zero level of the oscillations to guide the eye.
The kink at $k=0.04$
Mpc$^{-1}$ is due to the joining of the large and small scale solutions.
}
\label{fig:5power}
\end{figure}

%% file: chap5/peak.tex
\begin{figure}[t]
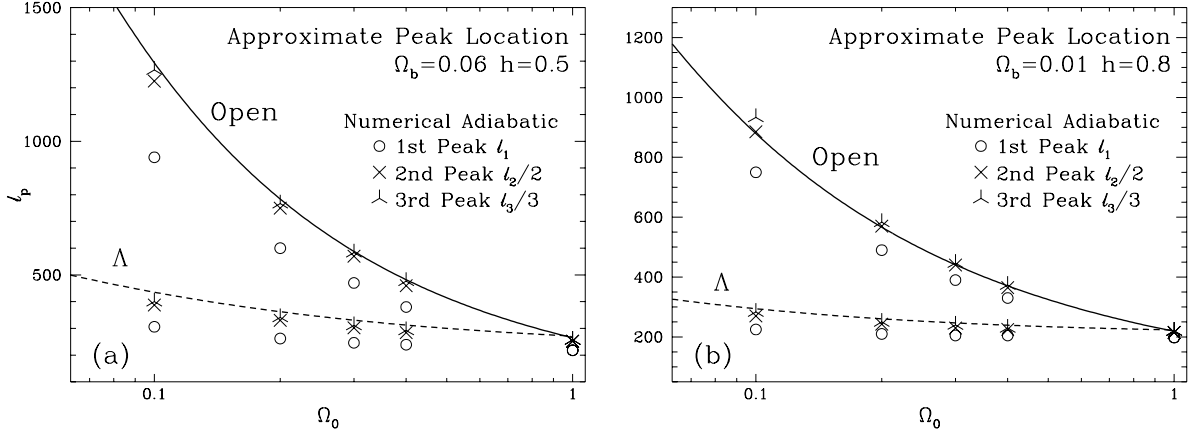

\centerline{ \hskip -0.25 truecm
\epsfxsize=3.0in \epsfbox{chap5/peaka.epsf} \hskip 0.25 truecm
\epsfxsize=3.0in \epsfbox{chap5/peakb.epsf}}
 \vskip -0.5truecm
\caption{Acoustic Peak Location}
\mycaption{The physical scale of the peaks is simply related
to the sound horizon
at last scattering and corresponds to
multiples of the angle 
that this scale subtends on the sky $\ell_p = p\pi r_\theta
/r_s$ for adiabatic models.
Varying $\Omega_0 h^2$ changes 
both the sound horizon at $\eta_*$ and the present horizon
$\eta_0$ leaving
little effect. For open models,
a given scale will correspond to a smaller angle by geodesic deviation.
This projection estimate for the peak location is valid
for pure acoustic contributions and underestimates the scale of
the first peak in low $\Omega_0 h^2$
models due to neglect of the early ISW effect.
}
\label{fig:5peak}
\end{figure}

%% file: chap5/heights.tex
\begin{figure}[t]
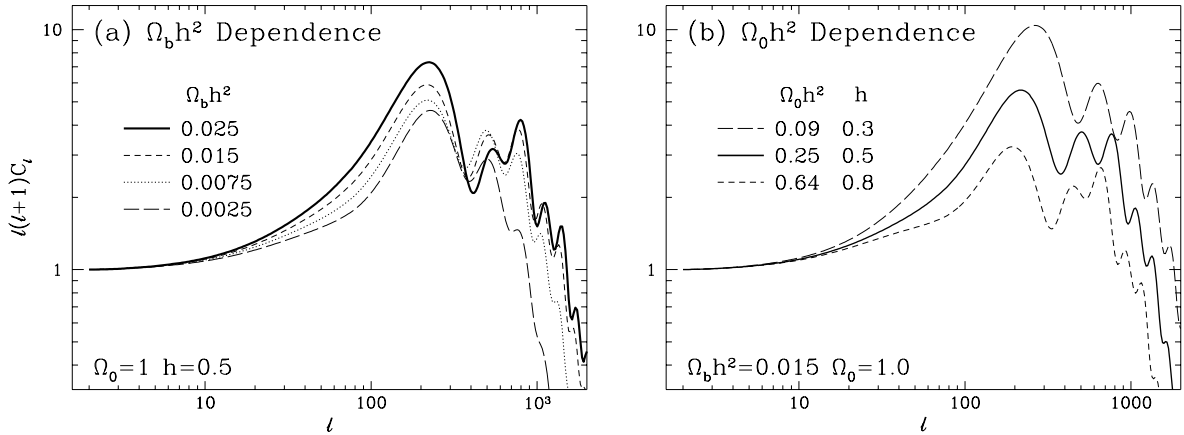

\centerline{ \hskip -0.15 truecm
\epsfxsize=3.0in \epsfbox{chap5/heightsa.epsf} \hskip 0.15 truecm
\epsfxsize=3.0in \epsfbox{chap5/heightsb.epsf}}
\vskip -0.5 truecm
\caption{Acoustic Peak Heights}
\mycaption{(a) The baryon-photon ratio $R\propto \Omega_b h^2$ determines
the balance between pressure and gravity and thus the zero point of
the oscillation.  Gravitational enhancement of compression leads to 
higher odd peaks as $\Omega_b h^2$ increases. For sufficiently 
high $\Omega_b h^2$, the even peaks cannot be distinguished at all. 
(b) Decay of the potentials
$\Psi$ and $\Phi$ due to radiation pressure inside the horizon during
radiation domination drives 
the oscillation to higher amplitude.  If  
matter radiation equality is delayed by lowering $\Omega_0 h^2$, this 
enhancement can boost the first few peaks.  The radiation also 
changes the expansion rate and shifts the location of
the peaks. 
}
\label{fig:5heights}
\end{figure}

%% file: chap5/diffusion.tex
\begin{figure}[t]
\centerline{\hskip -0.5truecm
\epsfxsize=3.5in \epsfbox{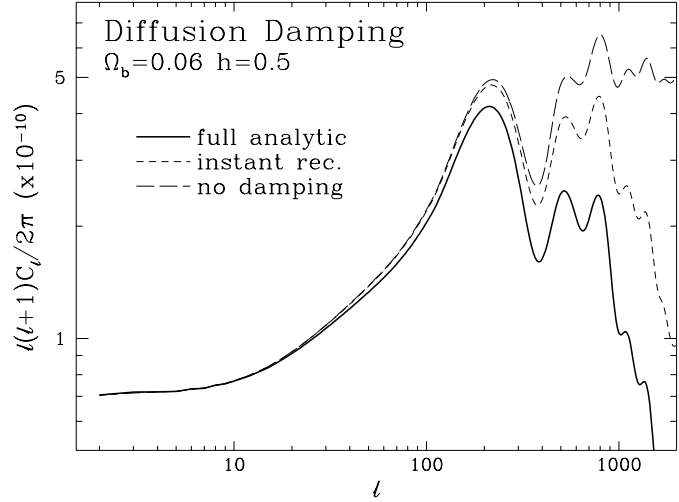}}
 \vskip -0.5truecm
\caption{Diffusion Damping}
\mycaption{The effect of the finite duration of
 last scattering from the results of Appendix \ref{ss-5analytic}.  
Estimating the damping
in the instantaneous recombination approximation leads to a significant
underestimate
of the damping scale.  It is however far better than neglecting
diffusion damping entirely.
}
\label{fig:5diffusion}
\end{figure}

%% file: chap5/damping.tex
\begin{figure}[t]
\centerline{ \hskip -0.25 truecm
\epsfxsize=3.0in \epsfbox{chap5/dampinga.epsf} \hskip 0.25 truecm
\epsfxsize=3.0in \epsfbox{chap5/dampingb.epsf}}
\vskip -0.5 truecm
\caption{Damping Scale}
\mycaption{The diffusion damping scale depends somewhat differently on
cosmological parameters than the acoustic scale.  It is more sensitive
to $\Omega_b$ (see Fig.~\ref{fig:5heights}) and less sensitive to $h$.
Its presence can also alter the pattern of heights expected from 
the acoustic peaks, \eg\ the small scale boost from dilation and
the alternating peak heights from infall.  
}
\label{fig:5damping}
\end{figure}

%% file: chap6.tex
\chapter{Secondary Anisotropies}
\label{ch-secondary}
\begin{verse}
\footnotesize\it\baselineskip=7pt
Mingled and merged, densely sprouting,

In the primaeval mass, there is no shape.

Spreading and scattering, leaving no trail behind,

In the darkness of its depths, there is no sound.
\vskip 0.1 truecm
\hskip 3.0truecm
{--Chuang-tzu, 14}
\end{verse}
\def\etau{e^{-\tau(\eta,\eta_0)}}

Between recombination and the present, astrophysical processes can alter 
the anisotropy spectrum.  In general, they may have two distinct effects:

\begin{enumerate}

\item Erasure or masking of primary anisotropies by rescattering and 
other filtering.

\item Generation of secondary fluctuations imprinting the mark of a much more
evolved and complex universe.

\end{enumerate}

\noindent
Indeed from the null result of the Gunn-Peterson test \cite{GP}, we know that
the universe is almost completely ionized out to redshift 4-5 \cite{SSG,WBCP}.
Although this alone would only have a percent or so affect on primary 
anisotropies,
it raises the possibility that reionization of the universe could have
occurred
at a much higher redshift.  In models with sufficient small scale power,
it is plausible that an early round of structure formation may have
released
the energy required to keep the universe ionized at high redshift
(see \eg\ \cite{GO,TS}).

Early reionization scenarios enjoyed a brief period of popularity following
the detection of puzzlingly small anisotropies at the $\sim 2^\circ$
scale by the SP91 experiment \cite{Schuster}, as the great number of
papers that it generated bears witness 
to \cite{GSJ,DJ93,CSS,HSSa,SSV,Models,DJ95}.  Although the status is
far from clear at the present, their popularity has declined due to
the steady stream of higher detections on roughly the same scale
\cite{WSS,Krauss}.  Still, some filtering of primary anisotropies must have
occurred. 
Indeed, for typical (primordial) isocurvature baryon (PIB)
models \cite{PeeblesPIBa,PeeblesPIBb}
significant reionization is both necessary and natural due to their excessive
amounts of small scale power.

Since secondary anisotropies 
depend on the astrophysical details of structure formation, 
they provide interesting constraints on models and clues to the process
of structure formation.  On the other hand, they
  do not have much power to measure background parameters
in a model independent manner.   In this sense, primary and secondary
anisotropies complement each other.  
If reionization is not too substantial, both mechanisms may
contribute.  In this case, the information
contained in the CMB increases and consequently so does the care needed
to extract it.  The problem of extraction alone would motivate the
study of secondary anisotropy formation (see also Appendix 
\ref{sec-5smallstuff}).  

In this chapter, we first discuss the general principles that govern
secondary anisotropy formation in linear theory. 
Since the results are quite model dependent, we will offer the CDM
and PIB models to illustrate their effect. 
Linear contributions
are generally cancelled at small scales.  It is therefore necessary to include
higher order effects.  We discuss second order calculations 
in detail and find
that the Vishniac, or second order Doppler contribution is the dominant
source at small angles.  We then briefly survey highly non-linear effects
and their importance for secondary anisotropy formation.  In this case,
even the qualitative sense of the effects can be model dependent.

\section{Linear Contributions}
\label{sec-6linear}

\subsection{Reionization Damping}
\label{ss-6damping}

Secondary anisotropy formation in linear theory follows the same basic
principles as primary anisotropy formation.  The main difference is
that the photons and baryons are no longer tightly coupled.  As 
shown in \S \ref{sec-4postrecombination},  the baryons
are released from Compton drag when the redshift falls below
\bel{eq:comptondragredshift}
z_d \approx 160 (\Omega_0 h^2)^{1/5} x_e^{-2/5},
\ee
where recall that $x_e$ is the ionization fraction.
Fluctuations in the matter
then are free to grow and follow the pressureless solution $D$ to the
evolution equations (see \S \ref{sec-4superhorizon}).  
Likewise, the photon diffusion length grows to be comparable
to the horizon size.  Last scattering effectively occurs when the
Compton scattering time becomes greater than the expansion time.  
More specifically, we can define it as the epoch when optical 
depth reaches unity. Since the optical depth
\beal{eq:opticaldepth}
\tau \eal 
%{\sigma_T H_0 \over 4\pi G m_p} 
4.61 \times 10^{-2} 
	(1-Y_p/2)x_e {\Omega_b h \over \Omega_0^2} \nonumber\\
&& \times \cases
{
  [2-3\Omega_0 + (1+\Omega_0 z)^{1/2}(\Omega_0 z+ 3\Omega_0 -2)] & 
	$\Omega_\Lambda=0$ \cr
  \Omega_0 [ 1-\Omega_0 + \Omega_0(1+z)^3]^{1/2} -\Omega_0 &
	$\Omega_0 + \Omega_\Lambda = 1$ \cr}
\eea
if $x_e$ is constant,  this occurs at
\bel{eq:lastscattering}
z_* \approx 98 \left( {\Omega_0h^2 \over 0.25} \right)^{1/3}
	       \left[ {(x_e \Omega_b h^2) \over 0.0125}{(1-Y_p/2) \over 0.885} 
	       \right]^{-2/3},
\ee
for both cases 
since  last scattering occurs before
curvature or $\Lambda$ domination.  Notice that last scattering occurs 
after the end of the drag epoch for sufficiently high ionization and
baryon fraction.  

\input chap6/pibrms.tex

In this limit, photons diffuse amongst the freely falling baryons inside
the horizon.  
Recall that diffusion damps intrinsic photon 
fluctuations as $e^{-\tau}$ due to 
streaming conversion of inhomogeneities to anisotropies and subsequent 
rescattering
isotropization.  Thus primary anisotropies are
sharply damped below the horizon scale implying that no acoustic
oscillations will survive (see Fig.~\ref{fig:6pibrms}).

\input chap6/cltau.tex

Features in the primary spectrum will be accordingly damped away as the
optical depth between recombination and the present increases.  For 
sufficiently high optical depth, the ability to measure fundamental
cosmological parameters through the location of the peaks may be lost 
(see Fig.~\ref{fig:6cltau}a).  Notice that for $\tau \simlt 1$, the oscillation
amplitudes are still high enough to make measurements possible.  Beyond
this value, the primary signal is likely to be lost in the noise, 
foreground contamination, and non-linear source contributions.  
For the low $\Omega_b h^2 = 0.0125$ standard CDM model, this only occurs
for an ionization redshift $z_i \simgt 100$.  This possibility is highly
unlikely since its $n=1$ primordial spectrum does not have enough power
for such early structure formation. 

\input chap6/trans.tex

\subsection{\COBE\ Constraints on PIB Models}
\label{ss-6PIB}

Reionization damping can on the other hand save models which would otherwise
predict too high an amplitude for small scale anisotropies. 
Such is the case for standard PIB models which have initial
isocurvature fluctuations $|S(0,k)|^2 \propto k^m$ in a 
baryon-dominated $\Omega_0 = \Omega_b$ universe
\cite{PeeblesPIBa,PeeblesPIBb}. 
Although $\Omega_0 = 0.1-0.3$ models, 
designed to satisfy dynamical estimates of 
the mass, consequently
fail to satisfy nucleosynthesis constraints on the baryon density,
astrophysical processes could alter light element abundances \cite{GO,GOR}. 
Moreover since there is no {\it ab initio} mechanism for generating the
required entropy perturbations, the index $m$ is fixed by measurements of
large scale structure today.  Recall from \S \ref{ss-4dampedacoustic} 
that isocurvature 
perturbations evolve
such that below the photon diffusion scale, the initial entropy
fluctuations become the density perturbations that seed large
scale structure (see Fig.~\ref{fig:6trans}).  
The observed power spectrum of approximately
$P(k) \propto k^{-1}$ at large scale structure scales \cite{Peacock} then
implies an $m \approx -1$ initial power law in the model.  Numerical simulations
which take into account non-linearities
confirm this result \cite{SugSuto}.
At the largest scales, however, isocurvature conditions prevent the
formation of potential perturbations leaving $k^3 |\Phi|^2 \propto k^{3+m}$
which is steeply rising for $m=-1$. When normalized to the \COBE\ DMR
measurement, this leads to a steeply rising spectrum of anisotropies
with effective slope $n_{\rm eff} \approx 2$.  
This model therefore
has three difficulties to overcome

\begin{enumerate}

\item Steeply rising \COBE\ slope.

\item Large degree scale anisotropies.

\item High matter power spectrum normalization, $\sigma_8$ the amplitude
at the $8h^{-1}$Mpc scale.

\end{enumerate}

\noindent They are all alleviated somewhat by reionization.
Since Silk damping \cite{Silk}
does not destroy entropy fluctuations, the large amount of small scale
power in the model allows for collapse of objects immediately following
recombination (see \S \ref{ss-4comptondrag}).  
This could lead to sufficient energy input to reionize
the universe as early as $z_i = 800$ \cite{PeeblesPIBa,PeeblesPIBb}.
As we have noted, reionization damps the steeply rising primary signal
(see Fig.~\ref{fig:6cltau}b) and can help the first two problems.
Furthermore, because Compton drag prevents the growth
of structure, the ionization history can be tuned to provide the right
ratio of matter to temperature fluctuations (see Figs.~\ref{fig:4comptondrag}, 
\ref{fig:6trans}).

\input chap6/pibcobe.tex

Unfortunately, reionization can only damp fluctuations under the horizon
scale at last scattering.  Thus it is difficult to lower the
effective slope $n_{\rm eff}$ 
at \COBE\ scales $\ell \approx 2 - 25$.  
Geodesic deviation carries the same physical scale
onto smaller angles for open universes.  Thus  open
models will thus be even less affected by reionization than
$\Lambda$ models. Smaller effects include raising the baryon
content through $\Omega_b h^2$ which
delays last scattering and increases
the physical scale of the horizon.  However even for flat models,
the projection from the last scattering surface
depends strongly
on $\Omega_0$ and counters the $\Omega_b$ dependence in these $\Omega_0=
\Omega_b$ baryonic models.
Furthermore, the late ISW effect 
boosts the low order multipoles slightly
as $\Omega_0$ decreases (see \S \ref{ss-5iso}).  
In the range of interest, decreasing $\Omega_0$
leads to a shallower \COBE\ slope.
High $x_e$, high $h$, low $\Omega_0$, $\Lambda$ models
therefore offer the best prospects of bringing
down the \COBE\ slope. 

The amount of reionization allowable is moreover constrained by 
the lack of spectral distortions in the CMB, $y \le 2.5 \times
10^{-5}$ (95\% CL) \cite{Mather}, 
where recall from \S\ref{ss-3comptonization} that 
$
y = \int d\tau \, k(T_e-T) / m_e c^2
$
measures the amount of upscattering in frequency from hot electrons.
For collisional
ionization, the electron temperatures must be quite high to overcome
the Boltzmann suppression factor, typically $T_e \simgt
15000$K \cite{GO,COP}.
For photoionization, there is no firm lower limit on $T_e$  since
we can always fine tune the photoelectron energy
to zero (\eg\ with a decaying neutrino that produces
13.6 eV photons).  
Yet, given the ionization potential, we
would generically expect electron energies of a few eV.
Compton cooling from energy transfer to 
the CMB (see \S \ref{ss-3electrontemp})
then suppresses the equilibrium
electron temperature to an average of $T_e \sim 5000$K \cite{TS}.
We will therefore adopt an electron temperature of $T_e =5000K$.
Since the collisionally
ionized model is to date the only
isocurvature scenario to successfully modify nucleosynthesis \cite{GO},
this is a very conservative choice.

Bunn, Scott, \& White \cite{BSW} find that the observational constraints
require $n_{\rm eff} = 1.3_{-0.37}^{+0.24}$ (with quadrupole) which 
indicates that $n_{\rm eff} = 2$ should be ruled out at greater than 
95\% confidence.  Since PIB spectra are not pure power laws in the 
effective slope (see Fig.~\ref{fig:6cltau}b), to quantify this constraint, we
employ a full likelihood analysis of the two-year
\COBE\ DMR sky maps for  open and $\Lambda$ isocurvature baryon models
fixed by $\Omega_0$, $h$, and $x_e$ \cite{HBS}.
We expand the two-year DMR data in a set of basis functions which
are optimized to have the maximum rejection power for incorrect
models (see \cite{Bunn} for a full discussion). 
To set limits on $m$ and the normalization
$Q$, the rms quadrupole, we assume a prior distribution
which is uniform for all $Q$ and $m \le 0$.  Spectra with $m > 0$ are
unphysical due to non-linear effects which regenerate an $m=0, P(k) \propto
k^4$ large
scale tail to the fluctuations \cite{PeeblesLSS}.
The constraint in
the crucial $m \approx -1$ regime is not sensitive to the details of this
cutoff.  It is furthermore not very sensitive to ambiguities in the
definition of power law initial conditions at the curvature scale
(see \S \ref{ss-4laplacian} and \S \ref{ss-5iso}
for a discussion).   Shown in Fig.~\ref{fig:6pibcobe} 
are the 95\% confidence upper
limits imposed on $m$ by integrating over the normalization $Q$ to form
the marginal likelihood in $m$.   As expected, all open models
with $m \approx -1$ are ruled out regardless of ionization fraction, whereas
highly ionized $\Lambda$ models remain acceptable.  Notice however that
the constraint tightens for the highest ionization fractions.  This is
because fluctuations are in fact regenerated at the new last scattering
surface if the optical depth is sufficiently 
high (see \S \ref{ss-6regeneration} below).

Since the PIB model is phenomenologically based, it is always possible
to add free parameters to adjust the model to fit observations.  Indeed 
an initial power spectrum with $m\approx-1$ is required only in
the large scale structure regime.  
Aside from simplicity arguments, we have no firm reason
to believe that the power law behavior extends to \COBE\ scales.  It 
is therefore worthwhile to consider smaller scale anisotropy formation
where CMB and large scale structure observations will overlap.  This 
will eventually provide powerful consistency tests for {\it any} model
since the two measure fluctuations at very different epochs in 
the evolution of structure (see \eg\ \cite{TegTegTeg}).  In the case of
early reionization, regeneration of small scale anisotropies can be
significant.  It is to this subject that we now turn.

\subsection{Anisotropy Regeneration}
\label{ss-6regeneration}

Fluctuations are not entirely damped away by reionization (see 
Fig.~\ref{fig:6pibrms}).  Since the baryons are in 
free fall after the drag epoch,
they possess a non-negligible bulk velocity.  Compton scattering
still attempts to isotropize 
the photons in the electron rest frame and couples the
photon and baryon bulk velocities
$V_\gamma$ and $V_b$.  Thus at each scattering event, the photons
are given a Doppler kick from the electrons.  Subsequent diffusion
over many wavelengths of the fluctuation damps away this contribution. 
Thus fluctuations will be on the order of $V_b \tau_k$
if the optical depth through a wavelength of the 
fluctuation, $\tau_k \approx \dot\tau/k \ll 1$.  
In the opposite regime, the photons are
still tightly coupled. Doppler fluctuations then go to $V_b$ and add 
to the undamped temperature fluctuations.  

We can employ analytic techniques to better understand these Doppler 
contributions.  Ignoring curvature, as is appropriate for these small
scales before last scattering, 
the formal solution to the Boltzmann equation is
\bel{eq:BoltSoln}
\left[\Theta + \Psi \right](\eta,k,\mu) = \left[\Theta+\Psi\right](\eta_d,
k,\mu)e^{ik\mu(\eta_d-\eta)}e^{-\tau(\eta_d,\eta)} +
[\Theta_{D}+\Theta_{ISW}](\eta,k,\mu) ,
\ee
where recall $k\mu = \bk \cdot \bg$ and 
the optical depth $\tau(\eta_1, \eta_2)=\int_{\eta_1}^{\eta_2} 
\dot \tau d\eta$.  Here  $\Theta_{D}$ and 
$\Theta_{ISW}$ represent the Doppler and the ISW effect respectively.  
The initial conditions are
taken at the drag epoch $\eta_d$ so that we can consider the matter
source $V_b$ as evolving independently.
As noted above, scattering rapidly
damps out the contributions from before the drag epoch as
$e^{-\tau}$, and we will hereafter ignore this term.
Thus the photon temperature perturbation is a function 
of the matter perturbations alone.
These source terms are explicitly given by
\beal{eq:SSFISW}
\Theta_{D}(\eta,k,\mu) \eal \int_{\eta_d}^\eta (\Theta_0 + \Psi
- i \mu V_b) \,\dot \tau e^{-\tau(\eta',\eta)}\,
e^{ik\mu(\eta'-\eta)}
d\eta', \nonumber\\
\Theta_{ISW}(\eta,k,\mu) \eal\int_{\eta_d}^\eta 2 \dot \Psi
e^{-\tau(\eta',\eta)} \,
e^{ik\mu(\eta'-\eta)}
d\eta', 
\eea
where we have neglected the small correction to the quadrupole from 
the angular dependence of Compton scattering (see \cite{HSa} for the 
justification) and recall that the plane-wave decomposition is
defined such that $\bg \cdot \V_b(\eta,\bx) = -i \mu V_b(\eta,k) \exp(i\bk
\cdot \bx)$.

To solve equation \eqn{BoltSoln} to the present, we must obtain an expression for
the effective temperature $\Theta_0+\Psi$ at last scattering.
Taking the zeroth moment of equation \eqn{BoltSoln}, we obtain
\bel{eq:zeromom}
[\Theta_0+\Psi](\eta,k,\mu) = \int_{\eta_d}^\eta 
\dot \tau e^{-\tau(\eta',\eta)}\,
\left\{ (\Theta_0 + \Psi + 2\dot\Psi) j_0[k(\eta-\eta')] 
- V_b j_1[k(\eta-\eta')] \right\} d\eta', 
\ee
where we have employed the identity
\bel{eq:besselidentity}
j_\ell(z) =  {i^\ell  \over 2} \int_{-1}^1 \exp(i\mu z) P_\ell(\mu) d\mu ,
\ee
with $P_\ell$ as the Legendre polynomial.  In the diffusion limit, the
optical depth across a wavelength is small and the sources do not vary
much over a time scale $\eta \sim 1/k$.  Taking these quantities out of the
integral and assuming $\eta \gg \eta_d$, we obtain
\bel{eq:isotropicsource}
[\Theta_0+\Psi](\eta,k,\mu) \approx [\Theta_0 + \Psi] {\dot \tau \over k}
{\pi \over 2} - V_b {\dot \tau \over k}  + 2{\dot \Psi \over k} {\pi \over 2},
\ee
where we have employed the relation
\bel{eq:besselintegral}
\int_0^\infty j_\ell(z)dz = {\sqrt{\pi} \over 2} {\Gamma[(\ell+1)/2]
	\over \Gamma[(\ell+2)/2]}.
\ee
As advertised, the contribution from the electron velocity is of order 
$\dot\tau/k$ or the optical depth through a wavelength.  It is thus 
suppressed
at short wavelengths.  Since last scattering occurs before curvature or
$\Lambda$ domination, the change in the potential across a wavelength is
negligibly small and we can neglect the ISW contribution at last scattering.
Therefore the effective temperature becomes
\bel{eq:effectivetemperature}
[\Theta_0+\Psi](\eta,k,\mu) \approx - V_b {\dot \tau \over k}
\ee
through last scattering. 

It may seem counterintuitive that a source to the dipole $\Theta_1$ 
creates an isotropic temperature fluctuation $\Theta_0$.  Mathematically,
it is clear from the Boltzmann hierarchy \eqn{Hierarchy} that the dipole
indeed sources the monopole as photons travel across a wavelength, 
$k\delta\eta \sim 1$.  Consider an observer at the origin of a sine
wave baryon velocity fluctuation in real space $v_b(x) = V_b \sin(kx)$.
The observer sees photons coming from both the crest at $kx = \pi/2$, where
$v_b > 0$, and the trough at $kx = -\pi/2$, where $v_b < 0$.   The scattered
photon distribution at these sights will be oppositely aligned dipoles.
Thus the scattered radiation observed at the origin will be {\it red}shifted
in both directions.  This leads to a net temperature fluctuation.  
Of course, the effect is not cumulative.  Radiation from further crests
and troughs have shifts that cancel leaving an effect only for 
the photons which scattered within a wavelength of the perturbation, 
$\Theta_0 = {\cal O}(V_b \dot \tau/k)$.

Although this contribution is suppressed at short wavelengths, it is
comparatively important since the dipole source $V_b$ itself is severely 
cancelled.  Inserting the effective temperature \eqn{effectivetemperature}
in equation \eqn{BoltSoln} 
 and integrating the dipole source by parts,
we obtain
\bel{eq:effectivedoppler}
[\Theta+\Psi](\eta_0,k,\mu) = \int_{\eta_d}^{\eta_0} {1 \over k}(\dot V_b \dot \tau
+ V_b \ddot\tau + 2k\dot\Psi) \etau\,
e^{ik\mu(\eta-\eta_0)}
d\eta. 
\ee
The multipole decomposition is then obtained from 
equation \eqn{besselidentity},
\bel{eq:multipolemoments}
{\Theta_\ell(\eta_0,k) \over 2\ell +1}
= \int_{\eta_d}^{\eta_0}  {1 \over k}
\left[ \dot V_b
\dot \tau 
+ V_b 
\ddot \tau 
+ 2 k {\dot \Psi} \right] 
e^{-\tau(\eta,\eta_0)} j_\ell[k(\eta_0-\eta)]d\eta,
\ee
where we have employed equation \eqn{besselidentity} and recall that
the multipole moments are defined such that $\Theta_\ell = 
i^\ell (2\ell+1) {1 \over 2} \int_{-1}^{1} 
P_\ell(\mu) \Theta d\mu$.  For the open universe generalization, 
replace $j_\ell$ with $X_\nu^\ell$.  

\input chap6/first.tex

We can further simplify the result by noting that in the small scale
limit the anisotropy is sourced over many wavelengths of the perturbation.
Contributions from crests and troughs of the perturbation cancel. 
In this case, $j_\ell(x)$ can be approximated as a $\delta$-function
at $x = \ell + 1/2$.  In fact, we have already used this approximation
for the late ISW effect of $\Lambda$ models in \S \ref{ss-5lambda}.
Employing equation \eqn{besselintegral} and the Stirling approximation of
$\Gamma(x)/\Gamma(x+1/2) \approx x^{-1/2}$ for $x \gg 1$, we obtain
\bel{eq:thetaellapprox}  
{\Theta_\ell(\eta_0,k) \over 2\ell+1} \approx \sqrt{\pi\over 2\ell}{1 \over k^2} 
\left[\left( \dot V_b
\dot \tau 
+ V_b \ddot \tau 
+ 2 k{\dot \Psi}\right)
e^{-\tau(\eta,\eta_0)} 
\right] \bigg|_{\eta=\eta_0 -\ell/k},
\ee
in a flat universe.  With the relations
\beal{eq:VPsiD}
k V_b \eal - {\dot D \over D_0} {\Delta_T(\eta_0,k)}, \nonumber\\
k^2 \Psi \eal - {3 \over 2} H_0^2 \Omega_0 {D \over D_0}{a_0 \over a} 
{\Delta_T(\eta_0,k)},
\eea
from the continuity and Poisson equations \eqn{BeqnContPois},
the final expression for $C_\ell$ becomes
\bel{eq:Cldoppler}
C_\ell^D =  {V \over \ell} \int {dk \over k} {1 \over (k\eta_0)^6}
 S_L^2(\eta_0 -\ell/k) k^3 P(k),
\ee
where the matter power spectrum is $P(k) = |\Delta_T(\eta_0,k)|^2$
and the linear theory source is
\bel{eq:Slinear}
S_L(\eta) = \left[ {\ddot D \over D_0} \dot \tau + {\dot D \over D_0} \ddot \tau
 + 3H_0^2\Omega_0 { a_0 \over a} \left( {\dot D \over D_0} - {D \over D_0}
{\dot a \over a} \right) \right] \eta_0^3
e^{-\tau(\eta,\eta_0)}.
\ee
This relation accurately describes the anisotropy on scales smaller than
the horizon at last scattering if last scattering occurs {\it well} after the
drag epoch (see Fig.~\ref{fig:6first}).  
For low baryon fraction models such as CDM or partially ionized
PIB models, these relations become less accurate.   
Notice that the amplitude of
the Doppler effect depends strongly on the epoch of last scattering.
This is due
to the presence of a cancellation scale $k\eta_* \sim 1$ as we shall now
see.   

\subsection{Cancellation Damping}
\label{ss-6cancellation}

It is instructive to consider 
the spatial power spectrum of the radiation
$k^3 |\Theta+\Psi|_{rms}^2$ as well as the anisotropy spectrum. 
With the projection deconvolved, the physical processes are 
easier to understand.
In fact, historically the above analysis was originally
presented in $k$-space \cite{Kaiser84}.  The photons illuminate a 
surface of thickness $\delta \eta$ of the source field, \ie\ the
line of sight electron velocity for the Doppler effect and 
the decaying potential for the ISW effect.  For perturbations 
with wavelength smaller than the thickness, the observer sees through
many crests and troughs if the wavevector is aligned parallel to the line
of sight.   Thus contributions will be severely cancelled for these modes
(see Figs.~\ref{fig:1projection} and \ref{fig:1cancelpict}).
A loophole occurs however if the wavevector is aligned perpendicular
to the line of sight.  In this case, all the contributions are additive
along the line of sight and cancellation does not occur.  For an isotropic
source field, the net effect after summing over both components 
is a suppression of power by $(k\delta\eta)^{-1}$ 
or approximately the inverse number
of wavelengths across the fluctuation.  

\input chap6/cancel.tex

For the Doppler effect, the source 
field is not isotropic.  Indeed, it is only the line of sight component
of the velocity that contributes at all.  In linear theory, the 
potential gradient
$\nabla \Psi$  generates an infall velocity.  
Thus gravitationally induced flows are
irrotational $\nabla \times \V(\bx) =0$ or $\bk \times \V(\bk) = 0$ and
the
velocity is parallel to the wavevector.  The line of sight component
of the electron velocity vanishes for the perpendicular mode.  In this
case, cancellation is much more severe.  Only if
the electron velocity or the probability of scattering changes across a
wavelength do the redshifts and blueshifts from crests and troughs
not entirely cancel.  The contributing sources are of order
$\dot V_b/k$ and $V_b \ddot\tau/\dot \tau$, as we have seen, 
and suppress the net effect
by an additional $(k\delta\eta)^{-2}$ in power. 
  
We can formalize these considerations by noting that equation 
\eqn{effectivedoppler} is approximately a Fourier transform in $\eta$
whose transform pair is $k\mu$ (with $k$ fixed).  This implies the
relation
\bel{eq:transformpair}
k^2 \eta_0^3 {[\Theta + \Psi](\eta_0,k,k\mu) \over \Delta_T(\eta_0,k)} 
\longleftrightarrow^{\kern-15pt\rm FT}\kern 2pt
S_L(\eta),
\ee
where $S_L$ is the linear theory source given by equation \eqn{Slinear}.
Thus the two mean squares are related by Parseval's theorem,
\bel{eq:parceval}
\int_0^{\eta_0} S_L^2(\eta)d\eta \approx {1 \over 2\pi} 
k^4 \eta_0^6 P^{-1}(k)
\int k d\mu |\Theta+\Psi|^2
\ee
or rearranging the terms,
\bel{eq:radiationpower}
|\Theta+\Psi|^2_{rms}(\eta_0,k) \approx \pi {P(k) \over (k\eta_0)^5}
\int_0^{\eta_0} S^2_L(\eta) d\eta/\eta_0.  
\ee
where we have employed the relation $|\Theta+\Psi|^2_{rms} = 
{1 \over 2} \int_{-1}^{1} d\mu |\Theta+\Psi|^2$. 

All the terms in equation \eqn{radiationpower} are easy to understand.  
The velocity power spectrum
is proportional to $P(k)/k^2$ and the potential power spectrum to 
$P(k)/k^4$.  The Doppler term suffers cancellation in power by $k^{-3}$ 
and the late ISW effect by $k^{-1}$.  This brings the contribution
to $P(k)/k^5$ for both effects and represents a significant small
scale suppression compared with the matter fluctuations.  
In Fig.~\ref{fig:6cancel}, we show an isocurvature
baryon examples compared with the numerical results.  
Notice that the late ISW effect can make a strong contribution to
this {\it spatial} power spectrum even
at small scales \cite{HSISW}.
Equation \eqn{radiationpower} is slightly less accurate 
for the $\Lambda$ late ISW effect since the potential
is still decaying at the present and Parseval's theorem begins to
break down because of the upper limit of the integral.

In fact, the radiation power spectrum can be approximated
by taking a projection of real space onto angles
\bel{eq:powerprojection}
{\ell (2\ell +1) \over 4\pi} C_\ell \approx {V \over 2\pi^2} k_{\rm proj}^3
|\Theta+\Psi|^2_{rms}(\eta_0,k_{\rm proj})
\ee
where $k_{\rm proj} \approx \ell/r_\theta(\eta_{\rm max})$, $\eta_{\rm max}$
is the epoch when the source $S_L$ peaks, and the angle-distance relation
$r_\theta$ is given by equation \eqn{angledistance}.  This is often useful for
open universes where the radial eigenfunctions at high wavenumber are
difficult to compute.  However, one must be careful to separate component
effects if $S_L$ is bimodal.  For example, since the $\Lambda$ late
ISW effect arises near the present time, spatial scales are carried
to larger angles by the projection than for the Doppler contributions.
In fact, even for the $\Omega_0=0.1$ $\Lambda$ model, the late ISW
effect
is not visible in $C_\ell$.  This exhibits one of the dangers of
naively working with spatial power spectra. 

\input chap6/minimal.tex

\subsection{Minimal PIB Anisotropies}

As an example of the regeneration of fluctuations through the Doppler
effect, let us consider the open PIB model.  It is particularly interesting
to construct one with minimal anisotropies.  
We have seen that the steeply rising spectrum of anisotropies in
this model can only be moderately mitigated by reionization because
of the angle to distance relation in open universes.  On the other hand,
the lack of information about the initial spectrum near the curvature scale
can be employed to evade the large angle constraint of \S \ref{ss-6PIB}.
Degree scale anisotropies can alternately be employed to constrain 
the model.  Since the observational state is still in flux, we
shall limit ourselves to stating rules of thumb which may be useful
to model builders in the future.  
For a concrete use of current data sets along
these lines, see \cite{Models}.

We might generalize the 
standard PIB model with a two dimensional parameterization of the
ionization history involving both the ionization fraction $x_e$ and 
the ionization redshift $z_i$.  
Since the fundamental scale for cancellation damping
is the horizon at last scattering, anisotropies will depend sensitively on
the epoch of last scattering.  Raising the ionization fraction delays
last scattering and makes the damping scale larger.  By allowing more
growth between the drag and last scattering epochs, it also increases
the amplitude of velocity perturbations at last scattering.  These
two effects oppose each other but are not of equal magnitude:
cancellation damping is more significant than growth 
(see Fig.~\ref{fig:6minimal}).  Thus minimal anisotropies will occur
for maximal ionization fraction $x_e$.

The ionization redshift has a more complicated effect.  Before
reionization, fluctuations can grow in pressureless linear theory. 
Thus the baryon velocity and correspondingly the Doppler effect
will be lowest for the latest reionization.
However, at scales near to and above the horizon at last scattering,
adiabatic growth of the temperature fluctuation dominates
(see \S \ref{sec-4superhorizon}).  For these scales, the 
latest reionization that still permits significant optical depth
between recombination and the present minimizes fluctuations 
(see Fig.~\ref{fig:6minimal}).
Since PIB models must have high optical depth $\tau \simgt 3$ 
between recombination and the present to damp the large primary
fluctuations \cite{Models}, the ionization redshift must be significantly
before last scattering.  However, it must also be low enough to avoid
Compton-$y$ constraints.  These constraints together
with degree scale anisotropy and large scale structure observations
will make PIB model building a real challenge in the future.

\section{Second Order Contributions}
\label{sec-6second}

The severe but in some sense ``accidental''
cancellation of the linear effect for reionized scenarios 
leads to the possibility that higher order effects may dominate 
sub-degree scale anisotropies. 
In this section, we will consider anisotropy
generation to second order in perturbation theory \cite{HSa,DJ95}. 
The fundamental equations and concepts necessary to understand these 
these effects have already been discussed in \S \ref{ss-2terms}.  
Applying them to the case of reionized models, we find that one source,
the so-called Vishniac term \cite{OV,Vishniac}, dominates over 
all other contributions.

\def\bq{{\bf q}}
\subsection{Generalized Doppler Effect}
\label{ss-6generaldoppler}

As we have seen, cancellation is a 
geometric effect and its severity for the Doppler effect is due irrotational
nature of flows in linear theory.
All modes except those for which $\bk$ is perpendicular to the line of sight
are cancelled as the photon streams through many wavelengths of the
perturbation to the observer.  However for the Doppler effect,
only the parallel component of the electron velocity yields an effect.  
Thus, for irrotational flows $\V_b \parallel \bk$, Doppler contributions
are severely suppressed.  Note however that the full Doppler source is
in fact $\dot \tau \V_b$, where recall $\dot \tau = x_e n_e \sigma_T a/a_0$,
since the probability of scattering must be 
factored in.  A photon is  more likely to scatter in regions of high 
density or ionization.  Thus perturbations in $x_e$ 
and $n_e$ will change the Doppler source.   The effective velocity is
therefore 
\beal{eq:velocityeffective}
{\bf q}(\bx) \eal[1 + \delta n_e(\bx) /n_e][1+ \delta x_e (\bx) /x_e] \V_b(\bx) \nonumber\\
 \eal[1 + \Delta_b(\bx)][1+ \delta x_e(\bx) /x_e] \V_b(\bx) .
\eea
If fluctuations in the electron density or ionization are
small, the additional contributions will be of second order.  
They can however escape the severe cancellation of the first order
term.  For example, there could be
a large scale bulk flow $\V_b(k_1)$ with $\bk_1 \parallel \bg$ 
and a small scale density fluctuation
$\Delta_b(k_2)$ with $\bk_2 \perp \bg$. In this
case, scattering will induce a small scale temperature fluctuations 
perpendicular
to the line of sight since more photons will have been scattered in
the overdense regions (see Fig.~\ref{fig:1vishpict}).  
In the extreme limit of high density fluctuations,
this is the kinetic Sunyaev-Zel'dovich effect for clusters (see
\S \ref{ss-6cluster} and \cite{SZ}). 

The solution of equation \eqn{BoltSoln} can be generalized to
\bel{eq:AeqnFormal}
[\Theta + \Psi](\eta_0,\bk,\bg) = \int_{\eta_d}^{\eta_0} \dot\tau 
\etau \bg \cdot \bq e^{ik\mu(\eta-\eta_0)}d\eta.
\ee
We have neglected the feedback term into the temperature fluctuation 
at last scattering since it is suppressed by the optical depth through
a wavelength.  Following Vishniac \cite{Vishniac},
let us decompose the solution 
into multipole moments,
\bel{eq:lmdecomp}
[\Theta + \Psi](\eta_0,\bk,\bg) = \sum_{\ell,m} 
a_{\ell m}(\bk) Y_{\ell m}(\Omega),
\ee
so that
\bel{eq:thetalm}
|a_{\ell m}|^2  = \bigg| {
        \int d\Omega\, Y_{\ell m}(\Omega)\, \int_0^{\eta_0} \dot \tau \etau
        (\bg \cdot \bq) e^{ik\mu(\eta-\eta_0)} } \bigg|^2.
\ee
Since the final result after summing over $\bk$ modes has no preferred  
direction, let us average over $m$ such that 
$|a_\ell|^2 = { 1 \over 2\ell +1 } 
	\sum_m |a_{\ell m}|^2,
$
which corresponds to $|a_\ell|^2 = 4\pi |\Theta_\ell/(2\ell+1)|^2$.
Choosing ${\bf \hat z} \parallel \bk$, we
note that the azimuthal angle dependence
separates out components of $\bq$ parallel and perpendicular to $\bk$ by 
employing the angular addition formula
\beal{eq:angularaddition}
{4\pi \over 2\ell +1} \sum_m Y_{\ell m}^*(\theta,\phi) Y_{\ell m}(\theta',\phi'
) \eal
P_\ell(\cos\theta)P_\ell(\cos\theta')  \\
&& + 2\sum_m {(\ell-m)! \over
(\ell + m)!} P_\ell^m(\cos\theta)P_\ell^m(\cos\theta') \cos[m(\phi-\phi')].
\nonumber
\eea
Since $\bg \cdot \bq = \cos\phi\, \sin\theta q_\perp + \cos\theta q_\parallel$,
the cross terms between the two components
vanish after integrating over azimuthal angles.  The two contributions
add in quadrature and may be considered as separate effects.

We have already noted that the $\bq \parallel \bk$ term is strongly suppressed
by cancellation.  Thus let us calculate the perpendicular component,
\bel{eq:associatedP}
|a_\ell(k)|^2 = 
		{\pi \over 2\ell(\ell+1)}\bigg|
                \int_{-1}^1 d\mu P^1_\ell (1-\mu^2)^{1/2}
                \int_0^{\eta_0} d\eta \dot\tau \etau q_\perp
                e^{ik\mu(\eta-\eta_0)} \bigg|^2.
\ee
The $\mu$ integral can be performed with the following identity
\bel{eq:identity}
\int_{-1}^1 d\mu (1-\mu^2)^{1/2} P_\ell^1(\mu) e^{iq\mu} = -2\ell (\ell+1)
(-i)^{-\ell+1} j_\ell(q)/q,
\ee 
so that
\bel{eq:thetalgen} 
|a_\ell(k)|^2 
	= {2\pi \ell(\ell+1)}\bigg| \int_0^{\eta_0} d\eta
        \dot \tau \etau q_\perp {j_\ell(k\Delta\eta) \over k\Delta\eta}
         \bigg|^2,
\ee
where $\Delta\eta = \eta_0-\eta$.
Notice that this has a simple physical interpretation.  We know from
the spherical decomposition that a plane wave perturbation
projects onto the shell at distance $\Delta \eta$ as $j_\ell(k\Delta\eta)$.
If the amplitude of the plane wave has an angular dependence,
the projection is modified.  In particular,
the perpendicular component suffers
less projection aliasing (see Fig.~\ref{fig:1projection}) and thus
the higher oscillations are damped as $\eta/k\Delta\eta$.

\subsection{Vishniac Effect}
\label{ss-6vishniac}

The Vishniac effect \cite{OV,Vishniac}
is the second order Doppler effect due to the density
enhancement $n_e(\bx) = \bar n_e [1+\Delta_b(\bx)]$ in linear theory, \ie\
$\bq(\bx) = [1+\Delta(\bx)]\V_b(\bx)$ to second order. 
The convolution theorem tells us that
\bel{eq:qvishniac}
{\bq}_\perp(\bk) = \left( I - {\bk \bk \over k^2} \right){1 \over 2}
\sum_{\bk'}
{\bf v}_b(\bk')\Delta_b(|\bk-\bk'|) + {\bf v}_b(\bk-\bk')\Delta_b(k').
\ee
Taking the ensemble average
of the fluctuation and assuming random phases for the underlying linear
theory perturbations, we obtain
\bel{eq:ensembleav}
\left< q_\perp^*(k,\eta) q_\perp(k,\eta') \right> = {1 \over 2}
\dot D(\eta)D(\eta) \dot D(\eta')D(\eta') \sum_{k'} d^2 P(k') P(|\bk-\bk'|),
\ee
where the projected vector
\bel{eq:dvishniac}
{\bf d} \equiv \left( I - {\bk \bk \over k^2} \right) \left[ {\bk' \over k^2}
 + {\bk - \bk' \over |\bk - \bk'|^2} \right].
\ee
A bit of straightforward but tedious algebra yields
\bel{eq:thetalvish}
\left< |a_\ell(k)|^2 \right> = 
{ 1 \over 4\pi}{V \over \eta_0^3}
{\ell (\ell+1) \over k\eta_0} M_V(k) I_\ell^2(k) P^2(k),
\ee
where the mode-coupling integral is
\bel{eq:modecoupling}
M_V(k) =
        \int_0^\infty dy \int_{-1}^{1} d(\cos\theta)
        {(1-\cos^2\theta)(1-2y\cos\theta)^2 \over (1+y^2-2y\cos
        \theta)^2} {P[k(1+y^2-2y\cos\theta)^{1/2}] \over P(k)}
        {P(ky) \over P(k)}, \qquad
\ee
and the time integral is
\beal{eq:AeqnTime}
I_\ell(k) \eal \int_0^{\eta_0} {d\eta \over \eta_0}
S_V (\eta) j_\ell(k\Delta\eta) \nonumber\\
          \eapp \sqrt{\pi \over 2\ell}{1 \over k\eta_0}S_V(\eta_0 - \ell/k),
\eea
with
\bel{eq:SVishniac}
S_V(\eta) =  \dot {D \over D_0}{D \over D_0}
{\eta_0^3 \over \eta_0 - \eta} \dot \tau e^{-\tau}.
\ee

\input chap6/vishniac.tex

The random phase assumption for the underlying linear perturbations assures
us that there are no cross terms between first and second order contributions
or different $k$ modes.
Thus total anisotropy is obtained by integrating over all $k$ modes 
\cite{HuWhite},
\beal{eq:Clvishniac}
C_\ell^V \eal {V \over 2\pi^2} \int {dk \over k} k^3  \left< | a_\ell(k) |^2 
\right> \nonumber\\
       \eal {\ell (\ell+1) \over (2\pi)^3} {V^2 \over \eta_0^6} \int 
	   {dk \over k}
           (k\eta_0)^2 M_V(k) I_\ell^2(k) P^2(k) \nonumber\\
       \eapp { \ell \over (4\pi)^2 } {V^2 \over \eta_0^6} 
	   \int{dk \over k}
           M_V(k) S^2_V(\eta_0-\ell/k) P^2(k). 
\eea 
In Fig.~\ref{fig:6vishniac}, we plot the Vishniac effect for standard CDM.  
Notice that since $S_V^2$ depends on the amplitude of fluctuations to the
fourth power, contributions are highly weighted toward late times and allows
extremely small scales to contribute to observable anisotropies.  
Thus even with minimal ionization of $z_i=5$, for which primary anisotropies
are only damped at the percent level, the Vishniac effect can dominate
the anisotropy at small scales.

Again it is useful to consider the $k$-space power spectrum.  Employing
the same Parseval approximation as for the first order contribution, 
we obtain
\bel{eq:Vishniack}
|\Theta + \Psi|^2 = {V \over \eta_0^3} {P^2(k) \over 16\pi} M_V(k)
 \int_0^{\eta_0}
 (1-\eta/\eta_0)^2 S_V^2(\eta) d\eta/\eta_0,
\ee
where the extra factor $1-\eta/\eta_0$ in the integrand is due to the 
projection effect for the perpendicular mode.  The $k$ factors come
from weak cancellation of $(k\delta\eta)^{-1}$,
the continuity equation conversion of velocity to density
$(k\eta)^{-2}$, and the volume in $k$ available for mode coupling $k^3$.
Although the exact nature of the mode coupling integral can change the
scaling, this simple power counting implies that the Vishniac effect will
have more power at small scales than the cancelled first order contribution.

\input chap6/pibvish.tex

The $k$-space power spectrum has often been used in the past to estimate
the anisotropy through a distance to angle conversion 
such as equation \eqn{powerprojection}.
The common assumption is that the Vishniac effect projects as if it all
arises from the last scattering surface \cite{Efstathiou,HSa,DJ95}.  
Given the strong weighting toward late times, this significantly underestimates
its coherence scale (see Fig.~\ref{fig:6vishniac}a).
The magnitude of this misestimation increases with the amount
of small scale power in the model.   Take for example, a PIB model with
a steeply blue $m=0$ spectrum (see Fig.~\ref{fig:6pibvish}).  In this
case, the $k$ space power keeps on rising to small scales.   When this
is projected onto $\ell$ space, it predicts a divergent anisotropy.
Of course, second order theory breaks down as the fluctuation amplitude
becomes comparable to unity so that the 
real spectrum would not continue to rise indefinitely.  By inserting
a cutoff at the non-linear scale, the anisotropy predicted by
equation \eqn{Clvishniac} or power projection is finite.  However, 
to calculate the effect precisely, one needs to go to $N$-body
simulations to accurately track the non-linear evolution.

\subsection{Other Second Order Effects}
\label{ss-6secondgeneral}

It is by no means obvious that the Vishniac effect dominates over all other
second order sources.  It is therefore worthwhile to consider the 
general Boltzmann equation to second order \cite{HSa}.  Indeed spatial
variations in the ionization fraction $\delta x_e(\bx)$ from patchy
reionization can have an effect comparable to the Vishniac source.  However
because it is strongly dependent on the model for structure formation 
and reionization, it is beyond the scope of this discussion. 

The second order Boltzmann equation is obtained by integrating the sources
calculated in \S \ref{ss-2terms} over frequency 
and is given in real space by
\beal{eq:brightness2}
\dot \Theta + \dot \Psi + \gamma^i \partial_i
(\Theta+\Psi) \eal \dot\tau (1+\Delta_b)\bigg[\Theta_0 + \Psi - \Theta 
	+  \gamma_i v_b^i - v_b^2 + 7(\gamma_i v_b^i)^2  \\
	&& + 2\dot\Psi  + {\cal O}([\Theta_0-\Theta]v_b) \bigg],
\eea
where we have again
neglected the small correction to the quadrupole \cite{HSa}. We also
assume that the ionization is uniform. 
Aside from the ${\cal O}(\Delta_b v_b)$ Vishniac
contribution, 
there are several new terms to consider here.

\subsubsection{${\cal O}(v_b^2)$ Quadratic Doppler Effect}

The kinetic energy of the electrons can be transferred to the photons
in a manner identical to the thermal energy transfer of the Sunyaev-Zel'dovich
effect (see \S \ref{ss-2terms} and \S \ref{sec-6higherorder}).  
Spatial variations in the kinetic
energy cause of order $v_b^2$ anisotropies in the CMB.  Note that
these anisotropies carry spectral distortions of the Compton-$y$ just as 
their thermal counterpart. 

These fluctuations
do not suffer the drastic cancellation of the linear Doppler effect since
the energy is direction independent.  At small scales, the power is
reduced by a factor $(k\delta\eta)^{-1}$ like the late ISW and Vishniac
effect.  Counting powers in $k$, we expect that aside from a 
spectrum-dependent mode-coupling integral, the contribution will consist of
$(k\delta\eta)^{-1}$ from cancellation, $(k\eta)^{-4}$ from the 
velocity to density conversion, and $k^3$ for the volume available to
mode coupling.  This gives a total of $k^{-2}$ and implies that the
Vishniac effect should be more important at small scales.   

The Parseval approximation to the power spectrum confirms this 
scaling relation,
\bel{eq:Qkpower}
|\Theta + \Psi|^2_{rms} = {1 \over 32\pi}{V \over \eta_0^3} {1 \over (k\eta_0)^2}
M_Q(k) P^2(k) \int_0^{\eta_0} S_Q^2(\eta) d\eta/\eta_0,
\ee
where the mode-coupling integral is 
\def\ct{\cos\theta}
\def\cst{\cos^2\theta}
\beal{eq:Qmodecoupling}
M_Q(k) \eal \int_0^\infty dy \int_{-1}^{+1} d(\ct)
{(y-\ct)^2 - 7(1-\cst)(y-\ct)y + {147 \over 8} (1-\cst)^2y^2 \over
(1+y^2-2y\,\ct )^2 } \nonumber\\
&& \times {P[k(1+y^2-2y\ct)^{1/2}] \over P[k]}
{P[ky] \over P[k]},
\eea
and the source is
\bel{eq:Qsource}
S_Q(\eta) = {\dot D \over D_0}{\dot D \over D_0} \dot \tau \etau \eta_0^3.
\ee 
Therefore, unless the mode-coupling integral behaves much differently
than its Vishniac counterpart, this contribution will be small in
comparison.  
In Fig.~\ref{fig:6quadratic}, we show a comparison for the
CDM model.  Note that since the quadratic Doppler effect carries a
spectral distortion of $(\Delta T/T)_{RJ} = -2y$, we have multiplied
the power by a factor of 4 to correspond to the case where the
Raleigh-Jeans temperature is measured.  The quadratic
Doppler effect is never dominant in this model.

\input chap6/quadratic.tex

\subsubsection{${\cal O}([\Theta_0-\Theta]v_b)$ Quadratic Doppler Suppression}

As discussed in \S \ref{ss-2terms}, the quadratic Doppler effect ceases
to operate once the photons are isotropic in the baryon rest frame.  If the
optical depth within a coherence scale of the baryon velocity $\V_b(\bx)$
is high, then the CMB will possess a dipole $\Theta-\Theta_0$
 of exactly $\V_b(\bx)$.  This will cancel any further contributions from
the quadratic Doppler effect.  However, in the small scale diffusion 
limit, by definition 
the optical depth never reaches unity in a coherence scale. 
The critical division is the horizon scale at 
optical depth unity, \ie\ last scattering.  
In the mode-coupling integral, if the source of the contributions
arise from larger wavelengths than this,
 they will be cancelled by the
${\cal O}([\Theta_0-\Theta]v_b)$ term.  This can only make the small
quadratic Doppler contribution
even smaller.

\subsubsection{${\cal O}([\Theta_0-\Theta]\Delta_b)$ Vishniac Suppression}

The same suppression mechanism works for the Vishniac effect.  Recall that
the Vishniac effect arises since small scale overdensities can possess 
bulk velocities along the line of sight.  The increased probability of
scattering off overdense regions causes a small scale temperature variation
from the Doppler shift.  If the optical depth across the coherence scale
of the bulk velocity is high, then all the photons will have scattered.
Since further scattering does not affect the distribution, the increased
probability of rescattering in overdense regions has no effect.  In
other words, a dipole $\Theta-\Theta_0$ has already been generated, such
that the  ${\cal O}([\Theta-\Theta_0]\Delta_b)$ term exactly cancels with the 
Vishniac $v_b\Delta_b$ term.  
Again one must check whether the Vishniac effect arises from bulk flows
smaller or larger than the horizon at last scattering.  By inserting
cutoffs in the mode coupling integral equation \eqn{modecoupling}, one
can show that they arise from smaller scales for the range of
power law spectra usually considered in the CDM and PIB models.

\subsubsection{Mixed Order Terms}

It is possible that first and third order terms couple in the rms.  
We have shown that the parallel and perpendicular components of the Doppler
effect separate and add in quadrature for $C_\ell$ 
(see \cite{HSa} for the $k$-space
proof).  Since the first order
contribution only possesses a parallel part, the mixed effect will only
couple with the parallel third order term.  However, this term is again
severely suppressed by cancellation.  The mixed order Doppler effect
can therefore be entirely ignored. 

\section{Beyond Perturbation Theory: A Survey}
\label{sec-6higherorder}

\begin{verse}
\footnotesize\it\baselineskip=7pt
To acknowledge, mark out, study, assess, 

Divide, discriminate, compete, and dispute.

These are our eight powers.

What is outside the cosmos, acknowledge but do not study.

What is within the cosmos, study but do not assess

What is a matter of record, assess but do not dispute.

\vskip 0.1truecm
\hskip 3truecm --Chuang-tzu, 2
\end{verse}
Beyond the realm of linear calculations lies a plethora of higher order
effects that are highly sensitive to assumptions about structure formation.
Modeling and N-body simulations are needed to estimate their effects.
Consequently, a full study of these individual effects is beyond 
the scope of this
chapter. Instead, we survey the literature on these subjects and provide
order of magnitude estimates where possible.  Most of these effects are
small in the degree to arcminute regime where one hopes
that primary anisotropies will yield important cosmological information.
Others such as the cluster Sunyaev-Zel'dovich effect and foreground
sources in the galaxy may be filtered out by spectral information and
object identification.  

\subsubsection{Cluster Sunyaev-Zel'dovich Effect}
\label{ss-6cluster}

As pointed out by Sunyaev and Zel'dovich \cite{SZ}
clusters can induce
anisotropies in the CMB from Compton scattering off electrons in the hot
cluster medium.
These hot electrons transfer energy to the CMB, leading
to temperature anisotropies {\it and} spectral distortions in the CMB 
(see \S\ref{ss-3comptonization}).  The frequency dependence can be
used to separate its signal from the primary anisotropy.  

For a typical cluster of $T_e \approx 1-10 $keV and a typical optical
depth of $\tau \approx 0.1-0.01$, the effect is of order 
$(\Delta T/T)_{RJ} = -2y \approx 10^{-5}-10^{-3}$.  Of course, the rms
fluctuation on a random patch of the sky will be much lower than this.
Much effort has been expended to estimate the fluctuations caused
by the SZ effect with varying results
(e.g.~\cite{MarBFJSa,MakSut,BarSil,ColMRV}).
Recently, empirical modelling of clusters 
has shown that the anisotropy at arcminutes 
is on the order of $(\Delta T/T)_{RJ} \simlt 10^{-7}$
\cite{CelBar}. 
Moreover, the signal is in large part due to bright and easily identifiable
clusters.  If such known clusters are removed from the sample, the anisotropy
drops to an entirely negligible level.
 
The peculiar velocity of a cluster also produces anisotropies via
a Doppler shift of the scattered photons.  This is the
non-linear analogue of the Vishniac and patchy reionization effects.
This process
leads to no spectral distortions to first order and yields a true
temperature fluctuation of 
$\Delta T/T = {\cal O}(\tau_c v_c)$ for an individual cluster,
where the optical depth through the cluster is
typically of order $\tau_c \simeq 0.1-0.01$ and its peculiar velocity $v_p
\simeq $ few $ \times 10^{-3}$.
Again there is hope that the signal can be removed by
identifying bright clusters and perhaps even the thermal effect.

\subsubsection{Rees-Sciama effect}

Higher order corrections to the density evolution cause time dependence
in the gravitational potentials from the Poisson equation.  As pointed
out by Rees \& Sciama \cite{RS}, 
this can cause a late ISW effect even in an $\Omega_0=1$ universe.
The second order contribution has been shown to be negligibly 
small \cite{MGSS}.  One can understand this by simple scaling arguments.
Just as the first order late ISW contribution, this term suffers 
cancellation in power by $(k\delta\eta)^{-1}$ where $\delta\eta$ is now
the time scale for change in the potential.  The Poisson equation
relates potentials to densities via a factor $(k\eta)^{-4}$ and the
mode coupling volume factor yields $k^3$.  Thus the effect scales
as $k^{-2}P(k)$ and will be small in comparison to even the
minimal Vishniac effect
if the mode coupling integrals behave similarly.

The fully non-linear case has been estimated using N-body simulations
and power spectrum techniques \cite{SeljakISW}.
In the standard CDM model, 
non-linear contributions dominate over the primary fluctuations
only at $\ell \simgt 5000$ and are thus smaller than the minimal Vishniac
effect.  Ray tracing techniques 
corroborate these results by showing that
fluctuations are at the $10^{-7}$ level at degree scales \cite{TL}. 

\subsubsection{Gravitational Lensing}

The presence of potential fluctuations gravitationally lenses the CMB
and changes the projection of temperature inhomogeneities into anisotropies.
Lensing 
neither generates or erases power but merely redistributes it in 
angles.   The magnitude and sense of the effect is somewhat
dependent on the model for structure formation, including the assumptions
for non-linear clustering.   This has led to some seemingly inconsistent
results in the literature
(e.g.~\cite{BlaSch,ColEfs,Sas,TomWat,Linder}). 
Recently Seljak \cite{SeljakLensing} has shown that for CDM, and indeed
most realistic scenarios of structure formation, the effect is small above
the arcminute scales and above.  At arcminute scales,
it smooths out features such as the acoustic
peaks at the few percent level in power. 

\subsubsection{Galactic Foreground Contamination}

Though not a part of the cosmic microwave {\it background}, 
galactic foreground contamination contributes to anisotropies at 
microwave frequencies.  This may make the extraction of information
from the primary signal extremely difficult at small angular scales.  Typical
sources such as synchotron, brems\-strahlung and dust emission can be 
identified by their spectral signature with multifrequency experiments
(see \eg\ \cite{Brandt,Bennett92}).  Near 100 GHz, one expects that synchotron
and bremsstrahlung will have already died away, whereas dust has not
yet reached its peak.  
However, a 
sensitivity in the $\Delta T/T \simlt 10^{-6}$ range will be necessary
to extract some of the information encoded in the primary signal
(see Appendix \ref{sec-5smallstuff}).
It may be however that even with full 
sky coverage from the next generation of satellite experiments only 
a small fraction containing the clean patches will be useful for observing
the structure of
primary anisotropies at this level.  
Clearly further work is needed on this important
subject, but it may be that we will only know the full story once the
next generation of CMB satellites have flown and taken data.

\section{Final Thoughts}

\begin{quote}
\footnotesize\it
What goes on being hateful about analysis is that it implies that the analyzed 
is a completed set.   The reason why completion goes on being hateful is that
it implies everything can be a completed set.
\vskip 0.1truecm
\centerline{--Chuang-tzu, 23}
\end{quote} 

We have endeavored to cover all of the major sources of primary and
secondary anisotropies in the CMB known to date.  
Still, there is no doubt that
nature will continue to surprise us with the unexpected.  In the end,
despite the theory developed here, the ultimate answers can only 
be obtained through observations.  Currently, several groups are
testing long duration balloon flights in the hope that 
they will be able to 
measure anisotropies across a substantial fraction of the sky at
degree resolutions.  The experimental challenge to eliminate atmospheric
noise and sidelobe contamination is formidable (see \eg\ \cite{Wilkinson}).
Space based missions, for which these problems can be avoided,
are now under consideration.  A mission of this kind can essentially obtain
cosmic variance limited measurements of the anisotropy spectrum down to
ten arcminutes with a wide frequency coverage.  
With such data, one can realistically hope to measure all the classical
cosmological parameters, the curvature $K$, the matter content $\Omega_0 h^2$,
the cosmological constant $\Omega_\Lambda$, the baryon content $\Omega_b h^2$ 
and possibly even the 
gravitational wave background and neutrino
mass (see Appendix \ref{ss-5gravitywave} and \ref{ss-5massive}). 
The frequency coverage could allow measurements of the thermal SZ effect
in a large number of clusters and yield a calibration of the
distance scale and so measure the expansion rate $h$ itself 
(see \eg\ \cite{BH}).   Combined with
large scale structure measurements, the anisotropy data would provide
important information on the model for structure formation as well
as consistency tests for the gravitational instability 
scenario itself.
Perhaps even more exciting is the chance that new phenomena, either
cosmological or astrophysical,  will
be detected with all 
sky maps in the new frequency bands.   Until such a mission flies, we can
only guess at the possibilities.  
\vfill
\begin{quote}
\footnotesize\it
Rather than go toward what suits you, laugh.
Rather than acknowledge it with your laughter, shove it from you.
Shove it from you and leave the transformation behind, then you will
enter the unity of the featureless sky.
\vskip 0.1truecm
\centerline{--Chuang-tzu, 6}
\end{quote}

%% file: chap6/pibrms.tex
\begin{figure}[t]
\centerline{ \hskip -0.5truecm
\epsfxsize=3.5in \epsfbox{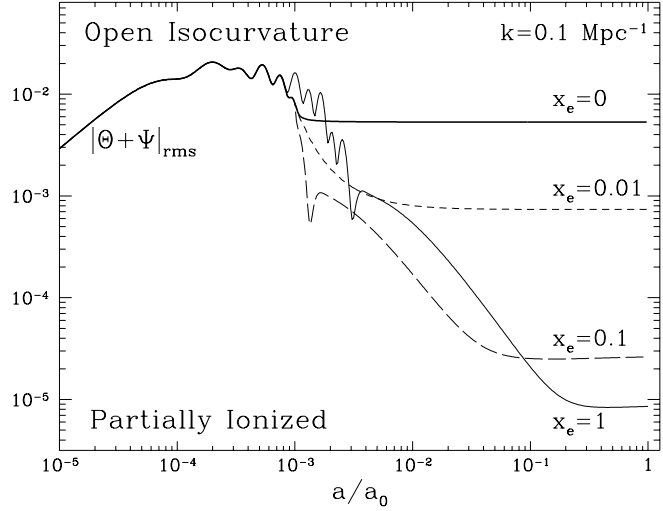}}
 \vskip -0.5truecm
\caption{Reionization Damping Evolution}
\mycaption{If the universe stays transparent after 
standard recombination at $z_* \approx a_0/a_* \approx 1000$, the
acoustic oscillations in the photon-baryon fluid will be frozen into
the rms temperature fluctuation.  For partially reionized models, the
diffusion length continues to grow and sharply damps the acoustic 
contributions.  Fluctuations are regenerated by scattering induced
Doppler shifts from the electrons. The model here is an open 
baryon isocurvature model with $\Omega_0=\Omega_b=0.2, h=0.5$.
}
\label{fig:6pibrms}
\end{figure}

%% file: chap6/cltau.tex
\begin{figure}[t]
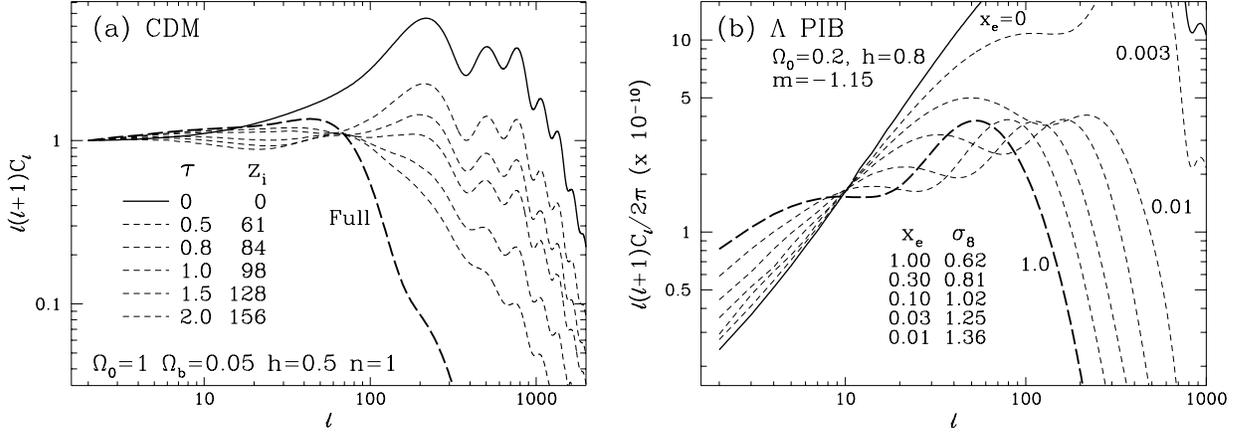

\centerline{ \hskip -0.5 truecm
\epsfxsize=3.0in \epsfbox{chap6/cltaua.epsf} \hskip 0.5 truecm
\epsfxsize=3.0in \epsfbox{chap6/cltaub.epsf}}
\vskip -0.5 truecm
\caption{Reionization Damped Spectrum}
\mycaption{(a) Standard CDM.  Reionization damps anisotropy power as
$e^{-2\tau}$ under the horizon (diffusion length) at last scattering.  
The models here are fully ionized $x_e=1.0$ out to a reionization 
redshift $z_i$.  Notice that with high optical
depth, fluctuations at intermediate scales are regenerated as 
the fully ionized (long-dashed) model shows.  (b)
$\Lambda$ PIB.  PIB models have excess small scale power and require
high optical depth to damp the corresponding anisotropy.  In this
case, both reionization damping and regeneration can be quite important
and the spectrum is sensitive to the details of the ionization history
not merely the optical depth.  Models here have constant ionization from
$z_i=800$ and are normalized to the COBE detection \cite{HBS}.  Note that
the amplitude of matter fluctuations $\sigma_8$ is also highly sensitive
to the ionization.  }
\label{fig:6cltau}
\end{figure}

%% file: chap6/trans.tex
\begin{figure}[t]
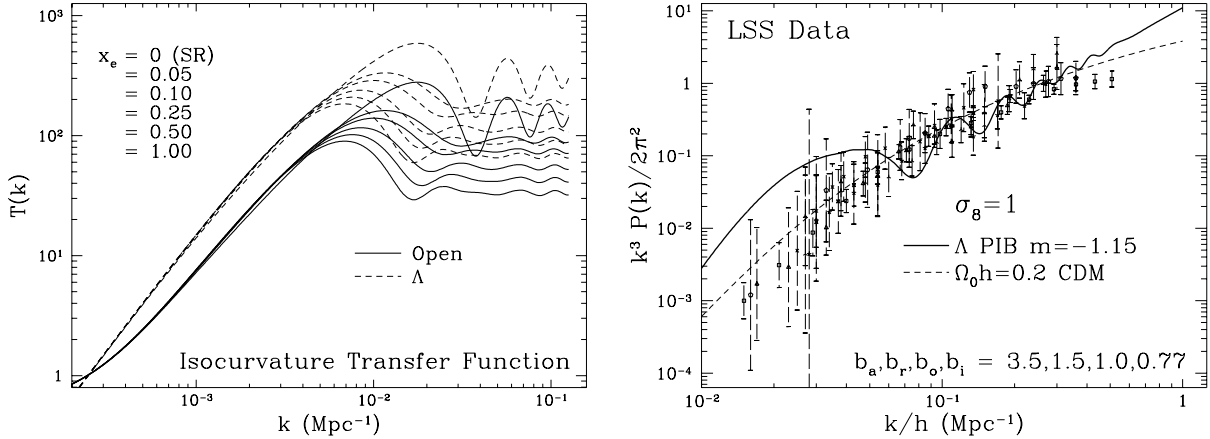

\centerline{ \hskip -0.5truecm
\epsfxsize=3.0in \epsfbox{chap6/trans.epsf} \hskip 0.5truecm
\epsfxsize=3.0in \epsfbox{chap6/transb.epsf}}
 \vskip -0.5truecm
\caption{PIB Matter Power Spectrum}
\mycaption{(a) Transfer function.
Baryon perturbations $\Delta_b(\eta_0,k) = T(k)S(0,k)$
have a prominent peak at the maximal Jeans scale. 
Silk damping of acoustic oscillations increases
with ionization leaving a constant small scale tail.  The amplitude of the
tail depends on the amount of time between the drag epoch and the present
for fluctuations to grow as $D(a)$.   The model here is
$\Omega_0=0.2$, $h=0.5$. (b) Large scale structure
data with optical bias unity and relative biases chosen to 
best reconstruct the power spectrum 
\cite{Peacock} (see also Appendix \ref{sec-Bnormalization}) require
that the isocurvature index $m \approx -1$. 
The model plotted is a $m=-1.15$ $\Lambda$ PIB model with $\Omega_0=0.2$,
$h=0.8$ and $x_e=0.1$ chosen to match $\sigma_8=1$ with a {\it COBE} 
normalization and not violate CMB constraints.  
A low $\Omega_0 h$ $\sigma_8$ normalized CDM
model is shown for comparison. 
}
\label{fig:6trans}
\end{figure}

%% file: chap6/pibcobe.tex
\begin{figure}[t]
\vphantom{MARKER}
%\vskip -6truecm
\centerline{ \hskip -0.15 truecm
\epsfxsize=3.0in \epsfbox{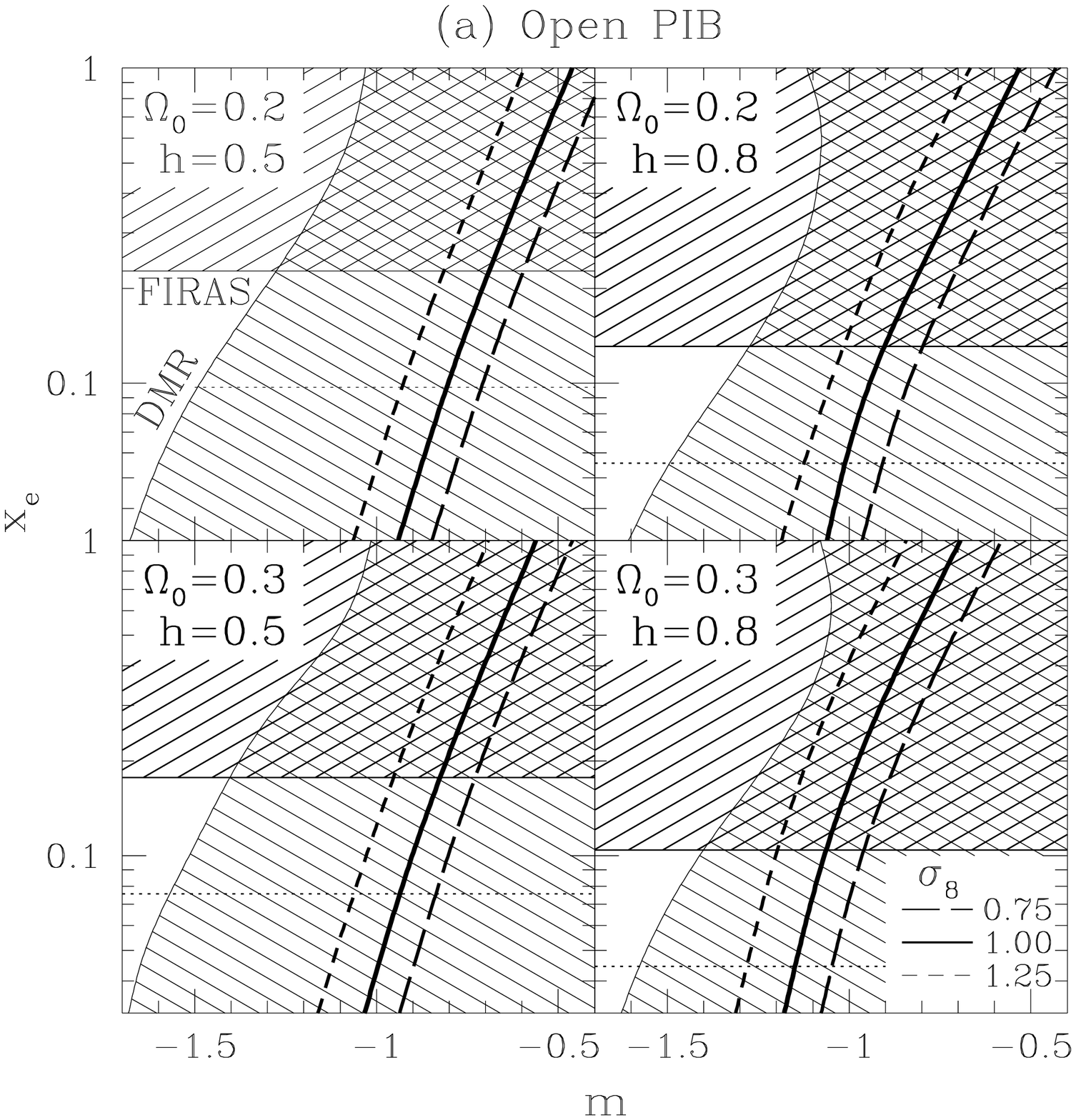} \hskip 0.15 truecm
\epsfxsize=3.0in \epsfbox{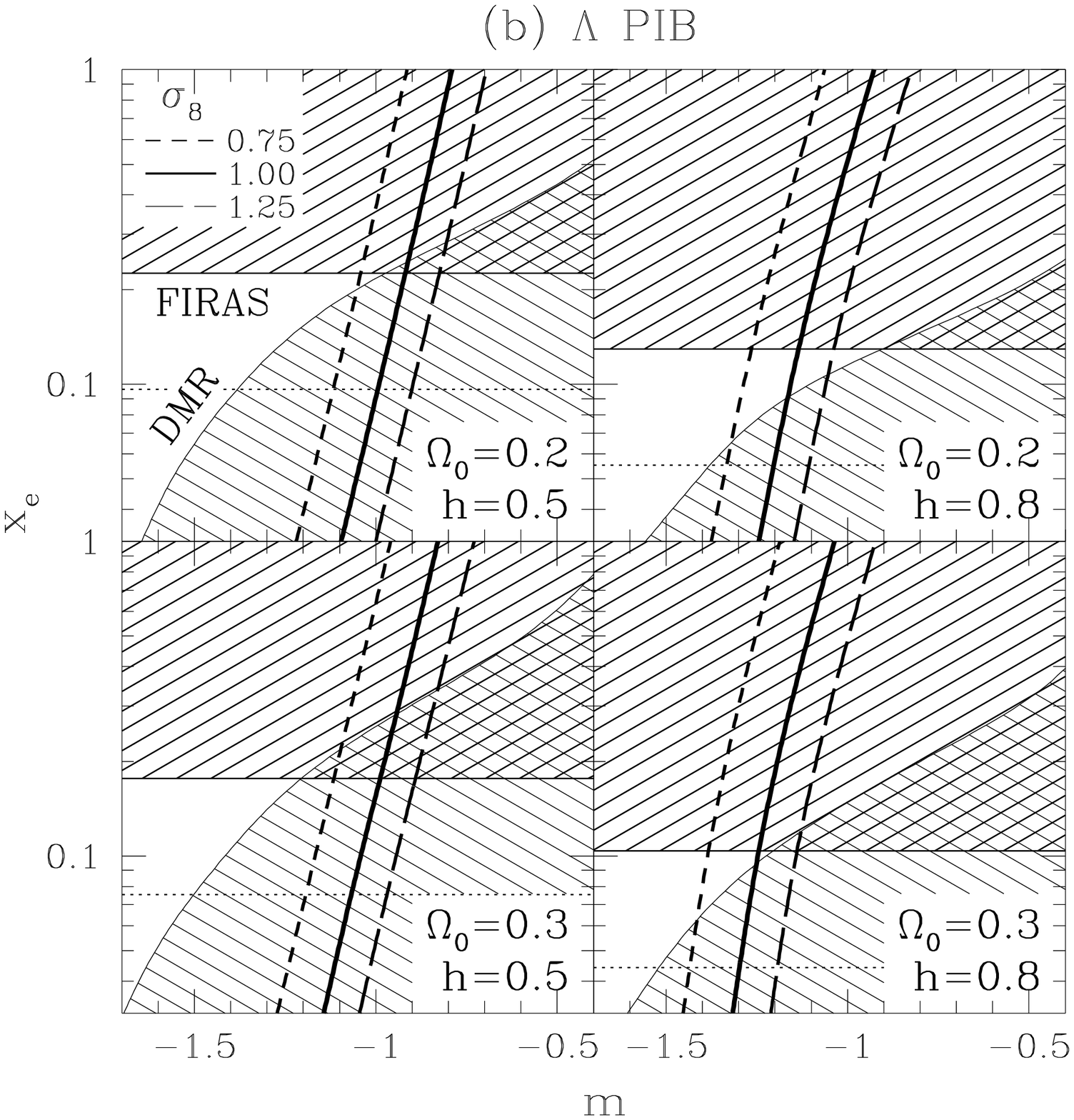}}
\caption{Constraints on PIB Models}
\mycaption{The \COBE\ DMR slope imposes
a upper (95\% confidence) limit on the initial spectral index $m$. 
The \COBE\ FIRAS constraint on spectral
distortions through the Compton-$y$ parameter sets an upper limit on the
ionization fraction.  Here a conservative $T_e=5000$K is assumed with
the more realistic $T_e=10000$K in dotted lines.
The \COBE\ DMR normalization also sets the level of matter fluctuations
at the 8 $h^{-1}$Mpc scale $\sigma_8$.  (a) No open model
simultaneously
satisfies all the observational constraints.
(b) For $\Lambda$ models, a small region of parameter space
is open for high $h$, low $\Omega_0$ models.
}
\label{fig:6pibcobe}
\end{figure}

%% file: chap6/first.tex
\begin{figure}[t]
\centerline{ \hskip -0.5truecm
\epsfxsize=3.5in \epsfbox{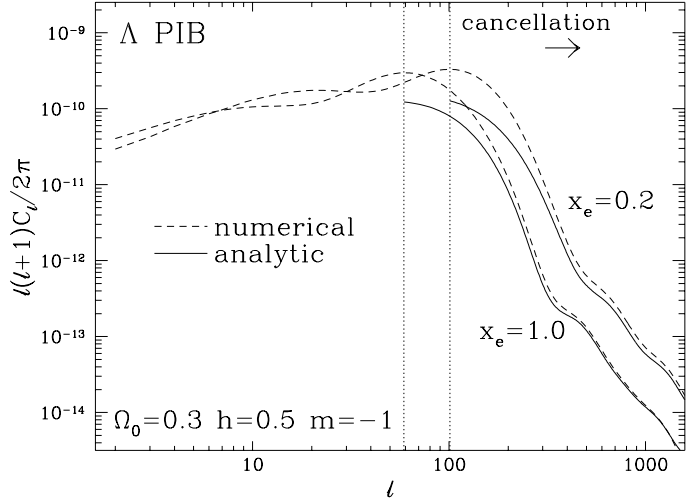}}
 \vskip -0.5truecm
\caption{First Order Doppler Effect}
\mycaption{Analytic calculations in the small scale cancellation regime
show that first order anisotropies are dominated by the cancelled 
Doppler effect.  Cancellation depends on the horizon scale
at last scattering which increases with the ionization fraction $x_e$.
As $x_e$ or $\Omega_b$ is lowered, last scattering approaches the 
drag epoch where the analytic estimate breaks down. 
}
\label{fig:6first}
\end{figure}

%% file: chap6/cancel.tex
\begin{figure}[t]
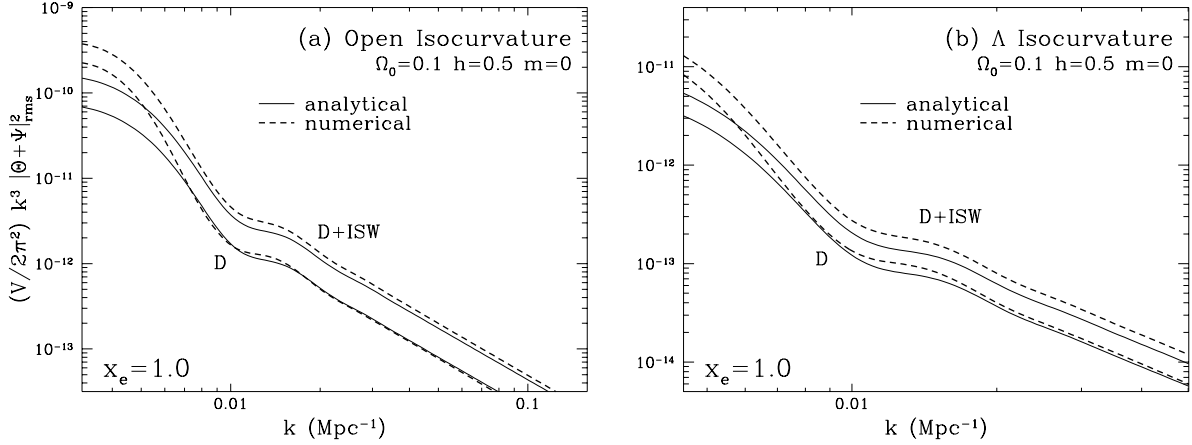

\centerline{ \hskip -0.25 truecm
\epsfxsize=3.0in \epsfbox{chap6/cancela.epsf} \hskip 0.25 truecm
\epsfxsize=3.0in \epsfbox{chap6/cancelb.epsf}}
\vskip -0.5 truecm
\caption{Cancellation Damping}
\mycaption{If the wavelength is much smaller than the thickness of the
surface upon which the anisotropy source lies, cancellation of contributions
as the photon streams over many wavelengths of the perturbation will 
damp the effect.  For the spatial power spectrum this implies mild 
cancellation of the late ISW effect and severe cancellation of the 
Doppler effect.  The two can be comparable at small scales.   For 
the $\Lambda$ model however the projection carries the late ISW
effect 
to larger angles where it is hidden by the Doppler effect in $C_\ell$. 
}
\label{fig:6cancel}
\end{figure}

%% file: chap6/minimal.tex
\begin{figure}[t]
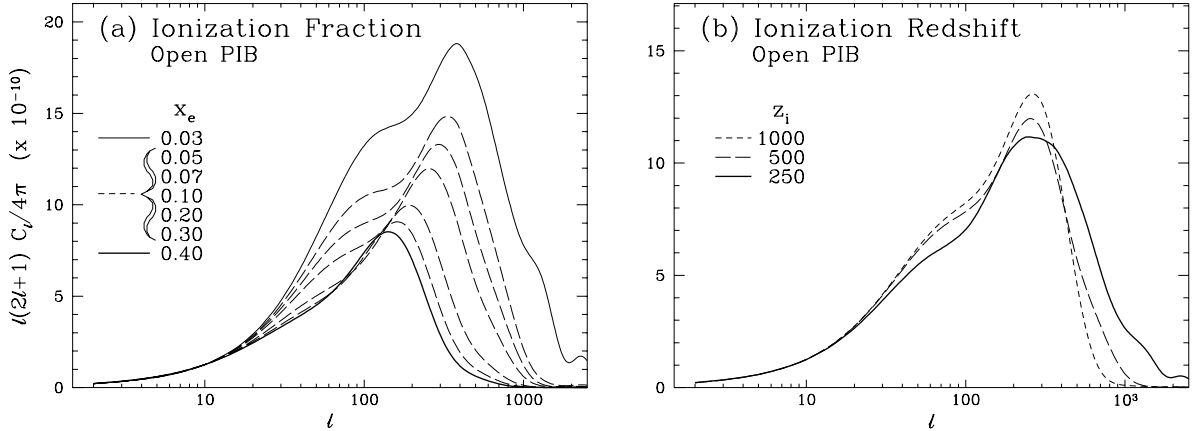

\centerline{ \hskip -0.25 truecm
\epsfxsize=3.0in \epsfbox{chap6/minimala.epsf} \hskip 0.25 truecm
\epsfxsize=3.0in \epsfbox{chap6/minimalb.epsf}}
\vskip -0.5 truecm
\caption{Minimal PIB Anisotropies}
\mycaption{Two free parameters in the standard PIB model are the ionization 
fraction $x_e$ and the ionization redshift $z_i$.  (a) The ionization 
fraction, assumed to be constant after redshift $z_i=500$,
fixes the epoch of last scattering and the amount of cancellation damping.
Aside from a small boost due to fluctuation growth, later last scattering
always leads to smaller anisotropies. (b) The ionization redshift
determines fluctuation growth before last scattering. Here $x_e=0.1$.  
Adiabatic photon growth at large scales and baryon velocity growth at small
scales yield opposite tendencies with $z_i$.  The model here is open
PIB with $\Omega_0=0.2$ $h=0.5$ and $m=-0.5$.  Ionization parameters are
chosen to avoid Compton-$y$ constraints.
}
\label{fig:6minimal}
\end{figure}

%% file: chap6/vishniac.tex
\begin{figure}[t]
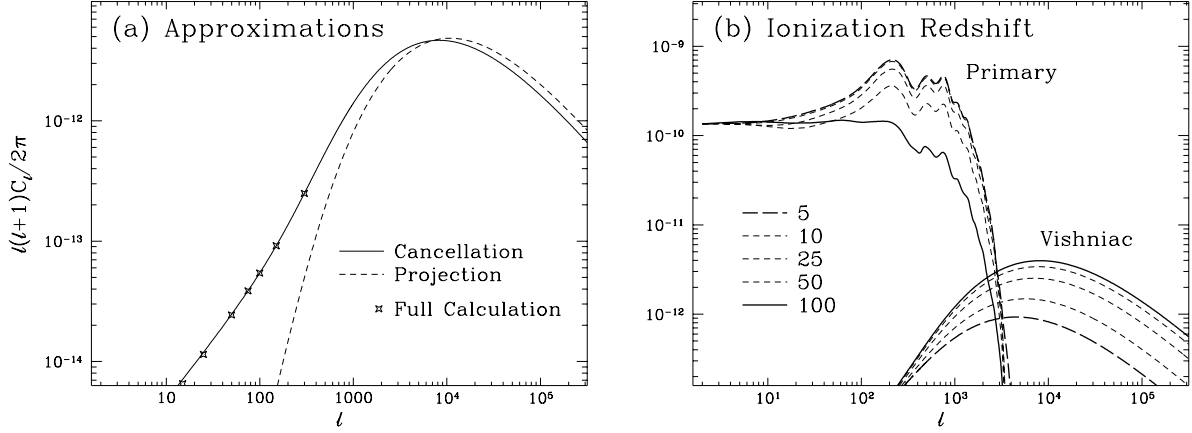

\centerline{ \hskip -0.25 truecm
\epsfxsize=3.0in \epsfbox{chap6/vishniaca.epsf} \hskip 0.25 truecm
\epsfxsize=3.0in \epsfbox{chap6/vishniacb.epsf}}
\vskip -0.5 truecm
\caption{Vishniac Effect}
\mycaption{The model is standard CDM $\Omega_0=1$, $\Omega_b=0.05$, $h=0.5$ with
a quadrupole normalization to \COBE\ of $20\mu$K.  (a) The cancellation
approximation to the Vishniac source is excellent.  Calculations in
$k$ space projected onto angles underestimates the coherence angle of
the Vishniac effect if fluctuations are all considered to come from
last scattering $\eta_{max}=\eta_*$ in equation \eqn{powerprojection}.
(b) The Vishniac effect originates mainly after last scattering.  
Therefore even if the optical depth is as low as its Gunn-Peterson
minimal value $z_i \approx 5$, the Vishniac effect contributes
a significant fraction of its total.  Both primary anisotropies
and the Vishniac effect may be present in the spectrum.  }
\label{fig:6vishniac}
\end{figure}

%% file: chap6/pibvish.tex
\begin{figure}[t]
\centerline{ \hskip -0.5truecm
\epsfxsize=3.5in \epsfbox{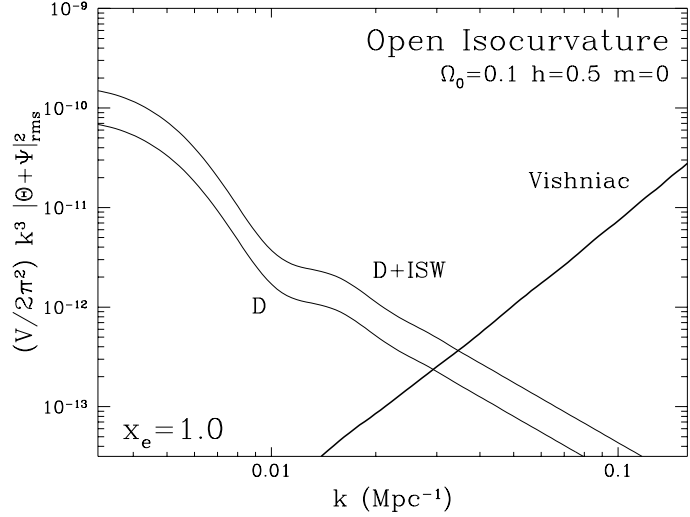}}
 \vskip -0.5truecm
\caption{PIB Vishniac Power Spectrum}
\mycaption{Analytic $k$ space power spectrum calculation of the Vishniac
effect in a PIB model.  Vishniac contributions dominate over
first order effects at small scales.  For this steeply small scale
weighted $m=0$ spectrum, high $k$ modes can contribute to lower
$\ell$ modes that one would naively think.  A full non-linear calculation
is needed to account for these high $k$ contributions.
}
\label{fig:6pibvish}
\end{figure}

%% file: chap6/quadratic.tex
\begin{figure}[t]
\centerline{ \hskip -0.5truecm
\epsfxsize=3.5in \epsfbox{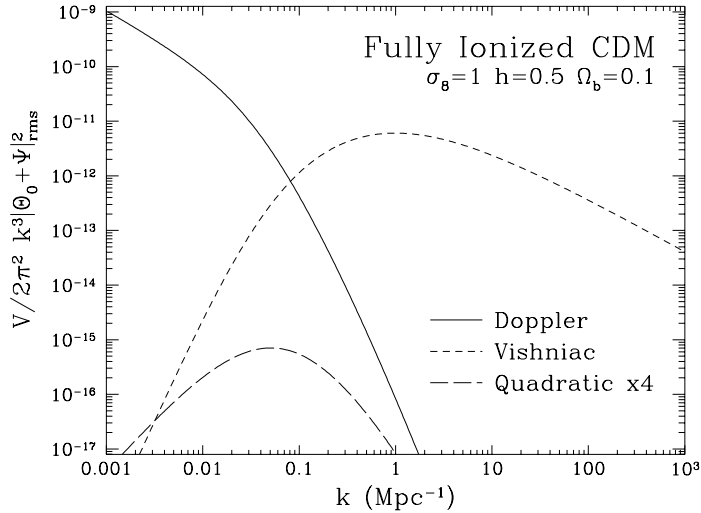}}
 \vskip -0.5truecm
\caption{Quadratic Doppler Effect}
\mycaption{Spatial power spectrum of the CMB for the first order
Doppler, Vishniac and quadratic Doppler effects in a CDM model. 
The quadratic effect is multiplied by $4$ to account for the
spectral distortion in the Rayleigh-Jeans regime but never dominates. 
}
\label{fig:6quadratic}
\end{figure}

%% file: appendixa.tex
\appendix
\chapter{Toward Higher Accuracy: A CDM Example} 
\label{sec-5CDM}

The scale invariant 
cold dark matter (CDM) model with $\Omega_0=1.0$ and $\Omega_b h^2$ 
near the nucleosynthesis value $\Omega_b h^2 \approx 0.01-0.02$ is
elegantly simple and succeeds in explaining the gross features of
both anisotropies in the CMB and large scale structure formation.  
It is therefore of value to study this model more closely. 
We will here refine our understanding of primary anisotropy formation first
to the 5\% level down to a fraction of
a degree.  It is possible and instructive to carry out this task through
analytic construction.  We then embark on the
quest of obtaining 1\% accurate results through the arcminute scale
by considering the numerical calculation of subtle effects.
This treatment should serve as an example of the types of consideration
necessary for accurate predictions in the general case.

Aside from the coupling to the baryons, photons only experience gravitational
effects from the other matter components.  
Primary anisotropy formation therefore depends sensitively on 
two quantities

\begin{enumerate}

\item The evolution of the metric perturbations.

\item The decoupling of the photons from the baryons.

\end{enumerate}

\noindent
We must therefore refine our understanding of both. As for the metric
perturbations $\Psi$ and $\Phi$, there are two
modifications we must make to the analysis of \S 
\ref{sec-4superhorizon} and \S \ref{sec-4subhorizon}.  At large scales
we must include the anisotropic stress 
contribution of the neutrinos.  Anisotropic stress serves to differentiate
the Newtonian potential $\Psi$ from the space curvature perturbation 
$\Phi$.  As we have seen in \S \ref{ch-primary}, they are both 
important in anisotropy formation.  At small scales,
we must be able to describe accurately
the pressure feedback effects from the radiation
onto the potentials. 
This in turn leads to some sensitivity to neutrino masses in the eV 
range.
 Finally, tensor metric perturbations, \ie\
gravity waves, can also produce gravitational redshifts and dilation
in the CMB.  Depending on the exact inflationary model, they can
perhaps significant at large scales but are almost
certainly small perturbations to the scalar spectrum 
near the acoustic peaks.  We shall
quantify this statement in section \ref{ss-5gravitywave}. 

We must also improve our understanding of recombination over the 
equilibrium Saha treatment presented in \S \ref{ss-5diffusion}. Last
scattering is delayed due to the high opacity to recombination photons
which keep the plasma ionized \cite{PeeblesRec,ZKS}.  This delay increases
the diffusion length and thus is responsible for further
damping of anisotropies.  Following the population of the first
excited state of hydrogen allows analytic construction of the 
anisotropies to 5\% through to the
damping scale.  Subtle effects can change the damping scale at the
several percent level.  Polarization feedback weakens photon-baryon coupling
by generating viscosity, \ie\ a quadrupole moment in the photons.
Helium ionization decreases the diffusion length before helium recombination.
It is quite possible that other subtle effects change the damping tail at
a comparable or even greater level.  We offer these two considerations
only as examples of the care that is required to obtain 1\% accurate
primary anisotropies under the damping scale. 
 
\bigskip
\goodbreak
\section{Refining the Gravitational Potentials}

\subsection{Neutrino Anisotropic Stress}
\label{ss-5largepotential}
\def\nufac{f_\nu}
The solution for the gravitational potentials given
by equation \eqn{BeqnUAD} must be corrected 
for the anisotropic stress $\Pi_T$.
Recall that the anisotropic stress is related to the quadrupole
moments of the radiation via equation \eqn{anisopi}, \ie\
\bel{eq:PiGN}
p_T \Pi_T = {12 \over 5}(p_\gamma \Theta_2 + p_\nu N_2).
\ee 
Due to the isotropizing effects of scattering,
the anisotropic stress of the photons is negligibly small before
recombination.
Hence the main contribution to $\Pi_T$ comes from the neutrino
quadrupole anisotropy $N_2$.
 
We can take it into account perturbatively.
Specifically,
we use the exact zeroth order growing and decaying solutions
\eqn{BeqnUAD} to obtain the anisotropic stress.  We then take this solution
to iteratively correct for anisotropic stress in the evolution 
equation~\eqn{Total}.
If we neglect higher order multipole components,
which is reasonable for superhorizon sized modes,
the second moment of the
the Boltzmann equation \eqn{Hierarchy} for
the neutrinos becomes
\bel{eq:AeqnNder}
\dot N_2 = {2 \over 3}k N_1 \approx {2 \over 3}k V_T \, ,
\ee
where recall 
from \S \ref{ss-4totalmattereqn}
that all fluid velocities are equal above 
the horizon
\ie\ $N_1 \equiv V_\nu \approx V_T$.

The exact zeroth order solution for $V_T$ 
is found by substituting the growing mode solution equation \eqn{BeqnUAD}
into the continuity equation \eqn{BeqnContPois}.
If the zeroth order solution is denoted $\Delta_T = C_G U_G$, then
the solution to equation~\eqn{AeqnNder} is 
\bel{eq:AeqnNint}
\bar N_2(a)/C_G \approx 2 \int_0^a {da' \over a'} {1 \over 3 a' + 4}
        \left( U_G  - (a' +1 )a' {d U_G \over d a'} \right),
\ee
where recall that $3w_T = 1/(1+a)$ with $a$ normalized at equality.
The overbar
represents the superhorizon solution since  
pressure growth suppression inside the horizon must be taken into account
(see \S 
\ref{ss-5smallpotential}).
Although it is possible to analytically
integrate equation \eqn{AeqnNint},
the expression is cumbersome.  Instead, we can employ
an approximate solution which is exact in the limit
$a \ll 1$ and $a \gg 1$,
\bel{eq:AeqnNfit}
\bar N_2(a)/C_G =  -{1 \over 10}{20a + 19 \over 3 a +4}U_G
        -{8 \over 3} { a \over 3 a +4} 
        + {8 \over 9}\ln\left({3a +4 \over 4}\right) \, .
\ee
We have checked that this approximation works quite well by comparing
it to equation~\eqn{AeqnNint} and
the full numerical solution.
 
Next, we employ the above solution for $\bar N_2$ in
equations \eqn{Total}.
These two first order equations may be rewritten as
one second  order equation for $\Delta_T$.
The particular solution including the source terms $\Pi_T$ and
$\dot \Pi_T$ can
be obtained from the homogeneous solutions $U_G$ and $U_D$ by
Green's method,
%\eqalign{
%\bar \Delta_T(a) = & \left(1 + {2 \over 5}\nufac\right)AU_G(a) \cr
%       & +{16\over 25}\nufac
%       \int_0^a {da' \over a'^2}{(a'+1)^{5/2} \over a' +4/3}
%       [U_G(a)U_D(a')-U_D(a)U_G(a')] G(a')\, , \cr
%}
\bel{eq:AeqnTotalit}
\bar \Delta_T(a)/C_G  =  \left(1 + {2 \over 5}\nufac\right)U_G(a)
        + {2 \over 5}\nufac [I_1(a)U_G(a) + I_2(a)U_D(a)] \, ,
\ee
where  $I_1(a) = \int_0^a da' F_\nu(a')U_D(a')$, $I_2(a) = \int_0^a da'
F_\nu(a')U_G(a')$,
\beal{eq:AeqnFa}
F_\nu(a) \eal {24 \over 5}{ (a+1)^{5/2} \over a^2(3a+4)}
\bigg\{ {2a \over 3a+4}{d \over da}U_G(a) 
        -{2 \over (3a+4)(a+1)}U_G(a) \\
         && \qquad
	 +\left[ {1 \over (a+1)^2} -{2 \over a+1} + {12 \over 3a+4}\right]
        \bar N_2(a)/C_G \bigg\} \nonumber\, , 
\eea
and recall $f_\nu$ is the ratio of neutrino to total radiation
density $f_\nu \equiv \rho_\nu /(\rho_\nu + \rho_\gamma)$.
If we assume three massless neutrinos and the standard thermal
history,
$\rho_\nu/\rho_\gamma = 3 (7/4) (4/11)^{4/3} / 2 =0.68$,
\ie\ $f_\nu = 0.405$.
The first term in equation \eqn{AeqnTotalit} comes from
the initial conditions for $\Delta_T$ which can be iteratively
established
by employing equation \eqn{AeqnNder} in \eqn{Total}.
All terms which are proportional to $f_\nu$ in the equation
\eqn{AeqnTotalit} come
from equation~\eqn{PiGN} since the anisotropic stress
$\Pi_T \approx (12/5)f_\nu N_2$.
The asymptotic behavior of equation \eqn{AeqnTotalit} is
\bel{eq:Totallim}
\bar \Delta_T(a) \rightarrow \cases{ 
	\left(1 + {2\over 5}\nufac \right)C_G  U_G(a)
%       = {10 \over 9} \left(1 + {2\over 5}\nufac \right)a^2 A
        & $(a \ll 1)$\cr
	\left(1 + {4\over 15}\nufac \right) C_G U_G(a).
%       = \left(1 + {2\over 5}\nufac (1-0.333) \right)a A
        & $(a \gg 1)$ \cr} 
\ee
Here we have used the fact that if $a \gg 1$, the decaying term $I_2 U_D$
may be ignored and $I_1 \rightarrow -{1 \over 3}$ approximately.
 
Therefore we may obtain a simple approximate expression for
the large scale density fluctuations,
\bel{eq:AeqnTotalfit}
\bar \Delta_T(a) \approx \left[1 + {2\over 5}\nufac \left(1-{1 \over 3}{a\over a+1}
\right)
         \right]C_G U_G(a) \, .
\ee
Again we have checked that this approximation works reasonably well by
comparing it to numerical calculations.
The potentials $\bar \Phi$ and $\bar \Psi$ are therefore written as
\beal{eq:AeqnPhiPsi}
\bar \Phi(a) \eal
 {3 \over 4}\left({k_{eq} \over k} \right)^2
 {a +1 \over a^2}\bar \Delta_T(a) \,, \nonumber\\
\bar \Psi(a) \eal
 -{3 \over 4}\left({k_{eq} \over k} \right)^2
 {a +1 \over a^2}\left(\bar \Delta_T(a) +{8\over 5}\nufac
        {\bar N_2(a) \over a +1}\right) \,  ,  
\eea
where recall $k_{eq} = \sqrt{2}(\Omega_0 H_0^2 a_0)^{1/2}$
is the scale that passes the horizon at matter-radiation equality.
By using the asymptotic form of $\bar \Delta_T$ and $\bar N_2$,
we easily obtain
the corresponding relation between $\bar \Phi$ and $\bar \Psi$,
\bel{eq:AeqnPhiAsympt}
\bar \Phi(a) = \cases{ - \bar\Psi(a) 
\left(1 + {2\over 5}\nufac \right)
& $(a \ll 1)$ \cr
- \bar \Psi(a). & $(a \gg 1)$}
\ee
Also of interest are the ratios of initial to final values of 
the gravitational potentials:
\beal{eq:InitialFinal}
\bar \Phi(a_0) \eal -\bar \Psi(a_0) = {9 \over 10} \left(1 + {4 \over 15}
	f_\nu \right) \left( 1 + {2 \over 5} f_\nu \right)^{-1} \bar\Phi(0), 
	\nonumber\\
\bar \Psi(a_0) \eal {9 \over 10} \left( 1 + {4 \over 15} f_\nu \right) 
	\bar\Psi(0) .
\eea
Thus
we see that the correction for anisotropic stress makes a 10\% difference
in $\bar \Psi$ during radiation domination. 
If recombination occurs near equality, this results in a small
correction to the standard
Sachs-Wolfe formula due to anisotropic stress.
 
The initial conditions for the perturbations may now be expressed
in terms of $\bar \Phi(0)$,
\beal{eq:AeqnInitial}
\Psi(0) \equiv \bar \Psi(0) \eal 
	-\left( 1+{2 \over 5}f_\nu \right)^{-1} \bar \Phi(0)\, , \nonumber\\
\Theta(0) \equiv \bar \Theta(0) \eal {1 \over 2} \left( 1 + {2 \over 5}
	f_\nu \right)^{-1} \bar \Phi(0)\, . 
\eea
Note that since all modes are superhorizon sized at the initial
epoch, the overbar is superfluous.
Moreover, even in the initial conditions, the anisotropic
stress represents a small but important correction to the $\Pi_T = 0$
solutions of \S \ref{sec-4superhorizon}, 
$\bar \Phi(0)=-\bar \Psi(0)=2\bar\Theta_0(0)$.
Finally, we can relate these quantities to the initial
power spectrum,
\bel{eq:AeqnNormalization}
k^3|\Phi(0,k)|^2 \equiv k^3|\bar \Phi(0,k)|^2 = \left[{5 \over 6}
\left( 1 + {2 \over 5} f_\nu \right) \right]^2
{\left(k_{eq} \over k\right) }^4 k^3 C_G^2(k) = Bk^{n-1} \, ,
\ee
where we have restored the implicit $k$ index. 
Note that $C_G$ is the normalization of the density fluctuations at
equality.  It is related to the matter power spectrum 
today $|\Delta(\eta_0,k)|^2 = Ak^n$ by
\beal{eq:NormRel}  
Ak^n \eapp \left[\left( 1 + {4 \over 15}f_\nu\right) C_G a_0/a_{eq} \right]^2.
\nonumber\\
\eapp \left( 1 + {4 \over 15}f_\nu \right)^2
      \left( 1 + {2 \over 5}f_\nu \right)^{-2}
      {36 \over 25} k_{eq}^{-4} (a_0/a_{eq})^2  Bk^n  \nonumber\\
\eapp \left( 1 + {4 \over 15}f_\nu \right)^2
      \left( 1 + {2 \over 5}f_\nu \right)^{-2}
      {9 \over 25} (\Omega_0 H_0^2)^{-2} Bk^n  ,
\eea
[{\it c.f.} equation \eqn{ABnormalization}]
where we have used equation \eqn{AeqnTotalfit}.
\bigskip
\goodbreak
\subsection{Small Scale Radiation Feedback}
\label{ss-5smallpotential}

%%%%%%%%%%%%%%%%%%%%%%%%%%%%%%%%%%%%%%%%%%%%%%%%
\smallskip
\input chap5/potential.tex

Next we need to obtain  solutions
of $\Psi$ and $\Phi$ in the small scale limit where pressure
cannot be neglected. Qualitatively speaking, we know that the
potentials decay inside the sound horizon in the radiation-dominated
epoch since pressure prevents $\Delta_T$ from growing.
However in general, it is impossible to obtain the exact solution
valid through matter-radiation equality
even if we neglect
the anisotropic stress term.
Only the asymptotic behavior in certain limits has
been found \cite{KS87}.
For the CDM scenario, it is well known that the {\it final} value of the
potential at small scales is obtained from the superhorizon solution
\eqn{AeqnPhiPsi} by the transfer function
 $\Phi(a_0)=-\Psi(a_0) =
T(k)\bar \Phi(a_0)$, where
\bel{eq:AeqnTrans}
T(k)={\ln(1+2.34q) \over 2.34q}[1+ 3.89q +(14.1q)^2+(5.46q)^3+(6.71q)^4]^{-1/4}
\, ,
\ee
with $q \equiv k/[\Omega_0 h^2 \exp(-2\Omega_b)]$ in units of
Mpc$^{-1}$ \cite{Peacock,BBKS}.
Note that $q \propto k/k_{eq}$ approximately, reflecting the fact that only
modes that cross the Jeans length before equality are suppressed.
This  implies that the potentials are larger in amplitude if equality occurs 
later,
\ie\ for high $\Omega_0 h^2$ models.
Equation \eqn{AeqnTrans} therefore
empirically accounts
for the lack of growth in the radiation-dominated era.
Now let us consider the time evolution of the potential.
We know that in the matter-dominated epoch the potentials are constant
on all scales.  Therefore,
we smoothly join the superhorizon scale solutions of equation \eqn{AeqnPhiPsi}
with a constant matter-dominated tail whose relative amplitude is given
by the transfer function.
Since the Jeans crossing epoch is approximately the same as the
horizon crossing time in radiation-dominated era,
we can take $(k/Ha) \sim a k/k_{eq} \sim 1 $ as the matching
epoch,
\beal{eq:AeqnFitt}
\Phi(a) \eal \bar \Phi(a) \left\{ [1-T(k)]
        \exp[-\alpha_1\left(a k/k_{\rm eq}\right)^\beta] +T(k)
\right\} \, ,\nonumber\\
\Psi(a) \eal \bar \Psi(a) \left\{ [1-T(k)]
        \exp[-\alpha_2\left(a k/k_{\rm eq}\right)^\beta]
        +T(k) \right\}
\, ,\
\eea
where $\alpha_1$, $\alpha_2$ and $\beta$ are fitting parameters.
We also need a small correction
to take into account the free streaming
oscillations of the neutrino quadrupole inside the Jeans scale.
A very simple approximation can be obtained by making
the replacement
$\bar N_2(a) \rightarrow  \bar N_2(a) \cos[0.5k/(Ha)]$ in 
equation~\eqn{AeqnPhiPsi}
for $\Psi(a)$.
 Here the factor $0.5$ is a best fit, and the Hubble parameter
$H(a)=(\dot a/a)(a_0/a)$.
Since it is a higher order correction, this 
crude approximation is sufficient for our
purposes.
Comparing this functional form \eqn{AeqnFitt} with
numerical results, we obtain a good fit
for $\alpha_1 = 0.11$, $\alpha_2 = 0.097$ and
$\beta = 1.6$.
%?? These fitting parameters are obtained by  employing
%simple approximate equations
%\Adis\AeqnTotalfit\ and \Adis\AeqnNfit\ for  the evolution of
%the total density perturbations  and the neutrino quadrupole, respectively.
In order to calculate the early ISW effect,
we take the direct derivative of
equations \eqn{AeqnFitt}.  In Fig.~\ref{fig:5potential}, we compare these
analytic approximations to the numerical results and find good agreement.

\section{Analytic Construction to 5\% Accuracy}

\subsection{Explicit Tight Coupling Solutions}

\label{ss-5explicit}

The first step in obtaining the explicit analytic solution for the anisotropy
is to calculate the photon fluctuation spectrum at last scattering.  
We have already seen in \S \ref{ss-4analyticacoustic} how this may
be obtained under the tight coupling approximation
once the potential evolution is known.
For calculational purposes, it is convenient to express the acoustic
solution of equation \eqn{PartSoln} in
a more explicit but cumbersome form.  
One advantage of the analytic
tight coupling
solutions is they do not require the use of time derivatives of the potentials
despite the appearance of equation~\eqn{PartSoln}.  Thus accuracy is
not compromised by our lack of a detailed description for $\dot \Phi$
and $\dot \Psi$. Integrating
equation~\eqn{PartSoln} by parts twice, we obtain
\bel{eq:DeqnMonopoleExplicit}
(1+R)^{1/4}[ \hat\Theta_0(\eta)+\Phi(\eta)] =
       \left[
         \cos kr_s(\eta) + J(0)
         \sin kr_s(\eta)
       \right] [\Theta_0(0) +\Phi(0)]
     +  I(\eta) \, ,
\ee
where the overhat denotes the undamped solution,
\bel{eq:DeqnJ}
J(\eta) \equiv - (1+R)^{3/4} {\sqrt{3} \over k}{ d \over d\eta}(1+R)^{-1/4} =
{\sqrt{3} \over 4k} {\dot R \over \sqrt{1+R}},
\ee
and
\bel{eq:DeqnI}
I(\eta) ={k \over\sqrt{3}}
\int_0^{\eta} d\eta' \Phi(\eta') G(\eta') \sin[k r_s(\eta)
- k r_s(\eta')] \, ,
\ee
with
\bel{eq:DeqnG}
G(\eta) =
        (1+R)^{-1/4}
          \left[1-(1+R){\Psi \over \Phi }
        +
          {3 \over 4k^2} \ddot R - J^2
          \right].
\ee
Here we have employed the identity $ \dot \Theta_0(0) = -\dot\Phi(0)$.
Since the ISW effect predicts constant $\Theta_0 +\Phi$ at superhorizon
scales, we
have written these expressions in terms of that quantity.
 
The dipole solution
$\hat \Theta_1$ can be similarly obtained from the photon
continuity equation $k\Theta_1 = -3(\dot \Theta_0 + \dot \Phi)$,
\beal{eq:DeqnDipoleExplicit}
(1+R)^{3/4}{\hat \Theta_1(\eta) \over \sqrt{3}} \eal
         [1+J(\eta)J(0)][\Theta_0(0) + \Phi(0)]
         \sin k r_s(\eta) \nonumber\\
&& + [J(\eta)-J(0)][\Theta_0(0) +\Phi(0)]
         \cos k r_s(\eta) \nonumber\\
&& +J(\eta)I(\eta) - {k \over \sqrt{3}} \int_0^{\eta} d\eta
         \Phi(\eta')G(\eta')
        \cos[ k r_s(\eta)- k r_s(\eta')] \, , 
\eea
where we have used the relation $\dot r_s = c_s =
 (1/\sqrt{3})(1+R)^{-1/2}$.
Notice that we do not need
$\dot \Phi$ even in the boundary terms in either
equation~\eqn{DeqnMonopoleExplicit} and \eqn{DeqnDipoleExplicit}.

At large scales, $k < 0.08 h^3$ Mpc$^{-1}$ 
this WKB solution fails because the oscillation rate
becomes comparable to rate at which the sound speed is changing (see
\S \ref{ss-4analyticacoustic}).
On the other hand, we know the large scale behavior is given by the
dilation effect $\Theta(\eta) = \Theta(0) - \Phi(\eta) + \Phi(0)$.  
Comparison with 
\eqn{DeqnMonopoleExplicit} suggests that an approximate matching onto
large scales can be obtained 
by dropping the explicit $R$ dependence,
\bel{eq:DeqnMonopoleLS}
[\hat \Theta_0(\eta)+\Phi(\eta)] = [\Theta_0(0)+\Phi(0)] \cos k r_s(\eta)
     + {k \over \sqrt{3}} \int_0^\eta d\eta' [\Phi(\eta')-\Psi(\eta')]
\sin[ k r_s(\eta)- k r_s(\eta')] \, .
\ee
Here we take the true $R \ne 0$ sound horizon $r_s$ in order to match
more smoothly onto the small scale solution. In the CDM model, the error
this causes at large scales is minimal.  The continuity equation
now implies
\bel{eq:DeqnDipoleLS}
(1+R)^{1/2} {\hat \Theta_1(\eta) \over \sqrt{3}}
 = [\Theta_0(0)+\Phi(0)]\sin k r_s(\eta)
- {k \over \sqrt{3}} \int_0^\eta d\eta' [\Phi(\eta')-\Psi(\eta')]
\cos[ k r_s(\eta)- k r_s(\eta')].
\ee
Finally, the following relations are useful for computation:
\bel{dq:DeqnRExplicit}
R = {1 \over 1-f_\nu} {3 \over 4}{\Omega_b \over \Omega_0} a,
\quad \dot R = \dot a R_{eq}= {k_{eq} \over \sqrt{2}}
                \sqrt{1+a} R_{eq},
\quad \ddot R = {1 \over 4} k_{eq}^2 R_{eq},
\ee
and recall
\bel{eq:DeqnPhase}
k_{eq} r_s
 =
{2 \over 3} \sqrt{6 \over R_{eq} }
\ln{
     { \sqrt{ 1 + R } + \sqrt{ R + R_{eq} } }
   \over
     { 1 + \sqrt{R_{eq}} }
    },
\ee
where $R_{eq} \equiv R(\eta_{eq})$ and we have employed the relation
$k_{eq} \eta = 2\sqrt{2} (\sqrt{1+a}-1)$.
Here $1+\rho_\nu/\rho_\gamma = (1-f_\nu)^{-1}=1.68$.
Note that $a$ is normalized at equality $a_{eq}/a_0 = a_0^{-1}
 = 2.38 \times 10^{-5} \To^4
 (\Omega_0 h^2)^{-1} (1-f_\nu)^{-1}$,
 and the scale which passes the horizon at equality
is $k_{eq} = 1.17/\eta_{eq}=
9.67 \times 10^{-2} \To^{-2} \Omega_0 h^2 (1-f_\nu)^{1/2}$
Mpc$^{-1}$. 
Evaluating these expressions at last scattering gives the solution in the
absence of diffusion damping.  To account for diffusion damping through
last scattering, one needs to know the ionization history through recombination. 

\input chap5/vis.tex

\subsection{Recombination Revisited}
\label{ss-5recombination}

\subsubsection{Atomic Considerations}

Some care must be taken in calculating the recombination history
of hydrogen.  In particular, hydrogen recombines more slowly than 
Saha prediction presented in \S \ref{ss-5diffusion}. Lyman
$\alpha$ and Lyman continuum photons from recombination to the
ground state immediately reionize another hydrogen atom leaving
no net effect.  It was realized long ago \cite{PeeblesRec,ZKS} that 
net recombination occurs through the forbidden 2-photon 
decay from the $2s$ level and by the loss of Lyman $\alpha$ photons
to the cosmological redshift.  The result is that the hydrogen
ionization fraction
\bel{eq:xH}
x_H \equiv n_e / n_H = (1-Y_p)^{-1} n_e/n_b = (1-Y_p)^{-1} x_e,
\ee
[see equation \eqn{heliumfractions}] obeys the differential equation
\cite{PeeblesRec},
\bel{eq:hydrogenrec}
{d x_H \over dt} = C_r \left[ \beta(1-x_H) -  n_H \alpha_B x_e^2 \right],
\ee
where 
\bel{eq:hydrobeta}
\beta = \left({ m_e k_B T_b \over 2\pi \hbar^2} \right)^{3/2} e^{-B_1 /k_B T_b}   
	\alpha_B
\ee
is the ionization rate out of the ground state,
with the ground state binding energy $B_1 = 13.6 $eV and
\bel{eq:alphab}
\alpha_B = 10^{-13} {a T_4^b \over 1+ cT_4^d}\, {\rm cm}^3 {\rm s}^{-1}
\ee
is the `case B' recombination rate which excludes those to the
ground state \cite{PPB}. Here the fitting constants are $a=4.309$, $b=-0.6166$,
$c=0.6703$, $d=0.5300$ with $T_4 = T_b/10^4$K.  The suppression 
factor 
\bel{eq:suppression}
C_r = {\Lambda_\alpha + \Lambda_{2s\rightarrow 1s}
	\over \Lambda_\alpha + \Lambda_{2s\rightarrow 1s} + 
	\beta e^{h\nu_\alpha/k_B T_b}},
\ee
takes into account the 2-photon decay rate 
$\Lambda_{2s \rightarrow 1s} = 8.22458$s$^{-1}$ \cite{Goldman} and the hydrogen production
rate through redshifting out of the line \cite{PeeblesRec}
\bel{eq:lambdaalpha}
\Lambda_\alpha  = {8\pi \over \lambda_\alpha^3 (1-x_H)n_H} H, \qquad
\lambda_\alpha = c/\nu_\alpha = {8\pi \hbar c \over 3B_1} = 1.216 \times 10^{-5} {\rm cm},
\ee
where recall $H$ is the Hubble parameter. 
Since helium recombination precedes hydrogen
we can assume that at the start of hydrogen recombination $x_e 
= (1-Y_p)$ or $x_H=1$.  We shall see below what effect helium recombination
has on the spectrum.  

From equation \eqn{electrontempev}, the baryon temperature evolution is
governed by
\bel{eq:electronTcoupling}
{d T_b \over dt} =
- {1 \over t_{cool}} (T_b - T) - 2{da \over dt}{1 \over a} T_b,
\ee
with 
\bel{eq:Tegagain}
t_{cool} = 7.66 \times 10^{19} 
	{ (1+x_e)/2-(3+2x_e)Y_p/8 \over 1-Y_p/2} x_e^{-1} \To^{-4}
(1+z)^{-4} \sec.
\ee
Since this short time scale implies that the electron temperature
tracks the photon temperature until late redshifts and low ionization,
we can determine its evolution away from the photon temperature 
iteratively by employing the $T_b=T$ solution for $x_e$.  The
two temperatures only start diverging at $z \simlt 100$ and thus
is irrelevant for CMB anisotropies \cite{PeeblesRec,HSSW}.  
We can therefore replace the baryon temperature with $T_b = T_0(1+z)$.   

\subsubsection{Ionization Fitting Formulae}

It is also useful to have fitting formula to the solutions of
equation \eqn{hydrogenrec}.
The total optical depth from the present to the critical recombination epoch $80
0 < z < 1200$
can be approximated as
\bel{eq:CeqnTau}
\tau(z,0) \approx \Omega_b^{c_1} \left({ z \over 1000} \right)^{c_2},
\ee
where $c_1=0.43$ and $c_2=16+1.8\ln\Omega_b$.  Since the range of reasonable
values for $h$ is limited to $0.5 \simlt h \simlt 0.8$, we have
ignored the small $h$ dependence.
For definiteness, we take last scattering to occur at $z_*$ where
the optical depth $\tau(z_*,0)=1$.
It immediately follows from \eqn{CeqnTau}
that this occurs at\footnote{A more general expression including
variations in $\Omega_0 h^2$ is given in \cite{HSsmall}}
\bel{eq:CeqnLS}
{z_* \over 1000} \approx \Omega_b^{-c_1/c_2} =
 \Omega_b^{-0.027/(1+0.11\ln\Omega_b)},
\ee
which is weakly dependent on $\Omega_b$.  The differential optical
depth $\dot \tau$ then becomes
\bel{eq:CeqnTaudot}
\dot\tau(z) = {c_2 \over 1000} \Omega_b^{c_1} \left( {z\over 1000} \right)^{c_2-
1}
{\dot a \over a} (1+z)\, ,
\ee
where $\dot \tau$ is by definition positive since $\dot \tau
\equiv {d[\tau(\eta',\eta)]/d\eta}$.
Finally, the ionization fraction is given by
$x_e(z) = \dot\tau a_0/n_e\sigma_T a$, where
\bel{eq:CeqnXe}
{(n_e \sigma_T a/a_0)}^{-1} = 4.3 \times 10^4
(1-Y_p/2)^{-1} (\Omega_b h^2)^{-1} (1+z)^{-2}  {\rm Mpc}.
\ee
Of course,
where the formula \eqn{CeqnTaudot} implies $x_e > 1$, we set
$x_e=1$, \ie\ $\dot \tau = n_e \sigma_T a/a_0$.  Or slightly better,
impose two step functions: from $x_e=1$ to $1-Y_p/2$ at $z=6000$ and
$1-Y_p$ at $z=2500$ to account for helium recombination.  To 
the level that we expect the analytic formulae to work, these corrections
are insignificant.
In Fig.~\ref{fig:5vis}, 
we show the numerical values for the visibility function
in redshift space
$-(d\tau/dz) e^{-\tau}$ compared with these analytic fits.
 
\input chap5/separation.tex

\subsection{Analytic Results}
\label{ss-5analytic}

The decrease in ionization fraction implies an increase in the Compton
mean free path and hence the diffusion length.  
Recall that the damping length is given by
\bel{eq:DampLengthAgain}
k_D^{-2}(\eta) = { 1 \over 6} \int_0^\eta  d\eta {1 \over \dot \tau}
{R^2 + 4(1+R)/5 \over (1+R)^2} .
\ee
and fluctuations are damped as $\exp[-(k/k_D)^2]$ assuming
$R\Psi \ll \Theta_0$ (see section \ref{ss-5polarization}).  
To account
for the evolution after last scattering, note that the Boltzmann
equation in flat space has the formal solution
\bel{eq:Formal}
[\Theta + \Psi](\eta_0,\mu) =
\int_0^{\eta_0}\left\{ [\Theta_0 + \Psi - i\mu V_b]\dot \tau
- \dot \Phi + \dot \Psi \right\}
e^{-\tau(\eta,\eta_0)}
e^{ik\mu(\eta-\eta_0)} d\eta \, .
\ee
For sufficiently large scales, we can take the slowly varying quantities
out of this integral.  Thus accounting for diffusion damping, the 
fluctuations at last scattering
become $[\Theta_0 +\Psi](\eta_*) = [\hat\Theta_0+\Psi](\eta_*) {\cal D}(k)$
and $\Theta_1(\eta_*) = \hat \Theta_1(\eta_*) {\cal D}(k)$, where
\bel{eq:AvgDampFact}
{\cal D}(k) = \int_0^{\eta_0} \dot \tau e^{-\tau(\eta,\eta_0)}
e^{-[k/k_D(\eta)]^2} d\eta.
\ee
Taking the multipole moments and setting $V_b = \Theta_1$,
we find for $\ell \ge 2$,
\beal{eq:FreeStreamAgain}
\Theta_\ell(\eta_0) \eapp [\Theta_0 + \Psi](\eta_*)
(2\ell+1) j_\ell(k\Delta\eta_*) 
+  \Theta_1(\eta_*)
[\ell j_{\ell-1}(k\Delta\eta_*) - (\ell+1) j_{\ell+1}(k\Delta\eta_*)] 
\quad\nonumber\\
&& 
+ (2\ell+1)\int_{\eta_*}^{\eta_0} [\dot \Psi - \dot \Phi] j_\ell(k\Delta\eta)
d\eta \, . 
\eea
Integrating over all $k$ modes of
the perturbation, we obtain
\bel{eq:Clagain}
{2 \ell + 1 \over 4\pi} C_\ell = {V \over 2\pi^2} \int {d k \over k}
{
k^3|\Theta_\ell(\eta_0,k)|^2
 \over 2\ell + 1} .
\ee
This completes the explicit construction of the anisotropy spectrum. 

\input chap5/analytic.tex

In Fig.~\ref{fig:5separation}, we show the analytic decomposition of
the spectrum into the effective temperature perturbation at last scattering
$[\Theta_0+\Psi](\eta_*)$, the dipole or Doppler term $\Theta_1/\sqrt{3}$,
the early ISW effect and diffusion damping.  Notice that without diffusion
damping the dilation boost of the acoustic oscillations for small scales
that enter during radiation domination is clearly evident.   The early
ISW
effect appears misleadingly small in power.  In fact it adds coherently 
with the SW effect,  whereas
the dipole roughly adds in quadrature.  
The 20\%
shift
in power spectrum normalization
from the monopole-only solution is entirely
due to the 1\% ISW effect.
Finally let us compare the analytic construction with the full numerical
results (see Fig.~\ref{fig:5analytic}).  The analytic approximation agrees
at the 5\% level to the damping scale for the range of parameters 
accessible to the CDM model.   By extending the analysis in this section
to other models, comparable accuracy can be obtained.

\section{Toward 1\% Accuracy}
\label{sec-5smallstuff}

The next generation of space based CMB anisotropy experiments have
the potential to measure all the $C_\ell$'s out to $\ell \sim 500$ to the
cosmic variance limit (\ie\ accuracy $\ell^{-1/2}$).  In this case,
the amount of information which may be retrieved from the CMB is
truly enormous.  If
the inflationary CDM cosmology turns out to be correct, there is
even a possibility that we can probe the physics of inflation through
tensor contributions (see \eg\ \cite{TWL,DKK}) and the shape of the 
initial power spectrum.  
A small difference between $C_\ell$'s for neutrinos with an eV scale
mass and the standard massless case appears near the damping scale
and provides the possibility of an indirect measure of the neutrino
mass through anisotropies.
To realize these goals, we must understand
the spectrum at the 1\% level.  Many secondary effects like those
discussed in \S \ref{ch-secondary} can contribute at this level.
As a first step toward the goal of 1\% accuracy, it is also necessary to refine
calculations of primary anisotropies.  The following
discussion draws results from \cite{HSSW}.

\input chap5/polar.tex
\subsection{Polarization Damping} 
\label{ss-5polarization}

The quadrupole
moment of the temperature distribution leads to linear polarization
in the microwave background (e.g.~\cite{Rees,Kaiser83}) and vice
versa \cite{BE84}.
The precise level of the temperature anisotropies therefore is not recovered
by neglecting polarization.
The Thomson cross section depends on angle
as $|\epsilon_f\cdot\epsilon_i|^2$,
where $\epsilon_f$ and $\epsilon_i$ are the final and initial polarization
vectors respectively
\cite{BE84,Kaiser83}.  A quadrupole temperature anisotropy therefore
sources polarization (see Fig.~\ref{fig:5polar}). Reversing the
arrow of time, polarization also feeds back to generate a temperature
quadrupole. 

To formally account for polarization, a separate Boltzmann hierarchy is
added for the temperature perturbation $\Theta^Q$ in 
the Stokes parameter $Q$
\cite{BE84,Kosowsky},  
\beal{eq:HierarchyQ}
\dot \Theta^Q_0 \eal 
	-{k \over 3}\Theta_1^Q
	-\dot \tau \left[ {1 \over 2}\Theta_0^Q - {1 \over 10}(\Theta_2 
		+ \Theta_2^Q) \right], \nonumber\\
{\dot \Theta^Q_1} \eal k\left[ \Theta^Q_0 -{2 \over 5}
        K_2^{1/2} \Theta^Q_2 \right] - \dot\tau\Theta^Q_1, \nonumber\\
{\dot \Theta^Q_2} \eal k \left[ {2 \over 3} K_2^{1/2} \Theta^Q_1
- {3 \over 7} K_3^{1/2} \Theta^Q_3 \right]
- \dot \tau \left( {9 \over 10}\Theta^Q_2 - {1 \over 10}\Theta_2 - {1 \over 2}
\Theta^Q_0 \right) , \nonumber\\
{\dot \Theta^Q_{\ell}} \eal k \left[ {\ell \over 2\ell-1}K_\ell^{1/2}
\Theta^Q_{\ell-1}
- {\ell+1 \over 2\ell+3} K_{\ell+1}^{1/2}
\Theta^Q_{\ell+1} \right] - {\dot \tau}\Theta^Q_\ell, \quad (\ell > 2)
\eea
where recall $K_\ell = 1 - (\ell^2+1)K/k^2$ and goes to unity if $K=0$.
Notice that as expected, it is the temperature quadrupole that sources
monopole and quadrupole polarization perturbations.  
Since the temperature quadrupole itself is suppressed in the tight 
coupling limit, we expect that polarization will yield only a higher
order correction for primary anisotropies.
Polarization feeds back to modify the quadrupole equation of the temperature
hierarchy \eqn{Hierarchy}
\bel{eq:PolarizationQuad}
{\dot \Theta_2} = k \left[ {2 \over 3} K_2^{1/2} \Theta_1
- {3 \over 7} K_3^{1/2} \Theta_3 \right]
- \dot \tau \left( {9 \over 10} \Theta_2 - {1 \over 10}\Theta^Q_2 -
{1 \over 2} \Theta^Q_0 \right).
\ee
Other multipole moments of the temperature hierarchy remain unmodified. 

\input chap5/polarization.tex

It is easy to see what effect polarization has on anisotropies. 
Let us expand these equations in the Compton scattering time $\dot\tau^{-1}$.
The polarization monopole $\ell=0$ and quadrupole $\ell=2$ equations
together
imply that
\bel{eq:PolarizationTC}
\Theta^Q_2 = \Theta^Q_0 = {1 \over 4}\Theta_2.
\ee
Putting these relations into equation \eqn{PolarizationQuad} for the 
feedback effect, we see that it changes the Compton coupling quadrupole
coefficient from ${9 \over 10} \rightarrow {3 \over 4}$.  
This affects the damping rate of acoustic oscillations as 
we shall now show.  

Diffusion damping occurs to second order in the tight coupling expansion
of the photon dipole and baryon Euler equations
[see equations \eqn{Hierarchy}, \eqn{Baryon}],
\beal{eq:CouplingExpansion}
\Theta_1 - V_b \eal \dot\tau^{-1} [k(\Theta_0+\Psi) - {2 \over 5} k\Theta_2 -
	\dot \Theta_1], \\
\label{eq:baryoncoupling}
\Theta_1 - V_b \eal \dot\tau^{-1} R [ \dot V_b + {\dot a \over a}V_b - k\Psi] 
\eea
where we have assumed that $K/k^2 \ll 1$.   Notice that a quadrupole
generated to first order in $\dot \tau^{-1}$ affects the evolution of
the dipole to second order.  
To lowest order, equation \eqn{PolarizationQuad} 
gives the quadrupole source as
\bel{eq:QuadSource}
\Theta_2 = \dot\tau^{-1} f_2^{-1} {2 \over 3} k\Theta_1,
\ee
where we have left the effect of polarization and the angular
dependence of Compton scattering implicit in 
\bel{eq:f2} 
f_2 =\cases { 9/10 & angular dependence \cr
	     3/4.  & polarization \cr }
\ee
The photon continuity or monopole equation yields
\bel{eq:PhotonContinuity}
\dot \Theta_0 = -{k \over 3} \Theta_1 - \dot\Phi.
\ee
To solve these equations to second order in $\dot \tau^{-1}$, let us
assume a solution of the form $\Theta_1 \propto \exp{i \int \omega d\eta}$ and
ignore variations on the expansion time scale compared with those
at the frequency of oscillation.  The electron velocity, obtained
by iteration is to second order
\bel{eq:Vbsecond}
V_b = \Theta_1 -\dot\tau^{-1} R[i\omega\Theta_1 - k\Psi] - \dot\tau^{-2} R^2
\omega^2\Theta_1.
\ee
Substituting this into the dipole equation \eqn{CouplingExpansion}
 and eliminating the zeroth
order term yields
\bel{eq:Theta1second}
i \omega (1+R)\Theta_1 
 =  k[\Theta_0 + (1+R)\Psi] - \dot\tau^{-1} R^2 \omega^2 \Theta_1
	- {4 \over 15} \dot\tau^{-1} f_2^{-1} k^2 \Theta_1.
\ee
The combination $\Theta_0 + (1+R)\Psi$ was shown in \S\ref{ss-4acousticdriven}
to oscillate acoustically around zero under the assumption of
a slowly varying $R$.  This is because of the baryonic infall contribution
$R\Psi$ and the photon blueshift $\Psi$ which displaces the zero point.   
It is therefore natural to try a solution where $\Theta_0 + (1+R) \Psi 
\propto \exp{i\int \omega d\eta}$, since its oscillations should match with
the dipole.  
Note also that after diffusion damping, the photon temperature retains
a contribution of order $R\Psi$ due to baryonic infall. 

Employing this relation in the photon continuity equation 
and ignoring slow changes in 
$R$, $\Phi$ and $\Psi$ yields the dispersion relation 
\bel{eq:dispersion}
(1+R)\omega^2 = {k^2 \over 3} + i\dot\tau^{-1}\omega \left( R^2 \omega^2 + 
{4 \over 15}
k^2 f_2^{-1} \right).
\ee
Using the lowest order solution to rewrite $\omega^3 = k^2 \omega/3(1+R)$ 
and solving the resultant quadratic equation, we obtain  \cite{Kaiser83}
\bel{eq:dispersionfinal}
\omega = \pm {k \over \sqrt{3(1+R)}} + {i \over 6}k^2 \dot\tau^{-1}
\left[ {R^2 \over (1+R)^2} + {4 \over 5} f_2^{-1}{ 1 \over {1+R}} \right].
\ee
In other words, the oscillations damp as $\exp[-(k/k_D)^2]$ and the
damping length becomes
\bel{eq:damplengthfull}
k_D^{-2} = {1 \over 6}
	\int d\eta {1 \over \dot \tau} {{R^2 + 4f_2^{-1}(1+R)/5} \over
(1+R)^2}.
\ee
In the photon-dominated $R\ll 1$ limit, 
the
damping length increases by 5\% ($f_2 = {9 \over 10}$) through
the angular dependence of Compton scattering and an additional 
10\% through polarization ($f_2 = {3 \over 4}$).  
Closer to baryon domination, the effect of
$f_2$ is less noticeable.  Qualitatively,  the polarization
sources the quadrupole and generates viscosity which is then
dissipated \cite{Kaiser83}.  Actual numerical results of the
effect of polarization are shown in in Fig.~\ref{fig:5polarization} 
and are in good agreement with these analytic estimates of the
relative effect.   The fractional
difference at small scales can be quite significant due to the near exponential
behavior of damping.

\def\Hei{He{\tencsc i}}
\def\Heii{He{\tencsc ii}}
\def\Heiii{He{\tencsc iii}}

\input chap5/helium.tex

\subsection{Helium Recombination}

One might naively expect helium recombination to have a negligible 
effect on the $C_\ell$'s
because helium recombines while the radiation and matter are still very
tightly coupled, at $z\simeq2500$ for singly ionized 
and $z\simeq6000$ for doubly ionized helium.
However the diffusion damping length grows continuously and is sensitive to
the full thermal history.
Inclusion of helium recombination affects the 2nd, 3rd and 4th
peaks at the 0.2\%, 0.4\% and 1\% levels, as shown in Fig.~\ref{fig:5helium}.
Hence it {\it is} important to follow the recombination of the helium in order
to obtain accurate $C_\ell$'s at the percent level.
Note that because of atomic collisions,
helium atoms are tightly coupled to the hydrogen through collisions
even after helium recombination,
Since they contribute to the inertia of the photon-baryon fluid, helium atoms
should be kept in the baryon evolution equations.
It has been shown that simple use of the Saha equation for 
helium is as accurate as treating helium atoms more fully \cite{HSSW}.
The trace of neutral hydrogen, present even at redshifts
$z\simeq2500$, can absorb the helium Ly$\alpha$ photons.  
This prevents helium recombination photons  from ionizing other helium atoms,
unlike their hydrogen counterparts. 

\input chap5/grav.tex

\subsection{Gravity Waves}
\label{ss-5gravitywave}

In addition to the scalar modes with which the previous discussion has
been involved, there is the possibility that inflation excites tensor
(\ie\ gravity wave) perturbations as well \cite{Staa}.
Early work on tensors and the CMB was performed 
by \cite{Stab,FabPol,AbbWis,Stab}.
There exist several semi-analytic approximations of varying accuracy, the
most recent and accurate being due to \cite{AllKor}.
To calculate the tensor spectrum numerically one uses the formalism of
\cite{Pol} as first worked out in detail by \cite{Crittenden}.
This leads to another set of Boltzmann equations, independent of those for
the scalars, which follow the temperature and polarization anisotropies of
the tensors.  The final result is then
$C_\ell^{\rm (tot)\vphantom{(S)}}=C_\ell^{(S)}+C_\ell^{(T)}$ where
the relative normalization of the tensor and scalar components depends on
the details of the perturbation generation scenario.
In Fig.~\ref{fig:5grav}, we plot the tensor 
contribution $C_\ell^{(T)}$, for a model with
the parameters of standard CDM.

The qualitative features of this spectrum are easy to understand.  
The CMB couples to gravity waves through its quadrupole moment.  
The amplitude of the metric perturbation induced by a gravity 
wave $h_\epsilon$, where $\epsilon$ represents the two possible
polarizations, evolves as
\bel{eq:gravitywave}
\ddot h_\epsilon + 2 {\dot a \over a} \dot h_\epsilon 
	+ (k^2 + 2K) h_\epsilon = 8\pi G \left( {a \over a_0} \right)^2 
	p_T \Pi_T^{(2)}
\ee
(see \cite{KS84} eqn.~4.15) where $\Pi_T^{(2)}$ is the {\it tensor}
contribution to the anisotropic stress.  
Ignoring the feedback effect through 
the radiation quadrupoles, this is a damped oscillator equation.  
Inflation predicts a spectrum of initial gravity waves $k^3 |h_\epsilon
(0,k)|^2 \propto k^{n_T-1}$.  For these initial conditions, $h_\epsilon$ 
remains
constant outside the horizon and feels the $k^2$ ``pressure'' force 
near horizon crossing.  The consequent changes in $h_\epsilon$ are
damped by the expansion.  Since $\dot a/a = \eta^{-1}$ in the 
radiation-dominated epoch and $2\eta^{-1}$ in the matter-dominated epoch, 
gravity waves are damped less rapidly in the former [$j_0(k\eta)$] 
than in the latter [$j_1(k\eta)/k\eta$].  

Just like changes in the  scalar spatial metric perturbation $\dot \Phi$,
$\dot h_\epsilon$ sources radiation perturbations through dilation from
the stretching of space.  The difference is that due to the spin two
nature of gravity waves, the deformation sources a quadrupole
rather than a monopole fluctuation in the matter.  This makes its effects
unimportant for density perturbations and structure formation.  
Contributions to the photon temperature perturbation
before last scattering
are rapidly damped away by Compton isotropization.
However during the free streaming epoch, the quadrupole source like the 
monopole projects onto higher multipoles as the photons free stream, causing
anisotropies in the CMB through the ISW effect.  This explains the three 
prominent features in the spectrum of Fig.~\ref{fig:5grav}. 
Modes that cross the horizon recently source mainly the quadrupole, 
boosting the low order multipoles. Smaller scales experience the 
full decay of $h_\epsilon$, leading to a small rise.  However
the smallest scales contribute negligibly since they enter the horizon
before last scattering.  Since these are exactly the scales on which
acoustic oscillations appear, it is very likely that gravity waves
are unimportant for small angle anisotropies.   On the other hand, the
ratio of large to small scale anisotropies may tell us something about
the relative amplitude of the tensor to scalar initial contributions
\cite{TWL,DKK}.

\input chap5/neutrinos.tex

\subsection{Massive Neutrinos}
\label{ss-5massive}

The radiation content determines the amount of dilation boost the
acoustic modes encounter at horizon crossing from the decay of the potential.  
Lowering the radiation content
lowers the boost.  Thus the CMB is sensitive to the number of 
effectively massless neutrino families $N_\nu$ before last scattering
(see Fig.~\ref{fig:5neutrinos}).  Massive neutrinos are a promising 
dark matter candidate and can solve some of the problems CDM models
have with large scale structure formation \cite{DSS,KHPR}.   
By assuming a neutrino
with a few eV mass and composing the rest of the critical density with
cold dark matter, one retains many of the features of the CDM model while also
lowering the excess of small scale power.  A low mass neutrino is 
relativistic when a galaxy-sized mass enters the horizon.  Thus
neutrino free streaming will collisionlessly damp power on these scales. 
However for the degree and larger scales that the current CMB 
experiments probe, such neutrinos are already non-relativistic
at horizon crossing and leave the same signature as CDM.  
The transition scale is around $\ell \approx 500$ for a neutrino mass
of a few eV.
Thus the CMB anisotropy spectrum should follow the $N_\nu=3$ CDM
prediction until roughly those scales and then decrease to the lower
$N_\nu$ prediction.  Extracting the neutrino mass will therefore
require a detailed understanding of the damping tail and any
secondary and foreground contributions -- an extremely challenging, 
but not unthinkable task. 

%% file: chap5/potential.tex
\begin{figure}[t]
\centerline{ \hskip -0.5truecm
\epsfxsize=3.5in \epsfbox{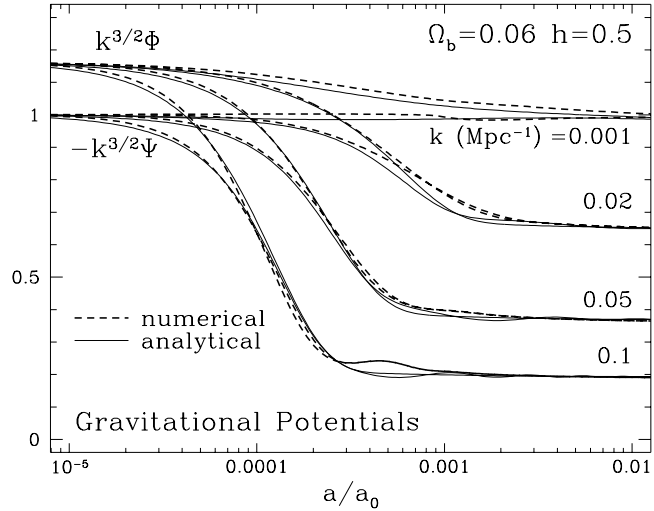}}
 \vskip -0.5truecm
\caption{Gravitational Potential Evolution}
\mycaption{Gravitational potentials in the Harrison-Zel'dovich
CDM model.  The potential decays after crossing the sound horizon in the
radiation dominated epoch and only flattens out {\it well} into the
matter dominated epoch.  Moreover $\Phi \ne -\Psi$ early on due to
anisotropic stress. 
The analytic approximations trace the numerical
potentials reasonably well.
}
\label{fig:5potential}
\end{figure}

%% file: chap5/vis.tex
\begin{figure}[t]
\centerline{ \hskip -0.5truecm
\epsfxsize=3.5in \epsfbox{chap5/vis.epsf}}
 \vskip -0.5truecm
\caption{Visibility Function}
\mycaption{The redshift visibility function.  Notice that
the weak dependence on $\Omega_b$ of the visibility function is adequately
described by the analytic fitting formula, whereas the Jones \& Wyse 
\cite{JW} fitting formula [their equation (23)] does not.
}
\label{fig:5vis}
\end{figure}

%% file: chap5/separation.tex
\begin{figure}[t]
\centerline{ \hskip -0.5truecm
\epsfxsize=3.5in \epsfbox{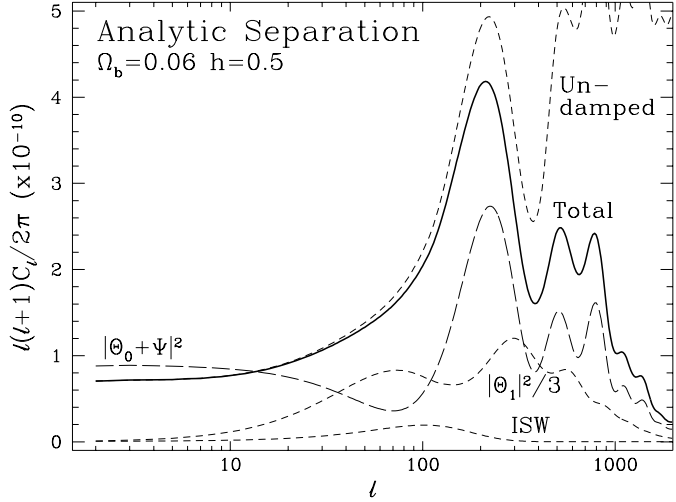}}
 \vskip -0.5truecm
\caption{Analytic Separation of Effects}
\mycaption{Individual contributions to the anisotropy in
the CDM model.
At the largest scales $(\ell \simlt 30)$,
the monopole $|\Theta_0 + \Psi|$ from the
ordinary Sachs-Wolfe effect dominates.  The 20\% correction
from the early
ISW effect on scales larger than the first Doppler peak
 appears misleadingly small in power (see text).
The ordinary
Sachs-Wolfe effect is overpowered by the acoustic
oscillations at small scales 
leading to a deficit at intermediate scales $(\ell \sim 70)$
which
is filled in by the adiabatic dipole $\Theta_1$
and the ISW effect. Although the dipole cannot be neglected,
the monopole is clearly responsible for
the general structure of the Doppler peaks.
Diffusion damping significantly reduces
fluctuations beyond the first Doppler peak and cuts off the anisotropies
at $\ell \sim 1000$.
}
\label{fig:5separation}
\end{figure}

%% file: chap5/analytic.tex
\begin{figure}[t]
\centerline{
\epsfxsize=4.5in \epsfbox{chap5/analytic.epsf}}
 \vskip -0.5truecm
\caption{Comparison of Analytic and Numerical Results}
\mycaption{The agreement between analytic and numerical results is
excellent on all scales. 
}
\label{fig:5analytic}
\end{figure}

%% file: chap5/polar.tex
\begin{figure}[t]
%\vphantom{marker} \vskip 0.5truecm
\centerline{ \hskip -1truecm
\epsfxsize=4.0in \epsfbox{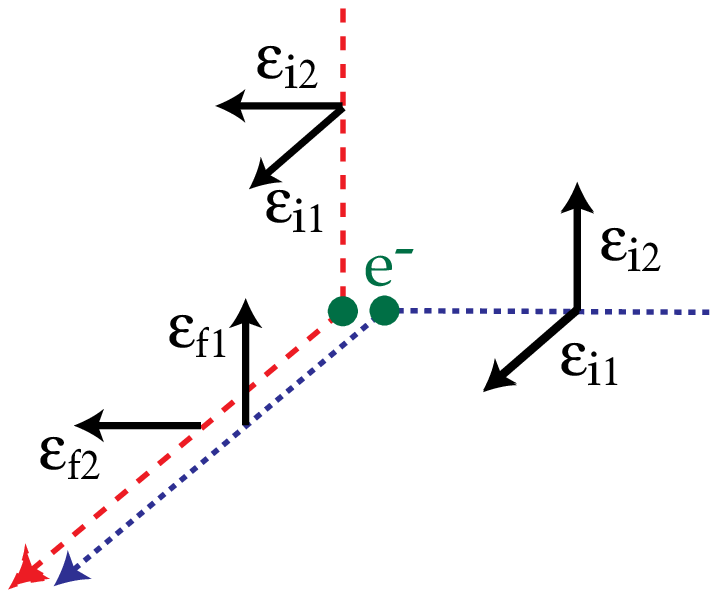}}
\vskip -0.5truecm
\caption{Polarization Generation}
\mycaption{A quadrupole moment in the temperature distributions sources
polarization.    Unless the temperature or mean energy of the radiation
at ninety degree angles is the same, polarization dependent preferential 
scattering will cause more intensity in one polarization than the other
in the outgoing scattered radiation.  Reversing the arrow of time, we
see that linear polarization sources a quadrupole anisotropy.  The length
of the dashes represents the wavelength of the photon.
}
\label{fig:5polar}
\end{figure}

%% file: chap5/polarization.tex
\begin{figure}[t]
\centerline{ \hskip -0.5truecm
\epsfxsize=3.5in \epsfbox{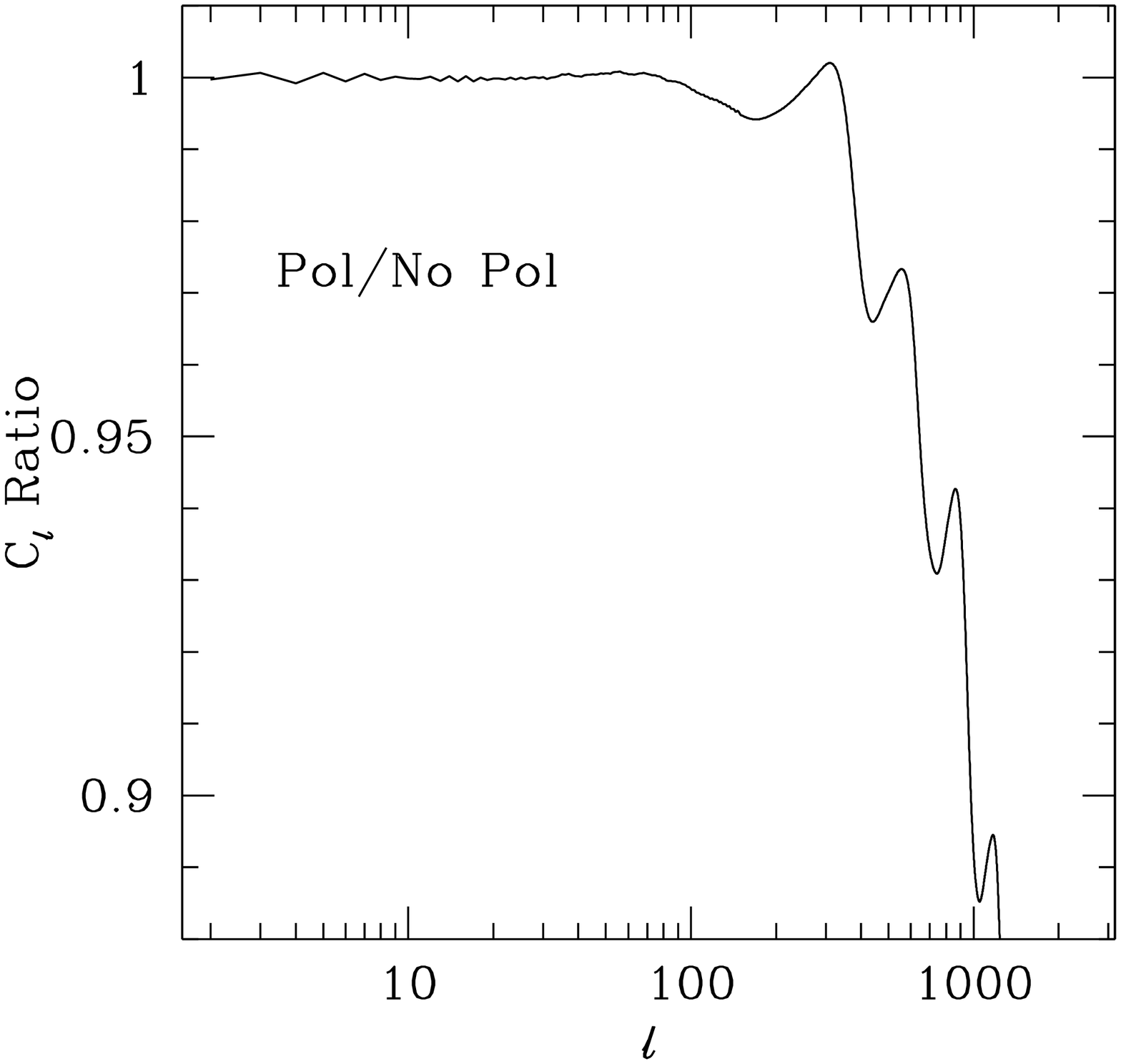}}
 \vskip -0.5truecm
\caption{Polarization Damping}
\mycaption{The ratio of $C_\ell$ for a standard CDM model where polarization is
explicitly followed, relative to a calculation where it is neglected.       
Polarization increases the damping scale of temperature anisotropies.
Calculations courtesy of M. White \cite{HSSW}.
}
\label{fig:5polarization}
\end{figure}

%% file: chap5/helium.tex
\begin{figure}[t]
\centerline{ \hskip -0.5truecm
\epsfxsize=3.5in \epsfbox{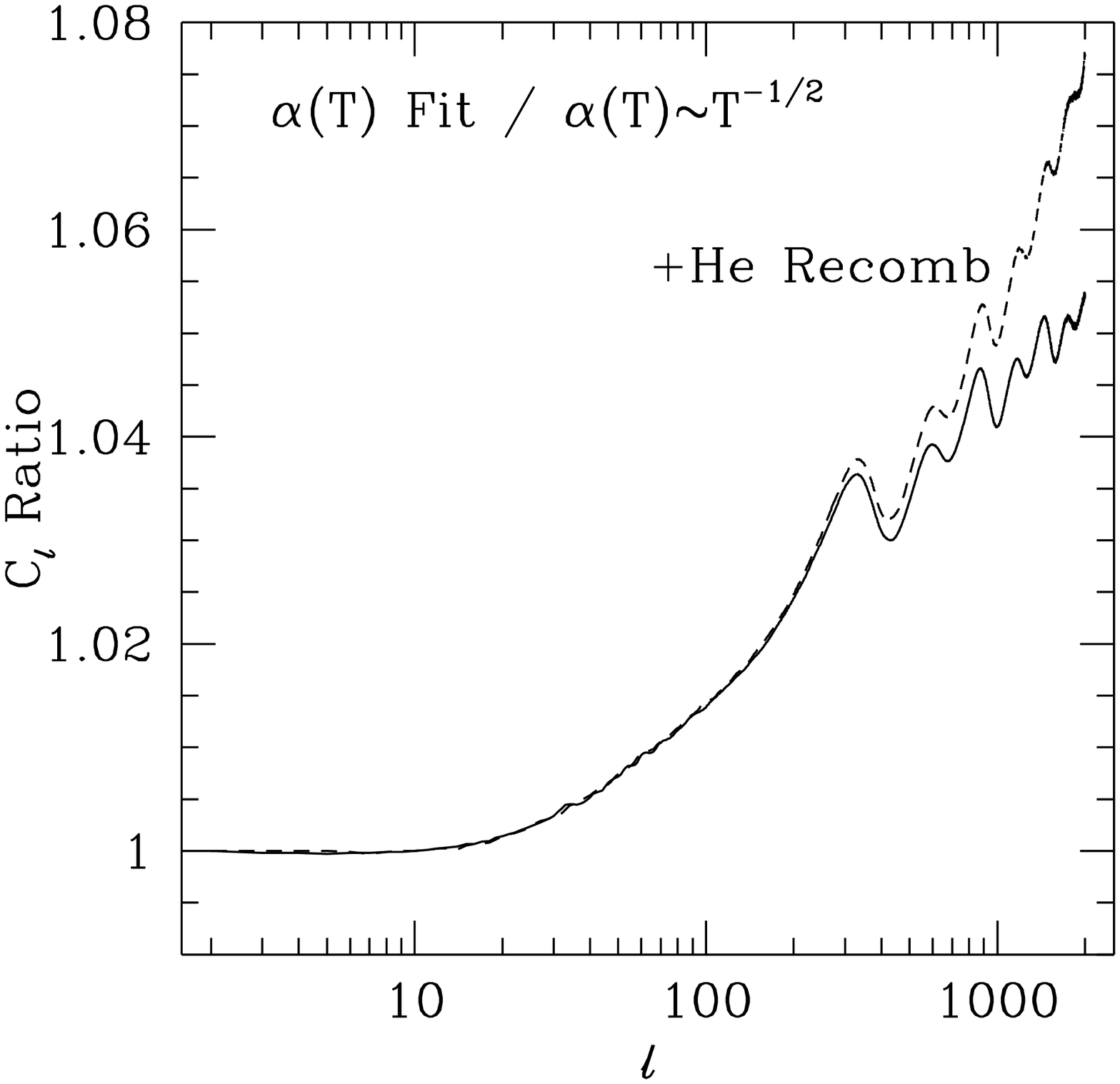}}
 \vskip -0.5truecm
\caption{Helium Recombination}
\mycaption{Ionized helium at high redshift contributes free electrons
to oppose photon diffusion and thus even at recombination has a small
effect in decreasing the damping scale.  Also shown here is the
effect of improving the fitting formula for the 
`case B' recombination coefficient
$\alpha_B$ of equation \eqn{alphab} from the $T^{-1/2}$ scaling of 
\cite{PeeblesRec}.
Calculations courtesy of M. White \cite{HSSW}.
}
\label{fig:5helium}
\end{figure}

%% file: chap5/grav.tex
\begin{figure}[t]
\centerline{ \hskip -0.5truecm
\epsfxsize=3.5in \epsfbox{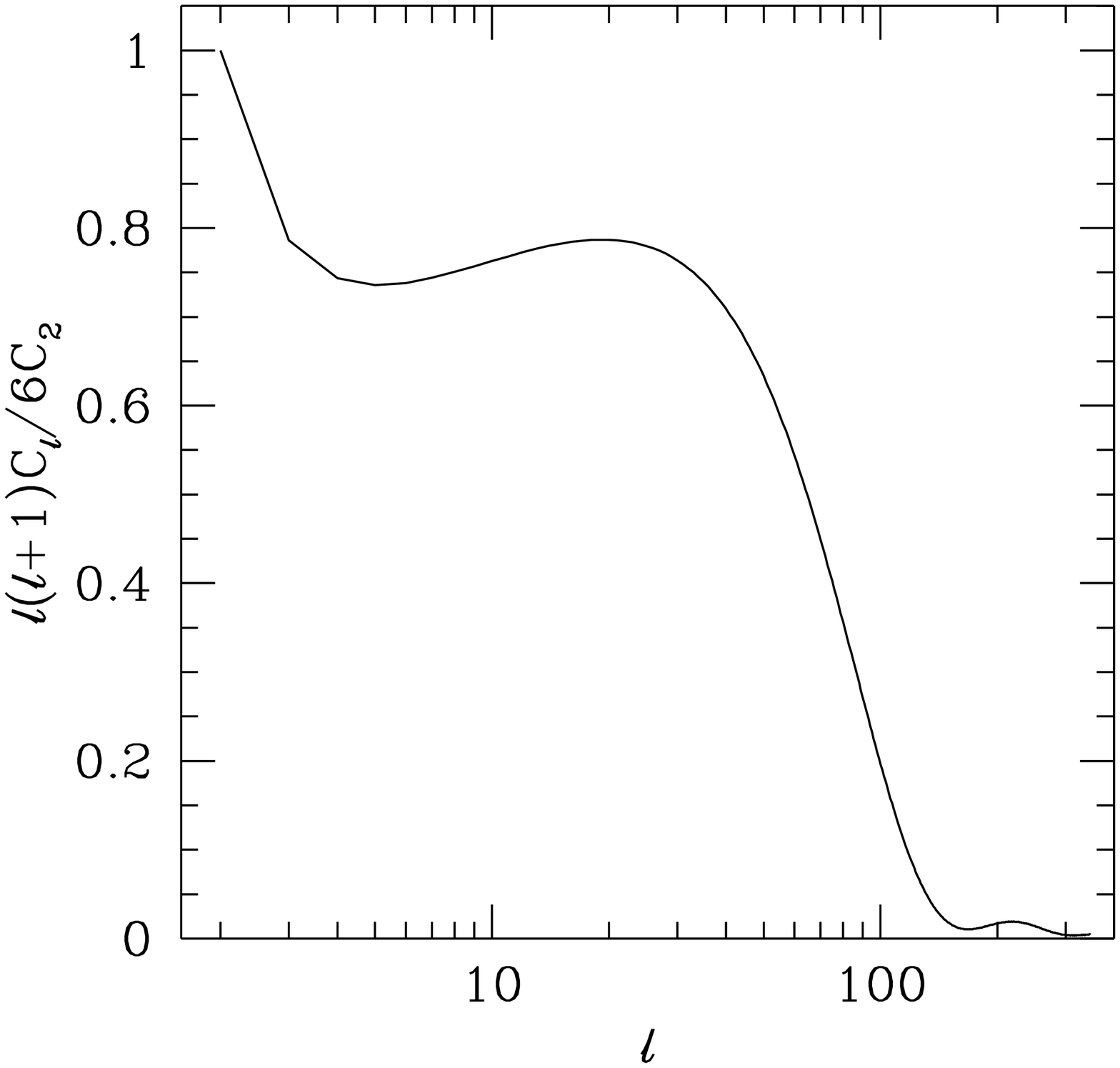}}
 \vskip -0.5truecm
\caption{Gravity Wave Spectrum}
\mycaption{A flat spectrum of tensor generated anisotropies with
for standard CDM $\Omega_0=1.0$, $h=0.5$, $\Omega_b =0.05$ $n_T=1$.  The
tensor and scalar spectrum add in quadrature.  Calculation courtesy
of M. White \cite{HSSW}. 
}
\label{fig:5grav}
\end{figure}

%% file: chap5/neutrinos.tex
\begin{figure}[t]
\centerline{ \hskip -0.5truecm
\epsfxsize=3.5in \epsfbox{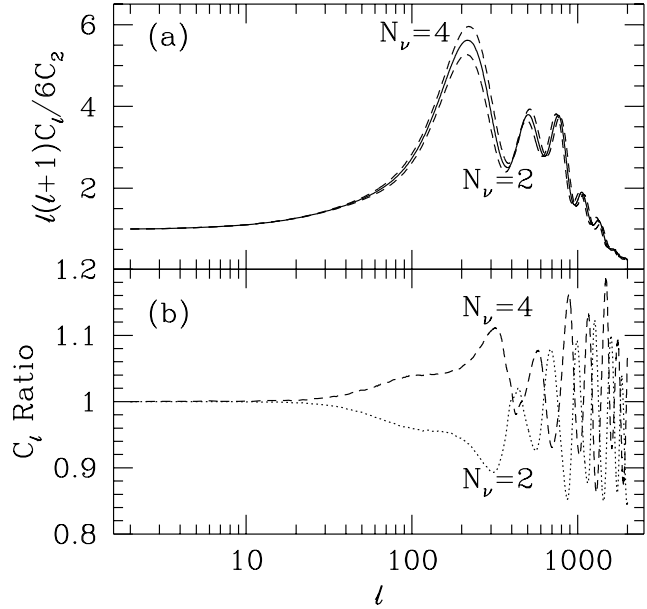}}
 \vskip -0.5truecm
\caption{Number of Massless Neutrinos}
\mycaption{Relativistic  neutrinos increase the amount of dilation boost
from the potential decay that  the acoustic mode encounters at horizon crossing.
The prediction for massive neutrino models 
depends on when the neutrinos become
non-relativistic.  For a mass of a few eV, the horizon at this epoch 
projects onto 
$\ell \sim 500$ and the resultant spectrum will be a smooth transition
from $N_\nu = 3$ to $N_\nu =2 $ near this scale. 
Calculation courtesy of M. White \cite{HSSW}. 
}
\label{fig:5neutrinos}
\end{figure}

%% file: appendixb.tex
\chapter{Useful Quantities and Relations} 

\input tables/conversions.tex
\section{FRW Parameters}

The expansion rate is given by the Hubble parameter 
\beal{eq:Hubbledefinition2}
H^2 \eadef \left( {1 \over a} {d a \over dt} \right)^2
 = \left( {\dot a \over a} {a_0 \over a} \right)^2 \nonumber\\
\eal \left( {a_0 \over a} \right)^4 { a_{eq} + a \over a_{eq} + a_0 }
\Omega_0 H_0^2
- \left({a_0 \over a}\right)^2 K
+ \Omega_\Lambda H_0^2,
\eea
where the curvature is $K = -H^2_0(1-\Omega_0-\Omega_\Lambda)$.
The value of the Hubble parameter today, for different choices of
the fundamental units (see Tab.~\ref{tab:conversions}), is expressed
as
\beal{eq:Hubblevalue}
H_0 \eal 100h\,{\rm km}{\rm s}^{-1} {\rm Mpc}^{-1} 
 \nonumber\\
    \eal 2.1331 \times 10^{-42} h\, {\rm GeV} 
 \nonumber\\
    \eal (2997.9)^{-1} h\, {\rm Mpc}^{-1}  
 \nonumber\\
    \eal (3.0857 \times 10^{17})^{-1} h\, {\rm s}^{-1} 
 \nonumber\\
    \eal (9.7778)^{-1} h\, {\rm Gyr}^{-1}.
\eea
Present day densities in a given particle 
species $X$ are measured in units of the critical density 
$\rho_X(a_0) = \Omega_X \rho_{\rm crit}$, where
\beal{eq:rhocrit}
\rho_{\rm crit} = 3H_0^2/8\pi G \eal 
	1.8788 \times 10^{-29} h^2\, {\rm g}\, {\rm cm}^{-3}\nonumber\\
	\eal 8.0980 \times 10^{-47} h^2\, {\rm GeV}^4 \nonumber\\
	\eal 1.0539 \times 10^4 h^2\, {\rm eV}\, {\rm cm}^{-3} \nonumber\\
	\eal 1.1233 \times 10^{-5} h^2 {\rm protons}\, {\rm cm}^{-3} \nonumber\\
	\eal 2.7754 \times 10^{11} h^2 {\rm M}_{\odot}\, {\rm Mpc}^{-3}.
\eea
For the CMB,
\beal{eq:CMBnumbers}
n_{\gamma 0} \eal 399.3 \To^3\, {\rm cm}^{-3}, \nonumber\\
\rho_{\gamma 0} 
\eal 4.4738 \times 10^{-34} \To^4\, {\rm g}\,{\rm cm}^{-3},\nonumber\\
\Omega_\gamma \eal 2.3812 \times 10^{-5} h^{-2} \To^{4} ,
\eea
and for the neutrinos
\beal{eq:Neutrinonumbers}
\rho_{\nu 0} \eal [(1-f_\nu)^{-1} - 1] \rho_{\gamma 0}, \nonumber\\
\Omega_\nu \eal [(1-f_\nu)^{-1} - 1]  \Omega_\gamma ,  
\eea
with $(1-f_\nu)^{-1} = 1.68$ for the standard model,
or for the total radiation
\beal{eq:radiationnumber}
\rho_{r 0} \eal (1-f_\nu)^{-1}\rho_{\gamma 0}, \nonumber\\
\Omega_r \eal (1-f_\nu)^{-1}  \Omega_\gamma  . 
\eea

\section{Time Variables}

Throughout the text we use four time variables interchangeably, they are
$a$ the scale factor, $z$ the redshift,
$\eta$ the conformal time, and $t$ the coordinate time.
In addition, three dimensionless time parameterizations are useful
to consider: $\chi$ the development angle in an open universe,
$D$ the relative amplitude of pressureless 
matter fluctuations, and $\tau$ the optical
depth to Compton scattering.

\input tables/redshifts.tex
\subsection{Scale Factor and Redshift}

The scale factor $a(t)$ describes the state of expansion and is the
fundamental measure of time in the Hubble equation \eqn{Hubbledefinition2}
since it controls the energy density of the universe.  
In this Appendix, we leave 
the normalization of $a$ free to preserve generality.  
However, the normalization
applied in \S\ref{ch-perturbation}, \S\ref{ch-evolution}, \S\ref{ch-primary},
and Appendix \ref{sec-5CDM} is $a_{eq} = 1$.  The conversion factor
between the more commonly employed normalization $a_{0} = 1$ is
\beal{eq:equality}
{a_{eq} \over a_0} \eal {\Omega_r \over \Omega_0 -\Omega_r} \nonumber\\
		   \eal 2.38 \times 10^{-5} (\Omega_0 h^2)^{-1} (1-f_\nu)^{-1}
			\To^{4}.
\eea
The redshift $z$ is defined by $(1+z) = a_0/a$ and serves the same
role as the scale factor normalized to the present.
We give the scale factor normalized to $3/4$ at
baryon-photon equality a special symbol $R$ given the frequency of its
appearance in equations related to Compton scattering.  More explicitly,
\beal{eq:rdef2}
R \eal {3 \over 4}{\rho_b \over \rho_\gamma} = 
	 (1-f_\nu)^{-1} {3 \over 4} {\Omega_b \over \Omega_0} 
	{a \over a_{eq}} \nonumber\\
\eal 31.5 \Omega_b h^2 \To^{-4} (z/10^3)^{-1}.
\eea
Epochs of interest for the CMB are listed 
in Tab.~\ref{tab:redshifts} by
their redshifts.  

\subsection{Conformal Time}

By definition, the conformal time $\eta = \int dt/a$ is related
to the scale factor as
\beal{eq:timedefinition}
\eta \eal \int {da \over a}{ 1 \over H} {a_0 \over a} .
\eea
Note that in these $c=1$ units, the conformal time doubles as the
comoving size of the horizon.
In an open universe, it is also related to the development angle by
\bel{eq:chidef}
\chi = \sqrt{-K}\eta.
\ee
Asymptotic relations are often useful for converting values.  Before
curvature or $\Lambda$ domination, the conformal time
\beal{eq:conformaleval}
\eta \eal {2 \sqrt{2} \over k_{eq} } \left(\sqrt{1+a/a_{eq}}  -1 \right) 
	\nonumber\\
     \eal 2 (\Omega_0 H_0^2)^{-1/2} (a_{eq}/a_0)^{1/2} 
				\left(\sqrt{1+a/a_{eq}} -1 \right),
\eea
and reduces to 
\bel{eq:conformalexpl}
\eta = \cases{ (\Omega_r H_0^2)^{-1/2} a/a_0 & RD \cr
    		  2(\Omega_0 H_0^2)^{-1/2} (a/a_0)^{1/2}, & MD \cr } 
\ee
where $\Omega_r/\Omega_0 \approx a_{eq} / a_0$.
In a $\Lambda=0$ universe, it also has an asymptotic solution 
for $a\gg a_{eq}$
\beal{eq:conformal2}
\eta \eal {1 \over \sqrt{-K}} \cosh^{-1} \left[ 1 + {2 (1-\Omega_0) \over 
\Omega_0}{a \over a_0} \right] \qquad {\rm MD/CD} \nonumber\\
\lim_{\Omega_0 \rightarrow 0} \eta_0 \!\!\!&\rightarrow&\!\!\!
	 (-K)^{-1/2} \ln(4/\Omega_0) ,
\eea 
and thus the horizon scale is larger than the curvature scale $(-K)^{-1/2}$
for low $\Omega_0$ universes.  
In a flat universe,
\bel{eq:angularlambda}
\eta_0 \approx 
2(\Omega_0 H_0^2)^{-1/2} (1+\ln\Omega_0^{0.085}), \qquad \Omega_0+\Omega_\Lambda=1
\ee
and the horizon goes to a constant $\eta = 2.8 H_0^{-1} \Omega_0^{-1/3}
(1-\Omega_0)^{-1/6}$ as $a/a_0 \rightarrow \infty$.

\subsection{Coordinate Time}

The coordinate time is defined in terms of the scale factor as,
\bel{eq:coordinatetimedef}
t = \int {da \over a} {1 \over H} .
\ee
It also takes on simple asymptotic forms, \eg\
\bel{eq:coordinatetime}
t = {2 \over 3} (\Omega_0 H_0^2)^{-1/2} a_0^{-3/2} [(a+a_{eq})^{1/2}
(a-2a_{eq}) + 2a_{eq}^{3/2}]. \qquad {\rm RD/MD}
\ee
Explicitly, this becomes
\beal{eq:timerd}
t \eal {1 \over 2} (\Omega_0 H_0^2)^{-1/2} (a_0/a_{eq})^{1/2} (a/a_0)^2 
 	\qquad {\rm RD}	\nonumber\\
  \eal 2.4358 \times 10^{19} \To^{-2} (1+z)^{-2} {\rm s}.
\eea
and 
\beal{eq:timemd}
t \eal {2 \over 3} (\Omega_0 H_0^2) (a/a_0)^{3/2} \qquad {\rm MD} \nonumber\\
  \eal 2.0571 \times 10^{17} (\Omega_0 h^2)^{-1/2} (1+z)^{-3/2} {\rm s} .
\eea
The expansion time, defined as $H^{-1}$ scales similarly
\beal{eq:timeexp}
t_{exp} \eal (\Omega_0 H_0^2)^{-1} (a/a_0)^2 a_0^{1/2} (a+a_{eq})^{-1/2} 
	\nonumber\\
	\eal 4.88 \times 10^{19} (z+z_{eq}+2)^{-1/2} \To^{-2} (1+z)^{-3/2}
	\rm{s.}
\eea
For $\Lambda=0$ universes, the coordinate time at late epochs when 
radiation can be neglected is given by
\bel{eq:curvaturetime}
t = H_0^{-1} \left[ { (1+\Omega_0 z)^{1/2} \over (1-\Omega_0) (1+ z)} 
- {\Omega_0 \over 2 (1-\Omega_0)^{3/2} } \cosh
\left( { 2(1-\Omega_0) \over  \Omega_0 (1+z)} + 1 \right) \right].
\qquad {\rm MD/CD}
\ee
In particular, the age of the universe today is
\bel{eq:ageopen}
t_0 = H_0^{-1} (1-\Omega_0)^{-1} \left[ 1 - {\Omega_0 \over 2} 
(1-\Omega_0)^{-1/2}\cosh (2/\Omega_0 - 1) \right], \qquad \Omega_\Lambda=0
\ee
where the factor in square brackets goes to unity as $\Omega_0 \rightarrow 0$.
This should be compared with the flat $\Omega_0 + \Omega_\Lambda = 1$ result
\bel{eq:agelambda} 
t_0 = {2 \over 3} H_0^{-1}  (1-\Omega_0)^{-1/2} \ln \left[ 
{1 + \sqrt{1-\Omega_0} \over \sqrt{\Omega_0}} \right],
	\qquad \Omega_0+\Omega_\Lambda=1,
\ee
which diverges logarithmically as $\Omega_0 \rightarrow 0$. 
Finally a microphysical time scale of interest for the CMB,
\beal{eq:comptontime}
t_C \eal (d\tau/dt)^{-1} = (x_e n_e\sigma_T)^{-1}  \nonumber\\
    \eal 4.4674 \times 10^{18} (1-Y_p/2)^{-1} (x_e \Omega_b h^2)^{-1} 
	(1+z)^{-3} {\rm s,} 
\eea
is the Compton mean free time between scatterings.

\subsection{Growth Function}

The amplitude of matter fluctuations undergoing pressureless growth 
is another useful parameterization of time. 
It is given by equation \eqn{BeqnCTotal} as
\bel{eq:Dgrowthagain}
D = {5 \over 2} \Omega_0 {a_0 \over a_{eq}} g(a) \int {da \over a} 
	{1 \over g^3(a)} \left({ a_0 \over a} \right)^2 ,
\ee
where the dimensionless, ``pressureless'' Hubble parameter is
\bel{eq:hubblepressureless}
g^2(a) = \left({a_0 \over a} \right)^3 \Omega_0 + \left({a_0 \over a}\right)^2
(1-\Omega_0-\Omega_\Lambda) + \Omega_\Lambda.
\ee 
In the matter or radiation-dominated epoch, $D = a/a_{eq}$ by construction.  
In a $\Lambda=0$ universe, $D$ becomes
\bel{eq:Dcurvature}
D = {5 \over 2 x_{eq}} \left[ 1 + {3 \over x} + 
	{3 (1+x)^{1/2} \over x^{3/2}} \ln[(1+x)^{1/2} - x^{1/2}] \right],
\ee
where $x = (\Omega_0^{-1} -1) a/a_0$.  
Fitting formulae for the growth factor, valid for the general case, are
occasionally useful \cite{CPT}:
\beal{eq:Dasymptotic}
{D_0 \over a_0} 
\eapp {5 \over 2} \Omega_0 \left[ \Omega_0^{4/7} - \Omega_\Lambda 
+ \left(1+ {1 \over 2}\Omega_0\right)\left(1-{1 \over 70}\Omega_\Lambda
\right) \right]^{-1}, \\
{d \ln D \over d \ln a} \eapp
\left[ {\Omega_0 (1+z)^3 \over \Omega_0(1+z)^3 - (\Omega_0+\Omega_\Lambda -1)
(1+z)^2 + \Omega_\Lambda} \right]^{4/7} .
\eea
The latter relation is often employed to relate the velocity to the
density field. 

\subsection{Optical Depth}

For the CMB, the optical depth $\tau$ to Compton scattering 
is a useful lookback time parameterization,
\beal{eq:tautime}
\tau(a,a_0) \eal \int_{\eta}^{\eta_0} d\eta' x_e n_e \sigma_T a' \nonumber\\
     \eal 6.91 \times 10^{-2} (1-Y_p/2)x_e \Omega_b h
	\int_{a}^{a_0} 
     {da' \over a'} {H_0 \over H} \left( {a_0 \over a'}\right)^3,
\eea
for constant ionization fraction.
If $a \gg a_{eq}$, this has closed form solution,
\beal{eq:tautimex} 
\tau(a,a_0) \eal 4.61 \times 10^{-2} (1-Y_p/2) x_e {\Omega_b h \over \Omega_0^2}
	\nonumber\\
&& \times \cases{ 2 - 3\Omega_0 + (1+\Omega_0 z)^{1/2} (\Omega_0 z + 3\Omega_0 -2)
 & $ \Omega_\Lambda = 0$ \cr
	       \Omega_0 [1-\Omega_0+\Omega_0(1+z)^3]^{1/2} - \Omega_0.
 & $ \Omega_0 + \Omega_\Lambda = 1$\cr}
\eea
Furthermore, since the optical depth is dominated by early
contributions the distinction between open and $\Lambda$ universes
for $\tau \simgt 1$
is negligible.  

\section{Critical Scales}

\subsection{Physical Scales}

Several physical scales are also of interest. We always use comoving
measures when quoting distances.  The most critical quantity is the horizon
scale $\eta$ given in the last section and the curvature scale
$(-K)^{-1/2} = 2997.9 h (1-\Omega_0 - \Omega_\Lambda)^{1/2} $Mpc.  
There are two related quantities of interest, the Hubble scale and the
conformal angular diameter distance to the horizon.
The Hubble
scale is often employed instead of the horizon scale because it 
is independent of the past evolution of the universe.  
The wavenumber
corresponding to the Hubble scale is 
\bel{eq:KHubble}
k_H ={\dot a \over a} =\cases{  (\Omega_r H_0^2)^{1/2} (a_0/a) & RD \cr
	       (\Omega_0 H_0^2)^{1/2} (a_0/a)^{1/2} & MD \cr 
	       (-K)^{1/2} & CD \cr
	       (\Omega_\Lambda H_0^2)^{1/2} a/a_0. & $\Lambda$D \cr}
\ee
Comparison with the relations for $\eta$ shows that $k_H \eta \sim 1$
during radiation and matter domination but not curvature or $\Lambda$
domination.  Indeed, due to the exponential expansion, the Hubble
scale goes to zero as $a/a_0 \rightarrow \infty$, reflecting the fact
that regions which were once in causal contact can no longer communicate.
This is of course how inflation solves the horizon problem.   Throughout
the main text we have blurred the distinction between the Hubble scale
and the horizon scale when discussing the radiation- and matter-dominated epochs.  

The distance inferred for an object of known spatial extent by
its angular diameter is known as the conformal angular diameter distance.  
It multiplies the angular part of the spatial metric.  Moreover, in 
an open universe, it is not
equivalent to the distance measured in conformal time.  
For an observer at the present, it is given by
\beal{eq:angulardiamdist}
r_\theta(\eta) \eal (-K)^{-1/2} \sinh[(\eta_0-\eta) (-K)^{1/2}].
\eea
Note that the argument of $\sinh$ is the difference in 
development angle $\chi$ in an open
universe.
Of particular interest is the angular diameter distance to the horizon
$r_\theta(0)$
since many features in the CMB are generated early
\bel{eq:angulardiamdistexpl}
r_\theta(0) \approx \cases{ 2(\Omega_0 H_0)^{-1}  & $\Omega_\Lambda=0$ \cr
	2(\Omega_0 H_0^2)^{-1/2} (1+\ln\Omega_0^{0.085}). & $\Omega_\Lambda
+ \Omega_0 = 1$ }
\ee
In the flat case, $r_\theta(0)=\eta_0$.

A microphysical scale, the mean free path of a photon to Compton scattering,
is also of interest for the CMB,
\bel{eq:comptonlength}
\lambda_C = (x_e n_e\sigma_Ta/a_0)^{-1}
    = 4.3404 \times 10^{4} (1-Y_p/2)^{-1} (x_e \Omega_b h^2)^{-1} 
	(1+z)^{-2} {\rm Mpc.} 
\ee 
The diffusion length is roughly the geometric mean of $\lambda_C$ 
and the horizon $\eta$.  More precisely, it is
given by equation \eqn{damplengthfull} as
\bel{eq:damplengthfull2}
\lambda_D^2 \sim  
	k_D^{-2} = {1 \over 6}\int d\eta {1 \over \dot \tau} {{R^2 + 4f_2^{-1}(1+R)/5} 
	\over (1+R)^2}.
\ee
where 
\bel{eq:definef2}
f_2 = \cases{ 1 & isotropic, unpolarized \cr
	      9/10 & unpolarized \cr
	      3/4 & polarized \cr }
\ee
where isotropic means that the angular dependence of Compton scattering
has been neglected,  and the polarization case accounts for feedback
from scattering induced polarization. 
Throughout the main text, we have used $f_2=1$ for simplicity. 
If the
diffusion scale is smaller than the sound horizon, acoustic oscillations
will be present in the CMB.  The sound horizon is given by
\bel{eq:SoundHorizon2}
r_s = \int_0^\eta c_s  d\eta'
= {2 \over 3} {1 \over k_{eq}} \sqrt{6 \over R_{eq}}
  \ln { \sqrt{1+R} + \sqrt{ R + R_{eq} }
    \over
        1 + \sqrt{ R_{eq}}},
\ee
which relates it to the horizon at equality $\eta_{eq} = (4-2\sqrt{2})
 k_{eq}^{-1}$, 
where
\beal{eq:keq}
k_{eq} \eal (2 \Omega_0 H_0^2 a_0/a_{eq})^{1/2} \nonumber \\
       \eal 9.67 \times 10^{-2} \Omega_0 h^2 (1-f_\nu)^{1/2}\To^{-2}{\rm Mpc}
	^{-1}, \nonumber  \\ 
\eta_{eq} \eal 12.1 (\Omega_0 h^2)^{-1} (1-f_\nu)^{1/2} \To^{2} {\rm Mpc},
\eea
with $k_{eq}$ as the wavenumber that passes the Hubble scale at equality.

\subsection{Angular Scales}

A physical scale at $\eta$ 
subtends an angle or equivalently a multipole on the
sky $\ell$ 
\bel{eq:anglesubtend}
\ell = kr_\theta(\eta) \approx \theta^{-1},  \qquad \ell \gg 1
\ee
where the angle-distance relation $r_\theta$ is given by equation 
\eqn{angulardiamdist}.  Three angular scales are of interest to the
CMB.  The sound horizon at last scattering determines the location
of the acoustic peaks
\beal{eq:acousticangle}
\ell_A \eal \pi {r_\theta(\eta_*) \over r_s(\eta_*)}, \nonumber\\
\ell_p \eal \cases{ m \ell_A & adiabatic \cr
		 ( m-{1 \over 2}) \ell_A, & isocurvature \cr}
\eea
where $\ell_p$ is the location of the $p$th acoustic peak.  
If $R_* \ll 1$, $\ell_A$ takes on a simple form
\bel{eq:ellA}
\ell_A = 172 \left( {z_* \over 10^3} \right)^{1/2} {f_G \over f_R},
\ee
where $f_R$ is the correction for the expansion during radiation domination
\beal{eq:fR}
f_R \eal (1+x_R)^{1/2} - x_R^{1/2}, \nonumber\\
x_R = {a_{eq}/a_*} 
	\eal 2.38 \times 10^{-2} (\Omega_0 h^2)^{-1} (1-f_\nu)^{-1} 
	\To^4 (z_*/10^3), 
\eea
and $f_G$ is the geometrical factor
\bel{eq:fG}
f_G \approx \cases { \Omega_0^{-1/2} & $\Omega_\Lambda=0$ \cr
	       1+\ln\Omega_0^{0.085}. & $\Omega_\Lambda+\Omega_0=1$ \cr}
\ee
The diffusion damping scale at last scattering subtends an angle given by
\bel{eq:ldamp}
\ell_D = k_D(\eta_*) r_\theta(\eta_*),
\ee
where $k_D(\eta_*)$ is the effective damping scale at last scattering
accounting for the recombination process.  From \S \ref{ss-5diffusion},
to order of magnitude it
is 
\bel{eq:ldampnumber}
\ell_D \sim 10^3 (\Omega_b/0.05)^{1/4}_{\vphantom{0}}
 \Omega_0^{-1/4} f_R^{-1/2} f_G, 
\ee 
if $\Omega_b h^2$ is low as required by nucleosynthesis.
The scaling is only approximate since the detailed physics of
recombination complicates the calculation of $k_D$ (see Appendix 
\ref{ss-5recombination}).
The curvature radius at the horizon distance (\ie\ early times)
subtends an angle given by
\beal{eq:lcurvature}
\ell_K \eapp \sqrt{-K}r_\theta(0)  \nonumber\\
       \eapp { 2 \sqrt{1-\Omega_0} \over \Omega_0  } .
\eea
This relation is also 
not exact since for reasonable $\Omega_0$, the curvature scale
subtends a large angle on the sky and the small angle approximation 
breaks down.  
Note also that at closer distances 
as is relevant for the late ISW effect, the curvature scale subtends an
even larger angle on the sky than this relation predicts.

\section{Normalization Conventions}
\label{sec-Bnormalization}

\subsection{Power Spectra}

There are unfortunately a number of normalization conventions used in 
the literature and indeed several that run through the body of this
work.  Perhaps the most confusing conventions are associated with
open universes.  The power in fluctuations is  expressed alternately 
per
logarithmic intervals of the Laplacian wavenumber $k$ or the eigenfunction
index $\nu = \tilde k/\sqrt{-K}$, 
$\tilde k = (k^2 + K)^{1/2}$.  The relation between the two follows
from the identity $kdk = \tilde k d\tilde k$,
\bel{eq:openmapping}
\tilde P_X (\tilde k) = { k \over \tilde k} P_X(k) ,
\ee
where $P_X$ is the power spectrum of fluctuations in $X$.  For example,
our power law spectra
\beal{eq:SPhiconv}
|\Phi(0,k)|^2 \eal Bk^{n-4},  \nonumber\\
|S(0,k)|^2 \eal Ck^m, 
\eea
become
\beal{eq:SPhimapping}
 |\tilde \Phi(0,\tilde k)|^2 \eal B (1-K/\tilde k^2)^{(n-3)/2} \tilde k^{n-4},
	\nonumber\\
 |\tilde S(0,\tilde k)|^2 \eal C(1-K/\tilde k^2)^{m/2} \tilde k^m .
\eea
To add to the confusion, adiabatic fluctuations are often expressed
in terms of the density power spectrum at present $P(k) = |\Delta_T(\eta_0,k)|^2$. 
The two conventions are related by the Poisson equation,
\bel{eq:Poissonnorm}
(k^2 - 3K) \Phi = {3 \over 2} \Omega_0 H_0^2 (1+ a_{eq}/a){a_0 \over a} 
		\Delta_T. 
\ee
To account for the growth between the initial conditions and the present,
one notes that at large scales ($k\rightarrow 0$) the growth function
is described by pressureless linear theory.   From equations
\eqn{AeqnTotalfit} and \eqn{AeqnPhiPsi}, 
\bel{eq:PsiDeltamapping}
\Delta_T(\eta_0,k) = {3 \over 5}(\Omega_0 H_0^2)^{-1} 
		     \left[1+ {4 \over 15} f_\nu \right]  
		     \left[1+ {2 \over 5} f_\nu \right]^{-1} 
		     (1-3K/k^2) {D \over D_{eq}}{a_{eq} \over a} \Phi(0,k).
\ee
If the neutrino anisotropic stress is neglected, drop the $f_\nu$ factors
for consistency.
Thus for a normalization convention of $P(k) =Ak^n$ at large scales
\bel{eq:ABmapping}
A = {9 \over 25}(\Omega_0 H_0^2)^{-2} \left[1+ {4 \over 15} f_\nu \right]^2  
    \left[1+ {2 \over 5} f_\nu \right]^{-2} (1-3K/k^2)^2
    \left( {D \over D_{eq}} {a_{eq} \over a} \right)^2 B.
\ee
Notice that in an open universe, power law conditions for the potential
do not imply power law conditions for the density,
\beal{eq:powerlaw}
P(k) \eaprop (k^2 - 3K)^2 k^{n-4}, \nonumber\\
\tilde P(\tilde k) \eaprop  \tilde k^{-1}(\tilde k^2 - K)^{-1} 
	(\tilde k^2 - 4K)^2 (\tilde k^2 - K)^{(n-1)/2}.
\eea
$\tilde P(\tilde k)$ is the form most often quoted in
the literature \cite{Wilson,HSa,RP}. 

The power spectrum may also be expressed in terms of the 
bulk velocity field.
At late times, pressure can be neglected and 
the total continuity equation \eqn{Total} reduces to
\beal{eq:Vnorm}
kV_T \eal - \dot \Delta_T \nonumber\\
     \eal - {\dot a \over a} {d \ln D \over d\ln a} \Delta_T,
\eea
and in particular
\bel{eq:Vnormnow}
kV_T(\eta_0,k) = -H_0 {d \ln D \over d\ln a}\bigg|_{\eta_0} \Delta_T(\eta_0,k),
\ee
or
\bel{eq:Vpower}
P_V(k) \equiv |V_T(\eta_0,k)|^2 = H_0^2 \left({d \ln D \over d\ln a}\right)^2
\bigg|_{\eta_0} P(k),
\ee
for the velocity power spectrum. 
Recall from equation \eqn{Dasymptotic} that 
$d\ln D / d\ln a \approx \Omega_0^{0.6}$ in an open universe.

\input tables/data.tex
\subsection{Anisotropies}

The anisotropy power spectrum $C_\ell$ is given by 
\bel{eq:Cltrans}
{2\ell+1 \over 4\pi } C_\ell = \int {dk \over k} T_\ell^2(k) 
\times \cases{ k^3 |\Phi(0,k)|^2 & adiabatic \cr
	       k^3 |S(0,k)|^2, & isocurvature \cr }
\ee 
where $T_\ell(k)$ is the radiation transfer function from the solution
to the Boltzmann equation.  Examples are given in \S\ref{ch-primary}.
The power measured by a given experiment with a window function $W_\ell$
has an ensemble average value of 
\bel{eq:dttrms}
\left({\Delta T \over T} \right)^2_{rms} 
= {1 \over 4\pi} \sum_{\ell} (2\ell+1)C_\ell
W_\ell.
\ee
Only if the whole sky is measured at high signal to noise
does the variance follow the ``cosmic variance'' prediction of 
a $\chi^2$ with $2\ell+1$ degrees of freedom.  Real experiments make
noisy measurements of a 
fraction of the sky and therefore require a more detailed
statistical treatment. 
To employ likelihood techniques, we must assume some underlying
power spectrum.
In order to divorce the measurement from theoretical
prejudice, experimental results are usually quoted with a model independent
choice.  The two most common conventions are the gaussian autocorrelation
function $C_{\rm gacf}(\theta) = C_0 \exp(-\theta^2/2\theta_c^2)$
and the ``flat'' power spectrum motivated by the Sachs-Wolfe tail of
adiabatic models (see \eg\ \cite{WSS}), 
\beal{eq:gacfflat}
C_{\ell \rm gacf} \eal
	2\pi C_0 \theta_c^2 \exp[-\ell(\ell+1)\theta_c^2/2], \nonumber\\
C_{\ell \rm flat} \eal {24\pi \over 5}\left( {Q_{\rm flat} \over T_0 } 
	\right)^2 [\ell(\ell+1)]^{-1} .
\eea
The two power estimates are thus related by
\bel{eq:gacfflatrelation}
Q_{\rm flat}^2 {6 \over 5} \sum_\ell {2 \ell + 1 \over \ell (\ell+1) } W_\ell 
 = C_0 \theta_c^2  {1 \over 2} \sum_\ell (2\ell + 1) \exp[-\ell(\ell+1)
	\theta_c^2/2] W_\ell .
\ee
The current status of measurements is summarized in Tab.~\ref{tab:data}
\cite{SSW}.

\subsection{Large Scale Structure}

Large scale structure measurements probe a smaller scale and have
yet another set of normalization conventions based on the two point
correlation
function of astrophysical objects
\bel{eq:correlation}
\xi_{ab}(\bx) = \left< \delta\rho_a(\bx'+\bx)\delta\rho_b(\bx')/
	\bar\rho_a\bar\rho_b \right> .
\ee
If all objects are clustered similarly, 
then all $\xi_{aa} = \xi$ and the two-point correlation
function is  related to the underlying power spectrum by
\beal{eq:twopt}
\xi(r) \eal {V \over 2\pi^2} 
	\int {dk \over k} k^3 P(k) X_\nu^0(\sqrt{-K}r)  \nonumber\\
       \eapp {V \over 2\pi^2} \int {dk \over k} k^3 P(k) {\sin(kr) \over kr},
\eea
where the approximation assumes that scales of interest are well
below the curvature scale.  
The normalization of the power spectrum is often quoted by the $N$th 
moment of the correlation function $J_N(r) = \int_0^r \xi(x) x^{(N-1)} dx$
which implies
\bel{eq:J3}
J_3(r) = {V \over 2\pi^2} \int {dk \over k} P(k) (kr)^2 j_1(kr).
\ee
For reference, $j_1(x) = x^{-2}\sin x - x^{-1}\cos x$.
Another normalization convention involves the rms density fluctuation in
spheres of constant radii
\bel{eq:sigma}
\sigma^2(r) = {V \over 2\pi^2} \int {dk \over k} k^3 P(k) \left( {3 j_1(kr)
	\over kr} \right)^2.
\ee
The observed galaxy distribution implies that 
\beal{eq:observedvalues}
J_3(10 h^{-1}{\rm Mpc}) \eapp 270 h^{-3} {\rm Mpc}^3 \\
\sigma_8 \equiv \sigma(8 h^{-1} {\rm Mpc}) \eal \cases{1.1 \pm 0.15 & 
	optical \cite{Loveday} \cr
		0.69 \pm 0.04.  & 
	IRAS  \cite{Fisher} \cr }
\eea
The discrepancy between estimates of the normalization obtained by
different populations of objects implies that they may all be biased
tracers of the underlying mass.
The simplest model for bias assumes $\xi_{aa} = b_a^2 \xi$ with 
constant $b$.  Peacock \& Dodds \cite{Peacock} find that the best fit
to the Abell cluster (A), radio galaxy (R), optical galaxy (O), 
and IRAS galaxy (I) data sets yields $b_A: b_R: b_O: b_I = 4.5: 1.9: 1.3: 1$.
\eject

\input tables/definitions.tex

%% file: tables/conversions.tex
\begin{table}[t]
\def\phantom{\vphantom{$\Big[$}}

\begin{center}
\begin{tabular}{|l l|}
\hline
$1\, {\rm s}$ & $= 9.7157 \times 10^{-15}\, {\rm Mpc}^{\vphantom{A}^
	{\vphantom{A}}}$  \\
$1\, {\rm yr}$ & $=3.1558 \times 10^{7}\, {\rm s}$ \\
$1\, {\rm Mpc}$ & $= 3.0856 \times 10^{24}\, {\rm cm}$ \\
$1\, {\rm AU}$ & $= 1.4960 \times 10^{13}\, {\rm cm}$ \\
$1\, {\rm K}$ & $= 8.6170 \times 10^{-5}\, {\rm eV}$ \\
$1\, {\rm M_\odot}$ &$ = 1.989 \times 10^{33}\, {\rm g}$ \\
$1\, {\rm GeV}$ & $= 1.6022 \times 10^{-3}\, {\rm erg}$ \\
 & $= 1.7827 \times 10^{-24}\, {\rm g}$ \\
 & $= (1.9733 \times 10^{-14}\, {\rm cm})^{-1}$ \\
 & $= (6.5821 \times 10^{-25}\, {\rm s})^{-1}_{\vphantom{A}_{\vphantom{A}}}$ \\ 
\hline
\end{tabular}
\end{center}

\begin{center}
\begin{tabular}{|l l|}
\hline
Planck's constant & $ \hbar  
	= 1.0546 \times 10^{-27}\, {\rm cm}^2 \, {\rm g} \, {\rm s}^{{-1}
	^{\vphantom{A}}}$ 
	\\
Speed of light &$ c  = 2.9979 \times 10^{10}\, {\rm cm}\, {\rm s}^{-1}$ \\
Boltzmann's constant &$
	 k_B  =  1.3807 \times 10^{-16}\, {\rm erg}\,{\rm K}^{-1}$ \\
Fine structure constant &$ \alpha = 1/137.036$ \\
Gravitational constant &$ G = 6.6720 \times 10^{-8}\, 
	{\rm cm}^3 \, {\rm g}^{-1} \, {\rm s}^{-2}$ \\
Stefan-Boltzmann constant &$ \sigma = ac/4 = \pi^2 k_B^4 / 60 \hbar^3 c^2$ \\
&$ a = 7.5646 \times 10^{-15}\, {\rm erg}\, {\rm cm}^{-3} {\rm K}^{-4}$ \\
Thomson cross section & $\sigma_T = 8\pi\alpha^2/3m_e^2 =
6.6524 \times 10^{-25}\, {\rm cm}^2$ \\
Electron mass & $m_e = 0.5110\, {\rm MeV}$ \\
Neutron mass & $m_n = 939.566\, {\rm MeV}$ \\
Proton mass & $m_p = 938.272\, {\rm MeV}_{\vphantom{A}_{\vphantom{A}}}$ \\
\hline
\end{tabular}
\end{center}

\caption{Physical Constants and Conversion Factors}
\label{tab:conversions}
\end{table}

%% file: tables/redshifts.tex
\begin{table}[p]
\def\phantom{\vphantom{$\Big[$}}
\begin{center}
\begin{tabular}{|l l|}
\hline
\dsp
Epoch & Definition \phantom\\ 
\hline
$z_*=10^3\Omega_b^{-0.027/(1+0.11\ln\Omega_b)}\qquad \Omega_0=1$ 
	& Last scattering (recomb.) \vphantom{$\Big[_A^A$}\\
$\hphantom{z_*}= 10^2 (\Omega_0 h^2/0.25)^{1/3} 
	(x_e\Omega_b h^2/0.0125)^{-2/3}$
	& Last scattering (reion.) \phantom\\
$z_d = 160(\Omega_0 h^2)^{1/5} x_e^{-2/5}$ 
	& Drag epoch \phantom\\
$z_{eq} = 4.20 \times 10^4 \Omega_0 h^2 (1-f_\nu) \To^{-4} $
	& Matter-radiation equality \phantom\\
$z_{b\gamma} = 3.17 \times 10^4 \Omega_b h^2 \To^{-4} $
	& Baryon-photon equality \phantom\\
$z_H = (1+z_{eq})\{4(k/k_{eq})^2/[1+(1+8(k/k_{eq})^2)^{1/2}]\} -1 $
	& Hubble length crossing \phantom\\
$z = (1-\Omega_0-\Omega_\Lambda)/\Omega_0 - 1$ 
	& Matter-curvature equality \phantom\\
$z = (\Omega_\Lambda/\Omega_0)^{1/3}-1  $
	& Matter-$\Lambda$ equality \phantom\\
$z = [\Omega_\Lambda/(1-\Omega_0-\Omega_\Lambda)]^{1/2} -1 $
	& Curvature-$\Lambda$ equality \phantom\\
$z_{cool} = 9.08 \To^{-16/5} f_{cool}^{2/5} (\Omega_0 h^2)^{1/5} -1$
	& Compton cooling era\phantom\\
$z> 4\sqrt{2} z_K$ 
	& Bose-Einstein era \phantom\\
$z < z_K/8 $
	& Compton-$y$ era \phantom\\
$z_K = 7.09\times 10^3 (1-Y_p/2)^{-1/2} (x_e\Omega_b h^2)^{-1/2} \To^{1/2}$
    	& Comptonization epoch \phantom\\
$z_{\mu,dc} = 4.09 \times 10^5 (1-Y_p/2)^{-2/5} (x_e\Omega_b h^2)^{-2/5} 
	\To^{1/5} $
	& Dbl. Compton therm.~epoch \phantom\\
$z_{\mu,br} = 5.60 \times 10^4 (1-Y_p/2)^{-4/5} (x_e\Omega_b h^2)^{-6/5}
	\To^{13/5} $
	& Bremss. therm.~epoch \phantom\\
\hline
$\To = T_0/2.7$K$\approx 1.01$ & Temperature Scaling \phantom\\ 
$Y_p = 4 n_{He}/n_b \approx 0.23$ & Helium mass fraction \phantom\\
$(1-f_\nu)^{-1} = 1+\rho_\nu/\rho_\gamma \rightarrow 1.68132$ & Neutrino
density correction  \phantom\\
$k_{eq} = (2\Omega_0 H_0^2 a_0/a_{eq})^{1/2}$
	 & Equality Hubble wavenumber 
	\phantom\\
$\hphantom{k_{eq}}
	= 9.67\times 10^{-2}\Omega_0 h^2 (1-f_\nu)^{1/2}\To^{-2}$Mpc$^{-1}$
	 & \phantom\\
$f_{cool} = x_e^{-1} [(1+x_e)/2 - (3+2x_e) Y_p/8]
(1-Y_p/2)^{-1}$&  Cooling correction factor \phantom\\
\hline
\end{tabular}
\end{center}
\caption{Critical Redshifts}
\mycaption{Critical epochs are also denoted as the corresponding
value in the coordinate time $t$, 
scale factor $a$, and conformal time $\eta$.  The neutrino fraction 
$f_\nu$ is given for three families of massless neutrinos and the
standard thermal history. The Hubble crossing redshift $z_H$ is given
for the matter and radiation dominated epochs. }
\label{tab:redshifts}
\end{table}

%% file: tables/data.tex
\begin{table}[t]
\def\phantom{\vphantom{$\Big[$}}

\begin{center}
\begin{tabular}{|l l l l l l|}
\hline
Experiment & $\ell_0$ & $\ell_1$ & $\ell_2$ & $Q_{\rm flat}(\mu$K) & Ref. \\
\hline
COBE	& --	& --	&$ 18$	& $ 19.9 \pm 1.6$& \cite{Gorski}\\
FIRS	& --	& --	&$ 30$ 	& $ 19 \pm 5 $ 	 & \cite{GPCM}  \\
Ten.	&$ 20 	$&$ 13 	$&$ 30	$&$ 26 \pm 6 $   & \cite{Hancock}\\
SP94	&$ 67	$&$ 32	$&$ 110	$&$ 26 \pm 6 $	 & \cite{SP94}\\
SK	&$ 69	$&$ 42	$&$ 100 $&$ 29 \pm 6 $   & \cite{Netterfield}\\
Pyth.	&$ 73	$&$ 50	$&$ 107	$&$ 37 \pm 12 $  & \cite{Dragovan}\\
ARGO	&$ 107 	$&$ 53	$&$ 180	$&$ 25 \pm 6 $	 & \cite{deBernardis}\\
IAB	&$ 125	$&$ 60	$&$ 205	$&$ 61 \pm 27 $	 & \cite{Piccirillo}\\
MAX-2 $(\gamma$UMi)
	&$ 158	$&$ 78	$&$ 263	$&$ 74 \pm 31 $  & \cite{Alsop}\\
MAX-3 $(\gamma$UMi)
	&$ 158	$&$ 78	$&$ 263	$&$ 50 \pm 11 $	 & \cite{Gundersen}\\
MAX-4 $(\gamma$UMi)
	&$ 158	$&$ 78	$&$ 263	$&$ 48 \pm 11 $  & \cite{Devlin}\\
MAX-3 $(\mu$Peg)
	&$ 158	$&$ 78	$&$ 263	$&$ 19 \pm 8 $	 & \cite{Meinhold}\\
MAX-4 $(\sigma$Her)
	&$ 158	$&$ 78	$&$ 263	$&$ 39 \pm 8 $ 	 & \cite{Clapp}\\
MAX-4 $(\iota$Dra)
	&$ 158	$&$ 78	$&$ 263	$&$ 39 \pm 11 $  & \cite{Clapp}\\
MSAM2	&$ 143	$&$ 69	$&$ 234	$&$ 40 \pm 14 $	 & \cite{Cheng}\\
MSAM3 	&$ 249	$&$ 152 $&$ 362	$&$ 39 \pm 12 $  & \cite{Cheng}\\
\hline
\end{tabular}
\end{center}

\caption{Anisotropy Data Points}
\mycaption{A compilation of anisotropy measurements from \cite{SSW}.
The experimental window function peaks at $\ell_0$ and falls to half
power at $\ell_1$ and $\ell_2$.  Points are plotted in Fig.~\ref{fig:1cdm}.}
\label{tab:data}
\end{table}

%% file: tables/definitions.tex
\section{Symbol Index}
\def\phantom{\vphantom{$A^{A^{A^A}}_{A_A}$}}
\begin{table}[h]
\vskip 0.5truecm
\begin{center}
\begin{tabular}{|c l c|}
\hline
Symbol & Definition & Equation \\
\hline
$\Gamma_i$ & Entropy fluctuation in $i$ & \eqn{entropygamma} 
	\hfill\phantom\\
$\Delta_i$ & T gauge density fluctuation in $i$ & \eqn{totalmatterpert} 
	\hfill\phantom\\
$\Lambda$ & Cosmological constant, $3H_0^2 \Omega_\Lambda$
         & \eqn{Hubbledefinition}
	\hfill\phantom\\
$\Theta$ &N Gauge CMB $\Delta T/T$ & \eqn{tempoperator} 
	\hfill\phantom\\
$\Theta_\ell$ & CMB $\ell$th multipole & \eqn{LDecomposition} 
	\hfill\phantom\\
$\To$ & $T_0/2.7$K & \eqn{BRt} 
	\hfill\phantom\\
$|\Theta+\Psi|_{rms}$ & CMB rms fluctuation & \eqn{RMSL} 
	\hfill\phantom\\
$\Pi_i$ & Anisotropic stress in $i$ & \eqn{stressenergy} 
	\phantom\\
$\Psi$ & Newtonian potential & \eqn{metric} 
	\phantom\\
$\Phi$ & Space curvature perturbation & \eqn{metric} 
	\phantom\\
$\Omega_i$ & Fraction of critical density in $i$ & \eqn{Hubbledefinition} 
	\phantom\\
$\gamma_i$ & Photon direction cosines & \eqn{pgamma} 
	\phantom\\
$\gamma_{ij}$ & Comoving three metric & \eqn{FRWmetric} 
	\phantom\\
$\delta_i^G$ & G gauge density fluctuation in $i$ & \eqn{gaugedensities} 
	\phantom\\
$\eta$ & Conformal time & \eqn{timedefinition}
	\phantom\\
$\mu$ & Dimensionless chemical potential & \eqn{BE} 
	\phantom\\
$\nu$ & Dimensionless eigenmode index & \eqn{AeqnHorribleFunc} 
	\phantom\\
$\rho_{crit}$ & Critical density $3H_0^2/8\pi G$ & \eqn{rhocrit}
	\phantom\\
$\rho_i$ & Energy density in $i$ & \eqn{stressenergy} 
	\phantom\\
$\sigma_T$ & Thomson cross section & \eqn{Thomson} 
	\phantom\\
$\tau$ & Thomson optical depth & \eqn{opticaldepthder} 
	\phantom\\
$\tau_K$ & Comptonization optical depth $4y$ & \eqn{ydef} 
	\phantom\\
$\tau_{abs}$ & Thermalization optical depth & \eqn{Ys} 
	\phantom\\
\hline
\end{tabular}
\end{center}
%\caption{Commonly Used Symbols}
%\label{common}
\end{table}

\begin{table}[t]
\begin{center}
\begin{tabular}{|c l c|}
\hline
$\chi$ & Curvature normalized radial coordinate & \eqn{radial} 
	\phantom\\
${\cal D}$ & Diffusion damping factor & \eqn{DampLS} 
	\phantom\\
${\cal G}$ & Drag growth factor & \eqn{DragGrow} 
	\phantom\\
$A$ & Final matter power spectrum norm. & \eqn{ABnormalization} 
	\phantom\\
$A^G$ & Time-time G gauge metric fluctuation & \eqn{generalmetric} 
	\phantom\\
$B$ & Initial power spectrum norm. & \eqn{ABnormalization} 
	\phantom\\
$B^G$ & Time-space G gauge metric fluctuation & \eqn{generalmetric} 
	\phantom\\
$C_\ell$ & Anisotropy power spectrum & \eqn{Cl} 
	\phantom\\
$D$ & Pressureless growth factor & \eqn{replacementD} 
	\phantom\\
$F$ & Gravitational driving force & \eqn{BeqnForceRepeat} 
	\phantom\\
$G$ & Gravitational constant & \eqn{einstein} 
	\phantom\\
$H$ & Hubble parameter & \eqn{Hubbledefinition} 
	\phantom\\
$H_L^G$ & Diagonal space-space G gauge metric pert. & \eqn{generalmetric} 
	\phantom\\
$H_T^G$ & Traceless space-space G gauge metric pert. & \eqn{generalmetric} 
	\phantom\\
$N_\ell$ & Neutrino $\ell$th multipole & \eqn{Hierarchy} 
	\phantom\\
$K$ & Curvature & \eqn{FRWmetric} 
	\phantom\\
$K_\ell$ & $1- (\ell^2-1)K/k^2$ & \eqn{AeqnMFact} 
	\phantom\\
$Q$ & General Laplacian eigenfunction & \eqn{eigenmodes} 
	\phantom\\
$R$ & Baryon-photon norm.~scale factor $3\rho_b/4\rho_\gamma$ & \eqn{Baryon}
	\phantom\\
$S$ & Matter-radiation entropy fluctuation & \eqn{Sdefinition} 
	\phantom\\
$T_\ell$ & Radiation transfer function & \eqn{Weight} 
	\phantom\\
$T$ & CMB temperature & \eqn{Y} 
	\phantom\\
$T_e$ & Electron temperature & \eqn{EDist} 
	\phantom\\
$U_A$ & Adiabatic mode growth function & \eqn{BeqnUAD} 
	\phantom\\
$U_I$ & Isocurvature mode growth function & \eqn{BeqnUAD} 
	\phantom\\
$V_i$ & N/T gauge velocity in $i$ & \eqn{stressenergy} 
	\phantom\\
$X_\nu^\ell$ & Radial eigenfunction & \eqn{AeqnHorribleFunc} 
	\phantom\\
$a$ & Scale factor & \eqn{FRWmetric} 
	\phantom\\
$c_i$ & Sound speed in $i$ & \eqn{entropygamma} 
	\phantom\\
\hline
\end{tabular}
\end{center}
\end{table}

\begin{table}[t]
\begin{center}
\begin{tabular}{|c l c|}
\hline
$c_s$ & Photon-baryon fluid sound speed & \eqn{SoundSpeed} 
	\phantom\\
$f$ & Photon distribution function & \eqn{collisionless} 
	\phantom\\
$h$ & Dimensionless Hubble constant &  \eqn{Hubblevalue}
	\phantom\\
$k$ & Laplacian eigenvalue and wavenumber & \eqn{eigenmodes} 
	\phantom\\
$\tilde k$ & Eigenmode number $\nu \sqrt{-K}$ & \eqn{decompconventionalt}
	\phantom\\	
$k_D$ & Diffusion damping wavenumber & \eqn{damplengthfull} 
	\phantom\\
$k_{eq}$ & Equality horizon wavenumber & \eqn{BeqnCTotal} 
	\phantom\\
$\ell$ & Multipole number & \eqn{LDecomposition} 
	\phantom\\
$\ell_A$ & Acoustic angular scale & \eqn{acousticangle}
	\phantom\\
$\ell_D$ & Damping angular scale & \eqn{ldampnumber}
	\phantom\\
$\ell_p$ & $p$th acoustic peak scale & \eqn{acousticangle}
	\phantom\\
$m$ & Isocurvature spectral index & \eqn{Weight} 
	\phantom\\
$n$ & Adiabatic spectral index & \eqn{Weight} 
	\phantom\\
$n_i$ & Number density in $i$ & \eqn{deltaoverw} 
	\phantom\\
$p$ & Photon frequency/momentum & \eqn{momentum} 
	\phantom\\
$p_i$ & Pressure in $i$ & \eqn{stressenergy} 
	\phantom\\
$r_\theta$ & Conf. angular diameter distance & \eqn{rtheta} 
	\phantom\\
$r_s$ & Sound horizon & \eqn{SoundHorizon} 
	\phantom\\
$x_e$ & Electron ionization fraction & \eqn{EDist}  
	\phantom\\
$x_p$ & Dimensionless frequency $p/T_e$ & \eqn{boltzmannagain} 
	\phantom\\
$v_i$ & Real space velocity in $i$ & \eqn{EDist} 
	\phantom\\
$v_i^G$ & G gauge velocity in $i$ & \eqn{gaugedensities} 
	\phantom\\
$w_i$ & Equation of state for $i$, $p_i/\rho_i$ & \eqn{zeroterms} 
	\phantom\\
$y$ & Compton-$y$ parameter & \eqn{ydef2} 
	\phantom\\
$z$ & Redshift & \eqn{FRWmetric}
	\phantom\\
$z_h$ & Early energy injection redshift & \eqn{Peak} 
	\phantom\\
$z_i$ & Ionization redshift & \eqn{GrowLate}
	\phantom\\
\hline
\end{tabular}
\end{center}
\caption{Commonly Used Symbols}
\mycaption{T $=$ total matter gauge, N $=$ Newtonian gauge, G $=$ arbitrary
gauge, $i = \gamma,b,e,\nu,c,m,r,T$ for photons, baryons, electrons, neutrinos,
cold collisionless matter, non-relativistic matter, radiation, and total
energy density. Other redshifts of interest are 
listed in Tab.~\ref{tab:redshifts}}
\label{common}
\end{table}

%% file: thesis.bbl
\begin{thebibliography}{1}

\def\etal {{\it et al.}}

\bibitem{AllKor} Allen, B. \& Koranda, S. PRD {\bf 50}, 3713 (1994)

\bibitem{Alsop} Alsop, D.C. \etal\ ApJ {\bf 395}, 317 (1992)

\bibitem{AS} Abbott, L.F. \& Schaefer, R.K. ApJ
{\bf 308}, 546 (1986)

\bibitem{AbbWis} Abbott, L.F. \& Wise, M.B. Nucl. Phys. {\bf B244}, 541 (1984)

\bibitem{Bardeen} Bardeen, J.M.  PRD {\bf 22}, 1882 (1980)

\bibitem{BBKS} Bardeen, J.M., Bond, J.R., Kaiser, N. \& Szalay, A.S.
ApJ {\bf 304}, 15 (1986)

\bibitem{BarSil} Bartlett, J.G. \& Silk, J. ApJ {\bf 423}, 12 (1994)


\bibitem{BS} Bartlett, J.G.  and Stebbins, A. ApJ {\bf 371}, 8 (1991)

\bibitem{Bennett92} Bennett, C.L. \etal\ ApJ {\bf 396}, L7 (1992)

\bibitem{Bennett94} Bennett, C.L. \etal\ ApJ {\bf 436}, 432 (1994)

\bibitem{Bernstein} Bernstein, J., {\it Relativistic Kinetic Theory},
(Cambridge University, 1988).  

\bibitem{BD} Bernstein, J. \& Dodelson, S. PRD {\bf 41}, 354
(1990).

\bibitem{BH} Birkinshaw, M. \& Hughes, J.P.
ApJ {\bf 420}, 331 (1994)

\bibitem{BlaSch} Blanchard, A. \& Scheider A\&A {\bf 184}, 1 (1987)

\bibitem{Bond} Bond, J.R. in {\it The Early Universe}, 
(eds Unruh, W.G. and Semenoff, G.W.) 
(Reidel, Dordrecht, 1988) p.~283 

\bibitem{BE84} Bond, J.R. \& Efstathiou, G. ApJL {\bf 285}, L45
(1984)

\bibitem{BE87} Bond, J.R. \& Efstathiou, G. MNRAS
{\bf 227}, 655 (1987)

\bibitem{Confusion} Bond, J.R. \etal\ PRL {\bf 72}, 13 (1994)

\bibitem{Brandt} Brandt, C.L. \etal\ ApJ {\bf 424}, 1 (1994) 

\bibitem{Bucher} Bucher, M., Goldhaber, A.S., \& Turok hepph-9411206

\bibitem{Bunn} Bunn, E. {\it Statistical Analysis of Cosmic Microwave Background
Anisotropies}, UCB thesis (1995)

\bibitem{BSW} Bunn, E.,  Scott, D.  \&  White, M. 
ApJL {\bf 441}, L9 (1995)

\bibitem{BS}  Bunn, E., \& Sugiyama, N.  ApJ {\bf 446}, 49 (1995)

\bibitem{BDDa}  Burigana, C., Danese, L. \& De Zotti, G., 
A\&A {\bf 246}, 49 (1991)

\bibitem{BDDb}  Burigana, C., Danese, L. \& De Zotti, G., 
ApJ {\bf 379}, 1 (1991)

\bibitem{CPT} Carroll, S.M., Press, W.H. \& Turner, E.L. ARA\&A
{\bf 30}, 499 (1992)

\bibitem{CelBar} Ceballos, M.T. \& Barcons, X. MNRAS {\bf 271}, 817 (1994)

\bibitem{COP} Cen, R., Ostriker, J.P. \&  Peebles, P.J.E.  ApJ
{\bf 415}, 423 (1993)

\bibitem{CJ}  Chan, K.L \& Jones, B.J.T.  ApJ {\bf 195},1 (1975)

\bibitem{Cheng} Cheng, E.S. \etal\ ApJL {\bf 422}, L37 (1994)

\bibitem{CSS} Chiba, T.,  Sugiyama, N. \&  Suto, Y.  ApJ
{\bf 429}, 427 (1994)

\bibitem{Clapp} Clapp, A.C. \etal\ ApJL {\bf 433}, L57 (1994)

\bibitem{ColMRV} Colafrancesco, S., Mazzotta, P., Rephaeli, Y. \&
Vittorio, N. ApJ {\bf 433}, 454 (1994)

\bibitem{ColEfs} Cole, S. \& Efstathiou, E. MNRAS {\bf 239}, 195 (1989)

\bibitem{CouCT} Coulson, D., Crittenden, R.G., \& Turok, N. PRL {\bf 73},
2390 (1994)

\bibitem{CritCT} Crittenden, R.G., Coulson, D. \& Turok, N. PRD {\bf 52} R5402 (1995) 

\bibitem{Crittenden} Crittenden, R. \etal\ PRL {\bf 71}, 324
(1994)

\bibitem{EvilOne} Daly, R.A. ApJ {\bf 371}, 14 (1991)

\bibitem{DD77} Danese, L. \& De Zotti, G. Riv. Nuovo Cimento {\bf 7},
277 (1977)

\bibitem{DD80} Danese, L. \& De Zotti, G. A\&A {\bf 84}, 364 (1980)

\bibitem{DD82}  Danese, L. \&  De Zotti, G. A\&A {\bf 107}, 39
(1982)

\bibitem{DSS} Davis, M., Summers, F.J. \& Schlegel, D. Nature 
{\bf 359}, 393 (1992)

\bibitem{deBernardis} de Bernardis, P. \etal\ ApJL {\bf 422}, L33 (1994)

\bibitem{Devlin} Devlin, M. \etal\ ApJL {\bf 430}, L1 (1994)

\bibitem{DJ93} Dodelson, S. \& Jubas, J.M. PRL {\bf 70}, 2224 (1993)

\bibitem{DJ95} Dodelson, S. \& Jubas, J.M. ApJ {\bf 439}, 503 (1995)

\bibitem{DKK} Dodelson, S., Knox, L. \& Kolb, E.W. PRD {\bf 72}, 3444 (1994)

\bibitem{DZS} Doroshkevich, A.G., Zel'dovich, Ya.B., \& Sunyaev, R.A.
Sov. Astron {\bf 22}, 523 (1978)

\bibitem{Dragovan} Dragovan, M. \etal\ ApJL {\bf 427}, 67 (1993)

\bibitem{Efstathiou} Efstathiou, G. {\it  Large Scale Motions in the
Universe: A Vatican Study Week}, eds. Rubin, V.C. and Coyne, G.V.,
(Princeton University, Princeton, 1988) pg. 299


\bibitem{EB}  Efstathiou, E. \& Bond, J.R. MNRAS {\bf 227}, 33p (1987)

\bibitem{EBW} Efstathiou, G., Bond, J.R. \& White, S.D.M.
MNRAS {\bf 258}, P1 (1992)

\bibitem{Ellis}  Ellis, J. \etal, Nuc. Phys. B {\bf 373}, 399 (1992)

\bibitem{FabPol} Fabbri, R. \& Pollock, M. Ap. Phys. Lett. {\bf B125}, 445 (1983)

\bibitem{Fisher} Fisher, K.B., Davis, M., Strauss, M., Yahil, A. \&
Huchra, J. MNRAS {\bf 266}, 50 (1994)

\bibitem{Freedman} Freedman, W.L., \etal\ Nature {\bf 371}, 27 (1994)

\bibitem{GPCM} Ganga, K., Page, L., Cheng, E. \& Meyer, S. ApJL {\bf 432}, 
	L15 (1994)

\bibitem{GO} Gnedin, N.Y. \& Ostriker, J.P. ApJ
{\bf 400}, 1 (1992)

\bibitem{GOR} Gnedin, N.Y., Ostriker, J.P., \& Rees, M.J. ApJ {\bf 438}, 
40 (1995)

\bibitem{Goldman} Goldman, S.P. Phys. Rev. A {\bf 40}, 1185 (1989)

\bibitem{GSJ} Gorski, K.M., Stompor, R. \& Juskiewicz, R. ApJL {\bf 410}
L1 (1993)

\bibitem{Gorski} Gorski, K.M. \etal\ ApJL {\bf 430}, L89 (1994)

\bibitem{GSS} Gouda, N., Sasaki, M., Suto, Y., ApJ
{\bf 341}, 557 (1989)

\bibitem{GSS91} Gouda, N., Sugiyama, N., \& Sasaki, M. Prog.
Theor. Phys. {\bf 85}, 1023 (1991)

\bibitem{Graham} Graham, A.C. {\it Chuang-tzu: The Inner Chapters}
(Mandala, London 1989)

\bibitem{GC} Gregory, P.C., \& Condon, J.J. ApJS {\bf 75}, 1011 (1991)

\bibitem{Gundersen} Gundersen, J.O. \etal\ ApJL {\bf 413}, L1 (1993)

\bibitem{SP94} Gundersen, J.O. \etal\ ApJL {\bf 443}, L57 (1994)

\bibitem{GP} Gunn, J.E. \& Peterson, B.A. ApJL {\bf 318}, L11 (1965)

\bibitem{Hancock} Hancock, S. \etal\ Nature {\bf 367}, 333 (1994)

\bibitem{Harrison} Harrison, E.L. Rev. Mod. Phys. {\bf 39}, 862 (1967)

\bibitem{Holtzman} Holtzman, J.A. ApJS {\bf 71}, 1 (1989)

\bibitem{Hu} Hu, W, in {\it CWRU CMB Workshop: 2 Years after COBE},
eds. L. Krauss \& P. Kernan, (World Scientific, Singapore 1994) p. 188

\bibitem{HBS} Hu, W., Bunn, E., \& Sugiyama, N. ApJL {\bf 447}, L59 (1995) 

\bibitem{HSSa} Hu, W., Scott, D. \& Silk, J. PRD
{\bf 49}, 648 (1994)

\bibitem{HSSb} Hu, W., Scott, D. \&  Silk, J. ApJL {\bf 430},
L5 (1994)

\bibitem{HSSW} Hu, W., Scott, D., Sugiyama, N. \& White, M. PRD 
{\bf 52} 5498 (1995)

\bibitem{HSl} Hu, W. \& Silk, J. PRL {\bf 70}, 2661 (1993)

\bibitem{HS} Hu, W. \& Silk, J. PRD {\bf 48}, 485 (1994)

\bibitem{HSISW} Hu, W. \&  Sugiyama, N. PRD {\bf 50}, 627 (1994)

\bibitem{Models} Hu, W. \& Sugiyama, N. ApJ {\bf 436}, 456
(1994)

\bibitem{HSa} Hu, W. \& Sugiyama, N. ApJ {\bf 444}, 489 (1995)
 
\bibitem{HSb} Hu, W. \& Sugiyama, N. PRD {\bf 51}, 2599
(1995)

\bibitem{HSsmall} Hu, W. \& Sugiyama, N. ApJ {\bf 471}, 542 (1996)

\bibitem{Color} Hu, W., Sugiyama, N., \& Silk, J. Nature
	{\bf 386} 37 (1997)

\bibitem{HuWhite} Hu, W. \& White, M. A\&A {\bf 315}, 33 (1996)

\bibitem{ISa}  Illarionov, A.F. \& Sunyaev, R.A. Sov. Astron.
{\bf 18}, 413 (1975)

\bibitem{ISb} Illarionov, A.F. \& Sunyaev, R.A. 
 Sov. Astron. {\bf 18}, 691 (1975)

\bibitem{Jacoby} Jacoby, G.H. \etal\ PASP {\bf 104}, 599 (1992)

\bibitem{Jones} Jones, M, \etal\ Nature {\bf 365}, 320 (1993)

\bibitem{JW} Jones, B.J.T. \& Wyse, R.F.G. A\&A
{\bf 149}, 144 (1985)

\bibitem{Jorgensen} J{\o}rgensen, H.E., Kotok, E., Naselsky, P., \& Novikov, I.
A\&A {\bf 294}, 639 (1995) 

\bibitem{Kaiser83} Kaiser, N. MNRAS {\bf 202}, 1169
(1983)

\bibitem{Kaiser84} Kaiser, N. ApJ {\bf 282}, 374 (1984)

\bibitem{KamSper} Kamionkowski, M. \& Spergel, D.N. ApJ {\bf 432},
7 (1994)

\bibitem{KSS} Kamionkowski, M., Spergel, D.N., \& Sugiyama, N. ApJL
{\bf 434}, L1 (1994)

\bibitem{KHPR} Klypin, A., Holtzmann, J., Primack, J. \& Regos, E. ApJ
{\bf 416}, 1 (1993)

\bibitem{Kofman} Kofman, L.A. \& Starobinskii, A.A. Sov. Astron. Lett. {\bf
9}, 643 (1985)

\bibitem{KS84} Kodama, H. \& Sasaki, M. Prog. Theor. Phys. Supp. {\bf 78}, 1
(1984)

\bibitem{KS86} Kodama, H. \& Sasaki, M. Int. J. Mod. Phys. {\bf
A1}, 265 (1986)

\bibitem{KS87} Kodama, H. \& Sasaki, M. Int. J. Mod. Phys. {\bf
A2}, 491 (1987)

\bibitem{Kompaneets} Kompaneets, A.S. Sov. Phys.--JETP {\bf 4}, 730 (1957)

\bibitem{Kosowsky} Kosowsky, A. Ann. Phys. {\bf 246}, 49 (1996) 

\bibitem{Krauss} Krauss, L.M. and Kernan, P., 
ed. {\it CMB Anisotropies Two Years After
COBE}, (World Scientific, Singapore 1994)

\bibitem{Larson}  Larson, E.W. \etal\ J. Comput. Phys. {\bf 61}, 359 (1985)

\bibitem{Liftshitz} Liftshitz, E.M. \& Khalatnikov, I.M. Adv.
Phys. {\bf 12}, 185 (1963)

\bibitem{Lightman}  Lightman, A.P.  ApJ {\bf 244}, 392 (1981)

\bibitem{Linder} Linder, V.E. MNRAS {\bf 243}, 362 (1990)

\bibitem{Loveday} Loveday, J., Peterson, B.A., Efstathiou, G. \& Maddox, S.J.
 ApJ {\bf 390}, 338 (1992)

\bibitem{LS} Lyth, D.H. \& Stewart, E.D.  Phys. Lett. B {\bf 252}, 336
(1990)

\bibitem{LW} Lyth, D.H \&  Woszczyna, A. PRD {\bf 52}, 3338 (1995)

\bibitem{MakSut} Makino, N. \& Suto, Y. ApJ {\bf 405}, 1 (1993)

\bibitem{MS} Mandl, F. \& Shaw, G., {\it Quantum Field Theory}
(Wiley, New York, 1984) pg. 157

\bibitem{MarBFJSa} Markevitch, M,  Blumenthal, G.R.,  Forman, W., Jones, C. \&
Sunyaev, R.A. ApJ {\bf 378}, L33 (1991)

\bibitem{MGSS} Martinez-Gonazlez, E., Sanz, J.L. \& Silk, J. PRD 
{\bf 46}, 4193

\bibitem{Mather} Mather, J.C. \etal\ ApJL {\bf 420}, 439 (1994)

\bibitem{Meinhold} Meinhold, P.R. ApJL {\bf 409}, L1 (1993)

\bibitem{MFB} Mukhanov, V.F., Feldman, H.A., \& Brandenberger, R.H.
Phys. Rep. {\bf 215}, 203 (1992)


\bibitem{Netterfield} Netterfield, C.B., Jarosik, N.C., Page, L.A., 
	Wilkinson, D., \& Wollack, E. ApJL {\bf 445}, L69 (1995) 

\bibitem{Ostriker} Ostriker, J.P. ARA\&A
{\bf 31}, 689 (1993)

\bibitem{OV} Ostriker, J.P. \& Vishniac, E.T.  ApJ  {\bf 306}, 51 (1986)

\bibitem{Peacock} Peacock, J.A. \&
Dodds, S.J. MNRAS {\bf 267}, 1020 (1994)

\bibitem{PeeblesRec} Peebles, P.J.E. ApJ {\bf 153}, 1 (1968)

\bibitem{PeeblesLSS} Peebles, P.J.E. {\it Large Scale Structure of the
Universe}, (Princeton University, Princeton 1980)

\bibitem{PeeblesPIBa} Peebles, P.J.E. ApJL {\bf 315}, L73
(1987) 

\bibitem{PeeblesPIBb} Peebles, P.J.E. Nature {\bf 327}, 210 (1987)

\bibitem{PY} Peebles, P.J.E. \& Yu, J.T. ApJ {\bf 162}, 815 (1970)

\bibitem{PW} Penzias, A.A. \& Wilson, R.W. ApJ {\bf 142}, 419 (1965)

\bibitem{PPB} Pequignot, D., Petitjean, P. \& Boisson, C. A\&A {\bf 251}, 690
(1991)

\bibitem{Peyraud} Peyraud, J. J. Physics {\bf 29}, 88 (1968)

\bibitem{Piccirillo} Piccirillo, L, \& Calisse, P. ApJ {\bf 411}, 529 (1993)

\bibitem{Pol} Polnarev, A.G. Sov. Astron. {\bf 29}, 607 (1985)

\bibitem{PV} Press, W. \& Vishniac, E.T. ApJ {\bf 239}, 1 (1980)  

\bibitem{RP} Ratra, B. \&  Peebles, P.J.E. ApJL {\bf 432}, L5
(1994)

\bibitem{Rees} Rees, M.J. ApJL {\bf 153}, L1 (1968)

\bibitem{RS} Rees, M.J \& Sciama, D.N. Nature {\bf 519}, 611 (1968)

\bibitem{RT} Ressel, M. \& Turner, M. Comm. Astrophys. {\bf 14}, 323
(1990)

\bibitem{SW} Sachs, R.K. \& Wolfe, A.M. ApJ {\bf 147}, 73 (1967)

\bibitem{Saha} Saha, A., \etal\ ApJ {\bf 438}, 8 (1995) 

\bibitem{Sandage} Sandage, A.R. AJ {\bf 106} 719 (1993)

\bibitem{Shi} Shi, X., Schramm, D.N., Dearborn, D.S.P. \& Truran, J.W.
Comm. Astrophys. {\bf 17}, 343 (1995)

\bibitem{SC} Sarkar, S. \& Cooper, A.M. Phys. Lett. B {\bf 148}, 347 (1983)

\bibitem{Sas} Sasaki, M. MNRAS {\bf 240}, 415 (1989)

\bibitem{SSG} Schneider, D.P., Schmidt, M. \& Gunn, J.E. AJ {\bf 98}, 1951 
(1989)

\bibitem{Schuster} Schuster, J. \etal\ ApJL {\bf 419}, L47 (1993)

\bibitem{SSW} Scott, D., Silk, J. \& White, M. Science {\bf 268}, 829 (1995)

\bibitem{Seljak} Seljak, U. ApJL {\bf 435}, L87 (1994)

\bibitem{SeljakLensing} Seljak, U. ApJL {\bf 463}, L1 (1996)

\bibitem{SeljakISW} Seljak, U. ApJ {\bf 460}, 549 (1996)

\bibitem{Silk} Silk, J.  ApJ {\bf 151}, 459 (1968)

\bibitem{Smith} Smith, M.S., Kawano, L.H., \& Malaney, R.A. ApJ {\bf 85}, 219 
1993

\bibitem{Smoot91} Smoot, G., \etal\ ApJL {\bf 371}, L1 (1991)

\bibitem{Smoot} Smoot, G., \etal\ ApJL {\bf 396}, L1 (1992)

\bibitem{Staa} Starobinskii, A.A. JETP Lett. {\bf 30}, 682 (1979)

\bibitem{Stab} Starobinskii, A.A. Sov. Astron. Lett. {\bf 11}, 113 (1985)

\bibitem{Strauss} Strauss, M.A. \& Willick, J.A. Phys. Rep. {\bf 261}, 271 (1995)

\bibitem{SugSuto} Suginohara, T. \& Suto,  Y., ApJ {\bf 387}, 431
(1992)

\bibitem{SG} Sugiyama, N. \& Gouda, N., Prog. Theor. Phys. {\bf 88}, 803
(1992)

\bibitem{SugSilk} Sugiyama, N. \& Silk, J. PRL {\bf 73},
509 (1994)

\bibitem{SSV} Sugiyama, N., Silk, J. \& Vittorio, N. ApJL {\bf 419}, L1
(1993)

\bibitem{SZ72} Sunyaev, R.A. \& Zel'dovich, Ya.B. Comments Ap.
Space Phys. {\bf 4}, 79 (1972)

 
\bibitem{SZ} Sunyaev, R.A. \& Zel'dovich, Ya.B. Ap. Space Sci.
{\bf 7}, 3 (1970)

\bibitem{SZb} Sunyaev, R.A. \& Zel'dovich, Ya.B. Ap. Space Sci.
{\bf 9}, 368 (1970)

\bibitem{TegTegTeg} Tegmark, M., Bunn, E. \&  Hu, W.  ApJ {\bf 434},
1 (1994)

\bibitem{TS} Tegmark, M. \& Silk, J. ApJ {\bf 423}, 529 (1994); errata 
{\bf 423}, 529 (1994)

\bibitem{TomWat} Tomita, K. \& Watanabe, K. Prog. Theor. Phys. {\bf 82}, 563

\bibitem{TL} Tuluie, R. \& Laguna, P. ApJL {\bf 445}, L73 (1995)

\bibitem{TWL} Turner, M.S., White, M., \& Lidsey, J.E. PRD {\bf 48},
4613 (1993)

\bibitem{Vishniac} Vishniac, E.T. ApJ {\bf 322}, 597 (1987)


\bibitem{VS} Vittorio, N. \& Silk, J. ApJL {\bf 285}, L39 (1984)

\bibitem{Walker} Walker, T.P., Steigman, G., Schramm, D.N., Olive, K.A.,
\& Kang, H.-S. ApJ {\bf 376}, 51 (1991)

\bibitem{WBCP} Webb, J.K., Barcons, X., Carswel, R.F., \& Parnell, H.C.
MNRAS {\bf 244}, 319 (1992)

\bibitem{Weinberg} Weinberg, S. 
{\it Gravitation and Cosmology}, (Wiley, New York, 1972)

\bibitem{WSS} White, M., Scott, D., \& Silk, J. ARA\&A
{\bf 32}, 319 (1994)

\bibitem{Wilson} Wilson, M.L. ApJ {\bf 273}, 2 (1983)

\bibitem{WS}  Wilson, M. \& Silk, J. ApJ {\bf 243}, 14 (1981)

\bibitem{Weynmann} Weynmann, R. ApJ {\bf 145}, 560 (1966)

\bibitem{Wilkinson} Wilkinson, D. in {\it Proceedings of the 9th Lake Louise
Winter Institute}, ed.~A. Astbury \etal\ (World Scientific, Singapore, 1995)

\bibitem{Wright} Wright, E., \etal\ ApJ {\bf 420}, 450 (1994)

\bibitem{YST} Yamamoto, K., Sasaki, M., \& Tanaka, T.
ApJ {\bf 455}, 412 (1995)

\bibitem{Zdziarski} Zdziarski, A.A. ApJ {\bf 335}, 768 (1988)

\bibitem{ZIS}  Zel'dovich, Ya. B., Illarionov, A.F. \& Sunyaev, R.A.
 Sov. Phys.-- JETP {\bf 33}, 643 (1972)

\bibitem{ZKS} Zel'dovich, Ya. B., Kurt, V.G., \& Sunyaev, R.A. 
Sov. Phys.-JETP {\bf 28}, 146 (1969)

\bibitem{ZL}   Zel'dovich, Ya. B. \& Levich, E.V. Sov. Phys.-JETP Lett.
{\bf 11}, 35 (1970)

\bibitem{ZS69} Zel'dovich, Ya. B. \& Sunyaev, R.A.  Ap. Space Sci. 
{\bf 4}, 301 (1969)

\end{thebibliography}
